\documentclass[a4paper,11pt]{article}
\pdfoutput=1 

\usepackage{jheppub} 
\usepackage[T1]{fontenc} 
\usepackage{lmodern}
\usepackage{tikz}
\usepackage{amsmath,amssymb,xfrac,framed,verbatim,amsthm}
\usepackage{mathrsfs,accents,bbm,bbold} 

\usepackage{microtype,ctable}
\usepackage{simplewick}
\usepackage{slashed,mathtools} 
\usepackage{sectsty} 
\usepackage{graphicx,enumerate,bm} 
\usepackage{comment}

\usepackage{young}
\usepackage[vcentermath]{youngtab}
\usepackage{ytableau}

\newcommand{\mc}[1]{\mathcal{#1}}

\newcommand{\mbb}[1]{\mathbb{#1}}
\newcommand{\tr}{\text{Tr}}

\newcommand{\m}{\bm{-}}
\newcommand{\p}{\bm{+}}

\newcommand{\ud}{\text{d}}
\newcommand{\tl}{\tilde}
\newcommand{\sgn}{\text{sgn}}
\renewcommand{\i}{\text{i}}

\def\cC{\mathcal{C}}

\def\cS{\mathcal{S}}

\def\cV{{\cal V}}

\renewcommand{\a}{\alpha}
\renewcommand{\b}{\beta}
\newcommand{\e}{\epsilon}

\newcommand{\g}{\gamma}

\newcommand{\D}{\partial}

\preprint{TIFR/TH/20-22} \title{\boldmath Fermi seas from Bose
  condensates in Chern-Simons matter theories and a bosonic exclusion
  principle}

\author[a,1]{Shiraz Minwalla,\note{minwalla@theory.tifr.res.in}}
\author[a,2]{Amiya Mishra,\note{amiya.mishra@theory.tifr.res.in}}
\author[a,3]{Naveen Prabhakar\note{naveensp@theory.tifr.res.in}}

\affiliation[a]{Department of Theoretical Physics, \\ Tata Institute
  of Fundamental Research, Homi Bhabha Rd, Mumbai 400005, India}

\abstract{We generalize previously obtained results for the (all
  orders in the 't Hooft coupling) thermal free energy of bosonic and
  fermionic large $N$ Chern-Simons theories with fundamental matter,
  to values of the chemical potential larger than quasiparticle
  thermal masses. Building on an analysis by Geracie, Goykhman and
  Son, we present a simple explicit formula for the occupation number
  for a quasiparticle state of any given energy and charge as a
  function of the temperature and chemical potential. This formula is
  a generalization to finite 't Hooft coupling of the famous
  occupation number formula of Bose-Einstein statistics, and implies
  an exclusion principle for Chern-Simons coupled bosons: the total
  number of bosons occupying any particular state cannot exceed the
  Chern-Simons level. Specializing our results to zero temperature we
  construct the phase diagrams of these theories as a function of
  chemical potential and the UV parameters. At large enough chemical
  potential, all the bosonic theories we study transit into a
  compressible Bose condensed phase in which the runaway instability
  of free Bose condensates is stabilized by the bosonic exclusion
  principle. This novel Bose condensate is dual to - and reproduces
  the thermodynamics of - the fermionic Fermi sea.}

\begin{document}
\maketitle

\section{Introduction}

There is now considerable evidence to support the conjectured
\emph{Bose-Fermi} duality between families of $2+1$ dimensional
Chern-Simons gauge theories coupled to scalars and (roughly
level-rank) dual Chern-Simons gauge theories coupled to fermions
\cite{Sezgin:2002rt, Klebanov:2002ja, Giombi:2009wh, Benini:2011mf,
  Giombi:2011kc, Aharony:2011jz, Maldacena:2011jn, Maldacena:2012sf,
  Chang:2012kt, Jain:2012qi, Aharony:2012nh, Yokoyama:2012fa,
  GurAri:2012is, Aharony:2012ns, Jain:2013py, Takimi:2013zca,
  Jain:2013gza, Yokoyama:2013pxa, Bardeen:2014paa, Jain:2014nza,
  Bardeen:2014qua, Gurucharan:2014cva, Dandekar:2014era,
  Frishman:2014cma, Moshe:2014bja, Aharony:2015pla, Inbasekar:2015tsa,
  Bedhotiya:2015uga, Gur-Ari:2015pca, Minwalla:2015sca,
  Radicevic:2015yla, Geracie:2015drf, Aharony:2015mjs,
  Yokoyama:2016sbx, Gur-Ari:2016xff, Karch:2016sxi, Murugan:2016zal,
  Seiberg:2016gmd, Giombi:2016ejx, Hsin:2016blu, Radicevic:2016wqn,
  Karch:2016aux, Giombi:2016zwa, Wadia:2016zpd, Aharony:2016jvv,
  Giombi:2017rhm, Benini:2017dus, Sezgin:2017jgm, Nosaka:2017ohr,
  Komargodski:2017keh, Giombi:2017txg, Gaiotto:2017tne,
  Jensen:2017dso, Jensen:2017xbs, Gomis:2017ixy, Inbasekar:2017ieo,
  Inbasekar:2017sqp, Cordova:2017vab, Charan:2017jyc, Benini:2017aed,
  Aitken:2017nfd, Jensen:2017bjo, Chattopadhyay:2018wkp,
  Turiaci:2018nua, Choudhury:2018iwf, Karch:2018mer, Aharony:2018npf,
  Yacoby:2018yvy, Aitken:2018cvh, Aharony:2018pjn, Dey:2018ykx,
  Chattopadhyay:2019lpr, Dey:2019ihe, Halder:2019foo, Aharony:2019mbc,
  Li:2019twz, Jain:2019fja, Inbasekar:2019wdw, Inbasekar:2019azv,
  Jensen:2019mga, Kalloor:2019xjb, Ghosh:2019sqf, Inbasekar:2020hla, Jain:2020rmw, Jain:2020puw,toappear1}.
In the simplest large $N$ examples with the matter fields in the
fundamental of the gauge group, the `elementary' bosonic fields on the
LHS of the duality create stable bosonic particles that map, under
duality, to the fermionic particles created by the `elementary'
fermionic fields on the RHS of the duality\footnote{In this sense the
  $2+1$ dimensional dualities at large $N$ are qualitatively different
  from famous examples of bosonization in $1+1$ dimensions, in which
  particles created by the elementary fermionic field $\psi$ map to
  particles created by a complicated composite (sometimes called
  solitonic) bosonic operator $e^\phi$. This 1+1 dimensional duality
  is, in some respects, more similar to the $N=1$ case of the 2+1
  dimensional dualities
  \cite{Karch:2016sxi,Seiberg:2016gmd,Murugan:2016zal} in which
  elementary fermions map under duality to vortices bound to a single
  boson. The role of the vortex is analogous to the role of the
  Jordan-Wigner `string' in the 1+1 dimensional context. We thank
  D. Radicevic for a discussion on this point.}.

That bosons and fermions can be dual to each other in this elementary
fashion may seem surprising. In most familiar contexts, fermions have
half-integer spin while bosons have integer spin\footnote{Bosonisation
  is less surprising in $1+1$ dimensions as there is insufficient room
  to rotate in a one dimensional space (more formally the massive
  little group, $SO(1)$, is trivial) and so massive excitations do not
  carry a spin quantum number.}. While this distinction is completely
sharp in $D \geq 4$ dimensions, it, however, blurs out in $D=3$ where
particle spins are allowed to take arbitrary value\footnote{Massive
  particles in $D$ dimensions transform in representations of the
  little group $SO(D-1)$ - or more precisely $Spin(D-1)$, the covering
  group of $SO(D-1)$. For $D \geq 3$, this group is non-abelian and
  its unitary representations are discrete. Fermions transform in
  spinorial representations (those in which a $2 \pi$ rotation is
  implemented by the operator $-1$) while bosons transform in
  discretely distinct faithful representations (those in which a
  $2 \pi$ rotation is implemented by the operator $1$). In $D=3$, this
  little group is $\mathbb{R}$ (the covering group of $SO(2)$). The
  unitary representations of $\mathbb{R}$ are labelled by one
  continuous parameter (the spin, an arbitrary real number) rather
  than a collection of discrete parameters as in higher
  dimensions.}. Indeed, it turns out that the coupling to a
Chern-Simons gauge field renormalizes the integer (resp.~half-integer)
spins of the bosons (resp.~fermions) to general real values. For
various specific conjectured Bose-Fermi dualities one can check that
the renormalized boson spin does indeed match the renormalized fermion
spin across the duality \cite{toappear1}.

The matching of single particle spins does not address all
paradoxes. Given the differences between Bose and Fermi statistics,
we must still understand how the spectrum of multi-particle states can
match across the duality. This question has already been investigated
in the context of two particle scattering states in
\cite{Jain:2014nza}. At least in large $N$ limit, it turns out that
the key to answer this lies in the colour quantum numbers of the
bosonic and fermionic states\footnote{We have ignored the fact that
  the coupling to the Chern-Simons gauge field modifies the statistics
  of both bosonic and fermionic excitations. This is consistent in the
  large $N$ limit as the effective anyonic phase in the scattering of
  two fundamental fermions or bosons is of order $1/N$
  \cite{Jain:2014nza}. At finite values of $N$ (in particular when
  $N=1$) the modification of statistics plays a key role in
  reconciling the difference between bosonic and fermionic statistics
  and cannot be ignored.\label{statfn}}.

Let us recall that the gauge groups are different on the bosonic and
the fermionic sides of the duality\footnote{$SU(N_B)$ or $U(N_B)$ for
  the bosons, and $SU(N_F)$ or $U(N_F)$ for the fermions in the
  situation studied by \cite{Aharony:2015mjs,Hsin:2016blu}. In general
  $N_F \neq N_B$.}. As the elementary bosons and fermions each
transform in the fundamental representations of their respective gauge
groups, it is clear that {\it all} single particle quantum numbers
cannot match across the duality. While the `physical' spacetime
quantum numbers (mass and spin) match precisely, the `unphysical'
gauge quantum numbers lie in different spaces and so cannot be
identified.

Of course, this `mismatch' does not necessarily falsify
duality. Suppose we work on three dimensional spacetime which is
$S^2 \times \mathbb{R}$.\footnote{This is one way of regulating the
  spatial $\mathbb{R}^2$ in the IR. Another way would be to replace
  $\mathbb{R}^2$ with a disk of large radius. This is a more
  complicated regulator as it introduces boundary degrees of freedom -
  roughly the boundary WZW modes. We will not study such a regulator
  in this paper.} Gauss' law then forces the physical states to be
colour singlets. Charged single-particle (or multi-particle) states
are not by themselves physical, but are instead the building blocks
for physical states. It was demonstrated in \cite{Jain:2014nza} that,
under duality, two-particle states that are symmetric (antisymmetric)
in colour indices map to two-particle states that are antisymmetric
(symmetric) in colour. Thus, while dual pairs of bosonic and fermionic
two-particle states have opposite statistics under the interchange of
{\it all} quantum numbers, they have identical statistics under the
interchange of all physical - i.e.  non-gauge - quantum
numbers\footnote{So, for example, a two-particle fermion state that is
  antisymmetric in colour indices is forced, by Fermi statistics, to
  be symmetric under the interchange of all physical (non-gauge)
  quantum numbers like momenta. Its dual two-particle boson state is
  symmetric in colour indices and so is also forced by Bose statistics
  to be symmetric in all physical (non-gauge) quantum numbers.}. We
see that, as promised, the `hidden' colour indices are the key to
reconciling duality with the difference between Bose and Fermi
statistics.

In this paper we study four families of quantum field theories: the
quasi-fermionic theories which are the regular fermion and the
critical boson, and the quasi-bosonic theories which are the regular
boson and the critical fermion\footnote{The names quasi-fermionic and
  quasi-bosonic \cite{Maldacena:2011jn,Maldacena:2012sf} indicate only
  the following: by definition the large $N$ single-trace operator
  spectrum of quasi-fermionic theories is identical to the same
  spectrum of operators in a free, gauged, large $N$ fermion
  theory. Similarly, the large $N$ single-trace operator spectrum of
  quasi-bosonic theories is identical to the same spectrum of
  operators in a free, gauged, large $N$ bosonic theory. The names
  carry no other significance. In particular, there is no sense in
  which all quasi-fermionic theories are `almost fermionic'. As an
  example, the ungauged large $N$ bosonic Wilson-Fisher theory -
  certainly a bosonic theory from every other point of view - is
  quasi-fermionic with our definitions. We hope this terminology will
  not prove confusing. We thank A. Karch for a discussion on this
  point.}.  In our study of the quasi-fermionic theories, we adopt the
conventions and regularization scheme of \cite{Choudhury:2018iwf} (see
Section $1.1$ of that paper including footnotes 5-8). Similarly, in
our study of quasi-bosonic theories, we adopt the conventions and
regularization scheme of \cite{Dey:2018ykx} (see Section $2.1$ of that
paper). The Euclidean actions for these theories are listed in Section
\ref{theoriessec}.

We will address three related but distinct questions. We first revisit
the question of how Bose-Fermi duality can be reconciled with the
difference between bosonic and fermionic statistics, but this time
working with thermodynamic systems with macroscopic numbers of
particles rather than few-particle scattering states. Specialising
this discussion to zero temperature, we then investigate how Bose
condensates can replicate the behaviour of Fermi seas. Finally, we
perform a detailed quantitative study of the zero temperature, finite
chemical potential phase diagrams of the critical boson and regular
boson theories (and hence, by duality, of the regular fermion and
critical fermion respectively). This study is related to the previous
question because Bose condensates (and their dual Fermi seas) play a
key role in these phase diagrams. In the rest of this introduction we
briefly describe our results on each of these counts in turn.

\subsection{Reconciling statistics with duality}

How do bosonic and fermionic thermodynamic ensembles\footnote{Our
  systems are defined on an $S^2$ with effective volume $\mc{V}_2$,
  where the volume is large ($\mc{V}_2= V N$). While $V$ is held fixed
  in the large $N$ limit in general, we also study the separate limit
  $V \to \infty$ in some parts of our analysis.}  manage to exhibit
the same physics despite their apparent difference in statistics? A
study of this question was initiated in the pioneering article of
Geracie, Goykhman and Son \cite{Geracie:2015drf}. The authors of
\cite{Geracie:2015drf} studied the analytically tractable theory of
regular fermions coupled to $SU(N)$ Chern-Simons theories in the large
$N$ limit\footnote{As we explain in Appendix \ref{bfdfr}, there are
  three simple Bose-Fermi dualities. The first of these asserts the
  equivalence between an $SU(N_F)$ gauged fermion theory and a
  $U(N_B)$ gauged bosonic theory. The second between a $U(N_F)$ gauged
  fermion theory and a $SU(N_B)$ gauged bosonic theory. The third
  between a $U(N_F)$ gauged fermion theory and a $U(N_B)$ gauged
  bosonic theory. On general grounds we expect $SU(N)$ and $U(N)$
  theories - and so all three of these distinct dualities - to become
  indistinguishable in the large $N$ limit. The reader may wonder how
  this expectation is consistent with the fact that $U(1)$ global
  symmetries of the $SU(N)$ theories are of the simple `phase all
  fields' sort, while the $U(1)$ global symmetries of the $U(N)$
  theories are the so called `topological' symmetries (see Appendix
  \ref{usu}). As explained in Appendix \ref{usu}, the way this works
  is the following. If we integrate out the dynamical $U(1)$ factor
  from a large $N$ $U(N)$ theory, the topological symmetry of the
  parent $U(N)$ theory turns into the usual symmetry of the resultant
  effective $SU(N)$ theory.}.  The effective excitations in these
grand canonical ensembles at finite temperature and finite chemical
potential had previously been demonstrated to be parametrically
long-lived fermionic excitations with a dynamically determined
mass\footnote{The mass of these quasiparticles is renormalized in the
  grand canonical ensemble; in other words it differs from the zero
  temperature and zero chemical potential pole masses at the same
  values of UV couplings.}.  The authors of \cite{Geracie:2015drf}
(see Section \ref{qon} below for a review and generalisations)
demonstrated that the large $N$ exact charge of the ensemble with
temperature $\beta^{-1}$ and chemical potential $\mu$ in the infinite
volume limit is that of a non-interacting gas of these fermionic
quasiparticles, with the average occupation number of each
quasiparticle state with energy $\epsilon$ and charge $q$ given by
\begin{align}\label{occuptnFintro} 
  \bar{n}_F({\epsilon},q)& =  \frac{1}{2 \pi |\lambda_F|}\int_{-\pi|\lambda_F|}^{\pi|\lambda_F|} d\alpha \  \frac{1}{e^{\beta ({\e} - q{\mu}) -\i q \a } + 1}\ , \\
                       &=\frac{1}{2}-\frac{1}{\pi|\lambda_F|}  \tan^{-1}\bigg(\frac{e^{\beta(\epsilon-q\mu)}-1}{e^{\beta(\epsilon-q \mu)}+1}\tan\frac{\pi|\lambda_F|}{2}\bigg)\ ,\quad\text{with}\quad\tan^{-1}(\zeta) \in \Big[-\frac{\pi}{2}, \frac{\pi}{2} \Big]\ .\nonumber
\end{align}
($\lambda_F$ is the 't Hooft coupling defined in
\eqref{tHooftcoupling}; the first line of \eqref{occuptnFintro} was
presented in \cite{Geracie:2015drf}; the second line of
\eqref{occuptnFintro} is derived in Section \ref{eoiv} of this
paper). The equation \eqref{occuptnFintro} is a finite-$\lambda_F$
generalisation of the famous occupation number formula of Fermi-Dirac
statistics
\begin{equation}\label{occferm}
  \bar{n}_F({\epsilon},q) =  \frac{1}{e^{\beta ({\e} - q{\mu})  } + 1}\ , 
\end{equation}
and \eqref{occuptnFintro} reduces to \eqref{occferm} in the limit
$\lambda_F \to 0$.

What is the bosonic analogue of \eqref{occuptnFintro}? In addressing
this question\footnote{And also of the fermionic analogue of the same
  formula that holds in a general phase of the theory, i.e. away from
  the infinite volume limit.}, we face the following obstacle. The
free energy of the bosonic theories (see \eqref{cblag} and
\eqref{rblag} for the Lagrangians) and the fermionic theories (see
\eqref{rflag} and \eqref{cflag} for the Lagrangians) have previously
been computed only for a range of temperatures and chemical
potentials: those values of these parameters for which the
quasiparticle thermal masses turn out to be larger than the chemical
potential\footnote{The same is true for fermionic theories in some
  phases that dominate away from the infinite volume limit.}. As the
first step in the analysis of this paper we remedy this technical
defect in (see Section \ref{thptnfn} below) by generalising previously
obtained results to all values of the chemical potential and
temperature. 

The method we employ is to analytically continue previously computed
results for the `off-shell free energy' of matter Chern-Simons theories
in the chemical potential.  This procedure makes sense because
off-shell free energy depends on the chemical potential only through a
finite temperature determinant; as this determinant is effectively two
dimensional we do not expect it to undergo a sharp phase transition as
$\mu$ exceeds the thermal mass, see Section
\ref{offshellsec}.\footnote{We thank O. Aharony and S. Wadia for
  discussions on this point.} This analytic continuation is
conveniently accomplished by modifying a contour of integration in the
space of holonomy eigenvalues. The final results of this exercise are
listed in detail in \eqref{CB2voffshellfecor} and
\eqref{RBoffshellfecor} for the bosonic theories in the infinite
volume limit. For general finite volume phases, the formulas are in
\eqref{CB2voffshellfecorn}, \eqref{RBoffshellfecorn} for the bosonic
theories, and \eqref{RF2voffshellfecorn} and \eqref{CFoffshellfecorn}
for the fermionic theories.

Armed with exact results for the relevant thermodynamical formulae, it
is not difficult to imitate the analysis of \cite{Geracie:2015drf} for
bosonic Chern-Simons-matter theories.  Once again we find that the
large $N$ exact charge of the ensemble of temperature $\beta^{-1}$ and
chemical potential $\mu$ in the infinite volume limit\footnote{See
  Section \ref{ofep} for the results that apply in more general phases
  that dominate away from the infinite volume limit.}  is that of a
non-interacting gas of these quasiparticles, with the average
occupation number of each quasiparticle state of energy $\epsilon$ and
charge $q$ given by
\begin{align}\label{nBugintro}
  &\bar{n}_B(\epsilon,q)\nonumber\\
  & = \frac{1}{2\pi|\lambda_B|}\int_{-\pi|\lambda_B|}^{\pi|\lambda_B|} d \alpha \frac{1}{e^{\beta(\epsilon-q\mu)-\i q \alpha}-1} + \frac{1}{|\lambda_B|} \ \Theta(q\mu-\epsilon)\ , \\
  &=\frac{1-|\lambda_B|}{2|\lambda_B|} - \frac{1}{\pi|\lambda_B|}  \tan^{-1}\bigg(\frac{e^{\beta(\epsilon-q \mu)}-1}{e^{\beta(\epsilon-q \mu)}+1}\cot\frac{\pi|\lambda_B|}{2}\bigg)\ ,\quad\text{with}\quad\tan^{-1}(\zeta) \in \Big[-\frac{\pi}{2}, \frac{\pi}{2}\Big]\ ,\nonumber
\end{align}
where $\Theta(x)$ is the Heaviside step-function. As above,
\eqref{nBfinalt} may be regarded as the generalisation to finite 't
Hooft coupling $\lambda_B$ of the standard formula of Bose-Einstein
statistics
\begin{equation}\label{stdbe}
  \bar{n}_B({\e},q) =  \frac{1}{e^{\beta ({\e} - q{\mu})  } - 1} \ .
\end{equation} 
For quasiparticle energies $\epsilon > q \mu$, \eqref{nBugintro}
reduces to \eqref{stdbe} in the limit $\lambda_B \to 0$. Unlike
\eqref{stdbe}, however, $\bar{n}_B(\epsilon,q)$ listed in
\eqref{nBugintro} is positive at all values of $\epsilon$ (even when
$\epsilon< q \mu$). In the limit $\lambda_B \to 0$,
$\bar{n}_B(\epsilon,q)$ given in \eqref{nBugintro} diverges; this is
the physically correct answer for a free Bose ensemble at
$q \mu > \epsilon$ (see below
for more discussion). 

The modified occupation numbers \eqref{occuptnFintro} and
\eqref{nBugintro} are one-parameter generalisations of the standard
formulae \eqref{occferm} and \eqref{stdbe} for fermions and bosons
respectively, with the parameter being the 't Hooft coupling
$\lambda_F$ or $\lambda_B$ respectively. As remarked in Footnote
\ref{statfn}, the anyonic phase obtained by exchanging two particles
with each other or two antiparticles with each other is proportional
to $1/N$ and hence is negligible in the large $N$ limit.\footnote{In
  contrast the anyonic phase obtained in taking a particle around an
  antiparticle in the so called singlet channel \cite{Jain:2014nza} is
  given $\pi|\lambda_F|$ or $\pi|\lambda_B|$ (for the fermions or
  bosons respectively), and so is of order unity in the large $N$
  limit.} However the ensembles we study in this paper contain a
thermodynamic number of particles since we are working in the large
$N$ and large volume limit of \cite{Jain:2013py}. Consequently, the
modified occupation number formulae \eqref{occuptnFintro} and
\eqref{nBugintro} are presumably a consequence of these anyonic
phases, whose net effect builds up to an $\mc{O}(1)$ number. It would
be very interesting to understand this better, perhaps in the context
of a Schrodinger type description of the non relativistic limit of
this system, see \cite{toappear1}.

Under the conjectured Bose-Fermi duality (see Section \ref{dualitysec}
and in particular \eqref{kappa}, \eqref{tHooftcoupling}
\eqref{level-rankmap} and \eqref{lambdamap}), the single particle
occupations numbers \eqref{nBugintro} and \eqref{occuptnFintro} of
corresponding states are not equal but instead turn out to be related
as

\begin{equation}\label{mspnd}
  N_F\bar{n}_F(\epsilon,q)  = N_B\bar{n}_B(\epsilon,q) \ .
\end{equation}

The fact that $\bar{n}_F(\epsilon,q)$ and $\bar{n}_B(\epsilon,q)$
are not simply equal to each other may, at first, seem to contradict
Bose-Fermi duality. In fact, the opposite is true. Equation
\eqref{mspnd} is {\it precisely} what is needed in order to ensure
that the net occupation number of a particular fermionic state matches
the net occupation number of bosonic state once we sum over the
`invisible' colour quantum numbers of these states. Effective
occupation numbers as a function of physical quantum numbers agree
exactly across the duality once we sum over the unphysical gauge
quantum numbers. As in the case of two-particle scattering states
reviewed earlier in this introduction, the `invisible' colour quantum
numbers play a key role in reconciling statistics with duality.

\subsection{Fermi seas from Bose condensates}

It is useful to focus attention on the ensemble in which the dichotomy
between Bose and Fermi statistics intuitively appears most pronounced
viz.~finite chemical potential and zero temperature.  In this limit we
expect the fermions to arrange themselves into a Fermi sea while the
bosons are expected to form a Bose condensate. We now explain how
these two rather different phases manage to be dual to each other.

In the zero temperature limit\footnote{All statements here are
  accurate when we first take $N$ to infinity and then take the
  temperature to zero. It is possible that the phase obtained by first
  taking $T$ to zero and then taking $N$ to infinity has additional
  complications. See Section \ref{discu} for additional remarks.}  the
formula for the Fermionic occupation number \eqref{occuptnFintro}
simplifies to
\begin{equation}\label{ztnF}
  \bar{n}_{F}(\epsilon,q) = \Theta(q \mu - \epsilon)\ ,
\end{equation}
demonstrating that the fermionic phase is a vanilla Fermi sea at every
value of $\lambda_F$ (this was already noted in
\cite{Geracie:2015drf}).

On the other hand the formula for the bosonic occupation number
\eqref{nBugintro} simplifies in the zero temperature limit to
\begin{equation}\label{avon}
\bar{n}_{B}(\epsilon,q) ={\bar n}\ \Theta(q \mu -\epsilon)\ ,\quad\text{with}\quad {\bar n}=\frac{1-|\lambda_B|}{|\lambda_B|}\ .
\end{equation}
In fact, the state described by \eqref{avon} is a regularized Bose
condensate. Recall that in a free bosonic theory with $\lambda_B = 0$,
the partition function ${\rm Tr}\,e^{-\beta(\epsilon - q\mu)}$ defines
an ensemble which infinitely populates\footnote{To forestall confusion we recall that the standard discussion of 
	Bose condensates in free theories involves taking the limit  $\mu q \to m_B$ ($m_B$ is the boson mass) from below. This limit 
	populates the boson zero mode in a macroscopic manner. 
	$q\mu$ is never taken above $m_B$, precisely because the free
	ensemble with $q \mu > m_B$ infinitely populates some states - including the zero mode - and so yields an ill defined ensemble. We thank G. Mandal for a related discussion.}
every single-particle state
with energy $\epsilon < q \mu$\footnote{Analytic continuation of this
  formula from values of $q\mu < \epsilon$ to values of
  $q \mu >\epsilon$ gets rid of the divergence, but evaluates to an
  unphysical negative number (see \eqref{stdbe}).}. As a consequence,
the ensemble with $q \mu$ strictly larger than the thermal mass $c_B$
is simply ill-defined in the free theory. In agreement with this fact,
$\bar{n}_{B}(\epsilon,q)$ diverges\footnote{The formula \eqref{avon}
  applies both to the regular boson theory and the critical boson
  theory. It follows that the critical boson self-interactions are
  insufficient to cure the $\mu > c_B$ divergence of the free boson.}
when $\epsilon < \mu q$ in the limit $\lambda_B \to 0$. This
divergence is regulated at nonzero $\lambda_B$. Quite remarkably, in
the theories studied in this paper the resultant `non-perturbative'
phase is remarkably simple; it can be described by a collection of
non-interacting quasi particles with occupation numbers given by
\eqref{avon}.

We see that the regulated Bose condensate \eqref{avon} is very similar
to a Fermi sea with one key difference; every state with energy less
than $q \mu$ in this phase is occupied not once (as in the case of a
Fermi sea) but ${\bar n}$ times (see \eqref{avon}). As in the previous
subsection, the difference in occupation numbers between the Fermi sea
and the Bose condensate is perfectly consistent with duality. It is
easily verified (using the definition of the 't Hooft couplings
\eqref{tHooftcoupling} and the duality map \eqref{level-rankmap}) that
\begin{equation}\label{sumn}
  N_B {\bar n}= N_F\ ,
\end{equation} 
and hence the occupation numbers agree once we sum on both sides over
`invisible' colour quantum numbers.

\subsection{A bosonic exclusion principle}\label{bep}

The discussion above suggests that the large $N$ bosonic matter Chern-Simons theory appears to obey an effective exclusion principle. As the upper limit on the occupation number of single particle states\footnote{We thank J. McGreevy for a very useful discussion on the topic of this subsection.}, 
\begin{equation}
{\bar n}= \frac{1-|\lambda_B|}{|\lambda_B|}=\frac{|k_B|}{N_B} \ ,
\end{equation}
(see around \eqref{kappa} for definitions) is not an integer, this exclusion principle does not operate at the level of single particle states with given colour quantum numbers. Instead, it asserts the total number of bosons one can add in any given single particle state after summing over all colour indices is the integer
$N_B {\bar n}=|k_B|$. We pause to explore this result a little further.

Consider a collection of $M$ bosons each of which occupies the same
(not counting colour quantum numbers) single particle state. As the
net wave function of these particles is symmetric under interchange,
and as all non-colour quantum numbers are identical for each of the
particles, the wave function of these $M$ particles must be symmetric
in colour. In other words, these $M$ particles transform in the
$SU(N_B)$ representation with a single row and $M$ columns in the
Young diagram as below
$$
\underbrace{\yng(2) \cdots \yng(1)}_{M ~ \text{boxes}}
$$
The exclusion principle above asserts that the length of the Young
diagram row cannot exceed $k_B$.

The restriction on representations described in the previous paragraph
is, of course, familiar.  Recall that all integrable representations
of $SU(N_B)_{k_B}$ WZW theory are described with Young diagram have
with no more than $k_B$ columns. It seems very likely to us that this
well-known property of WZW theory will turn out to be the underlying
explanation for the bosonic exclusion principle.

The discussion of the previous paragraph fits very neatly with the
ideas of level-rank duality. Recall that $M$ fermions in a given
single particle state (ignoring colour quantum numbers) transform in
an $SU(N_F)$ representation with a single column with $M$ rows as
below
$$
M ~ \text{boxes} \begin{cases} \ytableausetup
{mathmode, boxsize=1.2em}
\begin{ytableau}
\\
\\
\none[\vdots]  \\
\\
\end{ytableau}  \end{cases}
$$
The fact that $M \leq N_F$ is just a statement of the Fermi exclusion
principle. However, the representation with a single column and $M$
rows maps, under level-rank duality, to a representation with a single
row and $M$ columns. The restriction $M \leq N_F$ is dual to
$M \leq |k_B|$, i.e. the restriction described in the previous
paragraph. We hope to return to this important and interesting point
in future work.

\subsection{Zero temperature phase diagrams at finite chemical
  potential}

The exact (to all orders in 't Hooft coupling) large $N$ results for
the thermal free energy at finite chemical potential described above
become completely explicit and fairly simple in the zero temperature
limit\footnote{At $\mu=0$ this fact was already observed in
  \cite{Choudhury:2018iwf,Dey:2018ykx,Dey:2019ihe}.}. We now describe
the detailed and quantitative phase diagrams for the bosonic and
fermionic theories (see \eqref{cblag}, \eqref{rflag}, \eqref{rblag}
and \eqref{cflag} for the Lagrangians) at zero temperature but
arbitrary chemical potential (see Sections \ref {pscbrf} and
\ref{pdrbcf} below) that follow from these results.

Before presenting our quantitative phase diagrams, it is useful to
first address a preliminary qualitative question. What is the precise
nature of the charged bosonic excitations that condense to form the
Bose condensate? This question is more subtle than it might first
seem, as we now explain.

At zero chemical potential and temperature, a $SU(N_B)_{k_B}$
Chern-Simons gauge theory coupled to fundamental bosons has two
distinct phases as a function of the UV parameters. The critical boson
theory \eqref{cblag} has one UV parameter: the mass deformation
$m_B^{\rm cri}$. At zero temperature and chemical potential, this
theory lies in the unHiggsed phase $m_B^{\rm cri}>0$ but in the
Higgsed phase for $m_B^{\rm cri}<0$ ($m_B^{\rm cri}$ is the mass
parameter defined in \eqref{cblag}). There is a sharp second order
phase transition between these two phases at $m_B^{\rm cri}=0$. The
situation is qualitatively similar but quantitatively more complicated
in the regular boson theory and will be addressed below.

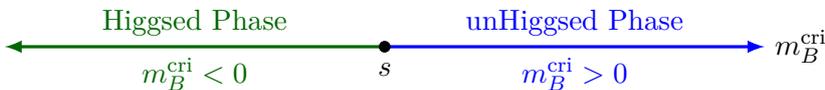
\begin{figure}[!h]
	\centering
	\begin{tikzpicture}
	\coordinate (L) at (-5,0);
	\coordinate (O) at (0,0) ; 
	\coordinate (R) at (5,0) ;
	\draw[latex-,black!60!green, very thick] (L) to [edge label={\text{Higgsed Phase}}, edge label'={$m_B^{\text{cri}}<0$}] (O) ;
	\draw[blue, very thick, -latex] (O) to [edge label={\text{unHiggsed Phase}}, edge label'={$m_B^{\text{cri}}>0$}] (R) ;
	\filldraw[black] (O) circle (2pt) ;
	\draw (O)+(0,-0.3) node {$s$}; 
	\draw (R) node[right] {$m_{B}^{\text{cri}}$} ; 
	\end{tikzpicture}
	\caption{Phase diagram of the critical boson theory as a
          function of $m_{B}^{\text{cri}}$ at $|\mu|=T=0$. Here, $s$
          marks the origin of the $m_{B}^{\text{cri}} $ axis at which point the theory undergoes a second order phase transition.}
	\label{CBphasediag|mu|=0int}
\end{figure}

The zero temperature, zero chemical potential phase diagram as a
function of $m_B^{{\rm cri}}$ takes the form depicted in Figure
\ref{CBphasediag|mu|=0int}.
In the ordinary or unHiggsed phase the excitations of the bosonic
theory are spin zero scalars created by the elementary bosonic field
$\phi$. In the Higgsed phase the gauge symmetry is Higgsed from
$SU(N_B)$ down to $SU(N_B-1)$ due to $\bar\phi\phi$ acquiring a
non-zero expectation value; the charged massive excitations in this
phase are the spin one $W$-bosons. The transmutation of degrees of
freedom from scalars to vectors is a consequence of the Higgs
mechanism.
  
Given that the charged excitations are qualitatively distinct in the
two phases depicted in Figure \ref{CBphasediag|mu|=0int}, it might
thus appear that there are two distinct Bose condensed phases; the
condensate of scalar excitations about the unHiggsed vacuum and the
condensate of spin one $W$-bosons around the Higgsed vacuum. At the
level of calculations this is true at intermediate steps; the
computations that determine the thermodynamics of these two phases
appear completely different. Remarkably, it turns out that the final
results for the thermodynamics in these two phases - at least at
leading order in large $N$ - turn out to be analytic continuations of
each other. This result holds in both the critical boson theory
\eqref{cblag} and the regular boson theory \eqref{rblag}. It follows
that these two apparently distinct Bose condensates are simply
different but equivalent descriptions of the same physical
state\footnote{A similar phenomenon occurs at non-zero temperature, at
  every value of the chemical potential including zero, as noted in
  e.g. \cite{Choudhury:2018iwf,Dey:2018ykx}.}.

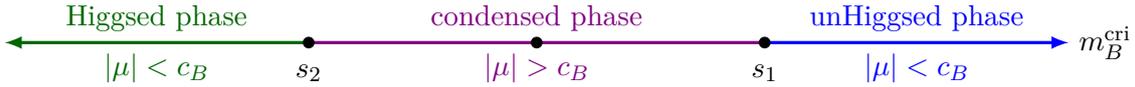
\begin{figure}[!ht]
	\centering
	\begin{tikzpicture}
	\coordinate (L) at (-7,0);
	\coordinate (s34) at (-3,0) ;
	\coordinate (s23) at (0,0) ;
	\coordinate (s12) at (3,0) ;
	\coordinate (R) at (7,0) ;
	\coordinate (O) at (0.3,0) ;
	\draw[latex-,black!60!green, very thick] (L) to [edge label={\text{Higgsed phase}}, edge label'={$|\mu|<c_B$}] (s34) ;
	\draw[violet, very thick] (s34) to [edge label={\text{condensed phase}}, edge label'={$|\mu|>c_B$}] (s12) ;
	\draw[blue, very thick, -latex] (s12) to [edge label={\text{unHiggsed phase}}, edge label'={$|\mu|<c_B$}] (R) ;
	\filldraw[black] (s34) circle (2pt) ;
	\draw (s34)+(0,-0.4) node {$s_{2}$}; 
	\filldraw[black] (s12) circle (2pt) ;
	\draw (s12)+(0,-0.4) node {$s_{1}$}; 
	\draw (R) node[right] {$m_{B}^{\text{cri}}$} ; 
	\filldraw[black] (s23) circle (2pt) ;
	\end{tikzpicture}
	\caption{Phase diagram of critical boson theory as a function of
          $m_B^{{\rm cri}}$ at fixed $\mu$. At the points $s_{1}=|\mu|$,
          $s_{2}=-|\mu|\big(\frac{2-|\lambda_B|}{|\lambda_B|}\big)$ the theory undergoes a second order phase transition. The
          point inside the condensed phase corresponds to
          $m_B^{\rm cri} =
          -|\mu|\big(\frac{1-|\lambda_B|}{|\lambda_B|}\big)$ and
          denotes a change in description from that a phase of
          condensed scalars to one of phase of condensed $W$-bosons
          though the condensed phase is itself unique.}
	\label{CBphasediagint|mu|>0}
\end{figure}

The phase diagram of the critical boson theory as a function of its
mass deformation parameter $m_B^{{\rm cri}}$ at zero temperature but
fixed nonzero $\mu$, is depicted in Figure \ref{CBphasediagint|mu|>0}.
As discussed in the previous subsection, the fermionic dual of the
Bose condensed phase is a Fermi sea phase. The absence of phase transitions within the Fermi sea phase as the effective fermion mass passes through zero is hardly a surprise. Thus, Bose-Fermi duality can be thought of
as a sort of `explanation' for the lack of an invariant distinction
between the Bose condensate of scalars and $W$ bosons.

\begin{figure}[htbp]
  \begin{center}
    \scalebox{1}{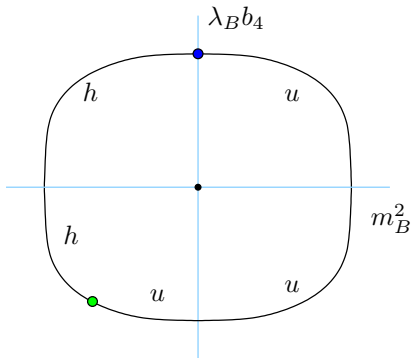}
    \caption{Phase diagram for the regular boson theory at zero
      temperature and chemical potential for the stable range
      $\phi_h < x_6^B < \phi_u$ of the marginal parameter $x_6^B$
      (refer to \eqref{phiuh} for the definitions of $\phi_h$ and
      $\phi_u$). The letter $u$ stands for the unHiggsed phase while
      the letter $h$ stands for the Higgsed phase. The blue dots are
      second order transitions while the green dots are first order
      transitions.}
    \label{RBphase}
  \end{center}
\end{figure}

In the rest of this introduction we describe the more intricate phase
diagram of the regular boson theory \eqref{rblag} (and so, of its
dual, the critical fermion theory \eqref{cflag}.) Let us first recall
that in the large $N$ limit, in addition to $\lambda_B$, the regular
theory is labelled by the dimensionless $UV$ parameter $x_6^B$, the
(mass-)dimension one parameter $b_4$ and the dimension two parameter
$m_B^2$. These correspond to the sextic, quartic and mass couplings
respectively.  At zero temperature and zero chemical potential, and at
fixed $\lambda_B$ and $x_6^B$, the phase diagram of this theory is a
function of $b_4$ and $m_B^2$ subject to the scaling equivalence
\begin{equation}\label{scaid} 
  (b_4, m_B^2) \sim (\alpha b_4, \alpha^2 m_B^2)\ ,\quad \alpha >0\ .
\end{equation} 
One way to incorporate the above equivalence while also keeping track
of the signs of $m_B^2$ and $b_4$ is to consider the following section
of the $(m_B^2, \lambda_B b_4)$ plane:
\begin{equation} \label{par}
  (m_B^2)^2 + (\lambda_B b_4)^4 = 1\ .
\end{equation}
The phase diagram is then the above ellipse-like curve. We present
this phase diagram (which was worked out in great detail in
\cite{Dey:2018ykx}, \cite{Dey:2019ihe}) in Figure \ref{RBphase}.
When the chemical potential is non-zero, the phase diagram of the
regular boson theory is two dimensional rather than one
dimensional. It is efficiently parametrized by the two dimensionless
variables $$\left(\frac{m_B^2}{|\mu|^2}, \frac{\lambda_B b_4}{|\mu|}\right)\ .$$
In the stable range of parameters ($ \phi_u <x_6<\phi_h$, see equation
\eqref{phiuh} for the definition of $\phi_u$ and $\phi_h$) the phase
diagram of our theory takes one of three qualitatively distinct forms
depending on the precise range in which $x_6^B$ lies. These three
phase diagrams have been sketched in Figure \ref{phaseintro} below.
\begin{figure}[!ht]
  \begin{center}
    \scalebox{0.55}{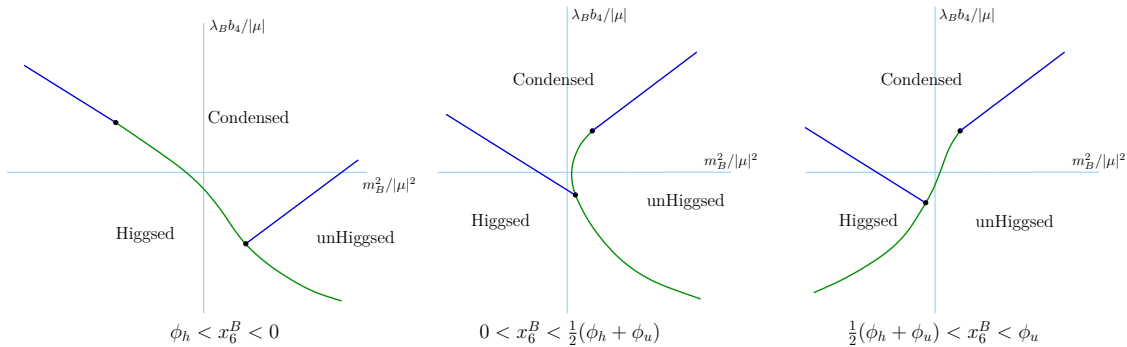}
    \caption{Phase diagram for the regular boson theory at zero
      temperature for various subranges of the stable range
      $\phi_h < x_6^B < \phi_u$. Refer to \eqref{phiuh} for the
      definitions of $\phi_u$ and $\phi_h$. The blue lines are second
      order transitions while the green lines are first order
      transitions. Actual numerical plots of the above phase diagrams
      can be found in Section \ref{numeric} in Figures
      \ref{phasecaseB}, \ref{phasecaseC1} and \ref{phasecaseC2}.}
    \label{phaseintro}
  \end{center}
\end{figure}

The phase diagrams in Figure \ref{phaseintro} allow one to determine
the thermodynamical behaviour of any given regular boson theory
(i.e. a theory at given fixed values of UV parameters) as a function
of the chemical potential (see Section \ref{vcp} for details). When
the chemical potential is small, the theory lies in the uncondensed
phase\footnote{Note that the quantity plotted on the $y$ axis scales like $\frac{1}{|\mu|}$ while the quantity plotted on the $x$ axis scales like $\frac{1}{|\mu|^2}$.
Going to small $\mu$ at fixed values of UV parameters 
therefore corresponds to moving along a parabola of the 
schematic form $y= \alpha \sqrt{x}$. Every such parabola moves into an uncondensed phase at infinity (corresponding to small $|\mu|$). See Section 
\ref{vcp} for details.}- unHiggsed or Higgsed - as depicted in Figure \ref{RBphase}
(b). As the (modulus of the) chemical potential is increased, for
every value of UV parameters, the theory undergoes a single phase
transition into the condensed phase and then stays in this phase at
every higher value of the chemical potential. The phase transition is
sometimes of second order and sometimes of first order depending on
the precise value of UV parameters.

\section{Theories and conjectured dualities}

\subsection{Theories}\label{theoriessec}
In this subsection, we present the Lagrangians for $SU(N)$
Chern-Simons theories coupled to bosonic and fermionic matter in the
fundamental representation of $SU(N)$. Analogous expressions exist for
Chern-Simons-matter theories for $U(N)$ gauge theories; see Appendix
\ref{bfdfr} for a detailed discussion. In particular, the conjectured
Bose-Fermi dualities involve theories with $U(N)$ gauge groups on at
least one side of the duality. However, the results of this paper are
accurate to leading order in $1/N$ in the 't Hooft large $N$ limit in
which case the distinction between $SU(N)$ and $U(N)$ appears only at
subleading orders in the $1/N$ expansion and hence ignored.

\subsubsection{The Regular Fermion}
The Regular Fermion theory is defined by the Lagrangian (see
e.g.~\cite{Choudhury:2018iwf})
\begin{align} \label{rflag} 
{S}_{\text{RF}}[y,\psi] & = \frac{\i \tl\kappa_F}{4\pi} \int \ud^3 x\ \epsilon^{\mu\nu\rho}\,\tr\big( y_\mu \partial_\nu y_\rho - \frac{2\i}{3} y_\mu y_\nu y_\rho\big)\nonumber \\ 
& +\int \ud^3 x \left(\bar\psi \gamma^\mu D_\mu \psi + m_F \bar\psi \psi\right)\ ,
\end{align}
where $y$ is an $SU(N)$ gauge field and $\psi$ is a Dirac spinor in
the fundamental representation of $SU(N)$. The Chern-Simons level
$\tl\kappa_F$ is given by
\begin{equation}
  \tl\kappa_F = \kappa_F - \frac{1}{2}\sgn(\kappa_F)\ .
\end{equation}
Note that the second term is subleading in the 't Hooft limit; hence,
we shall ignore this $\mc{O}(1)$ shift in all subsequent
calculations. The same comments apply for the critical fermion action
below.


\subsubsection{The Critical Fermion}
The critical fermion theory is defined by the Lagrangian
(see e.g.~\cite{Dey:2018ykx})
\begin{align}\label{cflag}
{S}_{\text{CF}}[y,\psi,\zeta_{F} ] & = \frac{\i \tl\kappa_F}{4\pi} \int \ud^3 x\ \epsilon^{\mu\nu\rho}\,\tr\big(y_\mu \partial_\nu y_\rho - \frac{2\i}{3} y_\mu y_\nu y_\rho\big)\nonumber \\
& +\int \ud^3 x \left(\bar\psi \gamma^\mu D_\mu \psi -\frac{4\pi}{\kappa_F}\zeta_{F}\bigg(\bar{\psi}\psi-\frac{\kappa_Fy_2^2}{4\pi}\bigg)-\frac{4\pi y_4}{\kappa_F} \zeta_F^2 +\frac{(2\pi)^2}{\kappa_F^2} x_6^F \zeta_F^3 \right) \ ,
\end{align}
where $y$ is an $SU(N)$ gauge field, $\psi$ is a Dirac spinor in the
fundamental representation of $SU(N)$ and $\zeta_{F}$ is an auxiliary
field which is a singlet under $SU(N)$.

\subsubsection{The Critical Boson}
The critical boson theory is defined by the Lagrangian
\cite{Choudhury:2018iwf}
\begin{align}\label{cblag} 
{S}_{\text{CB}}[y,\phi,\sigma] & = \frac{\i \kappa_B}{4\pi} \int \ud^3 x\ \epsilon^{\mu\nu\rho}\,\tr\big( y_\mu \partial_\nu y_\rho - \frac{2\i}{3} y_\mu y_\nu y_\rho\big)\nonumber \\ 
& + \int \ud^3 x\,\Big( \overline{D_\mu \phi} D_\mu \phi + \sigma  \big( \bar\phi \phi + \frac{N_B}{4 \pi} m_B^{\text{cri}} \big) \Big)\ ,
\end{align}
where $y$ is an $SU(N)$ gauge field, $\phi$ is a scalar field in the
fundamental representation of $SU(N)$ and $\sigma$ is a gauge-singlet
auxiliary field.


\subsubsection{The Regular Boson}
The regular boson theory is defined by the Lagrangian
\cite{Dey:2018ykx}
\begin{align}\label{rblag} 
{S}_{\text{RB}}[y,\phi] &  = \frac{\i \kappa_B}{4\pi }\int d^3 x~ \epsilon^{\mu\nu\rho } ~\text{Tr}\big(y_\mu \partial_\nu y_\rho -\frac{2\i }{3} y_\mu y_\nu y_\rho \big)\nonumber \\
&+\int d^3 x ~\Big((D_\mu \bar{\phi})(D^\mu \phi)+m_B^2 \bar{\phi}\phi+\frac{4\pi b_4}{\kappa_B}(\bar{\phi}\phi)^2+\frac{(2\pi)^2}{\kappa_B^2}(x_6^B+1) (\bar{\phi}\phi)^3\Big)\ ,
\end{align}
where $y$ is an $SU(N)$ gauge field and $\phi$ is a scalar field in
the fundamental representation of $SU(N)$.

The dimensionless parameters $x_6^F$ and $x_6^B$ run under the
renormalization group. However the beta functions of these parameters
are of order $\frac{1}{N}$ (see \cite{Aharony:2018pjn} for a detailed
study) and so vanish in the strict large $N$ limit. In the strict
large $N$ limit it is thus consistent to treat $x_6^F$ and $x_6^B$ as
freely tunable parameters - which one can later set equal to any
desired values (e.g. values at the fixed points of beta
functions). This is the attitude we adopt in this paper. See
\cite{Aharony:2018pjn} and \cite{Dey:2018ykx} for further discussion.

We have presented the actions \eqref{rflag}, \eqref{cflag},
\eqref{cblag} and \eqref{rblag} for the special case of $SU(N)$ gauge
theories. The generalization of these actions to $U(N)$ theories is
reviewed in detail in Appendix \ref{bfdfr}. At leading order in the
large $N$ limit, however, $SU(N)$ and $U(N)$ theories are identical,
so the subtle distinction reviewed in Appendix \ref{bfdfr} will play
no role in the discussions presented in the main text of this paper.

\subsection{Symmetries and Operators}\label{symop}

Each of the theories described above enjoys invariance under a single
$U(1)$ global symmetry. The precise nature of this symmetry depends on
whether the gauge group of the theory in question is $SU(N)$ or
$U(N)$. In the case that the gauge group is $SU(N)$, the symmetry is
simply the `baryonic' symmetry which corresponds to a uniform global
phase rotation of all fundamental fields and a simultaneous uniform
global rotation of all antifundamental fields by the inverse phase. In
the case that the gauge group is $U(N)$, the global symmetry is the
`topological $U(1)$ symmetry' corresponding to the $U(1)$ part of the
$U(N)$ gauge group. In this paper we will turn on a chemical potential
corresponding to the unique global symmetry the theory enjoys
(i.e. the baryonic symmetry if the theory is $SU(N)$ and the
topological symmetry if the theory is $U(N)$). See Appendix \ref{usu}
for more details about these symmetries.

At first sight the global symmetries of the $SU(N)$ and $U(N)$
theories appear to be rather different from each other. In Appendix
\ref{usu} we argue, however, that this distinction goes away in the
large $N$ limit. For this reason, the thermodynamical formulae
presented in this paper apply equally well to both the $SU(N)$ and the
$U(N)$ theories.

At large $N$ the gauge-invariant operators for each of the theories
listed above falls into two classes\footnote{We thank T. Senthil and
  A. Vishwanath for discussions on this point.}. The first class
consists of products of single-trace (or, more accurately, single-sum)
operators. A single-sum operator is given by the colour contraction of
(derivatives acting on) a fundamental field and (derivatives acting
on) an antifundamental field. Such operators generically have
dimensions of order unity and are all uncharged under the $U(1)$
global symmetries that each of these theories enjoy.

In the case of the $SU(N)$ theories, the second class of operators is
given by products of `baryons' (operators with (derivatives of) $N$
fundamental fields whose colour indices are contracted with an $SU(N)$
Levi-Civita tensor). In the case of $U(N)$ theories the second class
of operators consists of monopole operators. In the simple case of a
conformal theory these monopole operators are best understood via the
state-operator correspondence. These operators correspond to states
with one unit of flux, in, say, one of the diagonal entries of the
$N \times N$ $U(N)$ gauge field matrix. The Chern-Simons equations of
motion forces this configuration to be dressed by $|k|$ quanta of the
fundamental field where $k$ is the Chern-Simons level (see the next
subsection). See Appendix \ref{ops} and \cite{Radicevic:2015yla,
  Aharony:2015mjs} for a detailed discussion of these operators.

The operators in this second class are `heavy': their scaling
dimensions are of order $N$ in the large $N$ limit. These operators -
unlike those of the first class - are charged under the global $U(1)$
symmetries. Throughout this paper we work with global $U(1)$
symmetries normalised so that (roughly speaking) fundamental fields
carry unit charge under this symmetry. This means, for instance, that
the simplest baryonic operators carry charge $N$ under this
symmetry. Note that the $U(1)$ charges of these operators, like their
scaling dimensions, are of order $N$ in the large $N$ limit.

\subsection{A note on the Chern-Simons levels}\label{CSlevel}

We pause here to elaborate on the notation used in the actions
\eqref{rflag}, \eqref{cflag}, \eqref{cblag} and \eqref{rblag} above.

When the fermions or bosons in the theories above are all massive, the
low energy effective dynamics of our system is governed by a pure
Chern-Simons theory. In the case of the bosonic (resp.~fermionic)
theories we use the symbols $k_B$ (resp.~$k_F$) to denote the level of
the $SU(N)$ part of the WZW theory dual to the pure Chern-Simons
theory obtained after integrating out the matter fields with a
positive mass. The symbols $\kappa_B$ and $\kappa_F$ denote the
`renormalized levels' defined by
\begin{equation}\label{kappa}
  \kappa_B=\sgn(k_B) (|k_B|+N_B)\ ,\quad\kappa_F=\sgn(k_F) (|k_F|+N_F)\ .
\end{equation}
In this paper we will study the theories above in the large $N$
limit. The true coupling constant of these theories in this limit are
the 't Hooft couplings $\lambda_B$ (resp.~$\lambda_F$) defined by
\begin{equation}\label{tHooftcoupling}
\lambda_F =\frac{N_F}{\kappa_F}\ , \quad \lambda_B =\frac{N_B}{\kappa_B}\ .
\end{equation}
See \cite{Choudhury:2018iwf} and \cite{Dey:2018ykx} for further
details of notations and conventions.

\subsection{The conjectured Bose-Fermi duality map} \label{dualitysec}

The quasi-fermionic theories i.e.~regular fermion \eqref{rflag} and
critical boson \eqref{cblag} theories have been conjectured to be dual
to each other. Similarly, the quasi-bosonic theories i.e.~the critical
fermion \eqref{cflag} and regular boson \eqref{rblag} theories also
have been conjectured to be dual to each other. For each of these
conjectured dualities, the map between the renormalized levels and
ranks of the CS gauge groups is given by
\begin{equation}\label{level-rankmap} 
N_{B}=|\kappa_F|-N_{F}\ ,\quad \kappa_B = -\kappa_F\ .
\end{equation}
Equivalently, and perhaps more simply, 
\begin{equation}\label{level-rankmappr} 
N_{B}=|k_F| ,\quad k_B = -{\rm sgn}(k_F) N_F.
\end{equation}

The above relationship \eqref{level-rankmap} is conjectured to be
exact. In particular, they are precisely the rules for the
well-established level-rank duality of WZW theories that are dual to
the pure Chern-Simons theories that we obtain after integrating out
the massive matter fields. From \eqref{tHooftcoupling} it follows that
the 't Hooft couplings of the dual pairs of theories are related by
\begin{equation}\label{lambdamap}
\lambda_{B} = \lambda_{F} -\sgn(\lambda_F) \ .
\end{equation}

In addition to the Chern-Simons levels and ranks, the quasi-fermionic
theories are each characterised by an additional mass parameter, $m_F$
and $m_B^{{ \rm cri}}$ respectively for the regular fermion and
critical boson theories. In the large $N$ limit these mass parameters
are conjectured to be related via the duality map
\begin{equation}\label{rfcbmap}
  m_{F}=-\lambda_{B} m_{B}^{\text{cri}}\ .
\end{equation}
The quasi-bosonic theories are similarly characterised by three UV
parameters apart from the Chern-Simons ranks and levels. These
correspond to the sextic, quartic and mass terms in the Lagrangian
(see \eqref{rblag}, \eqref{cflag}). In the large $N$ limit these are
conjectured to be related via the duality maps
\begin{equation}\label{rbcfmap} 
x_{6}^F = x_{6}^B\ ,\quad y_{4}=b_{4}\ ,\quad y_{2}^2=m_{B}^2 \ .
\end{equation}
As mentioned above, while the relations \eqref{level-rankmap} and
\eqref{lambdamap} are expected to hold even at finite $N$, the
relations \eqref{rfcbmap} and \eqref{rbcfmap} are conjectured to hold
exactly at only infinite $N$ and may receive finite-$N$ corrections.

See e.g.~\cite{Dey:2018ykx, Choudhury:2018iwf} for more details of the
conjectured duality maps, including a detailed discussion of the
levels of $U(1)$ factors, as well as the regulation scheme in which
the duality map above is conjectured to hold.

\section{Large $N$ thermal partition functions} \label{thptnfn}

All four theories described in Section \ref{theoriessec} are
effectively solvable in the large $N$ limit. In particular, the free
energy of each of these theories at finite temperature and finite
chemical potential (in a range of chemical potentials, see below) has
previously been computed at every value of the 't Hooft coupling in
the large $N$ limit. In this section we first review previously
obtained results and then extend them to the range of chemical
potentials of interest to this paper.

The partition function $\mc{Z}$ on $S^2 \times S^1$ of each of these
theories is given by an expression of the sort
\begin{equation}\label{S2S1partnfunctn} 
  \mathcal{Z}_{S^2\times S^1} = \int [dU]_{\rm CS}\, e^{-\mc{V}_{2}T^2v[\rho]} \ ,
\end{equation}
where $[dU]_{\rm CS}$ is the Chern-Simons modified Haar measure over
$U(N)$ \cite{Aharony:2012ns, Jain:2013py}. At the physical level, the matrix $U$ is
the zero mode of the gauge field holonomy around the thermal circle
$S^1$. $\mathcal{V}_{2}$ is the volume of the two dimensional space
$S^2$ and $T$ is the temperature which is related to the radius
$\beta$ of the thermal circle $S^{1}$ as $T=\beta^{-1}$. The quantity
$\rho$ is the holonomy eigenvalue distribution function. As explained
in e.g.~Section 1.2.1 in \cite{Choudhury:2018iwf}, the zero mode of
the holonomy $U$ of the gauge field around $S^{1}$ is completely
specified (upto a permutation) by its eigenvalues, $e^{\i \alpha_{j}}$
where, $j=1,\cdots, N$ and $\alpha_{j} \in (-\pi, \pi]$. In the large
$N$ limit, the eigenvalue locations on the circle are effectively
specified by a continuous distribution function $\rho(\alpha)$ defined
by
\begin{equation}\label{rho(alpha)def} 
  \rho(\alpha) = \lim_{N\to \infty}  \frac{1}{N} \sum_{j=1}^{N}  \delta(\alpha-\alpha_{j})\ .  
\end{equation}
The quantity $v[\rho]$ that appears in \eqref{S2S1partnfunctn} is
simply the result of path integral over all non-holonomy modes
including the gauge field and matter fields. Indeed,
\begin{equation}\label{R2S1partnfunctn} 
  e^{-\mc{V}_{2}T^2v[\rho]} =\int_{\mathbb{R}^2 \times {S}^1} [d\phi]\,  e^{-S[\phi,\rho]}\ , 
\end{equation}
where $[d\phi]$ in \eqref{R2S1partnfunctn} is the path integral
measure over all matter and gauge degrees of freedom except the
holonomy of the gauge field around the thermal circle. The subscript
$\mathbb{R}^2 \times S^1$ indicates that the path integral is actually
computed for the theory on $\mathbb{R}^2 \times S^1$. The information
that we are in fact on $S^2 \times S^1$ goes into the Chern-Simons
modified Haar measure $[d U]_{\rm CS}$ in \eqref{S2S1partnfunctn}.


Partition functions at finite temperature and chemical potential are
physical observables, and so should be the same for two dual
theories. From the definition of the partition function in
\eqref{S2S1partnfunctn} it might at first seem that the requirement
that thermal partition functions match across a duality requires that
the quantities $v[\rho]$ match across the duality. However that is not quite the case; recall that unitary matrices are of different ranks on the bosonic and fermionic sides of the duality, and on each side one also has to
take into account the contribution from the Chern-Simons modified Haar measure over the unitary matrices in \eqref{S2S1partnfunctn}, which depends on the 't Hooft couplings $\lambda_B$ and $\lambda_F$ in a
non-trivial way. Nonetheless the bosonic and fermionic $v[\rho]$ functions must clearly be related in {\it some} way 
in order for the partition functions to match. It was demonstrated in \cite{Jain:2013py} that the thermal partition
functions given by \eqref{S2S1partnfunctn} of the bosonic (regular or
critical) and fermionic (critical or regular) theories {\it will} in
fact agree with each other in the large $N$ limit provided the
functions $v[\rho]$ obey the more subtle relationship
\begin{equation}\label{freenrgymap} 
  v_{\rm RB}[\rho_{B}] = v_{\rm CF}[\rho_F]\ ,\quad v_{\rm CB}[\rho_B] = v_{\rm RF}[\rho_F]\ ,
\end{equation}
where the holonomy distribution functions $\rho_B$ and $\rho_F$ are not equal to each other but are instead 
related by the equation
\begin{equation}\label{holonomymap}
|\lambda_{B}| \rho_{B}(\pi -\alpha) +|\lambda_{F}|\rho_{F}(\alpha) =\frac{1}{2\pi} \ .
\end{equation}

\subsection{The off-shell free energy}\label{offshellsec}

In the large $N$ 't Hooft limit, the path integral
\eqref{R2S1partnfunctn} is very effectively computed by the saddle
point approximation when the volume $\cV_2$ of the spatial $S^2$ on
which we work is taken to be $\cV_2 = N V$, and $V$ and all other
parameters (temperatures, chemical potentials, masses etc) are held
fixed in the t' Hooft limit. This is the limit we work with all
through this paper. In Sections \ref{thermo} onwards we also further
specialize to limit $V \to \infty$. In this case we first take the
limit $N \to \infty$ and then the limit $V \to \infty$.

 The final
results for $v[\rho]$, obtained this way in earlier work, are most
conveniently presented in terms of a so-called `off-shell free
energy' \cite{Dey:2018ykx}. The off-shell free energy is a quantity that depends on some
auxiliary variables in addition to the UV parameters of the theory,
the temperature, chemical potential and eigenvalue distribution
$\rho$. It turns out that once we perform the path integral
\eqref{R2S1partnfunctn} via saddle point approximation and obtain an
answer for $v[\rho]$ at each of the various large $N$ saddle points,
the whole procedure can be replaced by a simpler one: that of
performing an ordinary integral over a few auxiliary variables,
denoted collectively as $\varphi_{\rm aux}$, with the integrand being
a function of these variables rather than being a functional of
fields:
\begin{equation}\label{offshellFE}
  e^{-\mc{V}_{2}T^2v[\rho]} =\int d\varphi_{\rm aux}\  e^{-\mc{V}_2 \beta F(\varphi_{\rm aux};\rho]}\ ,
\end{equation}
where, the object $F(\varphi_{\rm aux};\rho]$ is the off-shell free
energy per unit volume. The funny brackets $(\cdots]$ indicate that
$F$ is still a functional of the holonomy distribution $\rho$ while
being an ordinary function of the auxiliary variables
$\varphi_{\rm aux}$. The off-shell free energy $F$ has an explicit
factor of $N$ in its expression and hence, in the large $N$ limit, the
integral above may be evaluated by saddle point approximation. The
quantity $v[\rho]$ is thus obtained from the off-shell free energy by
extremizing the latter w.r.t.~its auxiliary variables \cite{Dey:2018ykx}.

The existing results in the literature for the free energies $v[\rho]$ \cite{Giombi:2011kc, Jain:2012qi, Yokoyama:2012fa, Aharony:2012ns, Jain:2013py, Takimi:2013zca, Jain:2013gza, Yokoyama:2013pxa, Geracie:2015drf, Choudhury:2018iwf, Dey:2018ykx, Aharony:2018pjn, Dey:2019ihe, Jensen:2019mga}
and $F(\varphi_{\rm aux}; \rho]$ in the presence of chemical potential
were carefully computed only for values of the chemical potential
smaller than the thermal masses of the bosonic or fermionic
excitations in the respective theories. We present these expressions
for the free energy in the next subsection. It turns out that some
interesting modifications are required in the expressions for the free
energies in order to extend their validity for chemical potentials
larger than the thermal masses. We describe these modifications in the
subsection after next.

\subsection{Previously known results} \label{mulcbfe}

\textbf{Note:} Throughout this paper, we use the convention that,
given a quantity $x$ of mass dimension $a$, the corresponding
dimensionless hatted quantity $\hat{x}$ is defined to be the quantity
$x$ in the units of temperature, i.e., $\hat{x}= (\beta)^a x$. For
example, the mass parameter $m_B^{\rm cri}$ of the critical boson
theory has its dimensionless counterpart
${\hat m}_B^{\rm cri} = \beta m_B^{\rm cri}$.

\subsubsection{The Regular Fermion}\label{osrf}

The off-shell free energy for the RF theory, presented as a function
of the two auxiliary variables $\tilde\cC$ and $c_F$, 
is  \cite{Dey:2018ykx, Dey:2019ihe}  %
\begin{align}\label{RF2voffshellfe}
  &F_{\rm RF}({c}_F,\tilde{\mathcal{C}})\nonumber\\
  & = \frac{N_F}{6\pi\beta^3} \bigg[ - 8\lambda_{F}^2 \tl{\mc{C}}^3 - 3\tl{\mc{C}} \left( \hat{c}_{F}^2 - \big(2\lambda_F \tl{\mc{C}} +\hat{m}_F \big)^2 \right)  - 6\lambda_{F} \hat{m}_F \tl{\mc{C}}^2 \nonumber  \\
  &\qquad\qquad+ \hat{c}_{F}^3 - 3 \int_{\hat{c}_F}^{\infty}  d\hat{\e}\ \hat{\e}  \int_{-\pi}^{\pi}  d\alpha\, \rho_F(\alpha)\left(\log\big(1+e^{-\hat{\e}-\hat{\mu}-\i \alpha }\big)+\log\big(1+e^{-\hat{\e}+ \hat{\mu}+\i \alpha }\big)  \right)  \bigg]  \ .
\end{align}
Extremization of \eqref{RF2voffshellfe} w.r.t.~${c}_F$ and $\tl\cC$
gives, respectively,
\begin{equation}\label{tcdef}
  \tl{\mc{C}} =  \mc{C}({c}_F,{\mu})\ ,\quad\text{and}\quad \hat{c}_F^2 =
  \big(2\lambda_F \tl{\mc{C}} +\hat{m}_F  \big)^2\ ,
\end{equation}
where $\mc{C}$ is a particular moment of the holonomy distribution
given by\begin{equation}\label{Cdef}
\mathcal{C}({\epsilon}, {\mu}) \equiv \frac{1}{2} \int_{-\pi}^{\pi}  d\alpha \ \rho_F(\alpha)   \left( \log \left(2 \cosh \frac{\hat{\epsilon}+ \hat{\mu} +\i \alpha  }{2} \right) + \log \left(2 \cosh \frac{\hat{\epsilon} - \hat{\mu} - \i \alpha}{2} \right) \right)\ .
\end{equation}
The extremum value of the variable $c_F$ gives the thermal mass of the
fermionic excitations which is also denoted by the same symbol $c_F$.
We do not know of  a physical interpretation for the variable $\tl{\cC}$
except that it becomes the function $\cC(c_F,\mu)$ on-shell
\footnote{The logarithm in the free energy \eqref{RF2voffshellfe} and
  the moment function \eqref{Cdef} - and indeed everywhere in this
  paper - is defined to have a branch cut for negative real arguments,
  so that the imaginary part of $\ln (x + \i y)$ is $+\i \pi$ just
  above the negative $x$ axis, but is $-\i \pi$ just below the
  negative $x$ axis. In deriving the equation of motion \eqref{tcdef}
  we have used that when $x>0$
\begin{equation}\label{manip}
\begin{split}
&\int_{-\pi}^{\pi}  d\alpha \ \rho_F(\alpha)  \ \log \Big(2 \cosh \frac{x +\i \alpha  }{2} \Big)\\
& = \int_{-\pi}^{\pi}  d\alpha \ \rho_F(\alpha) \Big( \frac{x+\i  \alpha}{2} +  \log \big(1+ e^{ -x-\i \alpha } \big) \Big)\\
&= \frac{x}{2} + \int_{-\pi}^{\pi}  d\alpha \ \rho_F(\alpha) \ \Big( \log \big(1+ e^{ -x-\i \alpha } \big) \Big) 
\end{split} 
\end{equation} 
We use this identity once with $x=\hat{c}_F+\hat{\mu}$ and once with
$x=\hat{c}_F-\hat{\mu}$. The condition $x>0$ is met in both cases
because we have assumed ${\hat c}_F > |\hat{\mu}|$.  In going from the
first to the second line of \eqref{manip} we use the fact that when
the $\log$-function is defined to be real on the positive real
axis and with branch cut along the negative real axis
\begin{equation}\label{logbc}
\log z_1 z_2= \log z_1 + \log z_2, ~~~{\rm provided}~~~ |{\rm Arg}(z_1)|< \frac{\pi}{2}, ~~~|{\rm Arg}(z_2)|< \frac{\pi}{2},
\end{equation}
The conditions on the arguments of $z_1$ and $z_2$ in \eqref{logbc}
are met in going from the first to the second line of \eqref{manip},
because we have assumed $x>0$.  In going from the second to the third
line of \eqref{manip} we have used the fact that $\rho_F(\alpha)$ is
an even function that integrates to unity.\label{coshlogdef}}.

\subsubsection{The Critical Fermion} \label{oscf}

The off-shell free energy for the CF theory was presented in
\cite{Dey:2018ykx} (see equation (A.3) of \cite{Dey:2018ykx}) and is
given by
\begin{align}\label{CFoffshellfe}
  &F_{\rm CF}({c}_F,{\zeta}_F,\tilde{\cC})\nonumber\\
  & = \frac{N_F}{6\pi \b^3} \bigg[ -8\lambda_{F}^2 \tl{\mc{C}}^3 - 3\tl{\mc{C}} \bigg( \hat{c}_{F}^2 - \Big(2\lambda_F \tl{\mc{C}} - \frac{4\pi \hat{\zeta}_F}{\kappa_F} \Big)^2 \bigg) + 6\lambda_{F} \tl{\mc{C}}^2 \Big(\frac{4\pi \hat{\zeta}_F}{\kappa_F} \Big) \nonumber  \\ 
  &\qquad\qquad   + 3 \bigg( \frac{\hat{y}_2^2}{2\lambda_F} \frac{4\pi \hat{\zeta}_F}{\kappa_F} - \frac{\hat{y}_4}{2\lambda_F} \Big(\frac{4\pi \hat{\zeta}_F}{\kappa_F} \Big)^2 +\frac{x_6^F}{8\lambda_F} \Big(\frac{4\pi \hat{\zeta}_F}{\kappa_F} \Big)^3 \bigg)\nonumber   \\
&\qquad\qquad +  \hat{c}_{F}^3 - 3 \int_{\hat{c}_F}^{\infty} d\hat{\e}\ \hat{\e} \int_{-\pi}^{\pi}d\alpha\ \rho_F(\alpha) \left(\log\big(1+e^{-\hat{\e}-\hat{\mu}-\i\alpha }\big)+\log\big(1+e^{-\hat{\e}+ \hat{\mu}+\i\alpha }\big)  \right)  \bigg]  \ .
\end{align}
In \eqref{CFoffshellfe}, ${c}_F$, ${\zeta}_F$ and ${\tilde\cC}$ are
auxiliary quantities (w.r.t.~which we have to extremize $F_{\rm CF}$
in order to obtain $v_{\rm CF}[\rho]$). The gap equations that follow
from this extremization process are derived and described in detail in
the Section A.1 of \cite{Dey:2018ykx}. We present the equations here
for completeness. The extremization of \eqref{CFoffshellfe}
w.r.t.~$\tl{\mc{C}}$, ${c}_F$ and ${\zeta}_F$ respectively gives the
following equations
\begin{align}\label{extremizeFCF}
  & \hat{c}_F^2 = \bigg(2\lambda_F\tl{\mc{C}}-\frac{4\pi\hat{\zeta}_F}{\kappa_F} \bigg)^2\ ,\nonumber\\
  & \tl{\mc{C}}=\mc{C}({c}_F,{\mu})\nonumber\ ,\\
  & \frac{3}{4} \Big(\frac{4\pi\hat{\zeta}_F}{\kappa_F} \Big)^2 x_6^F +\frac{8\pi\hat{\zeta}_F}{\kappa_F} \big(2\lambda_F\tl{\mc{C}}- \hat{y}_4 \big) -4\lambda_F^2\tl{\mc{C}}^2+ \hat{y}_2^2 = 0\ .
\end{align}
As for the regular fermion, the extremum value of the variable $c_F$
gives the thermal mass of the fermionic excitations and the extremum
value of the variable $\tl{\cC}$ is $\cC(c_F, \mu)$. The variable
$\zeta_{F}$ in the critical fermion free energy \eqref{CFoffshellfe}
is precisely the (constant) expectation value of the auxiliary field
$\zeta_{F}(x)$ that appears in the critical fermion Lagrangian
\eqref{cflag}. If we (classically) integrate out the variables
${ c}_F$ and $\tl{\mathcal{C}}$ from the free energy
\eqref{CFoffshellfe} and perform the integral over the holonomy, the
resultant function of $\zeta_F$ is the (large $N$ exact) quantum
effective potential of the theory as a function of $\zeta_F$.


\subsubsection{The Critical Boson} \label{oscb}

The off-shell free energy for the critical boson theory, presented in
terms of the two auxiliary variables $c_B$ and ${\tilde \cS}$, is
given by \cite{Dey:2018ykx}
\begin{align}\label{CB2voffshellfe}
  &F_{\rm CB}({c}_B, \tilde{\cS})\nonumber\\
  & =\frac{N_B}{6\pi \b^3} \bigg[\frac{3}{2} \hat{c}_B^2\hat{m}_B^{\text{cri}} - 4\lambda_B^2\left(\tilde{\mc{S}}-\tfrac{1}{2}\hat{m}_{B}^{\text{cri}} \right)^3  +6|\lambda_B| \hat{c}_B\left(\tilde{\mc{S}}-\tfrac{1}{2}\hat{m}_{B}^{\text{cri}}\right)^2\nonumber \\
&\qquad\qquad -\hat{c}_B^3 +
3\int_{\hat{c}_B}^{\infty} d\hat{\e}\ \hat{\e} \int_{-\pi}^{\pi} d\alpha \ \rho_B(\alpha) \ \big(\log(1-e^{-\hat{\e}+\hat{\mu}+\i \a})+\log(1-e^{-\hat{\e}-\hat{\mu}-\i \alpha})\big)\bigg]\ .
\end{align}
Extremizing the free energy \eqref{CB2voffshellfe}
w.r.t.~$\tl{\mc{S}}$ and $c_B$ respectively gives
\begin{align}\label{ofeom}
  &\left(\tilde{\mc{S}}-\tfrac{1}{2}\hat{m}_{B}^{\text{cri}}\right) \left(\frac{\hat{c}_B}{|\lambda_B|} - \Big( \tilde{\mc{S}}-\tfrac{1}{2} \hat{m}_{B}^{\text{cri}}\Big)\right) = 0\ ,\nonumber\\
  &\left(\mc{S}({c}_B,{\mu})-\tfrac{1}{2}\hat{m}_{B}^{\text{cri}}\right)\hat{c}_B-|\lambda_B| \left( \tilde{\mc{S}}-\tfrac{1}{2}\hat{m}_{B}^{\text{cri}}\right)^2 = 0\ ,
\end{align}
where the function $\cS$ is defined by
\begin{equation}\label{Sdef}
\mathcal{S}({\epsilon}, {\mu}) \equiv \frac{1}{2} \int_{-\pi}^{\pi}  d\alpha \ \rho_B(\alpha)   \ \left( \log \left(2 \sinh \frac{\hat{\epsilon}+ |\hat{\mu}| +\i\alpha  }{2} \right) + \log \left(2 \sinh \frac{\hat{\epsilon}  - |\hat{\mu}| - \i\alpha}{2} \right) \right)\ .
\end{equation}
(the logarithm in \eqref{Sdef} is defined in exactly the same way as
the logarithm in \eqref{Cdef} in Footnote \ref{coshlogdef}).

The variable ${ c}_B$ has a direct physical interpretation: its
extremum value is the thermal mass of the bosonic excitations. We do
not have a similar physical interpretation for the variable $\tl\cS$
except that it becomes the function $\cS(c_B,\mu)$ above.

\subsubsection{The Regular Boson} \label{osrb}

As demonstrated in \cite{Dey:2018ykx}, the off-shell thermal free
energy (per unit volume $\beta\mathcal{V}_2$) of the RB theory at a finite
chemical potential $\mu$ is given by
\begin{align}\label{RBoffshellfe}
  &F_{\rm RB}({c}_B,{\sigma}_B,\tilde\cS)\nonumber\\
  & =\frac{N_B}{6\pi \b^3} \bigg[-3\hat{c}_B^2\hat{\sigma}_B+\lambda_B^2\hat{\sigma}_B^3 -4\lambda_B^2(\tilde{\cS}+\hat{\sigma}_{B})^3 +6|\lambda_B| \hat{c}_B(\tilde{\cS}+\hat{\sigma}_B)^2 \nonumber \\
&\qquad\qquad +3(\hat{m}_B^2\hat{\sigma}_B+2\lambda_B\hat{b}_4\hat{\sigma}_B^2+(x_6+1) \lambda_B^2 \hat{\sigma}_B^3)\nonumber \\
&\qquad\qquad  -\hat{c}_B^3 +
3\int_{\hat{c}_B}^{\infty} \hat{\e} \ d\hat{\e} \int_{-\pi}^{\pi} d\alpha \ \rho_B(\alpha) \ \big(\log(1-e^{-\hat{\e}+\hat{\mu}+\i \a})+\log(1-e^{-\hat{\e}-\hat{\mu}-\i \alpha})\big)\bigg]\ .
\end{align}
In \eqref{RBoffshellfe}, ${c}_B$, ${\sigma}_B$ and $\tilde{\mc{S}}$
are auxiliary quantities (w.r.t.~which we have to extremize
$F_{\rm RB}$ in order to obtain $v_{\rm RB}[\rho]$ ). The gap
equations that follow from this extremization process are derived in
detail in equation (4.3) of \cite{Dey:2018ykx}, and for completeness,
we list these here. Extremization of \eqref{RBoffshellfe}
w.r.t. $\tl{\mathcal{S}}$, $\hat{c}_B$ and $\hat{\sigma}_B$
respectively gives the following equations
\begin{align}\label{extremizeFRB}
& \big(\tilde{\mc{S}}+\hat{\sigma}_{B}\big) \big(\hat{c}_B-|\lambda_B| \big( \tilde{\mc{S}}+\hat{\sigma}_{B}\big)\big) = 0\ ,\nonumber \\
& \big(\mc{S}({c}_B,{\mu})+\hat{\sigma}_{B} \big)\hat{c}_B-|\lambda_B| \big( \tilde{\mc{S}}+\hat{\sigma}_{B} \big)^2 = 0\ ,\nonumber \\
& \hat{c}_B^2 -\hat{m}_B^2 -4|\lambda_B|\hat{c}_B \big( \tilde{\mc{S}}+\hat{\sigma}_{B}\big) + \lambda_B \Big( 4\tl{\mc{S}}^2 \lambda_B -4\hat{b}_4 \hat{\sigma}_B +8\lambda_B \hat{\sigma}_B \tl{\mc{S}} -3\lambda_B x_6^{B}\hat{\sigma}_B^2 \Big) = 0\ .
\end{align}
The quantity ${ \sigma_B}$ in the regular boson free energy
\eqref{RBoffshellfe} also has a simple physical interpretation. It is
related to the expectation value of the lightest gauge-invariant
operator, ${\bar \phi} \phi$, of the regular theory as
\begin{equation}\label{physintsig}
{\sigma}_{B} = \frac{2\pi }{N_B} \langle \bar{\phi}\phi \rangle \ .
\end{equation}
As for the critical boson, the variable $c_B$ has the physical
interpretation that, on shell, it is the thermal mass of the bosonic
excitations. The variable $\sigma_B$ also has an important physical
origin. In fact, if we (classically) integrate out the variables
${ c}_B$ and $\tl{\cS}$ in the regular boson free energy
\eqref{RBoffshellfe}, and then substitute \eqref{physintsig} for
$\sigma_B$ and finally perform the integral over the holonomy, the
resultant function of $\langle{\bar \phi} \phi\rangle$ is simply the
large-$N$-exact quantum effective potential of the theory as a
function of its `lightest' gauge invariant observable
$\langle{\bar \phi} \phi\rangle$.

The objects $\tl{\mc{S}}$ and $\cS({ c}_B, { \mu})$ are different
off-shell: one is a variable while the other is a function of a
different variable. However, on shell, we have
\begin{equation}
  \tl{\mc{S}} = \mc{S}({c}_B,{\mu})\ ,
\end{equation} 
which is easy to see by comparing the first two equations in
\eqref{extremizeFRB}.

\subsubsection{Duality of off-shell variables and off-shell free
  energies}

Under duality, the off-shell variables have to be appropriately mapped
to each other in addition to the duality identifications between the
coupling constants given in Section \ref{dualitysec}. The duality map
between the off-shell variables is as follows:
\begin{equation}\label{offshvarmap}
  c_B = c_F\ ,\quad \lambda_B \tl{\cS} = \lambda_F \tl\cC - \tfrac{1}{2} \sgn(\lambda_F) c_F\ ,\quad 2\lambda_B \sigma_B = -\frac{4\pi\zeta_F}{\kappa_F}\ .
\end{equation}
It can be shown that the off-shell free energies of the critical
fermion and regular boson map to each other under the duality map
\eqref{offshvarmap}. Similarly, it can be shown that the off-shell
free energies of the regular fermion and critical boson map to each
other under the first two relations of the map \eqref{offshvarmap}
(the third relation in \eqref{offshvarmap} is not required in this
case). We relegate the demonstration of the matching of off-shell free
energies to Appendix \ref{dualityapp}.

It is interesting that the off-shell free energies of the
(conjecturally dual) bosonic and fermionic theories map precisely to
each other at the full off-shell level rather than only on shell
(i.e. only on the solutions to the equations of motion)\footnote{The
  latter weaker agreement would have been sufficient to ensure the
  equality of the free energies of the bosonic and fermionic theories
  under the duality map of Section \ref{dualitysec}}. This stronger
duality invariance suggests that the off-shell free energy is not just
a computational device but is physically meaningful. It has previously
been partially demonstrated that this is indeed the case: the authors
of \cite{Dey:2018ykx} demonstrated that the off-shell free energy of
the regular boson theory as a function of $\sigma_B$ (obtained after
integrating out the variables $c_B$ and ${\tilde \cS}$) is closely
related to the quantum effective action as a function of its lightest
gauge invariant scalar operator ${\bar \phi \phi}$. It is possible
that the full off-shell free energy is physical in a similar manner
(i.e. it can be recast as the quantum effective action as a function
of three suitable physical operators). The details of such a
reformulation - if it exists - is an interesting problem for future
research.

\subsection{Conjecture for the off-shell free energy for
  \texorpdfstring{$|\mu|$}{mu} greater than thermal
  mass}\label{mugcfe}

In Section \ref{mulcbfe}, we presented the expressions for the
off-shell thermal free energies of the four classes of theories of
interest to us in this paper. The results of Section \ref{mulcbfe}
apply whenever the thermal mass of the bosonic or fermionic
excitations is larger than the modulus of the chemical potential. In
this subsection, we will present a conjecture for the extension of
these results to the situation in which the chemical potential is
larger than the quasiparticle thermal masses.

In order to motivate our conjecture, consider the case of the bosonic
theories. The effective squared mass of the boson $(c_B^2-\mu^2)$, vanishes when $\mu=c_B$ and becomes negative when $\mu^2 >c_B^2$, encouraging the boson to `condense'. As we will explain in detail in the next subsection, the off-shell free energies of these theories depend on the chemical potential only through a bosonic determinant. As this finite temperature determinant is  two
dimensional at long distances, however, we expect this `condensation' to occur without a sharp phase transition (recall that spontaneous symmetry breaking of continuous symmetries is forbidden in two dimensions). Note this expectation continues to hold even in the large $N$ limit, as the logarithm of a large $N$ determinant depends on $N$ only 
through an overall multiplicative factor of $N$ which does not affect its analyticity properties. \footnote{We thank S. Wadia for discussions on this point.}

The expectation on the fermionic side is similar: Fermi
sea formation, while sharp at zero temperature, is a crossover at
finite temperature.

At finite temperature, we then expect the off-shell free energies of
both the bosonic and fermionic theories to be analytic functions of
${\mu}$. {\it We thus conjecture that the correct formulae for the
  off-shell free energies for $|{\mu}| \geq c_B$ are simply the
  analytic continuation (in the variable ${\mu}$) of the relevant
  formulae for $|{\mu}| < {c}_B$ (presented in Sections
  \ref{osrf}$-$\ref{osrb}.}).

The conjecture presented in this section matches the
expectations of Bose-Fermi duality more or less trivially. Since the
fermionic and bosonic free energies have already been shown to match
when the chemical potential is smaller than the quasiparticle masses,
the analytic continuations of these two expressions necessarily agree
at all values of the chemical potential.

\subsection{Implementing the analytic continuation}

\subsubsection{Non-analyticity of the naive expressions}

The off-shell free energies of the previous section were computed in
the previous literature by evaluating the field theory partition
functions \eqref{offshellFE} via the following (schematic)
procedure. One first integrates out the gauge field (this is possible
because it appears in the action in a quadratic manner). One then
introduces various singlet Lagrange multiplier and
Hubbard-Stratanovich fields chosen so as to ensure that the action is
quadratic in the matter fields. The matter fields are then integrated
out, yielding a determinant (which exponentiates to a
log-determinant). The resultant action is a complicated non-local
functional of the auxiliary singlet fields.

One then derives the large $N$ saddle point equations for these
singlet fields. Remarkably, it turns out to be possible to exactly
solve the resultant non-linear integral equations. What is more
remarkable is that, these complicated non-linear integral equations
for the singlet fields can be recast into much simpler algebraic or
transcendental equations for a few $c$-number variables rather than
fields. These $c$-number variables are the `auxiliary' variables that
appear in the off-shell free energies \eqref{CB2voffshellfe},
\eqref{RBoffshellfe}, \eqref{RF2voffshellfe} and \eqref{CFoffshellfe},
and the algebraic or transcendental equations involving these
variables are the ones obtained by extremizing these simple off-shell
free energies listed in these four equations w.r.t.~their auxiliary
variables as discussed in detail in Section \ref{mulcbfe}.

What is important for the current discussion is that the
log-determinant described in the first paragraph above is the only
part of the effective action that had its origin in a term in the
Lagrangian involving derivatives acting on the matter
fields. Therefore, it is the only part of this effective action that
is sensitive to the chemical potential. This fact is also visible in
the explicit expressions \eqref{CB2voffshellfe}, \eqref{RBoffshellfe},
\eqref{RF2voffshellfe} and \eqref{CFoffshellfe} (the log-determinant
is the last line in each of those equations).

In summary, the general structure of the off-shell free energies \eqref{CB2voffshellfe}, \eqref{RBoffshellfe}, \eqref{RF2voffshellfe} and \eqref{CFoffshellfe} of the previous is 
\begin{equation}\label{offshstr}
  F = F_{\rm int} + F_{\rm det}\ ,
\end{equation}
where $F_{\rm det}$ is the log-determinant term and $F_{\rm int}$
(which is the rest of the free energy) is independent of $\mu$. As
reflected in the final answers \eqref{CB2voffshellfe},
\eqref{RBoffshellfe}, \eqref{RF2voffshellfe} and \eqref{CFoffshellfe},
the determinants $F_{\rm det}$ take a simple universal form,
correspond to the free energy of a single massive boson (of a mass
$c_B$ that remains to be determined) or a single massive fermion (of a
mass $c_F$ that remains to be determined) modified by its coupling to
the holonomy distribution $\rho$ \footnote{Recall that this holonomy
  distribution function $\rho_B$ or $\rho_F$ which appears in the
  determinant is itself determined at the saddle-point by extremizing
  its effective action obtained by integrating out the matter
  fields.}.

Let us first focus on the bosonic determinant. At small enough values
of $\mu$, $F_{\rm det}$ is, of course, an analytic function of
$\mu$. At $|\mu|=c_B$ something special happens.  At this value the
effective mass of the bosonic field - which is proportional to
$c_B^2-\mu^2$ - vanishes. After Kaluza-Klein reduction on the thermal
circle, this effectively massless scalar field reduces to an infinite
collection of two dimensional fields of squared masses
$$ \left(\frac{2 \pi n - \alpha}{\beta} \right)^2\ ,$$
where the Kaluza-Klein momentum $n$ runs from $-\infty$ to $\infty$,
$\alpha$ is the holonomy and $\beta$ is the inverse temperature. When
$\alpha=0$ the Kaluza-Klein mode with $n=0$ is a massless two
dimensional scalar field. The determinant over such a field has a
logarithmic divergence. As a consequence we should expect our
determinant to have a potential non-analyticity of the schematic form
$$ \int d \alpha \rho(\alpha) \log \left( \beta(c_B-\mu -i\alpha) \right)\ , $$
in the neighbourhood of $\alpha=0$. Indeed the exact expression for
the Bosonic determinant
\begin{align}\label{bosdet}
&\beta \mc{V}_2\, F_{B,\text{det}} (c_B, \mu)\nonumber\\
&= \frac{N_B \mc{V}_2}{2\pi} \int_{c_B}^{\infty} d\epsilon\,\epsilon \int_{-\pi}^\pi d\alpha\, \rho_B(\alpha) \left( \log \left(2 \sinh \frac{\hat{\epsilon}+ \hat{\mu} +\i\alpha  }{2} \right) + \log \left(2 \sinh \frac{\hat{\epsilon}  - \hat{\mu} - \i\alpha}{2} \right) \right)\ ,\nonumber\\
&= \frac{N_B \mc{V}_2}{2\pi} \left[ - \frac{\beta c_B^3}{3} + \int_{c_B}^{\infty} d\epsilon\,\epsilon \int_{-\pi}^\pi d\alpha\, \rho_B(\alpha) \left( \log \Big( 1 -  e^{-\hat{\epsilon} - \hat{\mu} - \i\alpha} \Big) + \log \Big( 1 -  e^{-\hat{\epsilon} + \hat{\mu} + \i\alpha} \Big) \right)\right]\ ,
\end{align}
does have precisely such a singularity. 

Similar remarks apply to fermionic fields. Once again the field is
effectively massless at $\mu=c_F$. Once again a Kaluza-Klein reduction
at this value of $\mu$ gives an infinite number of two dimensional
fields whose squared masses, this time, are
$$ \left(\frac{2 \pi (n+ \frac{1}{2}) - \alpha}{\beta} \right)^2\ .$$
(the shift of $n$ by $\frac{1}{2}$ is a consequence of the
antiperiodic boundary conditions of fermions around the thermal
circle). This time the mode with $n=0$ is effectively massless at
$\alpha=\pi$ and so we should expect the fermionic determinant to have
a potential non-analyticity of the form
$$ \int d \alpha \rho(\alpha) \ln \left( \beta(c_F-\mu -\i \alpha) \right)\ , $$
in the neighbourhood of $\alpha=\pi$. Indeed, the exact expression for
the fermionic determinant
\begin{align}\label{ferdet}
&\beta \mc{V}_2\, F_{F,\text{det}} (c_F, \mu)\nonumber\\
&=- \frac{N_F \mc{V}_2}{2\pi} \int_{c_F}^{\infty} d\epsilon\,\epsilon \int_{-\pi}^\pi d\alpha\, \rho_F(\alpha) \left( \log \left(2 \cosh \frac{\hat{\epsilon}+ \hat{\mu} +\i\alpha  }{2} \right) + \log \left(2 \cosh \frac{\hat{\epsilon}  - \hat{\mu} - \i\alpha}{2} \right) \right)\, ,\nonumber\\
&=-\frac{N_F \mc{V}_2}{2\pi} \left[- \frac{\beta c_F^3}{3} + \int_{c_F}^{\infty} d\epsilon\,\epsilon \int_{-\pi}^\pi d\alpha\, \rho_F(\alpha) \left( \log \Big( 1 +  e^{-\hat{\epsilon} - \hat{\mu} - \i\alpha} \Big) + \log \Big( 1 +  e^{-\hat{\epsilon} + \hat{\mu} + \i\alpha} \Big) \right)\right]\, .
\end{align}
does have precisely such a singularity\footnote{The second equality in
  both \eqref{bosdet} and \eqref{ferdet} holds in the dimensional
  regulation scheme.}.

In summary, the bosonic determinant \eqref{bosdet} is non-analytic
at $|\mu|=c_B$ if $\rho_B(\alpha)$ is non-vanishing in a neighbourhood
of $\alpha=0$. In a similar way the fermionic determinant
\eqref{ferdet} is potentially non-analytic at $|\mu|=c_F$ if
$\rho_F(\alpha)$ is non-vanishing in a neighbourhood of
$\alpha=\pi$. In the rest of this subsection we will explore the
nature of these non-analyticities and explain how we can analytically
continue around them.

\subsubsection{Analytic continuation in bosonic upper cap phases and
  fermionic lower gap phases} \label{ofebugp}

In this subsection, we consider phases of the fermionic theory
\cite{Jain:2013py} in which the eigenvalue distribution
$\rho_F(\alpha)$ vanishes in an interval around
$\alpha=\pi$. Correspondingly, the dual bosonic eigenvalue
distribution $\rho_B(\alpha)$ saturates the upper cutoff
$1 / 2\pi|\lambda_B|$ in an interval around $\alpha=0$. Concretely, we
focus on phases in which
\begin{equation}\label{fhgo}
  \rho_F(\alpha)= 0\quad{\rm when }\quad |\alpha| > \pi-b\ ,\quad\text{with}\quad b > 0\ .
\end{equation} 
From the duality map relating $\rho_B$ and $\rho_F$
\eqref{holonomymap}, it follows that the dual eigenvalue distribution
obeys
\begin{equation}\label{fhgob}
  \rho_B(\alpha)= \frac{1}{2 \pi |\lambda_B|}\quad{\rm when }\quad |\alpha| < b\ .
\end{equation} 
When the large $N$ saddle-point eigenvalue distributions $\rho_B$ and
$\rho_F$ satisfy \eqref{fhgo} and \eqref{fhgob}, the fermionic theory
is said to be in a lower gap phase while the bosonic theory is said to
be in an upper cap phase \cite{Jain:2013py}. We return to the study of
more general phases in the next subsection.

\subsubsection*{Fermions in the lower gap phase}
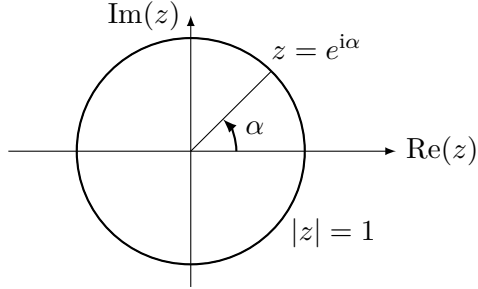
\begin{figure}[!h]
	\begin{center}
		\begin{tikzpicture}[scale=0.6]
		\coordinate (L) at (-4,0) ; 
		\coordinate (R) at (4.5,0) ;
		\coordinate (B) at (0,-3) ;
		\coordinate (U) at (0,3) ;
		\coordinate (P) at (1.2,0) ;
		\draw[dashed] (0,0) circle (2.5cm);
		\draw[dashed] (5:2.5cm) arc(5:355:2.5cm) ;
		\draw[-latex] (L) -- (R) node[right] {$\mathrm{Re}(z)$};
		\draw[-latex] (B) -- (U) node[left] {$\mathrm{Im}(z)$};
		\draw[thick] (0:2.5cm) arc(0:360:2.5cm) ;
                \draw (40:3.6cm) node {$z=e^{\i\alpha}$};
		\draw (-30:3.6cm) node {$|z|=1$};
                \draw[thick,-latex] (1cm,0) arc(0:45:1cm) ;
		\draw (0,0) -- (45:2.5cm) ; 
		\draw (20:1.5cm) node {$\alpha$};
		\end{tikzpicture} 
	\end{center} 
	\caption{The contour of integration in the $z$-plane for the
          integral \eqref{mudep}. The angle $\alpha$ is related to $z$
          as $z = e^{\i\alpha}$.}
	\label{alphavsz}
\end{figure}

\begin{figure}[!h]
	\begin{minipage}{.3\textwidth}
		\begin{tikzpicture}[scale=0.6]
		\coordinate (L) at (-5,0) ; 
		\coordinate (R) at (4,0) ;
		\coordinate (B) at (0,-3) ;
		\coordinate (U) at (0,3) ;
		\coordinate (P) at (-3,0) ;
		\draw[dashed] (0,0) circle (2.5cm);
		\draw[-latex] (L) -- (R) node[right] {$\mathrm{Re}(z)$};
		\draw[-latex] (B) -- (U) node[above] {$\mathrm{Im}(z)$};
		\draw[violet, thick] (120:2.5cm) arc(120:-120:2.5cm) ;
		\filldraw[black] (P) circle (2pt) ;
		\draw[green, very thick] (P) -- (L) ; 
		\draw (40:3.7cm) node {$\hat{\epsilon}>|\hat{\mu}|$};
		\draw[thick,-latex] (0.7cm,0) arc(0:120:0.7cm) ;
		\draw (0,0) -- (120:2.5cm) ; 
		\draw (45:1.35cm) node {$\pi-b$};
		\draw[thick,-latex] (0.8cm,0) arc(0:-120:0.8cm) ;
		\draw (0,0) -- (-120:2.5cm) ; 
		\draw (-45:1.35cm) node {$-\pi+b$};
		\draw (0,-3.5) node {(a)} ;
		\end{tikzpicture} 
	\end{minipage}
	\hspace{4.5cm}
	\begin{minipage}{.2\textwidth}
		\begin{tikzpicture}[scale=0.6]
		\coordinate (L) at (-4,0) ; 
		\coordinate (R) at (4,0) ;
		\coordinate (B) at (0,-3) ;
		\coordinate (U) at (0,3) ;
		\coordinate (P) at (-1.2,0) ;
		\draw[dashed] (0,0) circle (2.5cm);
		\draw[-latex] (L) -- (R) node[right] {$\mathrm{Re}(z)$};
		\draw[-latex] (B) -- (U) node[above] {$\mathrm{Im}(z)$};
		\draw[violet, thick] (120:2.5cm) arc(120:-120:2.5cm) ;
		\filldraw[black] (P) circle (2pt) ;
		\draw[green, very thick] (P) -- (L) ; 
		\draw (40:3.7cm) node {$\hat{\epsilon}<|\hat{\mu}|$};
		\draw (0,-3.5) node {(b)} ;
		\end{tikzpicture} 
	\end{minipage}
	\caption{Fermions in a lower gap phase: Branch point is at
          $z=-e^{\hat{\epsilon}-|\hat{\mu}|}$ and is denoted by the
          black dot on the negative real axis. The green line is the
          branch cut for the logarithm appearing in the fermionic free
          energies. The dashed curve is the unit circle and the purple
          curve is the counterclockwise contour. }
	\label{flg}
\end{figure}
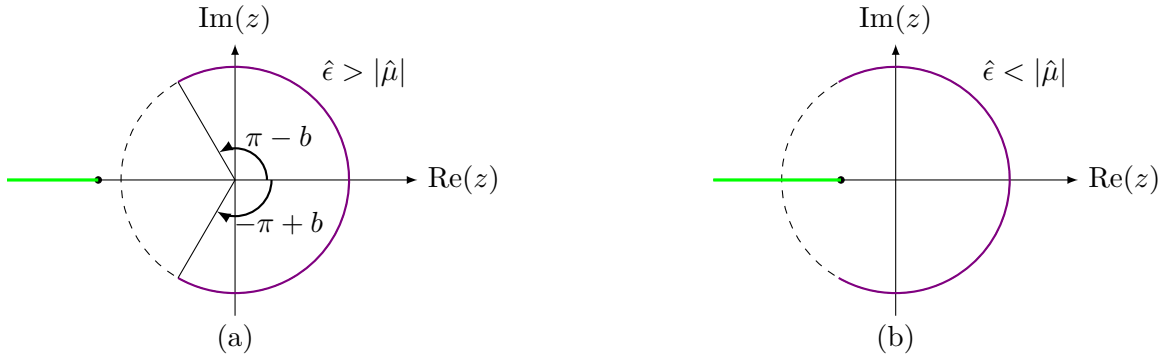

Let us first examine the fermionic determinant \eqref{ferdet}. The
term that depends on $\mu$ is the integral
\begin{equation}\label{mudep}
  \int_{{c}_F}^{\infty}  d{\epsilon}\ {\epsilon} \int_{-\pi}^{\pi} d\alpha \ \rho_F(\alpha)\   \Big(\log\big(1+e^{-\hat{\e}-\hat{\mu}-\i\alpha }\big)+\log\big(1+e^{-\hat{\e}+ \hat{\mu}+\i\alpha }\big)  \Big)  \ .
\end{equation} 
We recast the integral over $\alpha$ into a contour integral over the
unit circle in the complex $z$-plane with $z = e^{\i\alpha}$, see
Figure \ref{alphavsz}.
Let us study the branch cut structure of the logarithms
\eqref{mudep}. When ${c}_F > |{\mu}|$, the branch cut that appears in
both logarithms does not intersect the unit circle in the $z$-plane
(see Figure \ref{flg}(a)). As a consequence, the expression
\eqref{mudep} is analytic in ${\mu}$ for $|{\mu}|< {c}_F$. On the
other hand, when $|{\mu}|> {c}_F$, the unit circle intersects the
branch cut of the second logarithm in \eqref{mudep} when ${\epsilon}$
lies in the range ${c}_F< {\epsilon}< |{\mu}|$. For a general
distribution $\rho_F(\alpha)$, this intersection implies a breakdown
of the analyticity of one of the logarithms in \eqref{mudep} as a
function of ${\mu}$ at ${\mu}=c_F$ or at $\mu = -c_F$, depending on
the sign of $\mu$ (recall that $c_F$ is a positive quantity by
definition).

However, when the fermion is in the lower gap phase \eqref{fhgo} the
integration contour is not the full unit circle but the open arc
corresponding to
\begin{equation}\label{alpharange}
 -(\pi - b) < \alpha < \pi -b\ .
 \end{equation}
The branch cut of the logarithms in \eqref{mudep} lies at $\alpha=\pi$
and so does not intersect the open contour specified by
\eqref{alpharange} (see Figure \ref{flg}(b)). It follows that when
$\rho_F(\alpha)$ obeys \eqref{fhgo}, the expression \eqref{mudep} is
analytic as a function of ${\mu}$ at all values of ${\mu}$ including
${|\mu|} \geq c_F$. Recall that \eqref{mudep} is the only term in the
fermionic free energies that depends on ${\mu}$. Now, since this term
is analytic for all ${\mu}$, we conclude that whenever the fermion is
in a lower gap phase \eqref{fhgo}, the formulae for the fermionic
off-shell free energies are fully analytic in $\mu$ and apply without
modification for all values of ${\mu}$ including the range
$|{\mu}|> {c}_F$.

\subsubsection*{Bosons in an upper cap phase}

\begin{figure}[!h]
	\begin{minipage}{.3\textwidth}
		\begin{tikzpicture}[scale=0.7]
		\coordinate (L) at (-3,0) ; 
		\coordinate (R) at (4.5,0) ;
		\coordinate (B) at (0,-3) ;
		\coordinate (U) at (0,3) ;
		\coordinate (P) at (3,0) ;
		\draw[dashed] (0,0) circle (2.5cm);
		\draw[-latex] (L) -- (R) node[right] {$\mathrm{Re}(z)$};
		\draw[-latex] (B) -- (U) node[above] {$\mathrm{Im}(z)$};
		\draw[orange, thick] (60:2.5cm) arc(60:-60:2.5cm) ;
		\filldraw[black] (P) circle (2pt) ;
		\draw[red, very thick] (P) -- (R) ; 
		\draw (40:3.7cm) node {$\hat{\epsilon}>|\hat{\mu}|$};
		\draw[thick,-latex] (0.7cm,0) arc(0:60:0.7cm) ;
		\draw (0,0) -- (60:2.5cm) ; 
		\draw (30:1.2cm) node {$b$};
		\draw[thick,-latex] (0.8cm,0) arc(0:-60:0.8cm) ;
		\draw (0,0) -- (-60:2.5cm) ; 
		\draw (-30:1.35cm) node {$-b$};
		\draw (0,-3.5) node {(a)} ;
		\end{tikzpicture} 
	\end{minipage}
	\hspace{4.0cm}
	\begin{minipage}{.2\textwidth}
		\begin{tikzpicture}[scale=0.7]
		\coordinate (L) at (-3,0) ; 
		\coordinate (R) at (4.5,0) ;
		\coordinate (B) at (0,-3) ;
		\coordinate (U) at (0,3) ;
		\coordinate (P) at (1.2,0) ;
		\draw[dashed] (5:2.5cm) arc(5:355:2.5cm) ;
		\draw[-latex] (L) -- (R) node[right] {$\mathrm{Re}(z)$};
		\draw[-latex] (B) -- (U) node[above] {$\mathrm{Im}(z)$};
		\draw[orange, thick] (60:2.5cm) arc(60:3:2.5cm) ;
		\filldraw[black] (P) circle (2pt) ;
		\draw[red, very thick] (P) -- (R) ; 
		\draw (40:3.7cm) node {$\hat{\epsilon}<|\hat{\mu}|$};
		\draw[rounded corners, orange, thick] (3:2.5cm) -- (7.1:1.1cm) ;
		\draw[orange, thick] (7.1:1.1cm) to[bend right] (-7.1:1.1cm) ;
		\draw[orange, thick] (-7.1:1.1cm) -- (-3:2.5cm) ;
		\draw[orange, thick] (-3:2.5cm) arc(-3:-60:2.5cm) ;
		\draw (0,-3.5) node {(b)} ;
		\end{tikzpicture} 
	\end{minipage}
	\caption{Bosons in the upper cap phase: Branch point is at
          $z=+e^{\hat{\epsilon}-|\hat{\mu}|}$ and is denoted by the
          black dot on the positive real axis. Dashed curve is the
          unit circle. Orange curve is the contour (counterclockwise)
          over which the holonomy is integrated over in this
          phase. Red line is the branch cut for the logarithm
          appearing in the bosonic free energies.}
	\label{bug}
\end{figure}
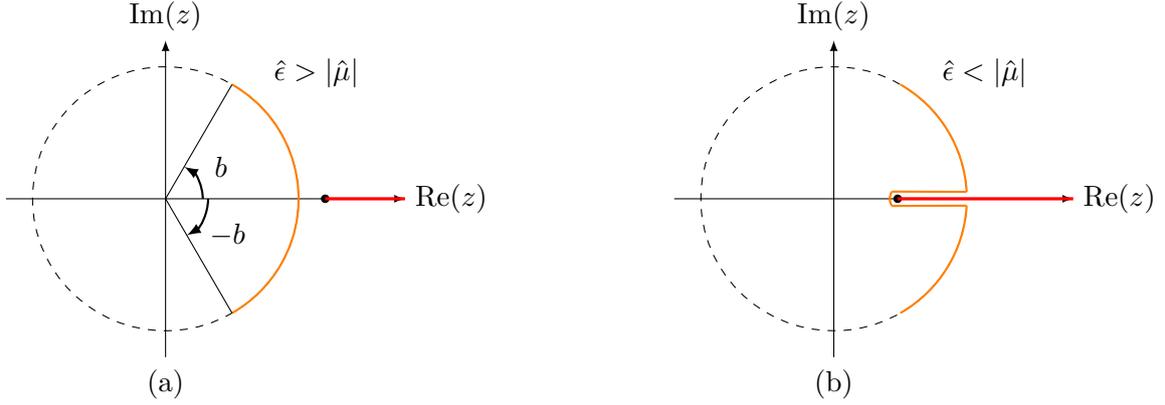

Let us now consider the bosonic determinant \eqref{bosdet}. The term
that depends on ${\mu}$ is
\begin{equation}\label{mudepb}
  \int_{{c}_B}^{\infty} {\epsilon} \ d{\epsilon}  \int_{-\pi}^{\pi} d\alpha \ \rho_B(\alpha)\  \Big(\log\big(1-e^{-\hat{\e}-\hat{\mu}-\i\alpha }\big)+\log\big(1-e^{-\hat{\e}+ \hat{\mu}+\i\alpha }\big)  \Big)\ .
\end{equation}
When the boson is in an upper cap phase \eqref{fhgob}, the integral
over $\alpha$ in \eqref{mudepb} can be split up into the integral
over the range $\alpha \in (-b, b)$ on the unit circle and the
integral over the rest of the unit circle. The integral over the rest
of the unit circle is manifestly analytic in ${\mu}$ for all values of
${\mu}$. The part of \eqref{mudepb} that is potentially non-analytic
in $\mu$ is
\begin{equation}\label{mudepbna}
  \frac{1}{2 \pi |\lambda_B|} \int_{{c}_B}^{|{\mu}| } {\epsilon} \ d{\epsilon} \int_{-b}^{b}  d\alpha \ \log\big(1-e^{-\hat{\e}+ |\hat{\mu}|+\i\alpha }\big)\ .  
\end{equation}
The potentially non-analytic part is one of the logarithms of
\eqref{mudepb} depending on the sign of $\mu$. In the above expression
\eqref{mudepbna}, we treat the two cases together by noticing that the
non-analyticity in either case depends only on the absolute value
$|\mu|$. As written, \eqref{mudepbna} is indeed non-analytic as a
function of ${\mu}$ at $|{\mu}|={\epsilon}$. However, it is easy to
find a modification of \eqref{mudepbna} that agrees with
\eqref{mudepbna} for $|{\mu}|< {\epsilon}$ but is analytic for all
$\mu$. We first note that \eqref{mudepbna} can be rewritten as an open
contour integral
\begin{equation}\label{mudepbnaz}
  \frac{1}{2 \pi |\lambda_B|} \int_{{c}_B}^{|{\mu}| } {\epsilon} \ d{\epsilon} \int_C \frac{dz}{\i z} \ \log\big(1-z\, e^{-\hat{\e}+ |\hat{\mu}| }\big)  \ ,
\end{equation}
where the contour $C$ starts at $e^{-\i b}$ and runs counterclockwise
along the unit circle to $e^{\i b}$ (see Figure \ref{bug}). The
modification of \eqref{mudepbnaz} that makes it analytic everywhere
replaces the contour $C$ with a contour $C'$ that begins and ends at
the same points, but no longer runs along the unit circle all the
way. It lies on the unit circle except in the neighbourhood of the
real axis where it makes a hairpin bend around the branch cut so that
it cuts the positive real axis at a point
$x_0 < e^{\hat{\epsilon} -|\hat{\mu}|}$ (see Figure
\ref{bug}(b)). This condition on the contour $C'$ is chosen to ensure
that this contour never passes through the branch cut of the logarithm
in \eqref{mudepbnaz} . With this choice of contour, \eqref{mudepbnaz}
defines a function that manifestly agrees with \eqref{mudepbna} when
$|{\mu}|<{c}_B$ and is analytic everywhere.

Note that when $z$ is slightly below (resp.~above) the branch cut, the
argument of the logarithm $1-z\,e^{-{\hat \epsilon} + |{\hat \mu}|}$,
has a small positive (resp.~negative) imaginary part. It follows that
the integral in \eqref{mudepbnaz} evaluates to
\begin{equation}\label{intoo}
  \frac{1}{2 \pi |\lambda_B|} \int_{{c}_B}^{|{\mu}| }  d{\epsilon}\ \epsilon  \int_{-b}^{b}  d\alpha \ \log\big(1-e^{-\hat{\e} + |\hat{\mu}| + \i \alpha }\big)   \ -\ \frac{1}{|\lambda_B|} \Theta(|{ \mu}|-{ c}_B ) \int_{{c}_B}^{|{\mu}| } \ d{\epsilon}\, \epsilon \int_{e^{{\hat \epsilon} - |{\hat \mu}|}}^1 \frac{dx}{x}\ .
\end{equation}
The first term in \eqref{intoo} is the integral over the arc on the
unit circle; the second term is the discontinuity across the branch
cut of the logarithm. Performing the integral over $x$ and then the
integral over ${ \epsilon}$ in the second term, we find that
\eqref{intoo} reduces to
\begin{equation}\label{intoofinal}
\frac{1}{2 \pi |\lambda_B|} \int_{{c}_B}^{|\mu| } d{\epsilon}\,{\epsilon}  \int_{-b}^{b}  d\alpha \ \log\big(1-e^{-\hat{\e}+ \hat{\mu}+\i \alpha }\big)  -\beta \Theta(|{ \mu}|-{ c}_B ) \ \frac{(|{ \mu}| -{ c}_B)^2(|{ \mu}| + 2{ c}_B)}{6|\lambda_B|}  \ .
\end{equation}
It follows from this analysis that whenever the boson is in the upper
cap phase (i.e.~when \eqref{fhgob} holds), the analytic continuation
of the log-determinant \eqref{bosdet} is
\begin{align}\label{bosdetuc}
  &\beta \mc{V}_2\, F_{B,\text{det}} (c_B, \mu)\nonumber\\
  &= \frac{N_B \mc{V}_2}{2\pi} \bigg[ - \frac{\beta c_B^3}{3} + \int_{c_B}^{\infty} d\epsilon\,\epsilon \int_{-\pi}^\pi d\alpha\, \rho_B(\alpha) \left( \log \Big( 1 -  e^{-\hat{\epsilon} - \hat{\mu} - \i\alpha} \Big) + \log \Big( 1 -  e^{-\hat{\epsilon} + \hat{\mu} + \i\alpha} \Big) \right)\nonumber\\
  &\hspace{6cm}  -\beta \Theta(|{ \mu}|-{ c}_B ) \ \frac{(|{ \mu}| -{ c}_B)^2(|{ \mu}| + 2{ c}_B)}{6|\lambda_B|}   \bigg]\ .
\end{align}
The log-determinants in the last line of the critical boson free
energy \eqref{CB2voffshellfe} and the regular boson free energy
\eqref{RBoffshellfe} have to be modified to include the additional
term in \eqref{bosdetuc} when $c_B < |\mu|$.  We conjecture that the
above procedure gives the correct expressions for the off-shell free
energies of the critical boson and regular boson theories for all
values of ${\hat \mu}$ in upper cap bosonic phases.
\begin{align}\label{CB2voffshellfecor}
  &F_{\rm CB}({c}_B, \tilde{\mc{S}})\nonumber\\
  &=\frac{N_B}{6\pi \b^3} \bigg[\frac{3}{2} \hat{c}_B^2\hat{m}_B^{\text{cri}} - 4\lambda_B^2\left(\tilde{\mc{S}}-\tfrac{1}{2}\hat{m}_{B}^{\text{cri}} \right)^3  +6|\lambda_B| \hat{c}_B\left(\tilde{\mc{S}}-\tfrac{1}{2}\hat{m}_{B}^{\text{cri}}\right)^2\nonumber \\
&\qquad\qquad -\hat{c}_B^3 +
3\int_{\hat{c}_B}^{\infty} d\hat{\e}\ \hat{\e} \int_{-\pi}^{\pi} d\alpha \ \rho_B(\alpha) \ \big(\log(1-e^{-\hat{\e}+\hat{\mu}+\i \a})+\log(1-e^{-\hat{\e}-\hat{\mu}-\i \alpha})\big)\nonumber\\
&\qquad\qquad - \Theta(|{ \mu}|-{ c}_B ) \frac{(|\hat{\mu}| - {\hat c}_B)^2(|\hat{\mu}| + 2 {\hat c}_B)}{2 |\lambda_B|} \bigg]\ .
\end{align}
\begin{align}\label{RBoffshellfecor}
  &F_{\rm RB}({c}_B,{\sigma}_B,\tilde\cS)\nonumber\\
  & =\frac{N_B}{6\pi \b^3} \bigg[-3\hat{c}_B^2\hat{\sigma}_B+\lambda_B^2\hat{\sigma}_B^3 -4\lambda_B^2(\tilde{\cS}+\hat{\sigma}_{B})^3 +6|\lambda_B| \hat{c}_B(\tilde{\cS}+\hat{\sigma}_B)^2 \nonumber \\
  &\qquad\qquad +3(\hat{m}_B^2\hat{\sigma}_B+2\lambda_B\hat{b}_4\hat{\sigma}_B^2+(x_6+1) \lambda_B^2 \hat{\sigma}_B^3)\nonumber \\
  &\qquad\qquad  -\hat{c}_B^3 +
    3\int_{\hat{c}_B}^{\infty}d\hat{\e}\  \hat{\e}  \int_{-\pi}^{\pi} d\alpha \ \rho_B(\alpha) \ \big(\log(1-e^{-\hat{\e}+\hat{\mu}+\i \a})+\log(1-e^{-\hat{\e}-\hat{\mu}-\i \alpha})\big)\nonumber\\
  &\qquad\qquad - \Theta(|{ \mu}|-{ c}_B ) \frac{(|\hat{\mu}| - {\hat c}_B)^2(|\hat{\mu}| + 2 {\hat c}_B)}{2 |\lambda_B|}\bigg]\ .
\end{align}
Note that the new term in \eqref{bosdetuc} is proportional to
$\frac{1}{\lambda_B}$ and hence is singular in the weak coupling limit
$\lambda_B \to 0$. As we will see below, this term will play a key
role in curing the runaway singularity of Bose condensation observed
in the free bosonic theories.

\subsubsection{Analytic continuation and off-shell free energies in general phases} \label{ofep}

Though the formulae of Section \ref{ofebugp} are all we will need
later in this paper, in this subsection we take the opportunity to
present the (conjecturally) correct expressions for the off-shell free
energy for both the bosonic and the fermionic theories with
$|{ \mu}|>{ c}_B$ and $|{ \mu}|>{ c}_F$ in more general phases than
the bosonic upper cap phases and dual fermionic theory lower gap
phases dealt with in the previous subsubsection.

When the boson is in an upper cap phase, the eigenvalue distribution
$\rho_B$ was constant in a finite interval around $\alpha = 0$ (see
equation \eqref{fhgob}; its analytic continuation to the interior of
the unit circle where the hairpin part of the contour was present (see
Figure \ref{bug}(b)) was trivially the same constant. However, when
the bosonic theory is not in a upper cap phase, this analytic
continuation has to be done more carefully.  For a general eigenvalue
distribution $\rho_B$, it is still true \cite{Jain:2013py} that there
is a sufficiently small interval, say $|\alpha| < c$, in which
$\rho_B(\alpha)$ is an analytic function.  Similarly, the dual
fermionic eigenvalue distribution $\rho_F$ is also analytic in the
interval $|\alpha| \geq \pi -c$. Within these intervals, these
eigenvalue distributions may be expanded into Fourier series as
\begin{align}\label{exprho} 
  \rho_B(\alpha)
  &= a_0 + \sum_{m=1}^\infty  a_m \left( e^{\i m \alpha} + e^{-\i m \alpha} \right)\ ,\quad |\alpha| < c\ , \nonumber\\
  \rho_F(\alpha)
  &= b_0 +\sum_{m=1}^\infty  b_m \left( e^{\i m \alpha} + e^{-\i m \alpha} \right)\ ,\quad |\alpha| > \pi - c\ .
\end{align}
We emphasize that the right hand sides of \eqref{exprho} - which are
well-defined and analytic everywhere on the unit circle - are
guaranteed to agree with the actual eigenvalue distributions only in
the restricted intervals described above, and not necessarily on the
whole unit circle.

We expect that the coefficients $a_n$ and $b_n$ in \eqref{exprho}
decay with $n$ at least as fast as $\varepsilon^n$ for some number
$\varepsilon < 1$ since the series are convergent on the unit circle
$z = e^{\i\alpha}$. It is then easy to see that the right hand sides
can be analytically continued at least in the annulus
\begin{equation}\label{annulus}
  \varepsilon < |z| < \frac{1}{\varepsilon}\ ,
\end{equation}    
with the simple replacement $e^{\i\alpha} \to z$. That is, consider
the functions ${\tilde \rho}_B(z)$ and ${\tilde \rho}_F(z)$
\emph{defined} by
\begin{align}\label{exprhocomp}
  {\tilde \rho}_B(z)&= a_0+\sum_m a_m \left( z^m+ \frac{1}{z^m}\right)\ ,\quad 
  {\tilde \rho}_F(z) = b_0 +\sum_m b_m \left( z^m+ \frac{1}{z^m}\right)\ .
\end{align}
These functions clearly agree with $\rho_B(\alpha)$ and
$\rho_F(\alpha)$ in the restricted intervals shown in
\eqref{exprho}. Hence, ${\tl \rho}_B(z)$ and ${\tl \rho}_F(z)$ are the
correct analytic continuations of the eigenvalue distributions
$\rho_B$ and $\rho_F$, at least in the strips
\begin{equation}
  \text{Bosons}:\  \varepsilon < |z| < \frac{1}{\varepsilon}\ ,\ |\text{Arg}(z)| < c\ ,\quad  \text{Fermions}:\  \varepsilon < |z| < \frac{1}{\varepsilon}\ ,\ |\text{Arg}(z)| < \pi - c\ .
\end{equation}
Of course, using the Bose-Fermi duality map of the holonomy
distributions \eqref{holonomymap} we can relate the $a_n$ and $b_n$
coefficients in \eqref{exprho} 
which can then be plugged back into the analytic continuations
${\tl \rho}_B$ and ${\tl \rho}_F$ to get the duality relation
\begin{equation}\label{tilderel}
  |\lambda_B| \tl{\rho}_B(z) + |\lambda_F| \tl{\rho}_F(-z) =\frac{1}{2\pi} \ .
\end{equation}
With all these definitions in place, we can now imitate the analysis
of the previous subsection to get expressions for the log-determinants
that are analytic for all $\mu$ by deforming the contours of
integration to go around the branch cuts as shown in Figures
\ref{bnug} and \ref{fnlg}.
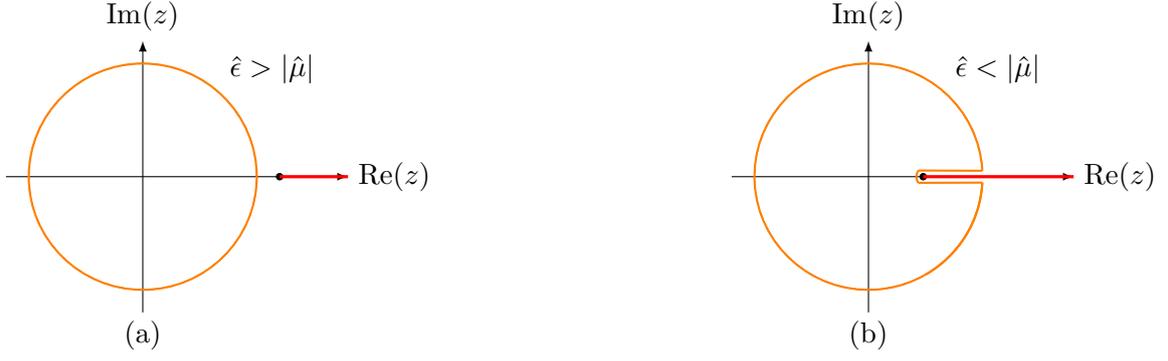
\begin{figure}[!h]
	\begin{minipage}{.35\textwidth}
		\begin{tikzpicture}[scale=0.6]
		\coordinate (L) at (-3,0) ; 
		\coordinate (R) at (4.5,0) ;
		\coordinate (B) at (0,-3) ;
		\coordinate (U) at (0,3) ;
		\coordinate (P) at (3,0) ;
		\draw[dashed] (0,0) circle (2.5cm);
		\draw[-latex] (L) -- (R) node[right] {$\mathrm{Re}(z)$};
		\draw[-latex] (B) -- (U) node[above] {$\mathrm{Im}(z)$};
		\draw[orange, thick] (0:2.5cm) arc(0:360:2.5cm) ;
		\filldraw[black] (P) circle (2pt) ;
		\draw[red, very thick] (P) -- (R) ; 
		\draw (40:3.7cm) node {$\hat{\epsilon}>|\hat{\mu}|$};
		\draw (0,-3.5) node {(a)} ;
		\end{tikzpicture} 
	\end{minipage}
	\hspace{4.0cm}
	\begin{minipage}{.2\textwidth}
		\begin{tikzpicture}[scale=0.6]
		\coordinate (L) at (-3,0) ; 
		\coordinate (R) at (4.5,0) ;
		\coordinate (B) at (0,-3) ;
		\coordinate (U) at (0,3) ;
		\coordinate (P) at (1.2,0) ;
		\draw[dashed] (5:2.5cm) arc(5:355:2.5cm) ;
		\draw[-latex] (L) -- (R) node[right] {$\mathrm{Re}(z)$};
		\draw[-latex] (B) -- (U) node[above] {$\mathrm{Im}(z)$};
		\draw[orange, thick] (3:2.5cm) arc(3:357:2.5cm) ;
		\filldraw[black] (P) circle (2pt) ;
		\draw[red, very thick] (P) -- (R) ; 
		\draw (40:3.7cm) node {$\hat{\epsilon}<|\hat{\mu}|$};
		\draw[rounded corners, orange, thick] (3:2.5cm) -- (7.1:1.1cm) ;
		\draw[orange, thick] (7.1:1.1cm) to[bend right] (-7.1:1.1cm) ;
		\draw[orange, thick] (-7.1:1.1cm) -- (-3:2.5cm) ;
		\draw[orange, thick] (-3:2.5cm) arc(-3:-60:2.5cm) ;
		\draw (0,-3.5) node {(b)} ;
		\end{tikzpicture} 
	\end{minipage}
	\caption{Bosons in non-upper cap phases: The branch point is
          at $z=+e^{\hat{\epsilon}-|\hat{\mu}|}$ and is denoted by the
          black dot on the +ve real axis. The orange curve is the
          contour (counterclockwise) over which the holonomy is
          integrated over in this phase. The red line is the branch
          cut for the logarithm appearing in the bosonic free
          energies.}
	\label{bnug}
\end{figure}
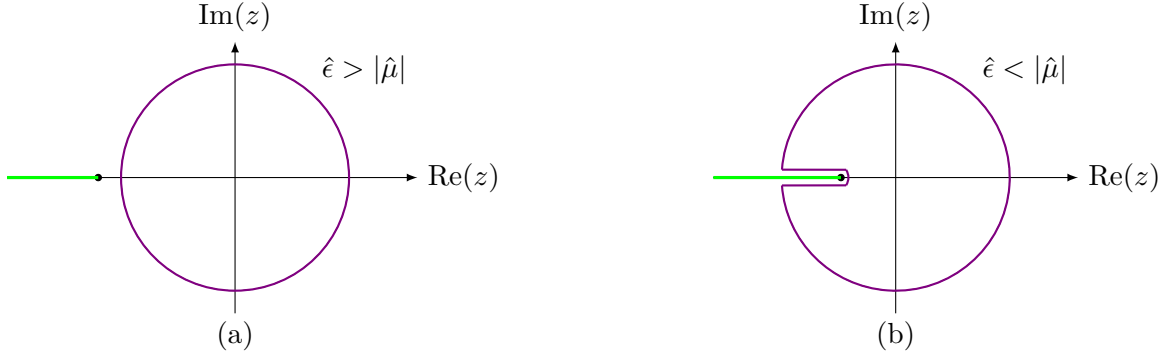
\begin{figure}[!h]
	\begin{minipage}{.3\textwidth}
		\begin{tikzpicture}[scale=0.6]
		\coordinate (L) at (-5,0) ; 
		\coordinate (R) at (4,0) ;
		\coordinate (B) at (0,-3) ;
		\coordinate (U) at (0,3) ;
		\coordinate (P) at (-3,0) ;
		\draw[dashed] (0,0) circle (2.5cm);
		\draw[-latex] (L) -- (R) node[right] {$\mathrm{Re}(z)$};
		\draw[-latex] (B) -- (U) node[above] {$\mathrm{Im}(z)$};
		\draw[violet, thick] (180:2.5cm) arc(180:-180:2.5cm) ;
		\filldraw[black] (P) circle (2pt) ;
		\draw[green, very thick] (P) -- (L) ; 
		\draw (40:3.7cm) node {$\hat{\epsilon}>|\hat{\mu}|$};
		\draw (0,-3.5) node {(a)} ;
		\end{tikzpicture} 
	\end{minipage}
	\hspace{4.5cm}
	\begin{minipage}{.2\textwidth}
		\begin{tikzpicture}[scale=0.6]
		\coordinate (L) at (-4,0) ; 
		\coordinate (R) at (4,0) ;
		\coordinate (B) at (0,-3) ;
		\coordinate (U) at (0,3) ;
		\coordinate (P) at (-1.2,0) ;
		\draw[dashed] (170:2.5cm) arc(170:-170:2.5cm) ;
		\draw[-latex] (L) -- (R) node[right] {$\mathrm{Re}(z)$};
		\draw[-latex] (B) -- (U) node[above] {$\mathrm{Im}(z)$};
		\draw[violet, thick] (176:2.5cm) arc(176:-176:2.5cm) ;
		\draw[rounded corners, violet, thick] (176:2.5cm) -- (171:1.1cm) ;
		\draw[violet, thick] (171:1.1cm) to[bend left] (-171:1.1cm) ;
		\draw[violet, thick] (-171:1.1cm) -- (-176:2.5cm) ;
		\filldraw[black] (P) circle (2pt) ;
		\draw[green, very thick] (P) -- (L) ; 
		\draw (40:3.7cm) node {$\hat{\epsilon}<|\hat{\mu}|$};
		\draw (0,-3.5) node {(b)} ;
		\end{tikzpicture} 
	\end{minipage}
	\caption{Fermions in non-lower gap phases: The branch point is
          at $z=-e^{\hat{\epsilon}-|\hat{\mu}|}$ and is denoted by the
          black dot on the -ve real axis. The purple curve is the
          contour (counterclockwise) over which the holonomy is
          integrated over in this phase. The green line is the branch
          cut for the logarithm appearing in the fermionic free
          energies.}
	\label{fnlg}
\end{figure}
\begin{align}\label{bosdetgen}
  &\beta \mc{V}_2\, F_{B,\text{det}} (c_B, \mu)\nonumber\\
  &= \frac{N_B \mc{V}_2}{2\pi} \bigg[ - \frac{\beta c_B^3}{3} + \int_{c_B}^{\infty} d\epsilon\,\epsilon \int_{-\pi}^\pi d\alpha\, \rho_B(\alpha) \left( \log \Big( 1 -  e^{-\hat{\epsilon} - |\hat{\mu}| - \i\alpha} \Big) + \log \Big( 1 -  e^{-\hat{\epsilon} + |\hat{\mu}| + \i\alpha} \Big) \right)\nonumber\\
  &\hspace{6cm}- 2 \pi \Theta(|{ \mu}|-{ c}_B ) \int_{{c}_B}^{|{\mu}| } d{\epsilon}\,\epsilon \int_{e^{{\hat \epsilon} -{|\hat \mu|}} }^1 dx \ \frac{{\tilde \rho}_B(x)}{x}   \bigg]\ ,
\end{align}
\begin{align}\label{ferdetgen}
  &\beta \mc{V}_2\, F_{F,\text{det}} (c_F, \mu)\nonumber\\
  &=-\frac{N_F \mc{V}_2}{2\pi} \bigg[- \frac{\beta c_F^3}{3} + \int_{c_F}^{\infty} d\epsilon\,\epsilon \int_{-\pi}^\pi d\alpha\, \rho_F(\alpha) \left( \log \Big( 1 +  e^{-\hat{\epsilon} - |\hat{\mu}| - \i\alpha} \Big) + \log \Big( 1 +  e^{-\hat{\epsilon} + |\hat{\mu}| + \i\alpha} \Big) \right)\nonumber\\
  &\hspace{6cm} - 2 \pi \Theta(|{ \mu}|-{ c}_F ) \int_{{c}_F}^{|{\mu}| } d{\epsilon}\,\epsilon \int_{e^{{\hat \epsilon} -{|\hat \mu|}} }^1 dx \ \frac{{\tilde \rho}_F(-x)}{x} \bigg]\ .
\end{align}
Finally, incorporating the modifications to the log-determinants into
the off-shell free energies given in Section \ref{mulcbfe}, we get
\begin{align}\label{RF2voffshellfecorn}
  &F_{\rm RF}({c}_F,\tilde{\mathcal{C}})\nonumber\\
  & = \frac{N_F}{6\pi\beta^3} \bigg[ - 8\lambda_{F}^2 \tl{\mc{C}}^3 - 3\tl{\mc{C}} \left( \hat{c}_{F}^2 - \big(2\lambda_F \tl{\mc{C}} +\hat{m}_F \big)^2 \right)  - 6\lambda_{F} \hat{m}_F \tl{\mc{C}}^2 \nonumber  \\
  &\qquad\qquad+ \hat{c}_{F}^3 - 3 \int_{\hat{c}_F}^{\infty}  d\hat{\e}\ \hat{\e}  \int_{-\pi}^{\pi}  d\alpha\, \rho_F(\alpha)\left(\log\big(1+e^{-\hat{\e}-\hat{\mu}-\i \alpha }\big)+\log\big(1+e^{-\hat{\e}+ \hat{\mu}+\i \alpha }\big)  \right) \nonumber\\
  &\qquad\qquad+6 \pi \Theta(|{ \mu}|-{ c}_F )  \int_{\hat{c}_F}^{|\hat{\mu}| } d\hat{\epsilon}\ \hat{\epsilon}  \int_{e^{{\hat \epsilon} -{\hat \mu}} }^1 dx \ \frac{{\tilde \rho}_F(-x)}{x} \bigg]   \ .
\end{align}
\begin{align}\label{CFoffshellfecorn}
  &F_{\rm CF}({c}_F,{\zeta}_F,\tilde{\cC})\nonumber\\
  & = \frac{N_F}{6\pi \b^3} \bigg[ -8\lambda_{F}^2 \tl{\mc{C}}^3 - 3\tl{\mc{C}} \bigg( \hat{c}_{F}^2 - \Big(2\lambda_F \tl{\mc{C}} - \frac{4\pi \hat{\zeta}_F}{\kappa_F} \Big)^2 \bigg) + 6\lambda_{F} \tl{\mc{C}}^2 \Big(\frac{4\pi \hat{\zeta}_F}{\kappa_F} \Big) \nonumber  \\ 
  &\qquad\qquad   + 3 \bigg( \frac{\hat{y}_2^2}{2\lambda_F} \frac{4\pi \hat{\zeta}_F}{\kappa_F} - \frac{\hat{y}_4}{2\lambda_F} \Big(\frac{4\pi \hat{\zeta}_F}{\kappa_F} \Big)^2 +\frac{x_6^F}{8\lambda_F} \Big(\frac{4\pi \hat{\zeta}_F}{\kappa_F} \Big)^3 \bigg)\nonumber   \\
  &\qquad\qquad +  \hat{c}_{F}^3 - 3 \int_{\hat{c}_F}^{\infty} d\hat{\e}\ \hat{\e} \int_{-\pi}^{\pi}d\alpha\ \rho_F(\alpha) \left(\log\big(1+e^{-\hat{\e}-\hat{\mu}-\i\alpha }\big)+\log\big(1+e^{-\hat{\e}+ \hat{\mu}+\i\alpha }\big)  \right) \nonumber\\
  &\qquad\qquad+6 \pi \Theta(|{ \mu}|-{ c}_F )  \int_{\hat{c}_F}^{|\hat{\mu}| }  d\hat{\epsilon}\ \hat{\epsilon} \int_{e^{{\hat \epsilon} -{\hat \mu}} }^1 dx \ \frac{{\tilde \rho}_F(-x)}{x}  \bigg]\ .
\end{align}
\begin{align}\label{CB2voffshellfecorn}
  &F_{\rm CB}({c}_B, \tilde{\mc{S}})\nonumber\\
  &=\frac{N_B}{6\pi \b^3} \bigg[\frac{3}{2} \hat{c}_B^2\hat{m}_B^{\text{cri}} - 4\lambda_B^2\left(\tilde{\mc{S}}-\tfrac{1}{2}\hat{m}_{B}^{\text{cri}} \right)^3  +6|\lambda_B| \hat{c}_B\left(\tilde{\mc{S}}-\tfrac{1}{2}\hat{m}_{B}^{\text{cri}}\right)^2\nonumber \\
&\qquad\qquad -\hat{c}_B^3 +
3\int_{\hat{c}_B}^{\infty} d\hat{\e}\ \hat{\e} \int_{-\pi}^{\pi} d\alpha \ \rho_B(\alpha) \ \big(\log(1-e^{-\hat{\e}+\hat{\mu}+\i \a})+\log(1-e^{-\hat{\e}-\hat{\mu}-\i \alpha})\big)\nonumber\\
&\qquad\qquad- 6 \pi \Theta(|{ \mu}|-{ c}_B ) \int_{\hat{c}_B}^{|\hat{\mu}| } d\hat{\epsilon}\ \hat{\epsilon}  \int_{e^{{\hat \epsilon} -{\hat \mu}} }^1 dx \ \frac{{\tilde \rho}_B(x)}{x} \bigg]\ .
\end{align}
\begin{align}\label{RBoffshellfecorn}
  &F_{\rm RB}({c}_B,{\sigma}_B,\tilde\cS)\nonumber\\
  & =\frac{N_B}{6\pi \b^3} \bigg[-3\hat{c}_B^2\hat{\sigma}_B+\lambda_B^2\hat{\sigma}_B^3 -4\lambda_B^2(\tilde{\cS}+\hat{\sigma}_{B})^3 +6|\lambda_B| \hat{c}_B(\tilde{\cS}+\hat{\sigma}_B)^2 \nonumber \\
&\qquad\qquad +3(\hat{m}_B^2\hat{\sigma}_B+2\lambda_B\hat{b}_4\hat{\sigma}_B^2+(x_6+1) \lambda_B^2 \hat{\sigma}_B^3)\nonumber \\
&\qquad\qquad  -\hat{c}_B^3 +
3\int_{\hat{c}_B}^{\infty}d\hat{\e}\  \hat{\e}  \int_{-\pi}^{\pi} d\alpha \ \rho_B(\alpha) \ \big(\log(1-e^{-\hat{\e}+\hat{\mu}+\i \a})+\log(1-e^{-\hat{\e}-\hat{\mu}-\i \alpha})\big)\nonumber\\
&\qquad\qquad- 6 \pi \Theta(|{ \mu}|-{ c}_B ) \int_{\hat{c}_B}^{|\hat{\mu}| } d\hat{\epsilon}\ \hat{\epsilon}  \int_{e^{{\hat \epsilon} -{\hat \mu}} }^1 dx \ \frac{{\tilde \rho}_B(x)}{x}\bigg]\ .
\end{align}
The expressions \eqref{CB2voffshellfecor} and \eqref{RBoffshellfecor}
for the bosonic free energies in an upper cap phase can be obtained by
setting ${\tilde \rho}_B(x)= {1}/{2 \pi |\lambda_B|}$ in the above
bosonic formulae \eqref{CB2voffshellfecorn} and
\eqref{RBoffshellfecorn}. Similarly, upon setting
${\tilde \rho}_F({-} x)=0$, the fermionic free energies
\eqref{RF2voffshellfecorn} and \eqref{CFoffshellfecorn} respectively
reduce to the original expressions \eqref{RF2voffshellfe} and
\eqref{CFoffshellfe} which are the correct free energies for a lower
gap phase. We conjecture that the above expressions
\eqref{RF2voffshellfecorn} - \eqref{RBoffshellfecorn} are the correct
expressions for the off-shell free energies of the bosonic and
fermionic theories for all values of ${\mu}$.

\subsubsection{Gap Equations} 

For chemical potentials less than the gap $c_B$ or $c_F$, the
equations that followed from varying the off-shell free energy were
presented for the regular fermions in \eqref{tcdef}, for the critical
fermions in \eqref{extremizeFCF}, for the critical bosons in
\eqref{ofeom} and for the regular bosons in \eqref{extremizeFRB}. In
general, these equations of motion are modified when the chemical
potential is greater than $c_B$ or $c_F$ since the off-shell free
energies are themselves modified in this case.

The modified equations of motion can be obtained by varying the
modified off-shell free energies presented in the previous
subsection. Alternatively, note that the only explicit occurrences of
$\mu$ in the equations are in the quantities $\cS(c_B, \mu)$ and
$\cC(c_F, \mu)$ which are the same factors which appear in the
log-determinants \eqref{bosdet} and \eqref{ferdet} under the
$\epsilon$-integral. Thus, the analytic continuation procedure for the
equations is identical to the one used for the off-shell free
energies. The final result is that the equations continue to hold for
$|\mu| > c_B$ and $|\mu| > c_F$ provided we make the replacements
\begin{align}\label{repoftildec}
  \mc{C}({c}_F,{\mu})
  &\rightarrow \mc{C}({c}_F,{\mu}) - \pi \Theta(|{ \mu}|-{ c}_F ) \int_{e^{\hat{c}_F-|{\hat \mu}|}}^1 dx \ \frac{{\tilde \rho}_F(- x)}{x}\ ,\nonumber\\
  \mc{S}({c}_B,{\mu})
  &\rightarrow \mc{S}({c}_B,{\mu})
    - \pi \Theta(|{ \mu}|-{ c}_B) \int^1_{e^{\hat{c}_B-|\hat{\mu}|}} dx \ \frac{{\tilde \rho}_B(x)}{x}\ ,
\end{align}
where, recall that $\tl\rho_F(x)$ and $\tl\rho_B(x)$ are the analytic
continuation of the holonomy distributions $\rho_F(\alpha)$ and
$\rho_B(\alpha)$ to a small neighbourhood of the unit circle. In an
upper cap phase for the bosons \eqref{fhgob}, we can use the second
equation in \eqref{repoftildec} after setting
${\tilde \rho}_B(x) = {1}/{2 \pi |\lambda_B|}$,
i.e.~make the replacement
\begin{equation}\label{repoftildecbosug}
\mc{S}({c}_B,{\mu}) \rightarrow \mc{S}({c}_B,{\mu})
- \Theta(|{ \mu}|-{ c}_B) \ \frac{|{\hat \mu}| -\hat{c}_B}{2|\lambda_B|} \ .
\end{equation}
The equation \eqref{repoftildecbosug} was already guessed in
\cite{Choudhury:2018iwf} (see equation A.5 of
\cite{Choudhury:2018iwf}).

\section{Quasiparticle occupation numbers} \label{qon} 

The usual thermodynamical formulae inform us that the charge of an
ensemble is obtained by varying the free energy $\mc{F}$ w.r.t.~the
chemical potential $\mu$ while keeping $\beta$ and the spatial volume
$\mc{V}_2$ fixed. In equations,
\begin{equation}\label{Qtherm}
  Q=-\left(\frac{\partial \mc{F}}{\partial \mu}\right)_\beta\ .
\end{equation}
The quantity $\mc{F}$ that appears in \eqref{Qtherm} is the
thermodynamic free energy related to the partition function $\mc{Z}$
as
\begin{equation}
  \mc{Z} = \exp\left(-\beta \mc{F}\right)\ .
\end{equation}
The free energy $\mc{F}$ is typically a rather complicated
function. As we have discussed in the previous section, in the large
$N$ limit, it is obtained by extremizing the effective action
functional $s[\rho]$ w.r.t.~the holonomy distribution $\rho$. This
action functional consists of two pieces. The first is the
contribution $v[\rho]$ that we discuss in this paper due to
integrating out all modes except the gauge field holonomy $\rho$ and
the second is the contribution due to the Chern-Simons modified path
integral measure (discussed extensively in \cite{Jain:2013py}).
Note, of course, that this measure is independent of $\mu$. 

We have seen in Section \ref{offshellsec}, the final
(usually complicated) answer for $v[\rho]$ can be obtained by
extremizing a rather simple `off-shell' free energy function
$F(\varphi_{\rm aux};\rho]$ with respect to the auxiliary variables
denoted collectively as $\varphi_{\rm aux}$. Thus the full free energy
$\mc{F}$ is obtained by extremizing the effective action functional
$\mc{S}(\varphi_{\rm aux};\rho]$ which consists of the `off-shell'
free energy $F(\varphi_{\rm aux};\rho]$ and the contribution due to
the measure which depends only on $\rho$ but not the other auxiliary
variables (and is also independent of $\mu$).

Let the full set of variables (auxiliary variables as well as
holonomies) be denoted by $\{\varphi_{i}\}$. The on-shell values
$\varphi^*_i$ of the variables $\varphi_i$ are determined by
extremizing the effective action $\mc{S}(\varphi_{\rm aux};\rho]$
w.r.t.~each of the $\varphi_i$:
\begin{equation}\label{actext}
  \frac{\partial \mc{S}}{\partial \varphi_{i}}\bigg|_{\varphi_i = \varphi^*_i} = 0\ .
\end{equation}
The final free energy is obtained by evaluating
$\mc{S}(\varphi_{\rm aux};\rho]$ at these extremum values.

In general, the extremum values of $\varphi_i$ are functions of
$\mu$. Hence, the final free energy $\mc{F}$ has two sources of
dependence on $\mu$: (a) the explicit dependence on $\mu$ coming from
the off-shell free energy $F$ (b) the implicit dependence coming from
the fact that the extremum values $\varphi^*_i$ are functions of
$\mu$. In other words,
\begin{equation}
  \mc{F}(\mu) \equiv  \mc{S}({\varphi^*_{i}(\mu)},\mu) \ .
\end{equation}
where $\mc{S}({\varphi^*_{i}(\mu)},\mu)$ is the effective action
evaluated at one of its extrema. It follows that
\begin{equation}\label{chargeint}
  -Q = \left(\frac{\partial \mc{F}(\mu)}{\partial \mu}\right)_\beta =  \frac{\partial \mc{S}}{\partial \mu}\bigg|_{\varphi_i = \varphi^*_i} + \sum_{i} \frac{\partial  \mc{S}}{\partial \varphi_{i}}\bigg|_{\varphi_i = \varphi^*_i} \frac{\partial \varphi^*_{i}}{\partial \mu}\ .
\end{equation}
Now, we use the equations of motion \eqref{actext} to set the second
term above to zero. Note that the only explicit dependence on $\mu$ in
the effective action $\mc{S}$ is through the off-shell free energy
$F$. Thus, the formula for the charge \eqref{chargeint} simplifies to
\begin{equation}\label{chargeexp}
  Q = -\frac{\partial F}{\partial \mu}\bigg|_{\varphi_i = \varphi^*_i}\ .
\end{equation}
Moreover the quantity $F$ in \eqref{chargeexp} can further be 
replaced by (the corrected form of) $F_{\rm det}$ as this is 
the only part of the off-shell free energy that depends on $\mu$. 

\subsection{Chemical potential smaller than quasiparticle 
thermal mass} \label{cpsqm} 

When the chemical potential is smaller than the thermal masses of the
excitations, the off-shell free energy formulae are given in Section
\ref{mulcbfe}. Recall that the only explicit dependence on ${\mu}$ of these
formulae is in the log-determinants in the last line of these
expressions. Evaluating the derivative w.r.t.~$\mu$ (see
\eqref{chargeexp}) we find that for both the regular
\eqref{RF2voffshellfe} and critical fermionic \eqref{CFoffshellfe}
theories\footnote{We restore the two-dimensional volume factor
  $\mc{V}_2$ and explicit factors of $\beta$ in the free energy
  expressions in this subsection.}
\begin{equation}\label{chargeonshellF}
  \boxed{ Q_{F}(c_F,\mu) = \frac{\mc{V}_2 N_F}{2\pi} \int_{{c}_F}^{\infty} d{\e}\ {\e}  \int_{-\pi}^{\pi}  d\alpha \, \rho_F(\alpha)  \left( \frac{1}{e^{\beta(\e-{\mu})-\i\alpha }+1}  -\frac{1}{e^{\beta({\e}+{\mu})+\i\alpha }+1} \right)\ .}
\end{equation}
Now recall that the number $dn$ of single-particle states of any fixed
colour is given by
\begin{equation}\label{nosps}
  dn = \frac{\mc{V}_2  d^2k}{(2\pi)^2}= \frac{\mc{V}_2 k dk}{2 \pi}
  = \frac{\mc{V}_2 \epsilon d \epsilon}{ 2 \pi}\ .
\end{equation} 
(where we have used the usual relativistic dispersion relation
$\epsilon^2= c_F^2 + \vec{k}^2$). It follows from \eqref{nosps} that
\eqref{chargeonshellF} can be rewritten as
\begin{equation}\label{chargeonshellFn}
  Q_{F}(c_F,\mu) =  N_F \int dn \int_{-\pi}^{\pi} d\alpha\,\rho_F(\alpha)\left( \frac{1}{e^{\beta({\e}-{\mu})-\i\alpha } + 1}  -\frac{1}{e^{\beta({\e} + {\mu}) + \i\alpha } + 1} \right)\ .
\end{equation}
The expression \eqref{chargeonshellFn} has an obvious
interpretation. At least as far as the charge goes, the system under
study consists of a collection of effectively non-interacting
particles where the occupation number of a single-particle state at
energy ${ \epsilon}$ and charge $q$ is given by
$\bar{n}_F({\e},q)$ where
\begin{equation}\label{occuptnF}
  \bar{n}_F({\e},q) \equiv  \int_{-\pi}^{\pi} d\a \ \rho_F(\a) \  \frac{1}{e^{\beta ({\e} - q{\mu}) -\i q \a } + 1}\ .
\end{equation}
where $q=1$ for fundamental excitations and $q=-1$ for antifundamental
excitations. The two terms under the integral in
\eqref{chargeonshellFn} then represent the occupation numbers of the
fundamental quasiparticle (charge $+1$) and antifundamental (charge
$-1$) states respectively.

\textbf{Note:} The formula \eqref{occuptnF} and the interpretation
proposed here, first appeared in the paper
\cite{Geracie:2015drf} in the form
\begin{align}\label{soncharge}
  n_{F}(\vec{p}) & =  \Big( \bar{n}_F({\e}_{\vec{p}},q=+1)  -\bar{n}_F({\e}_{\vec{p}},q=-1)  \Big)\ ,\nonumber\\
  & = \frac{1}{2} \int_{-\pi}^\pi d\alpha \rho_F(\alpha) \Bigg[ \tanh \frac{\beta({\epsilon}_{\vec{p}} +
    {\mu}) + \i \alpha }{2} - \tanh \frac{ \beta({\epsilon}_{\vec{p}}
    - {\mu}) - \i \alpha }{2} \Bigg]\ ,
\end{align}
where the notation $n_F(\vec{p})$ stands for the sum of the charges of
a fundamental quasiparticle state and its antifundamental counterpart
with momentum $\vec{p}$ and energy $\epsilon_{\vec{p}}$. It is easy to
rewrite the formula \eqref{chargeonshellF} for the charge above to
match the one in \cite{Geracie:2015drf} using the trigonometric
identity
\begin{equation}
\frac{1}{e^{x-y}+1} - \frac{1}{e^{x+y}+1} = \frac{1}{2} \Big( \tanh\frac{x+y}{2} - \tanh\frac{x-y}{2} \Big) \ .
\end{equation}

The analysis of the bosonic theories proceeds along completely
identical lines. Differentiating the free energies for the critical
boson \eqref{CB2voffshellfe} and the regular boson
\eqref{RBoffshellfe} w.r.t~$\mu$, we find a formula for the net charge
that admits an interpretation in terms of effectively non-interacting
particle states of energy ${ \epsilon}$ and charge $q$ with average
occupation number given by the formula
\begin{equation}\label{occuptnB}
  \bar{n}_B({\e},q) \equiv  \int_{-\pi}^{\pi} d\a\, \rho_B(\a)\, \frac{1}{e^{\beta({\e} - q {\mu}) - \i q \a } - 1}\ .
\end{equation}
The charge of the bosonic ensemble is then given by
\begin{equation}\label{chargeonshellB}
Q_{B}(c_B,\mu) = N_{B}  \int d n  \int_{-\pi}^{\pi} d\alpha\,\rho_B(\alpha) \left( \frac{1}{e^{\beta({\e}-{\mu})-\i\alpha } - 1} - \frac{1}{e^{\beta({\e}+{\mu}) + \i\alpha} - 1} \right)\ ,
\end{equation}
or, in terms of the occupation number $\bar{n}(\epsilon,q)$,
\begin{equation}\label{BchargenBbar} 
  Q_{B}(c_B,\mu) =  N_{B}  \int dn  \Big( \bar{n}_B({\e},q=+1)  -\bar{n}_B({\e}, q=-1)  \Big)\ .
\end{equation}

\subsection{Chemical potential larger than quasiparticle mass} 
\label{cplqm} 

The generalisation of the fermionic \eqref{occuptnF} and bosonic
\eqref{occuptnB} occupation numbers to $|{\mu}| \geq {c}_F$ and
$|{\mu}| \geq {c}_B$ may be obtained in one of two ways. First, we
could recompute the charge of the ensemble by differentiating the
corrected off-shell free energy formulae \eqref{RF2voffshellfecorn},
\eqref{CFoffshellfecorn} for the fermions and
\eqref{CB2voffshellfecorn}, \eqref{RBoffshellfecorn} for the bosons
w.r.t.~$\mu$ and then reinterpret the result in terms of a modified
occupation number for each single-particle state. Alternatively (and
more simply as well as more informatively), we could simply
analytically continue \eqref{occuptnF} and \eqref{occuptnB} to values
of ${\mu}$ larger than ${c}_F$ and ${c}_B$. We briefly outline this
second method (which proceeds along the lines of Section
\ref{ofep}).

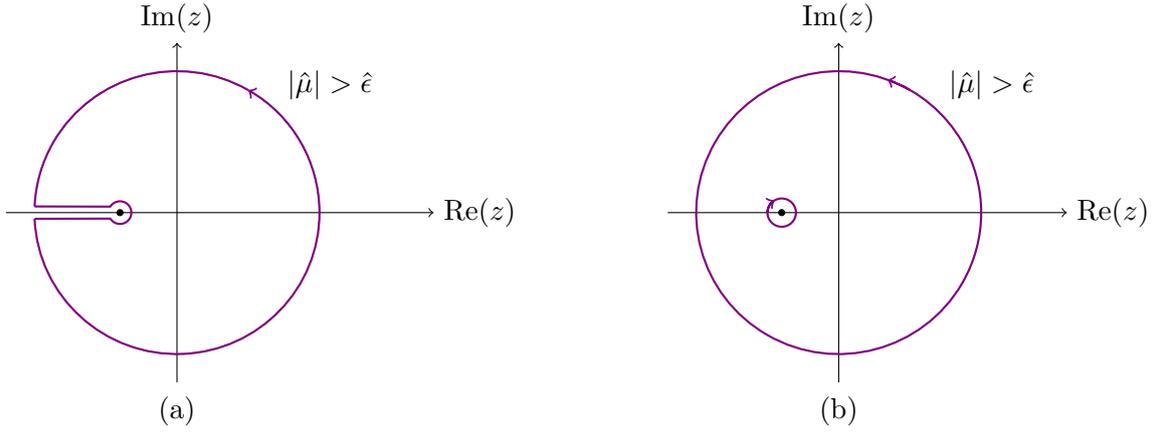
\begin{figure}[!h]
  \begin{minipage}{.3\textwidth}
    \begin{tikzpicture}[scale=0.75]
      \coordinate (L) at (-3,0) ; 
      \coordinate (R) at (4.5,0) ;
      \coordinate (B) at (0,-3) ;
      \coordinate (U) at (0,3) ;
      \coordinate (P) at (-1,0) ;
      \def\R{2.5cm};
      \def\r{0.2cm};
      \draw[dashed] (170:2.5cm) arc(170:-170:2.5cm) ;
      \draw[->] (L) -- (R) node[right] {$\mathrm{Re}(z)$};
      \draw[->] (B) -- (U) node[above] {$\mathrm{Im}(z)$};
      \draw[violet, thick,->] (0:\R) arc(0:60:\R) ;
      \draw[violet, thick] (60:\R) arc(60:180-2.5:\R) ;
      \draw[rounded corners, violet, thick] (180-2.5:\R) -- (180-5.2:1.17cm) ;
      \draw[violet, thick] (180+5.2:1.17cm) -- (180+2.5:\R) ;
      \draw[violet, thick] (180+2.5:\R) arc(180+2.5:360:\R) ;
      \filldraw[black] (P) circle (1.5pt) ;
      \draw[violet, thick] (P) +(150:\r) arc (150:-150:\r);
      \draw (40:3.5cm) node {$|\hat{\mu}|>\hat{\epsilon}$};
      \draw (B)+(0,-0.5) node {(a)};
    \end{tikzpicture}
  \end{minipage}
  \hspace{4.0cm}
  \begin{minipage}{.5\textwidth}
    \begin{tikzpicture}[scale=0.75]
      \coordinate (L) at (-3,0) ; 
      \coordinate (R) at (4,0) ;
      \coordinate (B) at (0,-3) ;
      \coordinate (U) at (0,3) ;
      \coordinate (P) at (-1,0) ;
      \def\R{2.5cm};
      \def\r{0.25cm};
      \draw[->] (L) -- (R) node[right] {$\mathrm{Re}(z)$};
      \draw[->] (B) -- (U) node[above] {$\mathrm{Im}(z)$};
      \draw[violet, thick,->] (60:\R) arc(60:360+70:\R) ;
      \filldraw[black] (P) circle (1.5pt) ;
      \draw[violet, thick,->] (P) +(180:\r) arc(180:-240:\r) ;
      \draw (40:3.5cm) node {$|\hat{\mu}|>\hat{\epsilon}$};
      \draw (B)+(0,-0.5) node {(b)};
    \end{tikzpicture}
  \end{minipage}
  \caption{Fermions: Pole is at $z=-e^{\beta(\epsilon-q\mu)}$ and is
    denoted by the black dot on the -ve real axis. The big circle is
    the unit circle. The purple curve is the contour $C'$
    (counterclockwise) over which the holonomy is integrated over.}
  \label{nFoc}
\end{figure}

The formula \eqref{occuptnF} for the fermionic occupation number is
analytic at all values of $\mu$ such that ${\rm sgn}(q \mu)=-1$. On
the other hand this expression has a potential
non-analyticity\footnote{However, the non-analyticity in the integrand
  of the holonomy integral in the expression of occupation number is a
  pole, whereas the analogous non-analyticity in the case of the free
  energy is a branch cut.} at $q\mu = { \epsilon}$ (recall $q= \pm 1$)
. In order see this non-analyticity, we rewrite the integral over
holonomies in \eqref{occuptnF} as a contour integral over
$z = e^{\i\alpha}$:
\begin{equation}\label{occuptnFn}
  \bar{n}_F({\e},q) \equiv  \oint_{C} \frac{dz}{\i z} \ {\tilde \rho}_F(z)   \frac{1}{ e^{\beta({\e} -  q{\mu})} z^{-1} + 1}\ .
\end{equation}
The contour $C$ in \eqref{occuptnFn} is initially the unit circle, but
with this definition \eqref{occuptnFn} is not analytic in $\mu$ at
$q\mu ={ \epsilon}$.  We obtain the analytic continuation of
\eqref{occuptnFn} by deforming the contour $C$ to the contour $C'$
(see Figure \ref{nFoc}(a)). $C'$ is chosen so that it continues to
circle the origin, but cuts the negative $x$ axis at a point closer to
the origin than $-e^{\beta( \epsilon- q \mu)}$. The formula
\eqref{occuptnFn}, with $C$ replaced by $C'$, is thus the correct
analytic continuation of \eqref{occuptnFn} to values of ${\hat \mu}$
s.t.~$q { \mu} > { \epsilon}$.
The contour $C'$ can be deformed to the sum of two contours without
changing the value of the contour integral: the first contour is the
original contour $C$ while the second is a second smaller loop that
runs clockwise around the pole at $z=-e^{\beta({ \epsilon}- q\mu)}$
(see Figure \ref{nFoc}(b)). The contribution of the small loop
encircling the pole is
 \begin{equation}\label{extrapieceF}
-2 \pi \Theta( q{ \mu}-{ \epsilon}) \ {\tilde \rho}_F(-e^{\beta( \epsilon - q \mu)})\ ,
\end{equation} 
where the Heaviside $\Theta$-function is to indicate that this
additional contribution is present only when $q \mu > \epsilon$. It
follows that the analytic continuation of \eqref{occuptnF} - and hence
the true fermionic occupation number of a quasiparticle state of
energy ${ \e}$ - is obtained by subtracting the contribution due to
the pole:
\begin{equation}\label{occuptnFfin}
  \bar{n}_F({\epsilon},q) = \int_{-\pi}^{\pi} d\alpha \ \Big(  \rho_F(\alpha) \  \frac{1}{e^{\beta(\e - q\mu) -\i q \alpha }+1} \Big) - \ 2 \pi \Theta( q{ \mu}-{ \epsilon}) \  {\tilde \rho}_F(-e^{\beta(\epsilon - q \mu)})\ .
\end{equation}
In the special case of a lower gap fermionic phase, ${\tilde \rho}_F$
vanishes, and so, \eqref{occuptnFfin} reduces to \eqref{occuptnF}.

The analysis of the bosonic case proceeds along similar lines. The
occupation number \eqref{occuptnB} can be rewritten as a contour
integral as
 \begin{equation}\label{occuptnBnz}
   \bar{n}_B({\e},q) =  \int_{C} \frac{dz}{\i z} \ {\tilde \rho}_B(z) \  \frac{1}{e^{\beta(\epsilon - q\mu)}z^{-1} - 1}\ .
\end{equation}
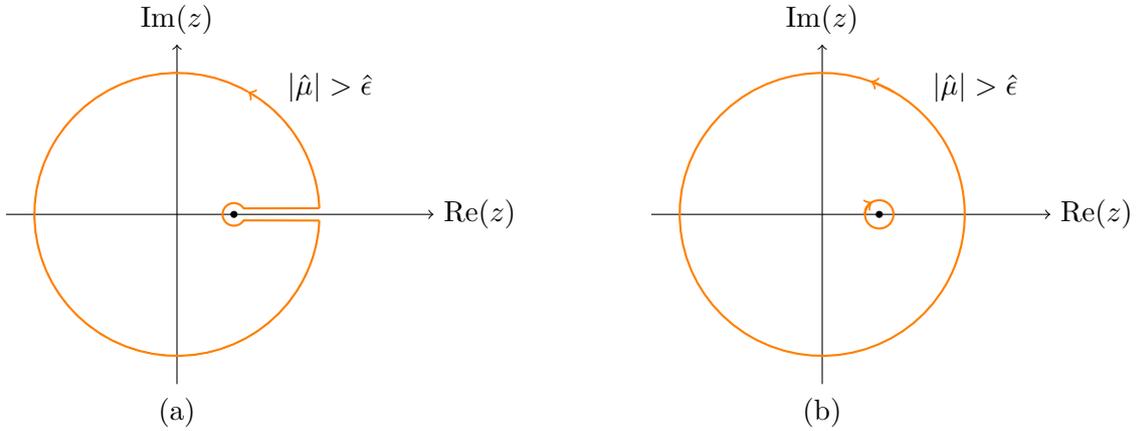
\begin{figure}[!h]
	\begin{minipage}{.28\textwidth}
		\begin{tikzpicture}[scale=0.75]
		\coordinate (L) at (-3,0) ; 
		\coordinate (R) at (4.5,0) ;
		\coordinate (B) at (0,-3) ;
		\coordinate (U) at (0,3) ;
		\coordinate (P) at (1,0) ;
		\def\R{2.5cm};
		\def\r{0.2cm};
		\draw[dashed] (5:2.5cm) arc(5:355:2.5cm) ;
		\draw[->] (L) -- (R) node[right] {$\mathrm{Re}(z)$};
		\draw[->] (B) -- (U) node[above] {$\mathrm{Im}(z)$};
		\draw[orange, thick,->] (2.5:\R) arc(2.5:60:\R) ;
		\draw[rounded corners, orange, thick] (2.5:2.5cm) -- (5.2:1.17cm) ;
		\draw[orange, thick] (-5.2:1.17cm) -- (-2.5:\R) ;
		\draw[orange, thick] (60:\R) arc(60:357.5:\R) ;
		\filldraw[black] (P) circle (1.5pt) ;
		\draw[orange, thick] (P) +(30:\r) arc (30:330:\r);
		\draw (40:3.5cm) node {$|\hat{\mu}|>\hat{\epsilon}$};
		\draw (B)+(0,-0.5) node {(a)};
		\end{tikzpicture}
	\end{minipage}
	\hspace{4.0cm}
	\begin{minipage}{.2\textwidth}
		\begin{tikzpicture}[scale=0.75]
		\coordinate (L) at (-3,0) ; 
		\coordinate (R) at (4,0) ;
		\coordinate (B) at (0,-3) ;
		\coordinate (U) at (0,3) ;
		\coordinate (P) at (1,0) ;
		\def\R{2.5cm};
		\def\r{0.25cm};
		\draw[->] (L) -- (R) node[right] {$\mathrm{Re}(z)$};
		\draw[->] (B) -- (U) node[above] {$\mathrm{Im}(z)$};
		\draw[orange, thick,->] (60:\R) arc(60:360+70:\R) ;
		\filldraw[black] (P) circle (1.5pt) ;
		\draw[orange, thick,->] (P) +(180:\r) arc(180:-240:\r) ;
		\draw (40:3.5cm) node {$|\hat{\mu}|>\hat{\epsilon}$};
		\draw (B)+(0,-0.5) node {(b)};
		\end{tikzpicture}
	\end{minipage}
	\caption{Bosons: the pole is at $z=+e^{\beta(\epsilon- q\mu)}$
          and is denoted by the black dot on the +ve real axis. Radius
          of the big circle is unity. The orange curve is the contour
          $C'$ (counterclockwise) over which the holonomy is
          integrated over.}
	\label{nBoc}
\end{figure}
Again, the analytic continuation of \eqref{occuptnB} (and hence the
correct formula for the occupation number of a bosonic quasiparticle
state at energy ${ \epsilon}$) is obtained by subtracting the
contribution due to the pole at $z=+e^{\beta(\epsilon- q\mu)}$. The
calculation can be followed pictorially in Figure \ref{nBoc}. The
modified occupation number is then
\begin{equation}\label{occuptnBfin}
\bar{n}_B({\e},q) = \int_{-\pi}^{\pi} d\alpha \ \Big(  \rho_B(\alpha) \  \frac{1}{e^{\hat{\epsilon} - q \hat{\mu}  -\i \alpha q}-1} \Big) + \ 2 \pi \Theta( q { \mu}-{ \e}) \ {\tilde \rho}_B(e^{-|{\hat \mu}|+{\hat \epsilon}})\ .
\end{equation} 
In the particular case of a bosonic upper cap phase,
\eqref{occuptnBfin} simplifies to
\begin{equation}\label{occuptnBfinug}
\bar{n}_B({\epsilon},q) = \int_{-\pi}^{\pi} d\alpha \ \Big(  \rho_B(\alpha) \  \frac{1}{e^{\hat{\epsilon} - q\hat{\mu} -\i q \alpha }-1} \Big) + \frac{1}{|\lambda_B|}\Theta( q { \mu}-{ \epsilon})\ .
\end{equation}

\section{The thermodynamic limit} \label{thermo} 

In the rest of the paper we specialise to the study of the
quasi-fermionic theories and quasi-bosonic theories in the
thermodynamic limit defined by taking the limit $V \to \infty$ (recall
that all through this paper we choose the volume of our $S^2$ to be
given by $\mc{V}_2 =N V$ and $N$ is taken to $\infty$ first before any
other limit). In this limit, the dominant phase of the holonomy
eigenvalues is one in which the bosonic and fermionic eigenvalue
distributions take the `tabletop' form 
which we reproduce below:
\begin{equation}\label{fhgon1}
\rho_F(\alpha)= \frac{1}{2\pi |\lambda_F|} \Theta(\pi|\lambda_F|-|\alpha|)\ ,\qquad \rho_B(\alpha)= \frac{1}{2\pi |\lambda_B|}\Theta(\pi|\lambda_B|-|\alpha|)\ .
\end{equation}
We will further be interested in the so-called 'thermodynamic,
zero-temperature limit' obtained by first taking the limit
$V \to \infty$ and only then taking the limit $T\to 0$.  The
importance of the above order of limits is the following. When we take
the limit $V \to \infty$ the eigenvalue distribution is frozen
to \eqref{fhgon1}. The zero-temperature limit is then taken with the
eigenvalue distribution frozen at \eqref{fhgon1}.

In this section, we first give expressions for the occupation numbers
in terms of explicit well-known functions in the thermodynamic
limit. We then consider the `thermodynamic, zero-temperature limit' of
those expressions. We also provide the expressions for the off-shell
free energies in the limit since they will be used extensively in the
subsequent sections.

\subsection{Occupation numbers in the infinite volume limit}\label{eoiv}

In this subsection we specialise the formulae for occupation numbers
presented in Section \ref{qon} to the $V \to \infty$ limit where the
eigenvalue distributions take the universal `tabletop' profile
\eqref{fhgon1}.

It turns out to be convenient to phrase the discussion in terms of the
variable $x$ defined as
\begin{equation}\label{xdef}
x=e^{\beta(\epsilon-q\mu)} \ .
\end{equation}
We shall encounter the multivalued function $\tan^{-1} \zeta$ in the
formulas of this section. This is defined unambiguously by choosing
the following branch:
\begin{equation}\label{tanidef}
  -\frac{\pi}{2}\ <\ \tan^{-1} \zeta\ <\ \frac{\pi}{2}\ .
\end{equation}
Note that with this definition of $\tan^{-1}(\zeta)$, we have
$\tan^{-1}(-\zeta) =-\tan^{-1}(\zeta).$

\subsubsection{The fermionic occupation number}

The fermionic occupation number is given by equation \eqref{occuptnF}
where $\rho_F(\alpha)$ takes the universal tabletop form
\begin{equation}
\rho_F(\alpha) = \frac{1}{2\pi|\lambda_F|} \ \Theta(\pi|\lambda_F|-|\alpha|)\ .
\end{equation}
In terms of the variable $x$ defined in \eqref{xdef}, the formula for
the occupation number of a quasiparticle state at energy $\epsilon$
and of charge $q=\pm 1$ simplifies to
\begin{align}\label{nFsim}
\bar{n}_F(\epsilon,q) & = \frac{1}{2\pi|\lambda_F|}  \int_{-\pi|\lambda_F|}^{\pi|\lambda_F|} d\alpha \ \frac{1}{x e^{-\i \alpha}+1}\ ,\nonumber \\
& = \frac{1}{2\pi|\lambda_F|}  \int_{-\pi|\lambda_F|}^{\pi|\lambda_F|} d\alpha \ \frac{(x\cos\alpha +1)}{1+x^2+2x\cos\alpha} \ ,
\end{align}
(an equivalent formula first appeared in \cite{Geracie:2015drf}).  In
going from the first to the second line of \eqref{nFsim} we have used
the fact that the integration limit is symmetric about $\alpha=0$. The
integral on the RHS of the second line of \eqref{nFsim} can be
performed explicitly and we find the relatively simple analytic
expression
\begin{align}\label{nFfin}
  \bar{n}_F(\epsilon,q) 
  & = \frac{1}{2}-\frac{1}{\pi|\lambda_F|}  \tan^{-1}\bigg(\frac{x-1}{x+1}\tan\frac{\pi|\lambda_F|}{2}\bigg)\ ,
\end{align}
where $x = e^{\beta(\epsilon - q\mu)}$ is defined in \eqref{xdef}. As
indicated in \eqref{tanidef}, the multivalued function
$\tan^{-1}(\zeta)$ that appears in \eqref{nFfin} is unambiguously
defined by choosing the branch
$-\frac{\pi}{2} < \tan^{-1}(\zeta) < \frac{\pi}{2}$. Substituting
\eqref{xdef} in \eqref{nFfin} it follows that
\begin{equation}\label{nFfin2} 
  \bar{n}_F(\epsilon,q) 
  = \frac{1}{2}-\frac{1}{\pi|\lambda_F|}  \tan^{-1}\bigg(\frac{e^{\beta(\epsilon-q\mu)}-1}{e^{\beta(\epsilon-q \mu)}+1}\tan\frac{\pi|\lambda_F|}{2}\bigg) \ .
\end{equation}
\eqref{nFfin} and \eqref{nFfin2} are the main results of this
subsubsection. In the rest of this subsubsection we will study various
limits of \eqref{nFfin2}.

\subsubsection*{Zero-temperature limit} 
In the zero temperature limit (i.e., $\beta\to \infty$), we have
\begin{align}\label{ztx}
x=e^{\beta(\epsilon-q\mu)} = \left\{ \begin{array}{cc}
0\ , & \quad \epsilon < q\mu\ ,  \\ 
1\ , & \quad \epsilon = q \mu\ , \\
\infty\ , & \quad \epsilon > q \mu\ .
\end{array}
\right.
\end{align}
It follows immediately from \eqref{nFfin2}, that in this limit,
\begin{equation}\label{ztnF1}
\bar{n}_{F}(\epsilon,q) = \left\{ \begin{array}{cc}
1\ , & \quad \epsilon < q \mu\ , \\ 
\frac{1}{2}\ , & \quad \epsilon = q \mu\ , \\ 
0\ , & \quad \epsilon > q \mu\ .
\end{array}
\right.
\end{equation}
Note in particular that all states with energy smaller than $q \mu$
are occupied with unit occupancy. In the zero temperature limit, in
other words, the fermions form a free Fermi sea at every value of
$\lambda_F$ (this fact was already noted in \cite{Geracie:2015drf}).

\subsubsection*{Small-$\lambda_F$ expansion}
Working at arbitrary temperature and $\mu$, \eqref{nFfin2} is easily
expanded in a power series expansion in $\lambda_F$ about zero; we
find
\begin{equation}\label{nFsla} 
\bar{n}_F(\epsilon,q) 
 =\frac{1}{e^{\beta(\epsilon-q\mu)} +1}-\frac{\pi^2\lambda_F^2}{6} \frac{e^{\beta(\epsilon-q\mu)}\big(e^{\beta(\epsilon-q\mu)}-1\big)}{\big(e^{\beta(\epsilon-q\mu)}+1\big)^3} \ + \  \mc{O}(\lambda_F^4)\ ,
\end{equation}
(as the expression \eqref{nFfin2} is an even function of $\lambda_F$,
the power series on the RHS of \eqref{nFsla} has only even powers in
$\lambda_F$). The expansion \eqref{nFsla} applies at all values of the
energy $\epsilon$.

Note that the first term on the RHS of \eqref{nFsla} is, of course,  the familiar formula of Fermi Statistics for the occupation number of a single particle state. Subsequent terms 
in this expansion represents universal `anyonic' corrections to this familiar formula.  
 
\subsubsection*{Small-$x$ expansion} 
When the energy of our quasiparticle state lies well below $q \mu$ in
units of the temperature, then $x =e^{\beta(\epsilon-q\mu)} \ll
1$. Equations \eqref{nFfin} or \eqref{nFfin2} takes the following form
\begin{align}\label{nFsxa}
  \bar{n}_{F}(\epsilon,q)
  & = 1 - \sum_{n=1}^{\infty} (-1)^{n+1} \ \frac{\sin(n\pi \lambda_F) }{n\pi \lambda_F}  x^n \ ,\nonumber\\
  &= 1 - \sum_{n=1}^{\infty} (-1)^{n+1} \ \frac{\sin(n\pi \lambda_F) }{n\pi \lambda_F} \ e^{n\beta(\epsilon-q \mu)} \ .
\end{align}
The first term on the RHS of \eqref{nFsxa} is unity indicating that at
leading order in this limit, all states are occupied with unit
occupancy. The first correction to unity on the RHS of \eqref{nFsxa},
$-\Big(\frac{\sin \pi \lambda_F}{\pi \lambda_F} \Big) x$, gives the
exponentially suppressed probability of this state becoming unfilled
(equivalently, of a `quasihole' being created). The intriguing factor
of $\frac{\sin \pi \lambda_F}{\pi \lambda_F}$ (which is unity in the
free limit $\lambda_F \to 0$ ) indicates that the formation at a hole
is less likely at nonzero $\lambda_F$ than in the free limit. It would
be very interesting to understand why this is the case directly from
the Schrodinger equation in an appropriate non-relativistic limit.

\subsubsection*{Large-$x$ expansion} 

On the other hand when the energy of our quasiparticle state lies well above 
$q \mu$ in units of the temperature, then $x =e^{\beta(\epsilon-q\mu)} \gg 1$. \eqref{nFfin} or \eqref{nFfin2} takes the  form 
 \eqref{nFfin2} takes the following form 
\begin{align}\label{nFlxa}
\bar{n}_{F}(\epsilon,q)& =  \sum_{n=1}^{\infty} (-1)^{n+1} \ \frac{\sin(n\pi \lambda_F) }{n\pi \lambda_F} x^{-n}\ ,\nonumber \\
& =  \sum_{n=1}^{\infty} (-1)^{n+1} \ \frac{\sin(n\pi \lambda_F) }{n\pi \lambda_F} \ e^{-n\beta(\epsilon-\mu)}\ .
\end{align}
\eqref{nFlxa} tells us that the occupation number of a state with such
energies is exponentially suppressed. Once again note the interesting
suppression of this probability by the factor of
$\frac{\sin \pi \lambda_F}{\pi \lambda_F}$, which once again calls out
for a non-relativistic explanation.

\subsubsection{The bosonic occupation number}
Recall the formula for the bosonic occupation number in an upper cap
phase \eqref{occuptnBfinug}:
\begin{equation}\label{nBug}
  \bar{n}_B(\epsilon,q) = \int_{-\pi}^{\pi} \rho_B(\alpha) d\alpha \ \frac{1}{e^{\beta(\epsilon-q\mu)-\i \alpha}-1} \ + \ \frac{1}{|\lambda_B|} \ \Theta(q\mu-\epsilon)\ . 
\end{equation}
Using the tabletop form of the bosonic eigenvalue distribution
\eqref{fhgon1}, we find the following formula for the bosonic
occupation number in the thermodynamic limit:
\begin{align}\label{nBsim}
    \bar{n}_B(\epsilon,q) &  = \frac{1}{2\pi|\lambda_B|}  \int_{-\pi|\lambda_B|}^{\pi|\lambda_B|} d\alpha \ \frac{1}{x e^{-\i \alpha}-1} \ + \ \frac{1}{|\lambda_B|} \ \Theta(q\mu-\epsilon)\ ,\nonumber \\
    & = \frac{1}{2\pi|\lambda_B|} \int_{-\pi|\lambda_B|}^{\pi|\lambda_B|} d\alpha \ \frac{(x\cos\alpha -1)}{1+x^2-2x\cos\alpha} \ + \
    \frac{1}{|\lambda_B|} \ \Theta(q\mu-\epsilon)\ .
\end{align}
As in the case of fermions, the integral on the RHS of the second line
of \eqref{nBsim} can be exactly evaluated and we find\footnote{As
  indicated in \eqref{tanidef}, the function $\tan^{-1}(\zeta)$ used
  in that equation is defined so that
  $-\frac{\pi}{2} < \tan^{-1}(\zeta) < \frac{\pi}{2}$.}
\begin{equation}\label{nBfin} 
\bar{n}_B(\epsilon,q) 
 = -\frac{1}{2}+\frac{1}{\pi|\lambda_B|}  \tan^{-1}\bigg(\frac{x+1}{x-1}\tan\frac{\pi|\lambda_B|}{2}\bigg) +  \frac{1}{|\lambda_B|} \ \Theta(q\mu-\epsilon)\ .
\end{equation}
Using the fact that 
\begin{align}
& x > 1\quad\text{for}\quad\epsilon > q\mu \ ,\nonumber\\ 
& x = 1\quad\text{for}\quad\epsilon = q\mu \ ,\nonumber\\ 
& x < 1\quad\text{for}\quad\epsilon < q\mu \ ,
\end{align}
the $\Theta$-function appearing in \eqref{nBfin} can also be rewritten
as $\Theta(1-x)$. So, \eqref{nBfin} can be rewritten as
\begin{equation}\label{nBfin2} 
  \bar{n}_B(\epsilon,q) 
  = -\frac{1}{2}+\frac{1}{\pi|\lambda_B|}  \tan^{-1}\bigg(\frac{x+1}{x-1}\tan\frac{\pi|\lambda_B|}{2}\bigg) + \frac{1}{|\lambda_B|} \ \Theta(1-x)\ .
\end{equation}
The $\tan^{-1}$ and the $\Theta$ terms on the RHS above discontinuous
at $x=1$. However these discontinuities cancel each other and the
final result is perfectly smooth at $x=1$. In order to see this we
proceed as follows. Using
\begin{equation}\label{thetasgn}
  \Theta(1-x) = \frac{1}{2} - \frac{1}{2} \sgn(x-1) \ ,
\end{equation}
and the identity
\begin{equation}\label{idde}
  \tan^{-1}(a) + \tan^{-1}\frac{1}{a} = {\rm sgn}(a) \frac{\pi}{2}\ ,
\end{equation} 
we see that \eqref{nBfin2} can be rewritten as 
\begin{equation}\label{nBfinaltx} 
\bar{n}_B(\epsilon,q) 
 =\frac{1-|\lambda_B|}{2|\lambda_B|} - \frac{1}{\pi|\lambda_B|}  \tan^{-1}\bigg(\frac{x-1}{x+1}\cot\Big(\frac{\pi|\lambda_B|}{2} \Big)\bigg)\ ,
\end{equation}
keeping in mind the choice of branch \eqref{tanidef} of the
$\tan^{-1}$ function.  Inserting the definition of $x$ \eqref{xdef}, we
obtain
\begin{equation}\label{nBfinalt} 
\bar{n}_B(\epsilon,q) 
 =\frac{1-|\lambda_B|}{2|\lambda_B|} - \frac{1}{\pi|\lambda_B|}  \tan^{-1}\bigg(\frac{e^{\beta(\epsilon-q \mu)}-1}{e^{\beta(\epsilon-q \mu)}+1}\cot\frac{\pi|\lambda_B|}{2}\bigg)\ .
\end{equation}
\eqref{nBfinaltx} and \eqref{nBfinalt} are the main results of this
subsection. In the rest of this subsection we study the various limits
of these formulae.

\subsubsection*{Zero-temperature limit} 
Using \eqref{ztx}, it follows that in the limit $\beta \to \infty$.
\begin{equation}\label{ztnB}
\bar{n}_{B}(\epsilon,q) = \left\{\renewcommand{\arraystretch}{2} \begin{array}{cc} \displaystyle
\frac{1-|\lambda_B|}{|\lambda_B|}\ , & \quad \epsilon < q\mu\ , \\ \displaystyle
  \frac{1-|\lambda_B|}{2|\lambda_B|}\ , & \quad \epsilon = q\mu\ , \\\displaystyle
0\ , & \quad \epsilon > q\mu\ .
\end{array}
\right.
\end{equation}
It follows from \eqref{ztnB}, in particular, that all states with
energies less than $q \mu$ (states that would have been infinitely
occupied by a Bose condensate in the free limit) have an occupation
number given by $$\frac{1-|\lambda_B|}{|\lambda_B|}\ .$$ Note that the
additional piece in the occupation number of the boson (the second
term in the formula for $\bar{n}_B$ in \eqref{nBug}) - plays a key
role in the formula \eqref{ztnB}. Without this term we would have
found $\bar{n}_B({\epsilon},q) = -1$; this is the nonsensical
`divergent' occupation number of a theory of free bosons.  The second
term in \eqref{nBug} `regulates' this free result into a physically
sensible answer, albeit one that does diverge in the limit
$\lambda_B\to 0$.

\subsection*{Small-$\lambda_B$ expansion}
Working at arbitrary values of $x$, \eqref{nBfinaltx} is easily expanded in a power series in $\lambda_B$. We find 
\begin{equation}\label{nBsla} 
\bar{n}_B(\epsilon,q) 
 = \frac{1}{|\lambda_B|} \Theta(q\mu-\epsilon) \ + \frac{1}{e^{\beta(\epsilon-q\mu)} -1}-\frac{\pi^2\lambda_B^2}{6} \frac{e^{\beta(\epsilon-q\mu)}\big(e^{\beta(\epsilon-q\mu)}+1\big)}{\big(e^{\beta(\epsilon-q\mu)}-1\big)^3} \ + \  \mc{O}(\lambda_B^4)\ .
\end{equation}
The above equation can be rewritten more explicitly as
\begin{equation}\label{nBslaexp} 
  \bar{n}_B(\epsilon,q) = \left\{\renewcommand{\arraystretch}{2}
    \begin{array}{cc} \displaystyle   \frac{1}{e^{\beta(\epsilon-q\mu)} -1}-\frac{\pi^2\lambda_B^2}{6} \frac{e^{\beta(\epsilon-q\mu)}\big(e^{\beta(\epsilon-q\mu)}+1\big)}{\big(e^{\beta(\epsilon-q\mu)}-1\big)^3} \ + \  \mc{O}(\lambda_B^4) \ , & \quad \epsilon> q\mu\ ,  \\ \displaystyle
 \frac{1}{|\lambda_B|} + \frac{1}{e^{\beta(\epsilon-q\mu)} -1}-\frac{\pi^2\lambda_B^2}{6} \frac{e^{\beta(\epsilon-q\mu)}\big(e^{\beta(\epsilon-q\mu)}+1\big)}{\big(e^{\beta(\epsilon-q\mu)}-1\big)^3} \ + \  \mc{O}(\lambda_B^4) \ , &\quad \epsilon<q\mu\ .\end{array}\right.
\end{equation}
The first term on the first line of \eqref{nBslaexp} is, of course,
the familiar expression for the occupation number of a single particle
state with $\epsilon >q \mu$ using the formulae of Bose statistics;
the remaining terms on this line are the universal\footnote{These
  corrections are universal in the following sense. They apply both to
  the critical and the regular boson theory, and all values of the UV
  parameters of these theories. Note, however, that these formulae
  only apply in the infinite volume limit where the holonomy
  eigenvalue distribution takes the tabletop form.} `anyonic'
corrections to this formula.

The second line of \eqref{nBslaexp} deals with the more interesting
case of a state whose energy is less than $q\mu $.  Any such state
would be infinitely occupied in a free Bose ensemble. The first terms
on the RHS of \eqref{nBslaexp} tell us that finite $\lambda_B$ effects
regulate this infinite occupation to a finite occupation number given
by $\frac{1}{|\lambda_B|}$. The remaining subleading terms in this
expansion give us the (temperature dependent) corrections to this
result.

The corrections to the leading order results in \eqref{nBslaexp}
are of order $\frac{\lambda_B}{\beta(\epsilon-q\mu)}$; at generic values of $\epsilon$, these corrections are small if $\lambda_B$ is small. No matter how small $\lambda_B$ is, however, these corrections are non-negligible when 
\begin{equation}\label{corwidth}
  \beta(\epsilon -q \mu) \sim \lambda_B\ .
\end{equation} 
At small $\lambda_B$, it follows that $\bar{n}_B(\epsilon,q)$ takes
the value $\frac{1}{|\lambda_B|}$ (for $\epsilon < q \mu$) and
$\frac{1}{e^{\beta(\epsilon-q\mu)} -1}$ for $\epsilon > q \mu$. The
transition between these two values takes place in the range of
energies $\eqref{corwidth}$. In fact at leading order in small
$\lambda_B$, $\bar{n}_B(\epsilon,q)$ is everywhere (i.e. at all
energies) well approximated by
 \begin{equation}\label{nBfinaltapp} \begin{split}
\bar{n}_B(\epsilon,q) 
& \approx \frac{1}{2|\lambda_B|} - \frac{1}{\pi|\lambda_B|}  \tan^{-1}\bigg(\frac{e^{\beta(\epsilon-q \mu)}-1}{\pi |\lambda_B|}\bigg) \\
\end{split} 
\end{equation}
Note that the RHS of tends to zero when $\beta(\epsilon - q\mu) \gg |\lambda_B|$ but tends to $\frac{1}{|\lambda_B|}$ for $\beta(\epsilon - q\mu) \ll -|\lambda_B|$.

\subsubsection*{Small-$x$ expansion} 
As in the case of fermions, $x$ is small for states whose energies lie
much below $q\mu$ in units of the temperature.  The occupation number
for such states is well approximated by the small-$x$ expansion of
\eqref{nBfinaltx}, given by
\begin{align}\label{nBsxa} 
\bar{n}_{B}(\epsilon,q)&= \frac{1-|\lambda_B|}{|\lambda_B|} - \sum_{n=1}^{\infty} \ \frac{\sin(n\pi \lambda_B) }{n\pi \lambda_B} \ x^n\ ,\nonumber \\
 &= \frac{1-|\lambda_B|}{|\lambda_B|} - \sum_{n=1}^{\infty} \ \frac{\sin(n\pi \lambda_B) }{n\pi \lambda_B} \ e^{n\beta(\epsilon-q\mu)}\ .
\end{align}
We see that at leading order in this expansion, all states are
occupied with occupation number $\frac{1-|\lambda_B|}{|\lambda_B|}$.
The probability of a quasihole appearing in this condensate is, at
leading order, proportional to
$\frac{\sin \pi \lambda_B}{\pi \lambda_B}\,x$. As in the case of
fermions it would be interesting to find a physical interpretation of
the prefactor $\frac{\sin \pi \lambda_B}{\pi \lambda_B}$.

\subsubsection*{Large-$x$ expansion} 

$x$ is large for states whose energies lie much above $q \mu$ in units of the temperature. The occupation number of such states is well approximated expanding \eqref{nBfinaltx} in a 
power series in $1/x$. This expansion is given by 
\begin{align}\label{nBlxa}
\bar{n}_{B}(\epsilon,q) & =  \sum_{n=1}^{\infty}  \ \frac{\sin(n\pi \lambda_B) }{n\pi \lambda_B} \ \frac{1}{x^n}\ ,\nonumber  \\ 
&=  \sum_{n=1}^{\infty}  \ \frac{\sin(n\pi \lambda_B) }{n\pi \lambda_B} \ e^{-n\beta(\epsilon-q\mu)} \ .
\end{align}
We see that the probability of occupation for any of these states is
given by $\frac{\sin \pi \lambda_B}{\pi \lambda_B}\,x^{-1}$ at
leading order. The factor of $x^{-1}$ is the familiar Boltzmann
suppression of a high energy state.  The universal prefactor is more
mysterious and requires an explanation.

\subsubsection{Duality of explicit forms of occupation number}

In this brief subsection we directly verify that the explicit
expressions \eqref{nFfin2} and \eqref{nBfinalt} are dual to each
other. Using \eqref{nFfin2} and \eqref{nBfinalt} we see that
\begin{multline}\label{ocndual}
|\lambda_B| \bar{n}_B  - |\lambda_F|\bar{n}_F = \frac{1-|\lambda_B|-|\lambda_F|}{2}  \\
 - \frac{1}{\pi} \Bigg( \tan^{-1}\bigg(\frac{e^{\beta(\epsilon-q\mu)}-1}{e^{\beta(\epsilon-q\mu)}+1}\cot\Big(\frac{\pi|\lambda_B|}{2} \Big)\bigg) - \tan^{-1}\bigg(\frac{e^{\beta(\epsilon-q\mu)}-1}{e^{\beta(\epsilon-q\mu)}+1}\tan\Big(\frac{\pi|\lambda_F|}{2} \Big)\bigg) \Bigg)\ .
\end{multline}
The duality map of the coupling constants\footnote{Throughout this
  section and in other sections, (unless mentioned explicitly) we have
  implicitly used the fact that $c_F=c_B$, so that the single particle
  energy spectra, $\epsilon=\sqrt{p^2+c_F^2}=\sqrt{p^2+c_B^2}$, are
  identical for both fermions and bosons.}
$|\lambda_F|=1-|\lambda_B|$ implies that
$\tan\big(\frac{\pi|\lambda_F|}{2} \big)=
\cot\big(\frac{\pi|\lambda_B|}{2} \big)$. It then follows that the
right hand side of \eqref{ocndual} vanishes. We conclude that
\begin{equation}\label{nBnFdual}
|\lambda_B| \bar{n}_B - |\lambda_F|\bar{n}_F  =0\ .
\end{equation}
as required by duality.

\subsection{Free energies at zero temperature}
The expressions for the free energies in Section \ref{ofebugp}
simplify in the limit $\beta \to \infty$ in that that the holonomy
integrals become almost trivial. Consider the holonomy integral in the
bosonic case:
\begin{equation}\label{holint}
  3\int_{{\hat c}_B}^\infty d\hat\epsilon\ \hat\epsilon  \int_{-\pi}^\pi d\alpha\,\rho_B(\alpha) \left(\log \Big(1 - e^{-\hat\epsilon + |\hat\mu| + \i\alpha}\Big) + \log \Big(1 - e^{-\hat\epsilon + |\hat\mu| + \i\alpha}\Big)\right)\ ,
\end{equation}
In the limit $\beta \to \infty$ with $\epsilon$ and $\mu$ fixed, the
quantities $\hat\epsilon$ and $\hat\mu$ go to $\infty$. It follows
that the second logarithm above in \eqref{holint} simply vanishes. The
first logarithm also vanishes when ${\epsilon}$ is larger than
$|{ \mu}|$. On the other hand, when ${ \epsilon}< { \mu}$, we have
\begin{equation}\label{expoflog}
  \log(1-e^{-\hat{\epsilon}+\hat{\mu}+\i \alpha}) = {\hat \mu} - {\hat \epsilon} + \i \theta_B(\alpha)\ , 
\end{equation} 
where 
\begin{equation}\label{deftheta}
\theta_B(\alpha)= \alpha - \pi\quad{\rm when}\quad\alpha>0\quad{\rm and}\quad\theta_B(\alpha)= \alpha + \pi\quad{\rm when}\quad\alpha<0\ .
\end{equation} 
Using the fact that $\rho_B(\alpha)$ is an even function of $\alpha$, it
follows that
\begin{equation}
  \int_{-\pi}^\pi d\alpha\, \rho_B(\alpha)\,\theta_B(\alpha) = 0\ .\nonumber
\end{equation}
The holonomy integral \eqref{holint} thus simplifies to
\begin{equation}\label{secondline} 
  3\,\Theta(|{ \mu}|-{ c}_B )  \int_{{\hat c}_B}^{|{\hat \mu}|} d {\hat \epsilon} \ {\hat \epsilon} \left( {\hat \mu}-{\hat \epsilon} \right)
  =\Theta(|{ \mu}|-{ c}_B ) \frac{(|{\hat \mu }| -{\hat c}_B )^2 (|{\hat \mu}|+2 {\hat c}_B) }{2}\ .
\end{equation} 
The same calculation works for the fermionic case except for one
simplification: there is no additional choice of branch for the
logarithm in \eqref{deftheta} since there is no additional minus sign
in the logarithm $\log (1 + e^{-\hat\epsilon + |\hat\mu| + \i\alpha})$
for the fermions.

Thus, in the zero-temperature limit, the off-shell free energies for
the bosonic and fermionic theories become
\begin{align}\label{RFzeroT}
  F_{\rm RF}({c}_F,\tilde{\mathcal{C}}) &= \frac{N_F}{6\pi\beta^3} \bigg[ - 8\lambda_{F}^2\tl{\mc{C}}^3 - 3\tl{\mc{C}} \bigg( \hat{c}_{F}^2 - \Big(2\lambda_F \tl{\mc{C}} + \hat{m}_F \Big)^2 \bigg)  - 6\lambda_{F} \hat{m}_F\tl{\mc{C}}^2 \nonumber  \\
  &\qquad\qquad\qquad\qquad\qquad+ \hat{c}_{F}^3 - \frac{1}{2}\Theta(|{ \mu}|-{ c}_F ) (|{\hat \mu }| -{\hat c}_F )^2 (|{\hat \mu}|+2 {\hat c}_F)  \bigg]  \ .
\end{align}
\begin{align}\label{CFzeroT}
  F_{\rm CF}({c}_F,{\zeta}_F,\tilde{\cC}) &= \frac{N_F}{6\pi \b^3} \bigg[ -8\lambda_{F}^2 \tl{\mc{C}}^3 - 3\tl{\mc{C}} \bigg( \hat{c}_{F}^2 - \Big(2\lambda_F \tl{\mc{C}} - \frac{4\pi \hat{\zeta}_F}{\kappa_F} \Big)^2 \bigg) + 6\lambda_{F} \tl{\mc{C}}^2 \Big(\frac{4\pi \hat{\zeta}_F}{\kappa_F} \Big) \nonumber  \\ 
  &\qquad\qquad   + 3 \bigg( \frac{\hat{y}_2^2}{2\lambda_F} \frac{4\pi \hat{\zeta}_F}{\kappa_F} - \frac{\hat{y}_4}{2\lambda_F} \Big(\frac{4\pi \hat{\zeta}_F}{\kappa_F} \Big)^2 +\frac{x_6^F}{8\lambda_F} \Big(\frac{4\pi \hat{\zeta}_F}{\kappa_F} \Big)^3 \bigg)\nonumber   \\
&\qquad\qquad\qquad\qquad\qquad +  \hat{c}_{F}^3 - \frac{1}{2}\Theta(|{ \mu}|-{ c}_F ) (|{\hat \mu }| -{\hat c}_F )^2 (|{\hat \mu}|+2 {\hat c}_F) \bigg]  \ .
\end{align}
\begin{align}\label{CBzeroT}
  F_{\rm CB}({c}_B, \tilde{\cS}) & = \frac{N_B}{6\pi \b^3} \bigg[\frac{3}{2} \hat{c}_B^2\hat{m}_B^{\text{cri}} - 4\lambda_B^2\left(\tilde{\mc{S}}-\tfrac{1}{2}\hat{m}_{B}^{\text{cri}} \right)^3  +6|\lambda_B| \hat{c}_B\left(\tilde{\mc{S}}-\tfrac{1}{2}\hat{m}_{B}^{\text{cri}}\right)^2\nonumber \\
&\qquad\qquad\qquad -\hat{c}_B^3 + \frac{1}{2}\left(1 - \frac{1}{|\lambda_B|}\right)\Theta(|{ \mu}|-{ c}_B ) (|{\hat \mu }| -{\hat c}_B )^2 (|{\hat \mu}|+2 {\hat c}_B)\bigg]\ .
\end{align}
\begin{align}\label{RBzeroT}
  F_{\rm RB}({c}_B,{\sigma}_B,\tilde\cS)
  & =\frac{N_B}{6\pi \b^3} \bigg[-3\hat{c}_B^2\hat{\sigma}_B+\lambda_B^2\hat{\sigma}_B^3 -4\lambda_B^2(\tilde{\cS}+\hat{\sigma}_{B})^3 +6|\lambda_B| \hat{c}_B(\tilde{\cS}+\hat{\sigma}_B)^2 \nonumber \\
&\qquad\qquad +3(\hat{m}_B^2\hat{\sigma}_B+2\lambda_B\hat{b}_4\hat{\sigma}_B^2+(x_6+1) \lambda_B^2 \hat{\sigma}_B^3)\nonumber \\
&\qquad\qquad  -\hat{c}_B^3 + \frac{1}{2}\left(1 - \frac{1}{|\lambda_B|}\right)\Theta(|{ \mu}|-{ c}_B ) (|{\hat \mu }| -{\hat c}_B )^2 (|{\hat \mu}|+2 {\hat c}_B)\bigg]\ .
\end{align}

\section{Quasi-fermionic theories} \label{pscbrf}

In this section, we analyse in detail the thermodynamical
zero-temperature phase structure of the quasi-fermionic theories
i.e.~the regular fermion and the critical boson. We mainly work in the
bosonic theory and appeal to Bose-Fermi duality in order obtain the
corresponding results for the fermions.

The off-shell free energy of the critical boson theory in the
zero-temperature limit was obtained in \eqref{CBzeroT} and is
reproduced here for convenience:
\begin{align}\label{CBzeroTmain}
  F_{\rm CB}({c}_B, \tilde{\cS}) & = \frac{N_B}{6\pi \b^3} \bigg[\frac{3}{2} \hat{c}_B^2\hat{m}_B^{\text{cri}} - 4\lambda_B^2\left(\tilde{\mc{S}}-\tfrac{1}{2}\hat{m}_{B}^{\text{cri}} \right)^3  +6|\lambda_B| \hat{c}_B\left(\tilde{\mc{S}}-\tfrac{1}{2}\hat{m}_{B}^{\text{cri}}\right)^2\nonumber \\
&\qquad\qquad\qquad -\hat{c}_B^3 + \frac{1}{2}\left(1 - \frac{1}{|\lambda_B|}\right)\Theta(|{ \mu}|-{ c}_B ) (|{\hat \mu }| -{\hat c}_B )^2 (|{\hat \mu}|+2 {\hat c}_B)\bigg]\ .
\end{align}
The equation of motion that follows by varying \eqref{CBzeroTmain}
w.r.t.~$\tl{\cS}$ is
\begin{equation}\label{cSeom}
 \frac{{c}_B}{|\lambda_B|} \left(\beta^{-1}\tilde{\mc{S}}-\tfrac{1}{2}{m}_{B}^{\text{cri}}\right)  - \Big(\beta^{-1}\tilde{\mc{S}}-\tfrac{1}{2} {m}_{B}^{\text{cri}}\Big)^2 = 0\ ,
\end{equation}
and the equation of motion from varying $c_B$ is
\begin{equation}\label{cbeom}
  \left\{ \renewcommand{\arraystretch}{2} \begin{array}{cc} \displaystyle \frac{c_B}{|\lambda_B|}\left(\frac{c_B}{2} -\tfrac{1}{2}m_B^{\rm cri}\right) -  \Big(\beta^{-1} {\tl \cS} - \tfrac{1}{2}m_B^{\rm cri}\Big)^2 = 0 &\quad c_B > |\mu|\ , \\ \displaystyle\frac{c_B}{|\lambda_B|}\left(\frac{|\mu|}{2} - \frac{1}{2|\lambda_B|} (|\mu|-c_B) -\tfrac{1}{2}m_B^{\rm cri} \right) -  \Big(\beta^{-1} {\tl \cS} -\tfrac{1}{2}m_B^{\rm cri}\Big)^2 = 0 &\quad c_B < |\mu|\ . \end{array}\right.
\end{equation}
Comparing \eqref{cSeom} and \eqref{cbeom}, we see that $\tilde{\cS}$ is
given on-shell by
\begin{equation}\label{cSonsh}
 \beta^{-1} \tl\cS = \left\{\renewcommand{\arraystretch}{2} \begin{array}{cc} \displaystyle\frac{{c}_B}{2} & \quad c_B > |\mu|\ , \\ \displaystyle\frac{ |\mu|}{2} - \frac{1}{2|\lambda_B|}(|\mu|-c_B) & \quad c_B < |\mu|\ .\end{array} \right.
\end{equation}
Plugging this back into the off-shell free energy \eqref{CBzeroTmain},
we get, for $c_B < |\mu|$,
\begin{multline}\label{fecond}
  F_{\rm condensed} = \frac{N_B}{12\pi}\Big[\lambda_B^2 (m_B^{\rm cri})^3 + 3 |\lambda_B|(1-|\lambda_B|) (m_B^{\rm cri})^2 |\mu| \\ + 3 (1 - |\lambda_B|)^2 m_B^{\rm cri} |\mu|^2 - (2-|\lambda_B)(1-|\lambda_B|) |\mu|^3\Big]\ ,
\end{multline}
and for $c_B > |\mu|$,
\begin{multline}\label{feuncond}
  F_{\rm uncondensed} = \frac{N_B}{12\pi}\Big[\lambda_B^2 (m_B^{\rm
    cri})^3 + 3 |\lambda_B|(1-|\lambda_B|) (m_B^{\rm cri})^2 c_B \\ +
  3 (1 - |\lambda_B|)^2 m_B^{\rm cri} c_B^2 -
  (2-|\lambda_B)(1-|\lambda_B|) c_B^3\Big]\ .
\end{multline}
Plugging the on-shell value of $\tl\cS$ into the equation of motion
for $c_B$ \eqref{cbeom}, we get
\begin{equation}\label{cbeom1}
  \left\{ \renewcommand{\arraystretch}{2} \begin{array}{cl} \displaystyle \left(\frac{2 - |\lambda_B|}{|\lambda_B|}c_B + m_B^{\rm cri}\right) \left(c_B - m_B^{\rm cri}\right) = 0 &\qquad c_B > |\mu|\ , \\ \displaystyle c_B^2 = |\lambda_B|^2 \left(\left(1 - \frac{1}{|\lambda_B|}\right)|\mu| - m_B^{\rm cri} \right)^2  &\qquad c_B < |\mu|\ . \end{array}\right.
\end{equation}
In the uncondensed case (corresponding to $c_B > |\mu|$), there are
two different solutions to the equation of motion for $c_B$. These two
saddle points correspond to the uncondensed Higgsed and uncondensed
unHiggsed phases of the free energy:
\begin{align}\label{uncond}
  \text{uncondensed, unHiggsed}:&\quad c_B = m_B^{\rm cri}\ ,\nonumber\\
  \text{uncondensed, Higgsed}:&\quad c_B = -\frac{|\lambda_B|}{2-|\lambda_B|} m_B^{\rm cri}\ .
\end{align}
These phases occur for the following ranges of $m_B^{\rm cri}$ that we
obtain from the restriction $c_B > |\mu|$:
\begin{align}\label{uncondmbcri}
  \text{uncondensed, unHiggsed}:&\quad m_B^{\rm cri} > |\mu|\ ,\nonumber\\
  \text{uncondensed, Higgsed}:&\quad m_B^{\rm cri} < -\frac{2-|\lambda_B|}{|\lambda_B|}|\mu|\ .
\end{align}
On the contrary, in the condensed phase (corresponding to
$c_B < |\mu|$), there seems to be a unique solution:
\begin{equation}\label{cond}
  \text{condensed}: \quad c_B = |\lambda_B|\left|\left(1-\frac{1}{|\lambda_B|} \right)|\mu| - m_B^{\rm cri}\right|\ .
\end{equation}
This phase exists for the following range of $m_B^{\rm cri}$ obtained
from the restriction $c_B < |\mu|$:
\begin{equation}\label{mbcricond}
  -\frac{2-|\lambda_B|}{|\lambda_B|}|\mu|\ <\ m_B^{\rm cri}\ <\ |\mu|\ .
\end{equation}
There are two possibilities of reaching the condensed phase from an
uncondensed phase by raising the chemical potential above $c_B$:
either from the uncondensed unHiggsed side or from the uncondensed
Higgsed side. Recall that the excitations in the unHiggsed phase are
scalars while those in the Higgsed phase are $W$-bosons. Depending on
the side we approach from, the description of the condensed phase is
either in terms of condensed $W$ bosons or condensed scalars. This can
be seen in the formula for $c_B$ \eqref{cond} which is non-analytic in
the coupling constant $m_B^{\rm cri}$ due to the absolute value on the
right hand side. The sign of the quantity inside the modulus on the
right hand side distinguishes between these two cases:
\begin{align}
  \text{condensed scalars}:&\quad m_B^{\rm cri}\ >\ -\frac{1 - |\lambda_B|}{|\lambda_B|}|\mu|\ ,\nonumber\\
  \text{condensed $W$ bosons}:&\quad m_B^{\rm cri}\ <\ -\frac{1 - |\lambda_B|}{|\lambda_B|} |\mu|\ .
\end{align}
Though the description of the condensed branch in terms of condensed
scalars or condensed $W$ bosons seems manifestly different, it turns
out that there is no real distinction between these two
descriptions. This can be seen from the free energy in the condensed
phase \eqref{fecond} which does not depend on the variable $c_B$ at
all! Thus, the non-analyticity in the equation of motion for $c_B$
does not affect the free energy in the condensed phase.

Plugging the solutions for each of the uncondensed phases into
\eqref{feuncond}, we find the free energy in the unHiggsed and Higgsed
phases to be
\begin{align}\label{uncondfin}
  F_{\rm unHiggsed} & =  \frac{N_B}{12\pi} (m_B^{\rm cri})^3\ ,\nonumber\\
  F_{\rm Higgsed} &= \frac{N_B}{12\pi} \frac{|\lambda_B|^2}{(2-|\lambda_B|)^2} (m_B^{\rm cri})^3\ .
\end{align}

\subsection{Phase diagram} 

\begin{figure}[!h]
	\centering
	\begin{tikzpicture}
	\coordinate (L) at (-7,0);
	\coordinate (s34) at (-3,0) ;
	\coordinate (s23) at (0,0) ;
	\coordinate (s12) at (3,0) ;
	\coordinate (R) at (7,0) ;
	\coordinate (O) at (0.3,0) ;
	\draw[latex-,black!60!green, very thick] (L) to [edge label={\text{Higgsed Phase}}, edge label'={$|\mu|<c_B$}] (s34) ;
	\draw[violet, very thick] (s34) to [edge label={\text{condensed Phase}}, edge label'={$|\mu|>c_B$}] (s12) ;
	\draw[blue, very thick, -latex] (s12) to [edge label={\text{un-Higgsed Phase}}, edge label'={$|\mu|<c_B$}] (R) ;
	\filldraw[black] (s34) circle (2pt) ;
	\draw (s34)+(0,-0.4) node {$s_{2}$}; 
	\filldraw[black] (s12) circle (2pt) ;
	\draw (s12)+(0,-0.4) node {$s_{1}$}; 
	\draw (R) node[right] {$m_{B}^{\text{cri}}$} ;
        \filldraw[black] (s23) circle (2pt) ;
	\end{tikzpicture}
	\caption{Phase diagram of critical boson theory as a function of
          $m_B^{{\rm cri}}$ at fixed $\mu$. At  $s_{1}=|\mu|$ and
          $s_{2}=-|\mu|\big(\frac{2-|\lambda_B|}{|\lambda_B|}\big)$ the system undergoes a sharp second order phase transition from a condensed to an uncondensed phase.
          The black dot inside the condensed phase marks the point
          where the description of the physically unique condensed
          phase changes from one of a phase of condensed scalars to a
          phase of condensed $W$ bosons; this occurs at $m_B^{\rm cri} =
          -|\mu|\big(\frac{1-|\lambda_B|}{|\lambda_B|}\big)$. }
	\label{mainCBphasediagint|mu|>0}
\end{figure}
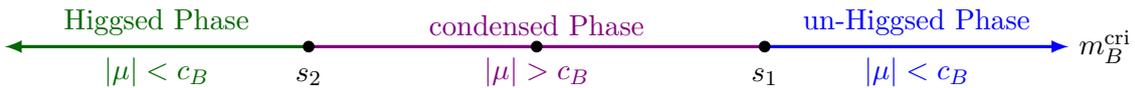

Note that the conditions \eqref{uncondmbcri} and \eqref{mbcricond} on
$m_B^{\rm cri}$ are mutually non-overlapping. Together they give a
single cover of the range of masses
\begin{equation}
  -\infty < m_B^{{\rm cri}} < \infty\ .\nonumber
\end{equation}
At any given value of $|\mu|$ and $m_B^{\rm cri}$, there exists
exactly one solution of the gap equations. At every value of
parameters, therefore, the system has a unique phase.  It follows that
the phase diagram of the system is as depicted in Figure
\ref{mainCBphasediagint|mu|>0}.
As depicted in Figure \ref{mainCBphasediagint|mu|>0}, our system
undergoes a sharp second order phase transition at the edges of the range
\eqref{mbcricond}. Within the range \eqref{mbcricond}, our system is
in a compressible phase, i.e.~the charge is a non-trivial function of
the chemical potential. Outside the range \eqref{mbcricond}, the
system is in an incompressible phase: its charge is independent of
$\mu$ (and vanishes in particular).

Within the compressible phase, the charge (per unit volume) as a
function of $\mu$ is given by the formula
\begin{equation}\label{cfm} 
Q=  \frac{N_B}{4\pi}\sgn(\mu)  \Big(\frac{1-|\lambda_B|}{|\lambda_B|}\Big) \big(\mu^2 - c_B^2 \big) \ .
\end{equation} 
Note that, the charge vanishes at both the edges of the condensed
range for $m_B^{\rm cri}$ \eqref{mbcricond}, but that its first
derivative is non-vanishing at these boundary points. This
discontinuity in the second derivative of the free energy demonstrates
that our system undergoes a second order phase transition at the edges
of the range \eqref{mbcricond}. The number of single particle states
(per unit volume) in the condensed phase, upto the energy $|\mu|$ is
given by
\begin{equation}\label{sps} 
\int \frac{d^2 k}{(2\pi)^2} =\int_{c_B}^{|\mu|} \frac{\epsilon d\epsilon}{2\pi} =\frac{1}{4\pi}(\mu^2-c_B^2)\ .
\end{equation}
Thus, we see that \eqref{cfm} can be thought of as the product of the
number of single particle states \eqref{sps} and the occupation number ${\bar n}$ listed in 
\eqref{avon}. 

Plugging in the expression for $c_B$ as a function of $\mu$ 
into this expression (\eqref{cond}) we find an explicit expression for the charge as a function of UV parameters and 
the chemical potential:  $\mu$-dependence, the charge can be
written as
\begin{align}\label{cfmexpl} 
Q 
& = \frac{N_B}{4\pi}\sgn(\mu)   \Theta(|\mu|-c_B)\, ( 1-|\lambda_B|) \Big( (2 -|\lambda_B|) \mu^2 -2(1-|\lambda_B| ){m}_B^{\text{cri}} |\mu|  -|\lambda_B|({m}_B^{\text{cri}} )^2 \Big)\ .
\end{align}
It is clear from \eqref{cfmexpl} that the charge $Q$ is a quadratic
function of $\mu$.  The compressibility of this phase is given by 
\begin{equation}\label{Qdervmu} 
  \frac{\partial Q}{\partial \mu} = \frac{N_B}{2\pi} \big( 1-|\lambda_B|\big)  \Theta(|\mu|-c_B)  \Big( (2 -|\lambda_B|) |\mu| -(1-|\lambda_B| ){m}_B^{\text{cri}} \Big)\ .
\end{equation} 
Note that $\frac{\partial Q}{\partial \mu} $ is always positive
in the condensed phase and vanishes at the edges.

\section{Quasi-bosonic theories}\label{pdrbcf}

In this section, we consider the detailed phase structure of the
quasi-bosonic theories viz.~the critical fermion and the regular
boson. We explicitly work in the bosonic setting and use the
Bose-Fermi duality map discussed previously to obtain the analogous
results for the critical fermion. The off-shell free energy for the
regular boson at zero temperature is given by \eqref{RBzeroT} and is
reproduced below:
\begin{align}\label{RBzeroTmain}
  F_{\rm RB}(c_B, \sigma_B,  \tl\cS)
  &= \frac{ N_B }{6\pi} \Bigg[-3 { c}_B^2 { \sigma}_B + 4\lambda_B^2 { \sigma}_B^3 +  3\left( { m}_B^2 { \sigma}_B  + 2\lambda_B { b}_4 { \sigma}_B^2 +  x^B_6  \lambda_B^2 { \sigma}_B^3\right) \nonumber\\
  &\qquad\qquad - 4\lambda_B^2 (\beta^{-1}{\tl \cS} + { \sigma}_B)^3 + 6|\lambda_B|{ c}_B (\beta^{-1}{\tl \cS} + { \sigma}_B)^2 \nonumber\\
  &\qquad\qquad -{ c}_B^3 + \frac{1}{2} \Theta(|{ \mu}| - { c}_B) \left(1 - \frac{1}{|\lambda_B|}\right)
    (|{\ \mu}| - { c}_B)^2 (|{ \mu}| + 2 { c}_B) \Bigg]\ .
\end{align}
The variation w.r.t.~$c_B$ and $\tl\cS$ give the equations
\begin{align}\label{fieldeq}
  &\beta\frac{\partial}{\partial {\tl \cS}}:\quad c_B (\beta^{-1} {\tl \cS} + \sigma_B) - |\lambda_B|(\beta^{-1}{\tl \cS} +  \sigma_B)^2 = 0\ ,\nonumber\\
  & \frac{\partial}{\partial c_B}:\ \left\{ \renewcommand{\arraystretch}{2} \begin{array}{cc} c_B\left(\frac{c_B}{2} + \sigma_B\right) - |\lambda_B| (\beta^{-1} \tl\cS + \sigma_B)^2 = 0 &\quad c_B > |\mu|\ , \\ c_B\left(\frac{c_B}{2} + \frac{1}{2}\left(1-\frac{1}{|\lambda_B|}\right) (|\mu|-c_B) + \sigma_B\right) - |\lambda_B| (\beta^{-1} {\tl \cS} + \sigma_B)^2 = 0 &\quad c_B < |\mu|\ . \end{array}\right.
\end{align}
Consistency of the two equations of motion gives the following
on-shell condition for ${\tl \cS}$ in the zero temperature limit:
\begin{equation}\label{defScb}
 \beta^{-1} \tl\cS = \left\{\renewcommand{\arraystretch}{2} \begin{array}{cc} \displaystyle\frac{{c}_B}{2} & \quad c_B > |\mu|\ , \\ \displaystyle\frac{ |\mu|}{2} - \frac{1}{2|\lambda_B|}(|\mu|-c_B) & \quad c_B < |\mu|\ .\end{array} \right.
\end{equation}
We substitute the above expression for $\tl\cS$ back into the
off-shell free energy \eqref{RBzeroTmain} and the $c_B$ equation of
motion in the second line of \eqref{fieldeq} respectively. The
off-shell free energy becomes
\begin{equation} \label{intefe}
F_{\rm RB}(c_B,\sigma_B) = \left\{\renewcommand{\arraystretch}{2}\begin{array}{cc} \renewcommand{\arraystretch}{1.2} \begin{array}{l} \displaystyle \frac{N_B}{2\pi } \Bigg[
x_6^B \lambda_B^2 {\sigma}_B^3 + 2\big( \lambda_B{b}_4 + |\lambda_B|(1-|\lambda_B|)c_B \big) {\sigma}_B^2  \\  + \big({m}_B^2 - (1-|\lambda_B|)^2 c_B^2){\sigma}_B - \tfrac{1}{6}(1-|\lambda_B|)(2-|\lambda_B|)c_B^3 \Bigg] \end{array} & \quad c_B >  |\mu|\ ,\\                                                                   \renewcommand{\arraystretch}{1.2} \begin{array}{l} \displaystyle \frac{N_B}{2\pi } \Bigg[
x_6^B \lambda_B^2 {\sigma}_B^3+2\big( \lambda_B{b}_4 +|\lambda_B|(1-|\lambda_B|)|\mu| \big) {\sigma}_B^2  \\  +\big({m}_B^2-(1-|\lambda_B|)^2|\mu|^2){\sigma}_B-\tfrac{1}{6}(1-|\lambda_B|)(2-|\lambda_B|)|\mu|^3 \Bigg] \end{array} & \quad  c_B <  |\mu|\ ,\end{array}\right.
\end{equation}
and the equation of motion for $c_B$ becomes
\begin{equation}\label{cbeq}
 \frac{\partial}{\partial c_B}:\ \left\{ \renewcommand{\arraystretch}{2.5} \begin{array}{cc} \displaystyle\left(\frac{2 - |\lambda_B|}{2|\lambda_B|}c_B -  \sigma_B\right) \left(\frac{c_B}{2} + \sigma_B\right) = 0 &\quad c_B > |\mu|\ , \\ \displaystyle\left(\frac{c_B}{2|\lambda_B|}\right)^2 - \left(\sigma_B + \frac{|\mu|}{2}\left(1 - \frac{1}{|\lambda_B|}\right)\right)^2 = 0 &\quad c_B < |\mu|\ . \end{array}\right.
\end{equation}
Each of the equations in \eqref{cbeq} is quadratic in $c_B$. It may
thus seem that the function $c_B(\sigma)$ obtained from the solution
of \eqref{cbeq} is multivalued with upto four branches. In actuality
this is not the case as we now explain. Solutions to the equation in
the second of \eqref{cbeq} are valid only when
\begin{equation}\label{whenisec}
  0 \leq c_B \leq |\mu|\ .
\end{equation} 
It follows that the second equation in \eqref{cbeq} has at most one
valid solution - the so called condensed phase solution - given by
\begin{equation}\label{cbbb}
  c_B = 2|\lambda_B| \left|\sigma_B + \frac{|\mu|}{2}\left(1 - \frac{1}{|\lambda_B|}\right)\right|\ .
\end{equation}
This solution is valid if and only if $\sigma_B$ satisfies
\begin{equation}\label{sigmacondensed}
\text{condensed}:\quad  - \frac{|\mu|}{2} < \sigma_B < \frac{2-|\lambda_B|}{2|\lambda_B|}|\mu|\ .
\end{equation}
The expression for $c_B$ in \eqref{cbbb} as a function of $\sigma_B$
is not analytic at
$\sigma_B = -\frac{|\mu|}{2}\left(1 -
  \frac{1}{|\lambda_B|}\right)$. Intuitively, this non-analyticity is
a consequence of the fact that the solution migrates from the
condensed unHiggsed to the condensed Higgsed phase at this value of
$\sigma_B$.  Remarkably enough, however, the free energy in the second
line of \eqref{intefe} is independent of $c_B$, and so the
non-analyticity of $c_B$ as a function of $\sigma_B$ does not
translate into a non-analyticity of the free energy as a function of
$\sigma_B$.

Notice that the above non-analytic point of $c_B$ is at $\sigma_B = 0$
when the chemical potential is zero. This was the point at which the
free energy was non-analytic and there was a phase transition between
the unHiggsed and Higgsed phases. The point
$\sigma_B = -\frac{|\mu|}{2}\left(1 - \frac{1}{|\lambda_B|}\right)$ is
the analog of such a transition when we turn on a chemical
potential. The discussion in the previous paragraph indicates that, at
least in the large $N$ limit, there is no distinction between the
condensed unHiggsed and the condensed Higgsed phases; we just have one
single condensed phase. The contribution of this phase to the free
energy as a function of $\sigma_B$ is simply the second line of
\eqref{intefe}. This expression for the free energy is valid when
\eqref{sigmacondensed} is obeyed.

There are clearly two different branches of solutions to the first
line of \eqref{cbeq}; these are the solutions
\begin{equation}\label{cbrange}
\text{unHiggsed}:\ c_B = -2\sigma_B\ ,\qquad \text{Higgsed}:\ c_B = \frac{|\lambda_B|}{2-|\lambda_B|} 2 \sigma_B\ .
\end{equation}
These solutions are only valid when $c_B > |\mu|$ which translates to
the following conditions on $\sigma_B$:
\begin{equation}\label{sigmarange}
\text{unHiggsed}:\ \sigma_B < -\frac{|\mu|}{2}\ ,\qquad \text{Higgsed}:\ \sigma_B > \frac{2-|\lambda_B|}{2|\lambda_B|} |\mu|\ .
\end{equation}
Plugging the solutions \eqref{cbrange} into the first line of
\eqref{intefe} yields the free energy as a function of $\sigma_B$ in
the ranges \eqref{sigmarange}.

Notice that the ranges of $\sigma_B$ in \eqref{sigmarange} are
non-overlapping, and are also exactly complementary to the range
\eqref{sigmacondensed}. It follows that the three ranges listed in
\eqref{sigmacondensed} and \eqref{sigmarange} together cover the
$\sigma_B$ line in a non-overlapping manner. The transitions between
the three ranges take place at the points
\begin{align}\label{sigmanonan}
\text{condensed $\leftrightarrow$ unHiggsed}&:\quad \sigma_B = -\frac{|\mu|}{2}\ ,\nonumber\\
\text{condensed $\leftrightarrow$ Higgsed}&:\quad \sigma_B = \frac{2-|\lambda_B|}{2|\lambda_B|} |\mu|\ .
\end{align}
We may set $|\mu| = 1$ in the above expressions since it just
corresponds to a choice of scale and we will do so in the intermediate
steps of the analysis below. However, it will be useful in the end to
restore the $|\mu$| dependence; for this we just rescale all
dimensionful quantities with appropriate powers of $|\mu|$:
\begin{equation}
  m_B^2 \to \frac{m_B^2}{|\mu|^2}\ ,\quad b_4 \to \frac{b_4}{|\mu|}\ ,\quad \sigma_B \to \frac{\sigma_B}{|\mu|}\ ,\quad F_{\rm RB} \to \frac{F_{\rm RB} }{|\mu|^3}\ .
\end{equation}
In summary, the off-shell free energy as a function of the single
variable $\sigma_B$ is given by piecewise analytic expressions in the
three ranges \eqref{sigmacondensed} and \eqref{sigmarange}:
\begin{align} \label{finalu} 
  &U_{\rm eff}(\sigma_B) = \frac{N_B}{2\pi} \times\nonumber \\
  & \left\{\renewcommand{\arraystretch}{2.5} \begin{array}{cc} (x_6^B - \phi_u) \lambda_B^2 \sigma_B^3 + 2\lambda_B b_4 \sigma_B^2 + m_B^2 \sigma_B &  \displaystyle\sigma_B < -\frac{1}{2}\ , \\ \renewcommand{\arraystretch}{1.4}\begin{array}{l} x_6^B \lambda_B^2 \sigma_B^3 + 2 \lambda_B \Big( b_4 + (\sgn(\lambda_B) - \lambda_B)\Big) \sigma_B^2 \\ + \Big(m_B^2 - (1-|\lambda_B|)^2 \Big) \sigma_B - \frac{1}{6} (1-|\lambda_B|) (2- |\lambda_B|) \end{array} &\quad  \displaystyle-\frac{1}{2} < \sigma_B < \frac{2 - |\lambda_B|}{2|\lambda_B|}\ , \\ (x_6^B - \phi_h) \lambda_B^2 \sigma_B^3 + 2\lambda_B b_4 \sigma_B^2 + m_B^2 \sigma_B & \ \displaystyle\sigma_B > \frac{2 - |\lambda_B|}{2|\lambda_B|}\ . \end{array} \right.
\end{align}
The quantities $\phi_h$ and $\phi_u$ are functions of $|\lambda_B|$
given by
\begin{equation}\label{phiuh}
  \phi_h = \frac{4}{3} \left(\frac{1}{(2-|\lambda_B|)^2} - 1\right)\ ,\quad \phi_u = \frac{4}{3} \left(\frac{1}{\lambda_B^2} - 1\right)\ ,
\end{equation}
and satisfy, for all $|\lambda_B| \leq 1$,
\begin{equation}
  \phi_h < 0 < \phi_u\ .
\end{equation}
Recall that the quantity $U_{\rm eff}(\sigma_B)$ has a clear physical
interpretation. The variable change in equation \eqref{physintsig} given by
$\sigma_B = \frac{2\pi}{N_B} \langle\bar\phi \phi\rangle$ turns the
free energy into the quantum effective potential for the lightest
gauge invariant operator ${\bar \phi} \phi$ of our theory (see
\cite{Dey:2018ykx}). We frequently refer to $U_{\rm eff}(\sigma_B)$ as
the effective potential while keeping in mind the above fact.

Notice that $U_{\rm eff}(\sigma_B)$ presented in \eqref{finalu} is a
piecewise cubic function of $\sigma_B$. Independently in each of the
three intervals in \eqref{finalu}, the equation
$U'_{\rm eff}(\sigma_B)=0$ is quadratic in $\sigma_B$. Real solutions
to the quadratic equations exist only when their discriminants are
positive. They first start existing on the curves along which the
discriminants vanish i.e.~the parabolas
\begin{align}\label{globaldisc}
  & D_u \equiv 4 \lambda_B^2 b_4^2 - 3 m_B^2 \lambda_B^2 (x_6^B - \phi_u) = 0\ ,\nonumber\\
  & D_c \equiv 4  \Big(\lambda_B b_4 + |\lambda_B|(1 - |\lambda_B|)\Big)^2 - 3 \lambda_B^2 x_6^B \Big(m_B^2 - (1-|\lambda_B|)^2\Big) = 0\ ,\nonumber\\
  & D_h \equiv 4 \lambda_B^2 b_4^2 - 3 m_B^2 \lambda_B^2 (x_6^B - \phi_h) = 0\ .    
\end{align}
The real solutions that are created are double zeroes of
$U_{\rm eff}'(\sigma_B)$\footnote{For a generic potential
  $U(\sigma)=A\sigma^3+B\sigma^2+C\sigma +D$, the extrema of this
  function are given by $U'(\sigma)=3A\sigma^2 + 2B\sigma +
  C=0$. The solutions of this quadratic equation are
  $$\sigma_{\pm}=\frac{-B \pm \sqrt{B^2-3AC}}{3A}\ .$$ When the
  discriminant $B^2-3AC$ vanishes, this simplifies to
  $\sigma_{\pm} = -\frac{B}{3A} = -\frac{C}{B}$.} and are located on
the $\sigma_B$ line respectively at
\begin{align}\label{globaldiscsol}
& D_u:\quad \sigma_B=  -\frac{2\lambda_B b_4}{3(x_6^B-\phi_u)\lambda_B^2}\ ,\nonumber\\
& D_c:\quad \sigma_B= -\frac{2\lambda_B \big(b_4+(\sgn(\lambda_B)-\lambda_B)\big)}{3x_6^B\lambda_B^2}\ ,\nonumber\\
& D_h:\quad \sigma_B=-\frac{2\lambda_B b_4}{3(x_6^B-\phi_h)\lambda_B^2}\ .    
\end{align}
Crossing any of the curves listed in \eqref{globaldisc} signals the
(dis)appearance of new extrema in the corresponding branch of the
effective potential $U_{\rm eff}(\sigma_B)$.  Of course the
discriminant curves \eqref{globaldisc} are physically relevant only
when the solutions \eqref{globaldiscsol} lie within their respective
domains of validity \eqref{sigmacondensed}, \eqref{sigmarange}. In
other words, only certain parts of the discriminant curves
\eqref{globaldisc} are relevant for the phase diagram.

\subsection{Strategy to obtain the phase diagram}\label{strategy}

In the rest of this paper, we work out the phase diagram of the theory
that follows from \eqref{finalu}. We split the analysis into four
cases depending on the range of the marginal parameter $x_6^B$:
\begin{alignat}{2}\label{x6case}
&\text{Case A}:\ x_6^B < \phi_h\ ,\qquad   &&\text{Case C}:\  0 < x_6^B < \phi_u\ ,\nonumber\\ 
&\text{Case B}:\ \phi_h < x_6^B < 0\ ,\qquad   &&\text{Case D}:\   \phi_u < x_6^B \ .
\end{alignat}
Looking at the potential \eqref{finalu}, it is clear that the
potential is unbounded below for large and positive $\sigma_B$ when
$x_6^B < \phi_h$ (Case A) and is unbounded below for large and
negative $\sigma_B$ when $x_6^B > \phi_u$ (Case D). When $x_6^B$ is in
the range $\phi_h < x_6^B < \phi_u$ corresponding to Cases B and C,
the potential is bounded below for both positive and negative
$\sigma_B$. Thus, the potential is stable for cases B and C while it
is unstable for cases A and D. We treat each of the cases in
\eqref{x6case} separately below. The stable cases are addressed in
Section \ref{phased} while the analysis for the unstable cases is
pushed to the appendix and is presented in Appendix \ref{unstable}.

The
phase diagram of the theory is two-dimensional and can be conveniently
described by the two dimensionless ratios
\begin{equation}
  \frac{m_B^2}{|\mu|^2}\quad\text{and}\quad  \frac{\lambda_B b_4}{|\mu|}\ .
\end{equation}
We have one phase diagram for every range of $x_6^B$ in
\eqref{x6case}. In drawing these diagrams, we place the variable
$\frac{\lambda_B b_4}{|\mu|}$ along the $Y$-axis and
$\frac{m_B^2}{|\mu|^2}$ along the $X$-axis. The phase of the theory at
any particular point on this diagram is decided by the dominant
minimum of the potential $U_{\rm eff}(\sigma_B)$ with the values of
the parameters $m_B^2$ and $\lambda_B b_4$ fixed by the coordinates of
the point on the phase diagram.

We obtain the detailed structure of the potential for each range of
$x_6^B$ at each point of the phase diagram. Our strategy is as
follows. First, recall that there are two values of $\sigma_B$ in the
neighbourhood of which the potential \eqref{finalu} is
non-analytic. These points, which we designate $\ell$ and $r$, are
\begin{equation}\label{defnonan}
\ell:\quad   \sigma_B = -\frac{1}{2}\ ,\qquad r:\quad  \sigma_B = \frac{2 - |\lambda_B|}{2|\lambda_B|}\ .
\end{equation}
The point $\ell$ is the junction between the unHiggsed and the
condensed branches of the potential, while the point $r$ is the
junction between the condensed and the Higgsed branches of the
potential.

In order to work out the full phase diagram we study the behaviour of
the potential near each of the non-analytic points above. To study the
potential near a non-analytic point, say $\sigma_B = \ell$, we shift
our origin in $\sigma_B$ to that point. Then, we study the behaviour
of the effective potential \eqref{finalu} as a function of $m_B^2$ and
$\lambda_B b_4$ by pretending that the effective potential has only
the unHiggsed and the condensed branches which meet at
$\sigma_B = \ell$ i.e.~ignoring the existence of the Higgsed branch
that, in reality, cuts off the condensed phase at $\sigma_B = r$. This
description is not expected to hold for all values of $m_B^2$ and
$\lambda_B b_4$ but only for a small region in the phase diagram where
the $\sigma_B = r$ cutoff is not important. Similarly, we study the
behaviour of the potential near $\sigma_B = r$ by ignoring the
existence of the unHiggsed branch of the potential which cuts off the
condensed branch at $\sigma_B = \ell$.

The two exercises above furnish two `local patches' of the behaviour
of the effective potential on the phase diagram. We then appeal to the
simplicity of the form of the effective potential to interpolate
between these two patches and obtain a global picture of the behaviour
of the effective potential. It is then a simple exercise to determine
the dominant minimum of the potential and thus the phase structure of
the theory. We now provide the expressions for the potential in the
neighbourhood of the two non-analytic points in $\sigma_B$ described
above in \eqref{defnonan}.

\subsection{Effective potential in the local patches around
  non-analytic points} \label{localpot}

\subsubsection{Near \texorpdfstring{$\ell$: $\sigma_B = -\tfrac{1}{2}$}{sigma=l}}\label{nearell}

First, let $\sigma_B'= \sigma_B + \frac{1}{2}$. Note that
$\sigma_B'=0$ when $\sigma_B=\ell $. In the local patch around
$\sigma_B' = 0$, we have
\begin{align}\label{potl}
&U_{\rm eff}(\sigma_B') = \frac{N_B}{2\pi} \times\left\{\renewcommand{\arraystretch}{2.7} \begin{array}{cc} \renewcommand{\arraystretch}{1.4} \begin{array}{r}(x_6^B - \phi_u) \lambda_B^2 \sigma_B'^{\,3} + 2\Big(\lambda_B b_4  - \tfrac{3}{4} \lambda_B^2 (x_6^B - \phi_u)\Big) \sigma_B'^{\,2} \qquad \\  + \Big(m_B^2 - 2 \lambda_B b_4 + \frac{3}{4} \lambda_B^2 (x_6^B - \phi_u) \Big)\sigma_B' \end{array} &\quad  \displaystyle\sigma_B' < 0\ , \\ \renewcommand{\arraystretch}{1.4}\begin{array}{r} x_6^B \lambda_B^2 \sigma_B'^{\,3} + 2\Big(\lambda_B b_4 - \tfrac{3}{4} \lambda_B^2 (x_6^B - \phi_u) - (1-|\lambda_B|) \Big) \sigma_B'^{\,2} \qquad \\  + \Big(m_B^2 - 2 \lambda_B b_4  + \frac{3}{4}  \lambda_B^2 (x_6^B - \phi_u)\Big)\sigma_B' \end{array} &\quad  \displaystyle \sigma_B' > 0\ . \end{array} \right.
\end{align}
There is an additional common constant to both the expressions above
which is unimportant for understanding the behaviour near
$\sigma_B' = 0$:
\begin{equation}\label{cprime}
 c' = \frac{1}{8} \left(-4 m_B^2 + 4 \lambda_B b_4  - \lambda_B^2 (x_6^B - \phi_u) \right)\ .
\end{equation}
The potential \eqref{potl} matches the general form of the potential
\ref{tmpot} considered in Appendix \ref{curveplot} which we reproduce
here for convenience:
\begin{equation}\label{genpot}
  U_{\rm eff}(\sigma) = \left\{\renewcommand{\arraystretch}{2}\begin{array}{cc} A \sigma^3 + B \sigma^2 + C \sigma &\qquad \sigma < 0\ , \\ (A + a) \sigma^3 + (B + b) \sigma^2 + C \sigma &\qquad \sigma > 0\ .\end{array}\right.
\end{equation}
The quantities $A$, $a$, $B$, $b$ and $C$ in the present case are given by
\begin{align}\label{ABCl}
  &A = x_6^B - \phi_u\ ,\quad a = \phi_u\ ,\quad B = 2(\lambda_B b_4 - \tfrac{3}{4} \lambda_B^2 (x_6^B - \phi_u))\ ,\nonumber\\
  &b = - (1 - |\lambda_B|)\ ,\quad C = m_B^2 - 2 \lambda_B b_4  + \frac{3}{4}  \lambda_B^2 (x_6^B - \phi_u)\ .
\end{align}

As we have mentioned above, we use $\lambda_B b_4= 0$ and $m_B^2 =0$
as the $X$-axis and $Y$-axis for the full phase diagram. However, the
local coordinate axes for the potential \eqref{potl} are different are
obtained by setting the quantities $B$ and $C$ in \eqref{ABCl} to
zero:
\begin{equation}\label{llocalaxes}
  \text{$X'$-axis}:\quad \lambda_B b_4 = \frac{3}{4}\lambda_B^2(x_6-\phi_u)\ ,
  \qquad \text{$Y'$-axis}:\quad m_B^2 - 2 \lambda_B b_4 + \frac{3}{4}\lambda_B^2(x_6-\phi_u) = 0\ .
\end{equation}
The origin of the local coordinate axes above lies at $B = C = 0$,
given by
\begin{equation}\label{lorigin}
O_\ell:\quad  m_B^2 = \lambda_B b_4 = \frac{3}{4}\lambda_B^2(x_6-\phi_u)\ .
\end{equation}
There is another point on this phase diagram in which the effective
potential \eqref{potl} becomes particularly simple: the point $L$ in
the phase diagram at which $B = -b$ and $C = 0$ in \eqref{ABCl}:
\begin{equation}\label{Ldef}
L:\left\{ \renewcommand{\arraystretch}{1.4}\begin{array}{c} m_B^2 = \tfrac{3}{4}\lambda_B^2 (x^B_6 - \phi_u) + 2(1 - |\lambda_B|)\ ,\\ \lambda_B b_4 = \tfrac{3}{4}\lambda_B^2 (x^B_6 - \phi_u) + (1 - |\lambda_B|)\ .\end{array}\right.
\end{equation}
Borrowing results from the general analysis of Appendix
\ref{curveplot}, we see that this `local patch' of the phase diagram
has the following special curves.
\begin{enumerate}
\item The $Y'$-axis \eqref{llocalaxes} is divided into three segments
  by the local origin $O_\ell$ \eqref{lorigin} and the point $L$
  \eqref{Ldef}. These segments correspond to $B > -b$, $0 < B < -b$
  and $B < 0$ in \eqref{ABCl}. These translate to the following
  conditions on $\lambda_B b_4$:
  \begin{align}\label{primedef}
    S':&\quad \tfrac{3}{4}\lambda_B^2 (x^B_6 - \phi_u) + (1 - |\lambda_B|)\ \leq\  \lambda_B b_4\ ,\nonumber\\
    N':&\quad \tfrac{3}{4}\lambda_B^2 (x^B_6 - \phi_u)\  \leq\ \lambda_B b_4 \ \leq\ \tfrac{3}{4}\lambda_B^2 (x^B_6 - \phi_u) + (1 - |\lambda_B|)\ ,\nonumber\\
    M':&\quad  \lambda_B b_4 \ \leq\ \tfrac{3}{4}\lambda_B^2 (x^B_6 - \phi_u)\ .
  \end{align}
  Please refer to Section \ref{phasean} for a detailed description of
  these segments and the role they play in understanding the structure
  of the potential.
  
\item We also have the discriminant parabolas \eqref{globaldisc} for
  the unHiggsed and condensed branches of the potential:
  \begin{align}
  D_u:&\quad 4\lambda_B^2 b_4^2  - 3 m_B^2 \lambda_B^2 (x_6^B - \phi_u) = 0\ ,\nonumber\\
  D_c:&\quad 4\left(\lambda_B b_4 + |\lambda_B|(1-|\lambda_B|)\right)^2
        - 3 \lambda_B^2 x_6^B \left(m_B^2 - (1-|\lambda_B|)^2\right) = 0\ .
  \end{align}
  Note that the curve $D_c$ passes through the point $L$ at the
  junction of $S'$ and $N'$; in fact, this parabola is tangent to the
  $Y'$-axis \eqref{llocalaxes} at $L$. In a similar manner the curve
  $D_u$ is tangent to the $Y'$-axis at the local origin $O_\ell$
  \eqref{lorigin} which is at the junction of $N'$ and $M'$.

\item In addition to the curves we have drawn above, there may be
  first order transition lines which have to be determined
  numerically.
\end{enumerate}

\subsubsection{Near \texorpdfstring{$r$: $\sigma_B = \tfrac{2-|\lambda_B|}{2|\lambda_B|}$}{sigma=r}}\label{nearr}

We look at the effective potential \eqref{finalu} in the neighbourhood
of $\sigma_B = \frac{2 - |\lambda_B|}{2|\lambda_B|}$ by setting
$\sigma_B'' = \sigma_B - \frac{2 - |\lambda_B|}{2|\lambda_B|}$ and
expanding the relevant expressions for $U_{\rm eff}$. Again, note that
$\sigma_B'' =0$ when $\sigma_B=r$. In a neighbourhood around
$\sigma_B'' =0$, we have
\begin{align}\label{potr}
  &U_{\rm eff}(\sigma_B'') = \frac{N_B}{2\pi} \times \nonumber\\
  &\left\{\renewcommand{\arraystretch}{2.7} \begin{array}{cc} \renewcommand{\arraystretch}{1.4}\begin{array}{r} x_6^B \lambda_B^2 \sigma_B''^{\,3} + 2\Big(\lambda_B b_4 + \tfrac{3}{4} |\lambda_B|(2-|\lambda_B|) (x_6^B - \phi_h) - \frac{|\lambda_B|(1-|\lambda_B|)}{2 - |\lambda_B|}\Big) \sigma_B''^{\,2} \qquad \\  + \Big(m_B^2 + 2 \frac{2-|\lambda_B|}{|\lambda_B|} \lambda_B b_4  + \frac{3}{4} (2-|\lambda_B|)^2 (x_6^B-\phi_h) \Big)\sigma_B'' \end{array} & \quad \sigma_B'' < 0\ , \\ \renewcommand{\arraystretch}{1.4} \begin{array}{r}(x_6^B - \phi_h) \lambda_B^2 \sigma_B''^{\,3} + 2\Big(\lambda_B b_4 + \tfrac{3}{4} |\lambda_B|(2-|\lambda_B|) (x_6^B - \phi_h)\Big) \sigma_B''^{\,2} \qquad \\  + \Big(m_B^2 + 2 \frac{2-|\lambda_B|}{|\lambda_B|} \lambda_B b_4  + \frac{3}{4} (2-|\lambda_B|)^2 (x_6^B-\phi_h) \Big)\sigma_B'' \end{array}  &\quad  \displaystyle \sigma_B'' > 0\ . \end{array} \right.
\end{align}
Again, there is an additional constant in both expressions which is
unimportant while looking at the behaviour near $\sigma_B'' = 0$:
\begin{equation}\label{cdprime}
  c'' = \frac{2 - |\lambda_B|}{2|\lambda_B|} \left(m_B^2 + 2 \lambda_B b_4 \frac{2 - |\lambda_B|}{2 |\lambda_B|} +  \left(\frac{2 - |\lambda_B|}{2|\lambda_B|}\right)^2 \lambda_B^2 (x_6^B - \phi_h)\right)\ .
\end{equation}
Again, the potential in \eqref{potr} is of the form of the general
potential \eqref{tmpot} in Appendix \ref{curveplot} (also reproduced
in \eqref{genpot} above) with the following identification of the
various parameters:
\begin{align}\label{ABCr}
  &A = x_6^B\ ,\quad a = -\phi_h\ ,\quad B = \lambda_B b_4 + \tfrac{3}{4} |\lambda_B|(2-|\lambda_B|) (x_6^B - \phi_h) - \frac{|\lambda_B|(1-|\lambda_B|)}{2 - |\lambda_B|}\ ,\nonumber\\
  &b =  \frac{|\lambda_B|(1-|\lambda_B|)}{2 - |\lambda_B|}\ ,\quad C = m_B^2 + 2 \frac{2-|\lambda_B|}{|\lambda_B|} \lambda_B b_4  + \frac{3}{4} (2-|\lambda_B|)^2 (x_6^B-\phi_h)\ .
\end{align}
The local coordinate axes for the above potential in the
$(m_B^2, \lambda_B b_4)$ plane are given by setting the quantities $B$
and $C$ to zero:
\begin{align}\label{rlocalaxes}
  &\text{$X''$-axis}:\ \lambda_B b_4 = -\frac{3}{4} |\lambda_B|{2-|\lambda_B|}(x_6^B - \phi_h)-\frac{|\lambda_B|(1-\lambda_B)}{2-|\lambda_B|}\ ,\nonumber\\
  &\text{$Y''$-axis}:\ m_B^2 + 2 \frac{2-|\lambda_B|}{|\lambda_B|} \lambda_B b_4  + \frac{3}{4} (2-|\lambda_B|)^2 (x_6^B-\phi_h) = 0\ .
\end{align}
The origin of the above local coordinate axes is at the point $R$
specified by $B = C = 0$ in \eqref{ABCr}:
\begin{equation}\label{Rdef}
  R:\left\{\ \renewcommand{\arraystretch}{1.4} \begin{array}{c}  m_B^2 = \tfrac{3}{4} (2-|\lambda_B|)^2 (x_6^B - \phi_h) - 2(1-|\lambda_B|)\ ,\\
  \lambda_B b_4 = -\tfrac{3}{4}|\lambda_B|(2-|\lambda_B|)(x_6^B - \phi_h) + \tfrac{|\lambda_B|(1-|\lambda_B|)}{2-|\lambda_B|}\ .\end{array}\right.
\end{equation}
There is another point on this phase diagram in which the effective
potential \eqref{potr} becomes particularly simple. This is the point
$P_r$ at which $B = -b$ and $C = 0$ in \eqref{ABCr}:
\begin{equation}\label{Pr}
  P_r:\left\{\ \renewcommand{\arraystretch}{1.4} \begin{array}{c}  m_B^2 = \tfrac{3}{4} (2-|\lambda_B|)^2 (x_6^B - \phi_h)\ ,\\
  \lambda_B b_4 = -\tfrac{3}{4}|\lambda_B|(2-|\lambda_B|)(x_6^B - \phi_h)\ .\end{array}\right.
\end{equation}
Appealing to the results of Appendix \ref{curveplot}, we see that this
`local patch' of the phase diagram has the following special curves.
\begin{enumerate}
\item The $Y''$-axis \eqref{rlocalaxes} is divided into three segments
  $S''$, $N''$ and $M''$ by the local origin $R$ \eqref{Rdef} and the
  point $P_r$ \eqref{Pr}. The importance of these segments are
  elucidated in Section \ref{phasean} in the Appendix. The segments
  are again specified by the following inequalities on
  $\lambda_B b_4$:
\begin{align}\label{dprimedef}
  &S'':\quad  -\tfrac{3}{4}|\lambda_B|(2-|\lambda_B|) (x_6^B - \phi_h) + \frac{|\lambda_B|(1 - |\lambda_B|)}{2-|\lambda_B|}\ \leq\ \lambda_B b_4 \ ,\nonumber\\
  &N'':\quad 0 \ \leq \ \lambda_B b_4 + \tfrac{3}{4}|\lambda_B|(2-|\lambda_B|) (x_6^B - \phi_h)  \ \leq \  \frac{|\lambda_B|(1 - |\lambda_B|)}{2-|\lambda_B|}\ ,\nonumber\\
  &M'':\quad \lambda_B b_4 \ <\ -\tfrac{3}{4}|\lambda_B|(2-|\lambda_B|) (x_6^B - \phi_h)\ .
\end{align}

\item The parabolas $D_c$ and $D_h$ were discussed around equation
  \eqref{globaldisc} and correspond to the discriminants in the
  condensed and the Higgsed branches.
  \begin{align}
    &D_c :\quad 4\left(\lambda_B b_4 + |\lambda_B|(1-|\lambda_B|)\right)^2 - 3 \lambda_B^2 x_6^B \left(m_B^2 - (1-|\lambda_B|)^2\right) = 0\ , \nonumber\\
    &D_h :\quad  4 \lambda_B^2 b_4^2 - 3 m_B^2 \lambda_B^2 (x_6^B - \phi_h) = 0\ .
  \end{align}
  Note that the curve $D_c$ passes through the local origin $R$ at the
  junction of $S''$ and $N''$; in fact, this parabola is tangent to
  the $Y''$-axis \eqref{rlocalaxes} at the point $R$. In a similar
  manner the curve $D_h$ is tangent to the $Y''$-axis at the point
  $P_r$ which is at junction of $N''$ and $M''$.

\item In addition to the curves we have described above, there may be
  first order transition lines which have to be determined
  numerically.
\end{enumerate}

\subsection{Some broad features of the phase diagram}

As we have mentioned above, we will present the phase diagram of our
theory on a plane whose $Y$-axis is $m_B^2=0$ and whose $X$-axis is
$\lambda_B b_4 = 0$. With these conventions we note the following
general properties of the curves we have defined above.

\subsubsection{Relative locations of the local patches}

The local patches of the phase diagram corresponding to the behaviour
of the potential near $\sigma_B = \ell$ and $\sigma_B = r$ are
described separately in the previous subsection. Now, we try to
understand their relative locations in the full phase diagram.
\begin{enumerate}
\item The $Y'$-axis in \eqref{llocalaxes} has slope
  $\tfrac{1}{2}$. Substituting the coordinates of the point $R$
  \eqref{Rdef} into the expression for $Y'$ in \eqref{llocalaxes}, we
  get
  \begin{equation}
    Y'(m_B^2, \lambda_B b_4) \bigg|_{(m_B^2, \lambda_B b_4) = R} = 3 x^B_6\ .
  \end{equation}
  Thus, the point $R$ lies to the right of the $Y'$-axis when $x^B_6$
  is positive, and to its left when $x^B_6$ is negative.

\item The slope of the $Y''$-axis in \eqref{rlocalaxes} is
  $-\frac{|\lambda_B|}{2(2-|\lambda_B|)}$. Similar to the previous
  case, substituting the coordinates of the point $L$ \eqref{Ldef}
  into the expression for $Y''$, we get
  \begin{equation}
    Y''(m_B^2, \lambda_B b_4) \bigg|_{(m_B^2, \lambda_B b_4) = L} = 3 x^B_6\ .
  \end{equation}
  Thus, the point $L$ lies to the right of $Y''$ when $x^B_6 > 0$ and
  to its left when $x^B_6 < 0$.

  
\item The points $L$ and $R$ are coincident when $x^B_6 = 0$ and are
  located at
  \begin{equation}
    m_B^2 = (1 - |\lambda_B|)^2\ ,\quad \lambda_B b_4 = - |\lambda_B| (1 - |\lambda_B|)\ .
  \end{equation}

\end{enumerate}

\subsubsection{The phase near the origin of the phase diagram}
Let us focus on the origin of the phase diagram
$(m_B^2, \lambda_B b_4) = (0,0)$. The effective potential
\eqref{finalu} becomes
\begin{align} \label{finaluzero}
  &U_{\rm eff}(\sigma_B) = \frac{N_B}{2\pi}\times \left\{\renewcommand{\arraystretch}{2.5} \begin{array}{cc} (x_6^B - \phi_u) \lambda_B^2 \sigma_B^3 & \displaystyle\sigma_B < -\frac{1}{2}\ , \\ \renewcommand{\arraystretch}{1.4}\begin{array}{l} x_6^B \lambda_B^2 \sigma_B^3 + 2 |\lambda_B| (1 - |\lambda_B|) \sigma_B^2 \\  - (1-|\lambda_B|)^2 \sigma_B - \frac{1}{6} (1-|\lambda_B|) (2- |\lambda_B|) \end{array} &\quad  \displaystyle-\frac{1}{2} < \sigma_B < \frac{2 - |\lambda_B|}{2|\lambda_B|}\ , \\ (x_6^B - \phi_h) \lambda_B^2 \sigma_B^3 & \ \displaystyle\sigma_B > \frac{2 - |\lambda_B|}{2|\lambda_B|}\ . \end{array} \right.
\end{align}
The potential is monotonic in the unHiggsed and Higgsed branches since
it contains only the cubic term. The only extrema possible are then in
the condensed branch. Depending on the choice of $x_6^B$, there may be
extrema or not in the range of validity of $\sigma_B$ of the condensed
branch. Thus, the origin of the phase diagram is either in the
condensed phase when an extremum exists or in a `no phase' region if
the potential is monotonic in the condensed branch as well. The same
logic applies if $m_B^2$ and $lambda_B b_4$ are non-zero but are small
when compared to the chemical potential. Thus, a neighbourhood of the
origin lies in the condensed phase or in a `no phase' region.

\subsubsection{The behaviour for
  \texorpdfstring{$|m_B^2| \gg |\mu|^2, |\lambda_B b_4| \gg
    |\mu|$}{mB^2 >> mu^2 and b4 >> mu}}\label{firstord}

When the parameters $m_B^2$ and $\lambda_B b_4$ are much larger in
modulus than $|\mu|^2$ and $|\mu|$ respectively, the effective
potential \eqref{finalu} takes the simple form
\begin{align} \label{largeparamU} 
  &U_{\rm eff}(\sigma_B) \approx \frac{N_B}{2\pi} \times\nonumber \\
  & \left\{\renewcommand{\arraystretch}{2.5} \begin{array}{cc} (x_6^B - \phi_u) \lambda_B^2 \sigma_B^3 + 2\lambda_B b_4 \sigma_B^2 + m_B^2 \sigma_B &  \displaystyle\sigma_B < -\frac{|\mu|}{2}\ , \\ x_6^B \lambda_B^2 \sigma_B^3 + 2 \lambda_B b_4 \sigma_B^2 + m_B^2 \sigma_B &\quad  \displaystyle-\frac{|\mu|}{2} < \sigma_B < \frac{2 - |\lambda_B|}{2|\lambda_B|}|\mu|\ , \\ (x_6^B - \phi_h) \lambda_B^2 \sigma_B^3 + 2\lambda_B b_4 \sigma_B^2 + m_B^2 \sigma_B & \ \displaystyle\sigma_B > \frac{2 - |\lambda_B|}{2|\lambda_B|}|\mu|\ . \end{array} \right.
\end{align}
Since the extrema of the potential scale like $\lambda_B b_4$ or
$\sqrt{m_B^2}$, they will not lie in the condensed branch of the
potential whose range of validity is $\mc{O}(|\mu|)$ around
$\sigma_B = 0$. Thus, when $m_B^2$ and $\lambda_B b_4$ are both large,
we can ignore the presence of the condensed branch. In this case the
analysis reduces to that of the regular boson without chemical
potential \cite{Dey:2018ykx} whose effective potential has only two
branches viz.~the unHiggsed and the Higgsed.

Suppose we go to large values of $m_B^2$ and $\lambda_B b_4$ along the
parabolas
\begin{equation}
  m_B^2 = \alpha (\lambda_B b_4)^2\ .
\end{equation}
Then the above analysis tells us that, after a certain point on
\emph{any} parabola of the type given above, the system is in one of
the two uncondensed phases i.e.~either the unHiggsed or the Higgsed
phase (or in a no-phase region if $x_6^B$ is in one of the unstable
ranges).

For the regular boson without chemical potential with $x_6^B$ in the
stable range $\phi_h < x_6^B < \phi_u$, there is a first order curve
for $\lambda_B b_4 < 0$ along the following parabola:
\begin{equation}
  m_B^2 = \nu(x_6^B,|\lambda_B|)\ (\lambda_B b_4)^2\ ,
\end{equation}
where $\nu(x_6^B, |\lambda_B|)$ is a numerically determined
function. The relevant information about $\nu$ that is needed is that
it is monotonically decreasing as a function of $x_6^B$ from $\phi_h$
to $\phi_u$ and is zero when $x_6^B = \tfrac{1}{2} (\phi_u +
\phi_h)$. It can be easily seen from the expressions for $\phi_h$ and
$\phi_u$ in \eqref{phiuh} that $\phi_h + \phi_u$ is always positive
for any value of $|\lambda_B| \leq 1$. Thus, the above parabola is in
the fourth quadrant when
$\phi_h < x_6^B < \tfrac{1}{2} (\phi_u + \phi_h)$ and is in the third
quadrant when $\tfrac{1}{2} (\phi_u + \phi_h) < x_6^B < \phi_u$. From
the discussion above, it is clear that we should see the above first
order curve even in phase diagrams of the current paper for
sufficiently large $\lambda_B b_4$ and $m_B^2$.

\subsection{The phase diagram}\label{phased}

In this subsection, we focus on the ranges of $x_6^B$ in
\eqref{x6case} that correspond to the potential being stable i.e.~it
is bounded below for large and positive $\sigma_B$ as well as for
large and negative $\sigma_B$. This corresponds to the cases B and C
in \eqref{x6case}. The unstable ranges of $x_6^B$ are addressed in
Appendix \ref{unstable}.

\subsubsection{Case B: \texorpdfstring{$\phi_h < x_6^B < 0$}{phi-h<x6<0}}

\begin{figure}[!ht]
  \begin{center}
    \scalebox{0.8}{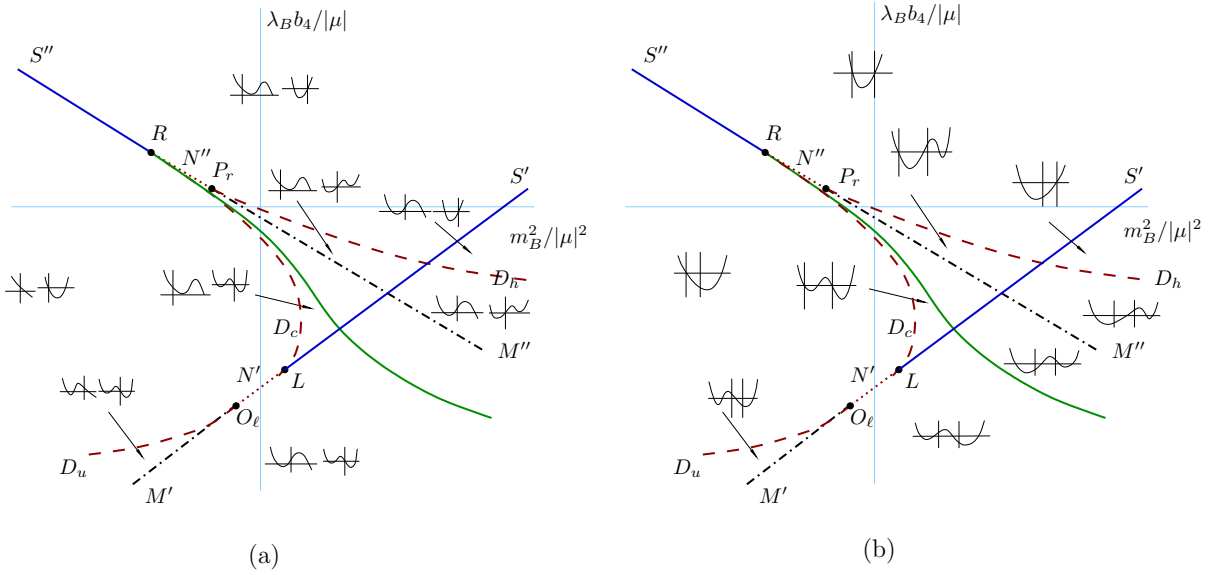}
    \caption{Structure of the potential for $\phi_h < x_6^B < 0$. The
      green curve is a first order transition while the blue curves
      are second order transitions. We have restored the $|\mu|$
      dependence of the phase diagram by rescaling $m_B^2$ and
      $\lambda_B b_4$ by appropriate powers of $|\mu|$.}
    \label{PhaseplotB}
  \end{center}
\end{figure}
\begin{figure}[!ht]
  \begin{center}
    \scalebox{0.8}{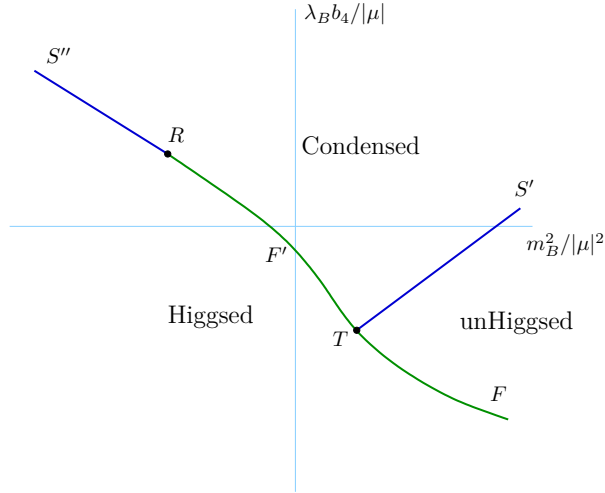}
    \caption{Phase diagram for $\phi_h < x_6^B < 0$. The blue lines
      are second order transitions and the green line is a first order
      transition. We have restored the $|\mu|$ dependence of the phase
      diagram by rescaling $m_B^2$ and $\lambda_B b_4$ by appropriate
      powers of $|\mu|$.}
    \label{PhasediagB}
  \end{center}
\end{figure}

Case III of Appendix \ref{curveplot} is pertinent near the
non-analytic point $\ell$ since the quantity $A$ in \eqref{ABCl}
satisfies $A < 0$ and $A + a < 0$, while Case II of Section
\ref{curveplot} is pertinent near the point $r$ since the quantity $A$
in \eqref{ABCr} satisfies $A < 0$ and $A + a > 0$. This implies that
the global potential is bounded for both positive and negative
$\sigma_B$. The unboundedness of the local potential near
$\sigma_B' = 0$ for large and positive $\sigma_B'$ is not very serious
since it is cutoff at $\sigma_B' = 1 / |\lambda_B|$.

The following conditions specify the segments
of $D_u$, $D_c$ and $D_h$ that appear in this phase diagram:
\begin{align}
  D_u: &\quad \lambda_B b_4\ \leq\ \tfrac{3}{4} \lambda_B^2 (x_6^B - \phi_u)\ ,\qquad   D_h: \quad \lambda_B b_4 \ \leq\ -\tfrac{3}{4}(2-|\lambda_B|)(x_6^B - \phi_h)\ ,
\end{align}
\begin{multline}\label{Dcbound}
D_c: \quad \tfrac{3}{4}\lambda_B^2 (x_6^B - \phi_u) - (1-|\lambda_B|)\ \leq\ \lambda_B b_4\ \leq \\ -\tfrac{3}{4}|\lambda_B|(2-|\lambda_B|)(x_6^B - \phi_h) + \frac{|\lambda_B|(1-|\lambda_B|)}{2-|\lambda_B|}\ .
\end{multline}
We display the detailed structure of the potential near each local
patch in Figure \ref{PhaseplotB}(a) and put together these two local
behaviours of the potential in Figure \ref{PhaseplotB}(b). The final
phase structure is displayed in Figure \ref{PhasediagB}. The first
order transition lines shown in solid green in Figures
\ref{PhaseplotB} and \ref{PhasediagB} are determined numerically. The
part of the first order line that separates the unHiggsed and Higgsed
phases is determined by comparing the value of the potential at the
respective minima in each branch. As was discussed in Section
\ref{firstord} (also see the discussion in \cite{Dey:2018ykx} and
\cite{Aharony:2018pjn}), this line goes to infinity in the fourth
quadrant when $x_6^B < \frac{1}{2}(\phi_u + \phi_h)$ and in the third
quadrant when $x_6^B > \frac{1}{2}(\phi_u + \phi_h)$. Since
$\phi_h + \phi_u$ is always positive, it follows that the first order
line always curves towards infinity in the fourth quadrant for all
values of $x_6^B$ such that $ \phi_h < x_6^B < 0$.

\subsubsection{Case C: \texorpdfstring{$0 < x_6^B < \phi_u$}{0<x6<phi-u}}

\begin{figure}[!htb]
  \begin{center}
    \scalebox{0.8}{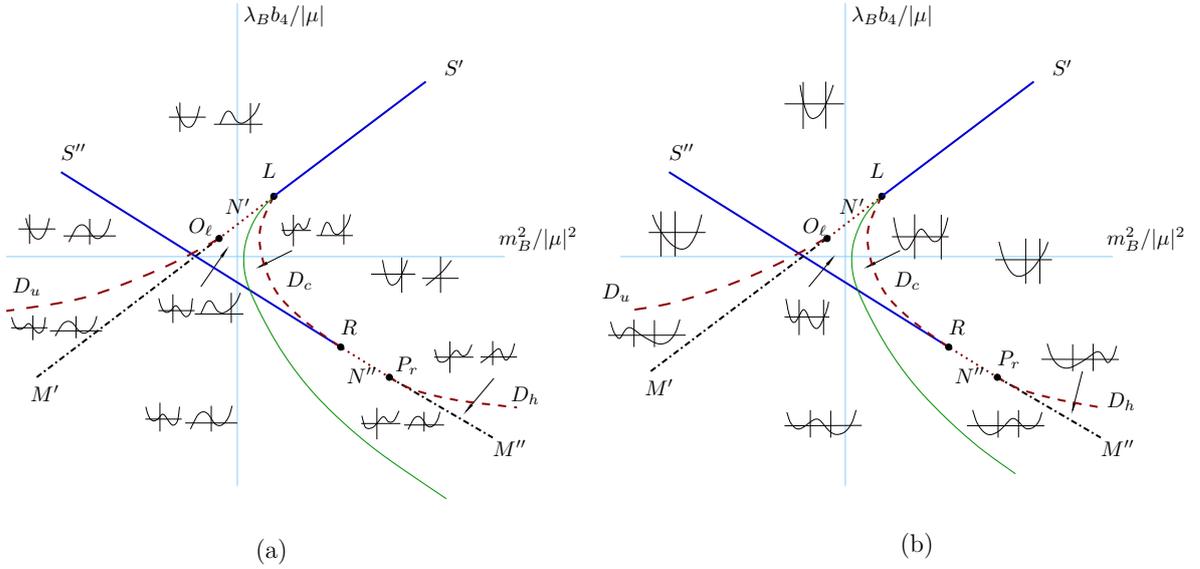}
    \caption{Structure of the potential for $0 < x_6^B < \phi_u$. We
      have restored the $|\mu|$ dependence of the phase diagram by
      rescaling $m_B^2$ and $\lambda_B b_4$ by appropriate powers of
      $|\mu|$. The green curve is a first order transition while the
      blue curves are second order transitions. The above plot is for
      the subrange $0 < x_6^B < \tfrac{1}{2}(\phi_h + \phi_u)$ where
      the first order curve goes to infinity in the fourth
      quadrant. The analysis proceeds similarly when the first order
      curve goes to infinity in the third quadrant for
      $ \tfrac{1}{2}(\phi_h + \phi_u) < x_6^B < \phi_u$.}
    \label{PhaseplotC}
  \end{center}
\end{figure}

\begin{figure}
  \begin{center}
    \scalebox{0.8}{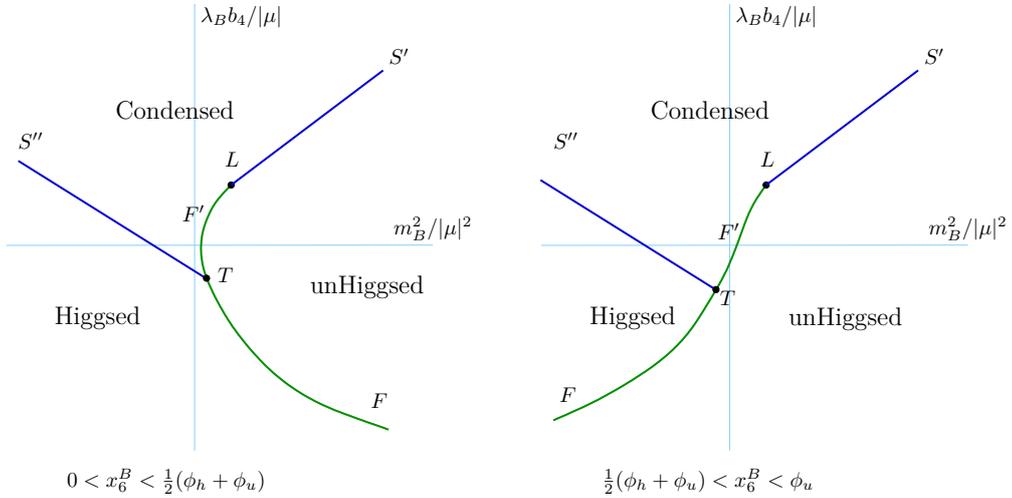}
    \caption{Phase diagram for $0 < x_6^B < \phi_u$. We have restored
      the $|\mu|$ dependence of the phase diagram by rescaling $m_B^2$
      and $\lambda_B b_4$ by appropriate powers of $|\mu|$. The blue
      lines are second order transitions and the green line is a first
      order transition. The figure on the left is for the range
      $0 < x_6^B < \tfrac{1}{2}(\phi_h + \phi_u)$ where the first
      order line goes to infinity in the fourth quadrant. The figure
      on the right corresponds to
      $ \tfrac{1}{2}(\phi_h + \phi_u) < x_6^B < \phi_u$ where the
      first order line goes to infinity in the third quadrant. At the
      special value $x_6^B=\tfrac{1}{2}(\phi_h + \phi_u)$ the green
      first order phase transition line from $T$ downwards runs down
      the $\lambda_B b_4$ axis.}
    \label{PhasediagC}
  \end{center}
\end{figure}

For this range of $x_6^B$, Case II of Section \ref{curveplot} applies
for the potential near $\sigma_B = \ell$ and Case I applies for the
potential near $\sigma_B = r$. The conditions specifying which
segments of the discriminant parabolas appear in this phase diagram
are
\begin{align}
  D_u: &\quad \lambda_B b_4\ \leq\ \tfrac{3}{4} \lambda_B^2 (x_6^B - \phi_u)\ ,\qquad   D_h: \quad \lambda_B b_4 \ \leq\ -\tfrac{3}{4}(2-|\lambda_B|)(x_6^B - \phi_h)\ ,
\end{align}
\begin{multline}
D_c: \quad -\tfrac{3}{4}|\lambda_B|(2-|\lambda_B|)(x_6^B - \phi_h) + \frac{|\lambda_B|(1-|\lambda_B|)}{2-|\lambda_B|} \ \leq\ \lambda_B b_4\ \leq \\ \tfrac{3}{4}\lambda_B^2 (x_6^B - \phi_u) - (1-|\lambda_B|)\ .
\end{multline}
The detailed structure of the potential is displayed in Figure
\ref{PhaseplotC}. In Figure \ref{PhaseplotC}(a), we provide a small
inset plot of the local potential near $\sigma_B = \ell$ and $r$ in
each region demarcated by the special lines of the potentials
\eqref{potl} and \eqref{potr}. In Figure \ref{PhaseplotC}(b) we show
the global potential in each of the above regions. Finally, we display
the phase structure of the theory in Figure \ref{PhasediagC}.

The part of the first order line that separates the unHiggsed and
Higgsed phases goes to infinity in the fourth quadrant when
$x_6^B < \frac{1}{2}(\phi_u + \phi_h)$ as explained in Section
\ref{firstord}. As indicated there, $\phi_u + \phi_h$ is always
positive for all values of $|\lambda_B|$. It follows that the first
order line curves towards infinity in the fourth quadrant for $x_6^B$
such that $0 < x_6^B < \frac{1}{2}(\phi_u + \phi_h)$ and goes to
infinity in the third quadrant when
$\frac{1}{2}(\phi_u + \phi_h) < x_6^B < \phi_u$. These two cases are
displayed in Figure \ref{PhasediagC}.

\clearpage

\subsection{A closer view at the triple point}\label{triple}

In the stable cases, Case B: $\phi_h < x_6^B < 0$ and Case C:
$0 < x_6^B < \phi_u$, there is an interesting point labelled $T$ on
the corresponding phase diagrams Figures \ref{PhasediagB} and
\ref{PhasediagC}. At this point, all three phases - the condensed, the
(uncondensed) unHiggsed and the (uncondensed) Higgsed - of the system
coexist. Equivalently, this `triple point' is the intersection of two
first order lines and one second order line. In this subsection, we
study the relative orientations of the two first order lines that meet
at the `triple point'.

The first order curve $F$ separating the unHiggsed and the Higgsed
phases takes the shape of a parabola as discussed in Section
\ref{firstord} which starts at the triple point $T$ and goes off to
infinity. The first order curve $F'$ which separates the condensed
phase and one of the uncondensed phases (Higgsed for Case B while
unHiggsed for Case C) has to be determined numerically and is
generally more complicated than a parabola. However, there are some
features which can be understood analytically.

\begin{figure}[!ht]
  \begin{center}
    \scalebox{1}{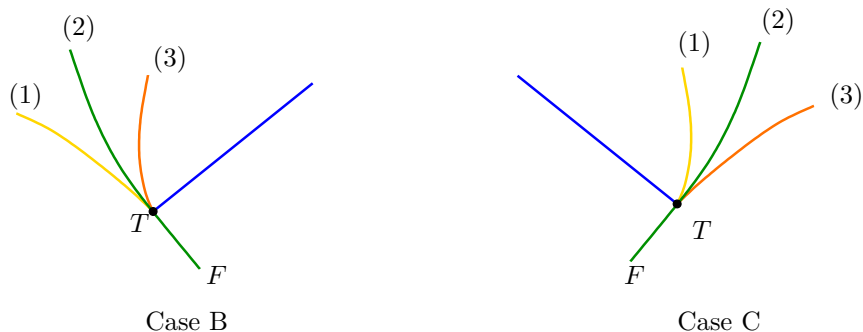}
    \caption{Different possibilities for the first order curve
      $F'$, labelled as $(1)$, $(2)$ and $(3)$, in relation to the
      first order curve $F$. The two subfigures differ in the
      relative placements of the second order line (blue) and the
      first order line.}
    \label{relFO}
  \end{center}
\end{figure}

Suppose we continue the first order curve $F$ that separates the
unHiggsed and Higgsed phases slightly further past the triple point
and closer to the origin. Then, the question is as follows: does the
other first order curve $F'$ go (1) to the left of, (2) along, or (3)
to the right of the extended curve $F$? We have illustrated the
possibilities in Figure \ref{relFO}.
It is in fact quite simple to figure out which of the above
possibilities actually occurs in our situation given the forms of the
effective potential \eqref{potl} and \eqref{potr} near the two
non-analytic points $\sigma_B = \ell$ and $\sigma_B = r$ respectively.

We study Case B for concreteness -- the results for Case C can be
obtained by swapping the unHiggsed and Higgsed phases. Recall that
$F'$ is the curve that separates the condensed phase and the Higgsed
phase. On the extended line $F$, the Higgsed and unHiggsed phases
coexist with equal free energy:
\begin{equation}
  U_{\rm Higgsed}(\sigma_1) = U_{\rm unHiggsed}(\sigma_2)\ ,
\end{equation}
where $\sigma_1$ is the minimum in the Higgsed branch of the potential
while $\sigma_2$ is the minimum in the unHiggsed branch. On this line
and very close to the triple point, we compare the free energy at the
condensed phase minimum and at the Higgsed phase minimum (which is the
same as the free energy unHiggsed phase minimum). If these free
energies are the same, then $F'$ coincides with the extended $F$. On
the other hand, if the condensed phase minimum has lower free energy,
the extended curve $F$ is in fact in the condensed phase, implying
that the first order curve $F'$ is to the left of the extended
$F$. Analogously, if the condensed phase minimum has the higher free
energy, then the first order curve $F'$ is to the right of the
extended $F$.

The triple point $T$ is on the second order line separating the
condensed and the unHiggsed phases. From the general analysis of
Section \ref{phasean} in the Appendix and the expression for the
potential \eqref{potl} in Section \ref{nearell}, we know that the
minimum in the condensed phase and the minimum in the unHiggsed phase
are both situated at the junction between the branches
$\sigma_B' = 0$. When we vary the parameters to go slightly into the
condensed phase along the extended $F$, both the condensed and
unHiggsed minima go to the right of $\sigma_B' = 0$,\footnote{Note
  that the unHiggsed minimum is not inside the range of validity
  $\sigma_B' < 0$ of the unHiggsed branch of the potential and is not
  useful for determining the phase. However, it is indeed useful for
  the ongoing analysis.} and if we are really close to the triple
point, then we can take these minima to be at the same value of
$\sigma_B' = \sigma_* \approx 0$. In this case, the free energies of
the unHiggsed and condensed minima are given by
\begin{align}
  U_{\rm unHiggsed}(\sigma_*) &=  A (\sigma_*)^3 + B (\sigma_*)^2 + C \sigma_*\ ,\nonumber\\
  U_{\rm condensed}(\sigma_*) &= (A + a) (\sigma_*)^3 + (B + b) (\sigma_*)^2 + C \sigma_*\ ,
\end{align}
where the parameters $A$, $a$, $B$, $b$ and $C$ are given in equation
\eqref{ABCl}. The difference between the condensed and unHiggsed phase
minimal free energies is then given by
\begin{equation}
  \Delta U =  U_{\rm condensed}(\sigma^*) - U_{\rm unHiggsed}(\sigma^*) = b (\sigma_*)^2\ ,
\end{equation}
where we have discarded the $(\sigma_*)^3$ term since $\sigma_*$ is
very close to $0$. Since the parameter $b = -(1- |\lambda_B|)$ is
negative for the potential \eqref{potl}, it follows that the condensed
phase has lower free energy than the unHiggsed phase and hence, also
lower free energy than the Higgsed phase since we are on the extended
$F$ line. Thus, possibility (1) in Figure \ref{relFO} is realised for
Case B.

In the same way, for Case C, one must compare the free energies at the
condensed and Higgsed phase minima
\begin{align}
  U_{\rm condensed}(\sigma'_*) &=  A (\sigma'_*)^3 + B (\sigma'_*)^2 + C \sigma'_*\ ,\nonumber\\
  U_{\rm Higgsed}(\sigma'_*) &= (A + a) (\sigma'_*)^3 + (B + b) (\sigma'_*)^2 + C \sigma'_*\ ,
\end{align}
where the parameters $A$, $a$, $B$, $b$ and $C$ are given in
\eqref{ABCr} corresponding to the potential \eqref{potr}. Again, the
sign of the parameter $b$ decides the phase with lower free energy. It
is easy to see from \eqref{ABCr} that the parameter $b$ is positive,
and hence the condensed phase has lower free energy than the Higgsed
phase, and consequently, than the unHiggsed phase since we are on the
extended $F$ line. Thus, possibility (3) in Figure \ref{relFO} is
realised for Case C.

\subsection{Varying the chemical potentials for fixed UV parameters} \label{vcp}

In the analysis above we have studied the phase diagram of our system
as a function of the three dimensionful parameters, $\lambda_B b_4$,
$m_B^2$ and $|\mu|$. While the first two of these parameters appear in
the Lagrangian of the theory (and so define its behaviour in the deep
ultraviolet), the third parameter (the chemical potential) specifies
the ensemble of the theory, and so the `vacuum' about which we work.

A question of clear physical interest is the following: what is the
phase structure of any particular theory as a function of the modulus
of the chemical potential? What phases, in other words, do we
encounter if we start at $|\mu|=0$ and then steadily raise the
chemical potential to $\infty$, all the while keeping the UV
parameters of the theory fixed?

The phase diagrams in Figures \ref{PhasediagB} and \ref{PhasediagC}
allow us to answer the questions posed in the previous paragraph
rather easily. Recall that that we have plotted
$\frac{m_B^2}{|\mu|^2}$ and $\frac{\lambda_B b_4}{|\mu|}$ on the
$X$-axis and $Y$-axis respectively of the phase diagrams referred to
above. Now consider the following two sets of parabolas
\begin{equation}\label{curn}
y = \alpha \sqrt{x}\ \ \text{with}\ \ x>0\ ,\qquad y = \gamma \sqrt{-x}\ \ \text{with}\ \ x<0\ .
\end{equation}
Note that the curve corresponding to the first equation in
\eqref{curn} lies in the first quadrant of the $x$-$y$ plane when
$\alpha$ is positive and in the fourth quadrant when $\alpha$ is
negative. Similarly the second curve in \eqref{curn} lies in the
second quadrant when $\gamma$ is positive but in the third quadrant
when $\gamma$ is negative.
\begin{alignat}{2}\label{alphgamdist}
  &\alpha>0:\quad \text{First quadrant}\ ,\qquad  &&\alpha<0:\quad \text{Fourth quadrant}\ ,\nonumber\\
  &\gamma>0:\quad \text{Second quadrant}\ ,\qquad  &&\gamma<0:\quad \text{Third quadrant}\ .
\end{alignat}
These four sets of (half-)parabolas foliate the $x$-$y$ plane and we
shall refer to them as \emph{foliating parabolas}.

By substituting the relations $y=\frac{\lambda_B b_4}{|\mu|}$
and $x=\frac{m_B^2}{|\mu|^2}$ we see that
\begin{equation}\label{alphgam}
\alpha = \frac{\sqrt{m_B^2}}{\lambda_B b_4}\ \text{with}\ m_B^2 > 0\ ,\quad \gamma = \frac{\sqrt{-m_B^2}}{\lambda_B b_4}\ \text{with}\ m_B^2 < 0\ .
\end{equation}
The important point here is that $\alpha$ and $\gamma$ depend only on
(and are fixed by) the UV parameters of the theory but are independent
of $|\mu|$. In other words, as we move along the foliating parabolas
\eqref{curn}, we are working in the same UV theory but at different
values of $|\mu|$.  Each of the parabolas run from the origin of the
phase diagram to infinity. When $|\mu|=0$ we start out at infinity. As
$|\mu|$ is increased we move in towards the origin, reaching the
origin only at $|\mu|=\infty$. Tracing this path in the phase diagrams
in Figures \ref{PhasediagA}, \ref{PhasediagB}, \ref{PhasediagC} and
\ref{PhasediagD} gives the phase diagram as a function of $\mu$ in any
particular theory given by the fixed UV parameters. In the rest of
this subsection we explain in more detail how this works for values of
$x_6^B$ that lie in the `stable range'.

\subsubsection{\texorpdfstring{$ \phi_h \leq x_6 \leq 0$}{phi-h<x6<0}}

\begin{figure}[!ht]
  \begin{center}
    \scalebox{0.8}{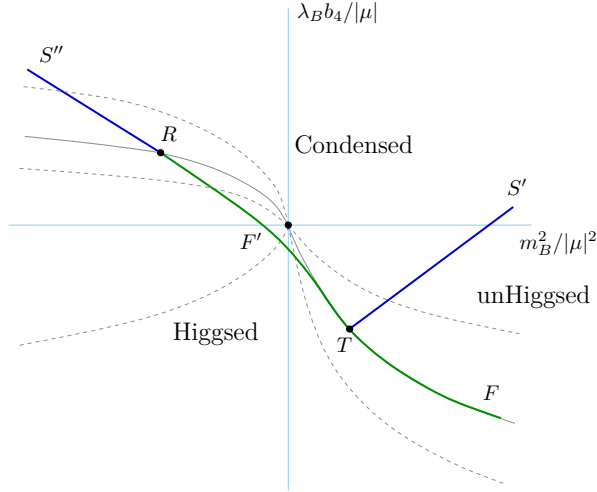}
    \caption{Foliating parabolas in gray for the phase diagram with
      $x_6^B$ in the range $\phi_h < x_6^B < 0$. Increasing the chemical potential at fixed values of UV parameters corresponds to moving along one of these parabolas (the values of the UV parameters determine which one). The parabolas from $T$ to $F$ (green line) and the parabola that pass through $R$ are distinguished. They separate parabolas across which the phase transition to a compressible phase is of second order from those along which the same transition is of first order.}
    \label{muchangeB}
  \end{center}
\end{figure}
In this case the phase diagram of our system is depicted in Figure
\ref{PhasediagB} and reproduced in Figure \ref{muchangeB}.  There are
two interesting parabolas of the form \eqref{curn} associated with
this phase diagram. The first of these passes through the triple point
$T$ in the fourth quadrant and goes along the green first order phase
transition line $F$ that starts at $T$ in Figure \ref{muchangeB}. This
curve is precisely a parabola of the form of the first of \eqref{curn}
with a particular value of $\alpha= \tl\alpha$, as explained in
Section \ref{firstord}. Note that $\tl\alpha$ is a negative
number\footnote{As explained in Section \ref{firstord}, this curve is
  a parabola because it represents a phase transition between the
  uncondensed Higgsed and the uncondensed unHiggsed phase, and so is
  completely insensitive to $\mu$. The value of $\tl\alpha$ is also
  insensitive to $\mu$ and was evaluated in \cite{Aharony:2018pjn}
  (see around Figures 25 and 26).}.

A second interesting parabola associated with the same phase diagram
is a parabola that passes through the point $R$ in the second quadrant
in Figure \ref{muchangeB}. This is of the the form of the second curve
in \eqref{curn} with a specific positive value for $\gamma$ given by
$\gamma = \gamma^*$. The value of $\gamma^*$ can be determined by
plugging in the coordinates of the point $R$ \eqref{Rdef} in the
formula for $\gamma$ in \eqref{alphgam}:
\begin{equation}
  \gamma^* = \frac{\left(2(1-|\lambda_B|) - \tfrac{3}{4}(2-|\lambda_B|)^2 (x_6^B - \phi_h)\right)^{1/2}}{-\tfrac{3}{4}|\lambda_B|(2-|\lambda_B|)(x_6^B - \phi_h) + \frac{|\lambda_B|(1-|\lambda_B|)}{2-|\lambda_B|}}\ .
\end{equation}
A glance at Figure \ref{muchangeB} will convince the reader that if
our UV parameters are such that we are on the foliating parabolas with
$\alpha^* < \alpha < \infty$, our theory starts out in the unHiggsed
phase at small $|\mu|$.  As $|\mu|$ is increased, the theory undergoes
a single order second order phase transition into a condensed phase
and then stays in this phase for all larger values of $|\mu|$.

If the UV parameters are such that we are on the foliating parabolas
with $\gamma^* < \gamma < \infty$, then our theory starts out in the
Higgsed phase at small $|\mu|$.  As $|\mu|$ is increased the theory
undergoes a single order second order phase transition into a
condensed phase and then stays in this phase for all larger values of
$|\mu|$.

Finally, if the UV parameters are such that we are on the parabolas
with $-\infty < \alpha < \alpha^*$ or quadrants with
$-\infty < \gamma < \gamma^*$, then our theory starts out in the
Higgsed phase at small $|\mu|$, and then undergoes a first order phase
transition to the condensed phase at a particular value of
$|\mu|$. The theory then stays in this phase for all larger values of
$|\mu|$.

\subsubsection{\texorpdfstring{$0 \leq x_6^B \leq
    \phi_u$}{0<x6<phi-u}}

\begin{figure}[!ht]
  \begin{center}
    \scalebox{0.8}{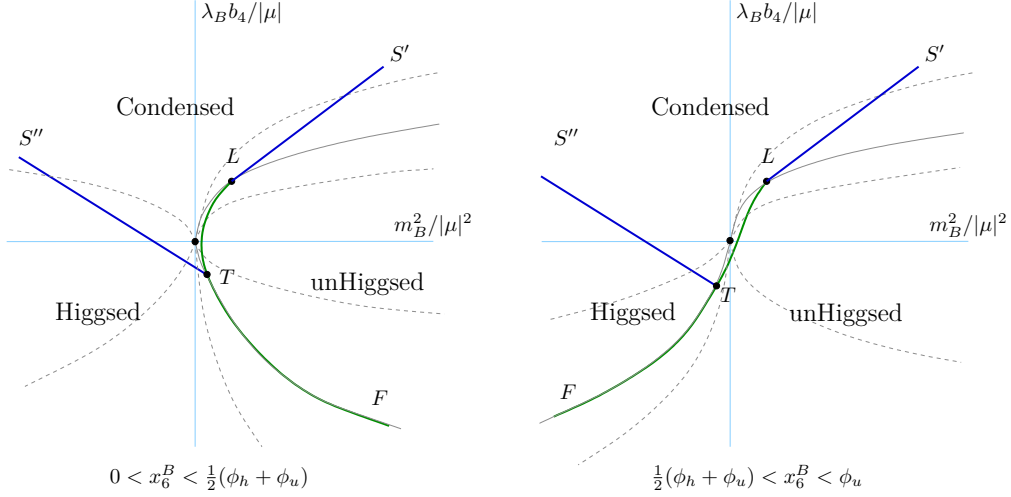}
    \caption{Foliating parabolas for the range $0 < x_6^B < \phi_u$. The significance of these parabolas was recounted in the caption to the previous figure. The distinguished parabolas (see caption to the previous figure) are the curves TF (green line) and the parabola that passes through $L$.}
    \label{muchangeC}
  \end{center}
\end{figure}

In this case the phase diagram of our system is depicted in Figure
\ref{PhasediagC} and reproduced in Figure \ref{muchangeC}.  Once
again, there are two interesting foliating parabolas \eqref{curn}
associated with this phase diagram. The first of these passes through
the triple point $T$ and continues along the first order curve $F$
that separates the unHiggsed and Higgsed phases. As above, this curve
is precisely a parabola. For
$0 < x_6^B < \tfrac{1}{2}(\phi_h + \phi_u)$, this curve is an
$\alpha$-parabola with a particular value of $\alpha = \tl\alpha$. For
the complementary range
$\tfrac{1}{2}(\phi_h + \phi_u) < x_6^B < \phi_u$, the curve is a
$\gamma$ parabola with a particular value of $\gamma =
\tl\gamma$. Note that both $\tl\alpha$ and $\tl\gamma$ are negative
are determined numerically in general.

A second interesting parabola associated with the phase diagram is an
$\alpha$-parabola which passes through the point $L$ in the first
quadrant of Figure \ref{muchangeC}. This corresponds to a positive
value of $\alpha$ given by $\alpha = \alpha^*$. We have an analytic
expression for $\alpha^*$ that we obtain by plugging in the
coordinates of the point $L$ \eqref{Ldef} in the formula for $\alpha$
in \eqref{alphgam}:
\begin{equation}
  \alpha^* = \frac{\left(\tfrac{3}{4}\lambda_B^2 (x_6^B - \phi_u) + 2 (1-|\lambda_B|)\right)^{1/2}}{\tfrac{3}{4}\lambda_B^2 (x_6^B - \phi_u) + (1-|\lambda_B|)}\ .
\end{equation}
For the range $0 < x_6^B < \tfrac{1}{2}(\phi_h + \phi_u)$, suppose our
UV parameters are such that we are on a $\gamma$-parabola with any
value of $\gamma$, or an $\alpha$-parabola with
$-\infty < \alpha < \tl\alpha$. Then, our theory starts out in the
Higgsed phase at small $|\mu|$.  As $|\mu|$ is increased the theory
undergoes a single second order phase transition into the condensed
phase and then stays in this phase for all larger values of $|\mu|$.

On the other hand if the UV parameters are such that we are on
$\alpha$-parabolas with $\alpha^* < \alpha < \infty$ then our theory
starts out in the unHiggsed phase at small $|\mu|$ and as $|\mu|$ is
increased it undergoes a single order second order phase transition
into a condensed phase and then stays in this phase for all larger
values of $|\mu|$. When $\tl\alpha < \alpha < \alpha^*$, the theory
starts out in the unHiggsed phase at small $|\mu|$, undergoes a first
order phase transition at an intermediate value of $|\mu|$ into the
condensed phase and stays in that phase for all higher values of
$|\mu|$.

A similar analysis can be performed in a straightforward way for the
case $\tfrac{1}{2}(\phi_h + \phi_u) < x_6^B < \phi_u$ depicted on the
right in Figure \ref{muchangeC}.

\subsection{Numerical plots}\label{numeric}

In this section we have so far presented only schematic phase diagrams
for the quasi bosonic theories. However the analysis that led up to
these phase diagrams was completely quantitative. Consequently, it is
not difficult, at any given value of $x_6$ $\lambda_B$ to plot fully
quantitative phase diagrams (the precise shapes of the first order
phase transitions in these diagrams can be determined only
numerically).

\begin{figure}[!b]
	\centering
	\scalebox{0.45}{\includegraphics{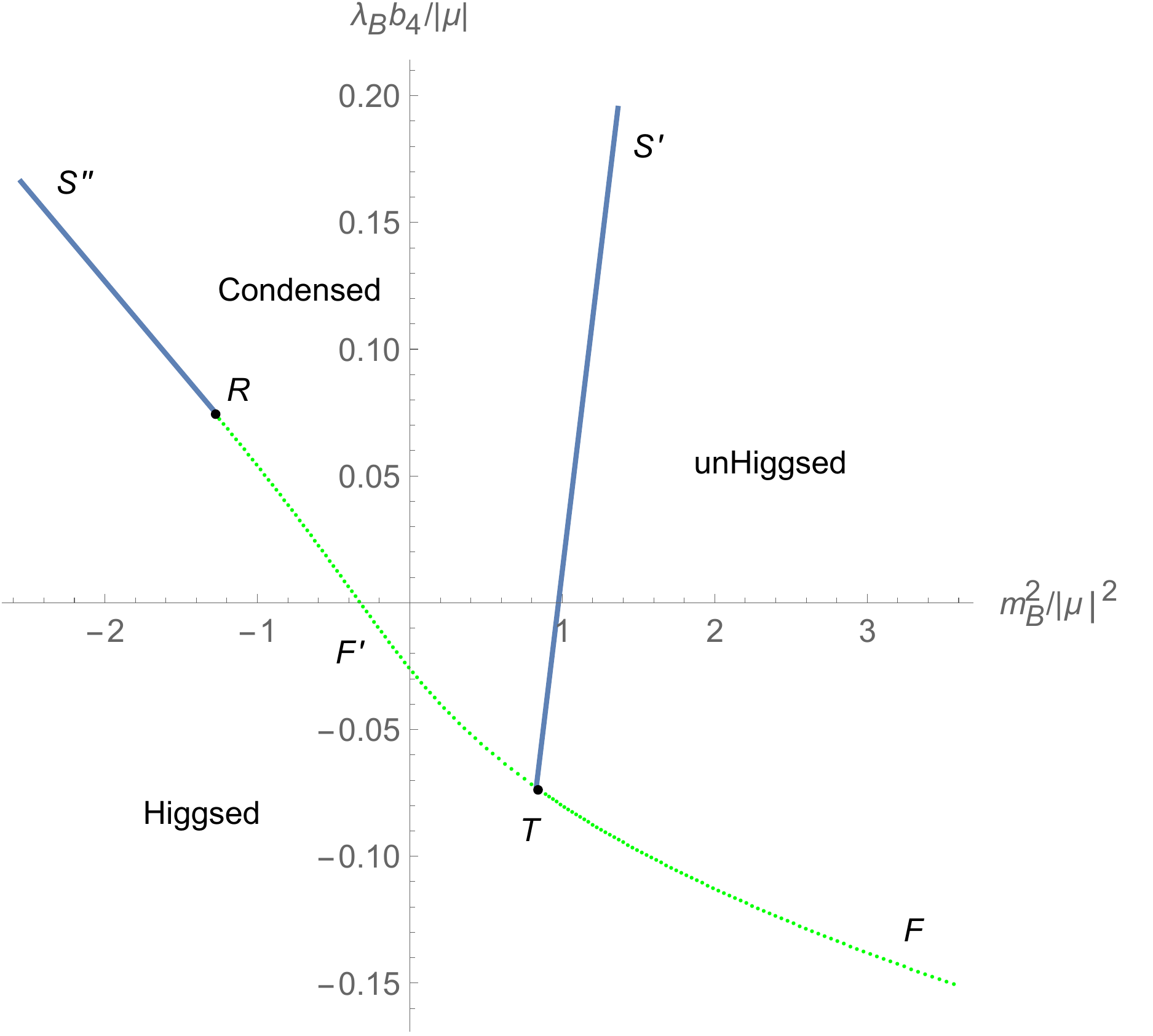}}
	\caption{The phase diagram of the regular boson theory for
          $|\lambda_B| = \tfrac{1}{4}$ and $x_6^B = -0.8$ such that it
          is in the stable range $\phi_h < x_6^B < 0$. }
	\label{phasecaseB}
\end{figure}

\begin{figure}[!htbp]
	\centering
	\scalebox{0.45}{\includegraphics{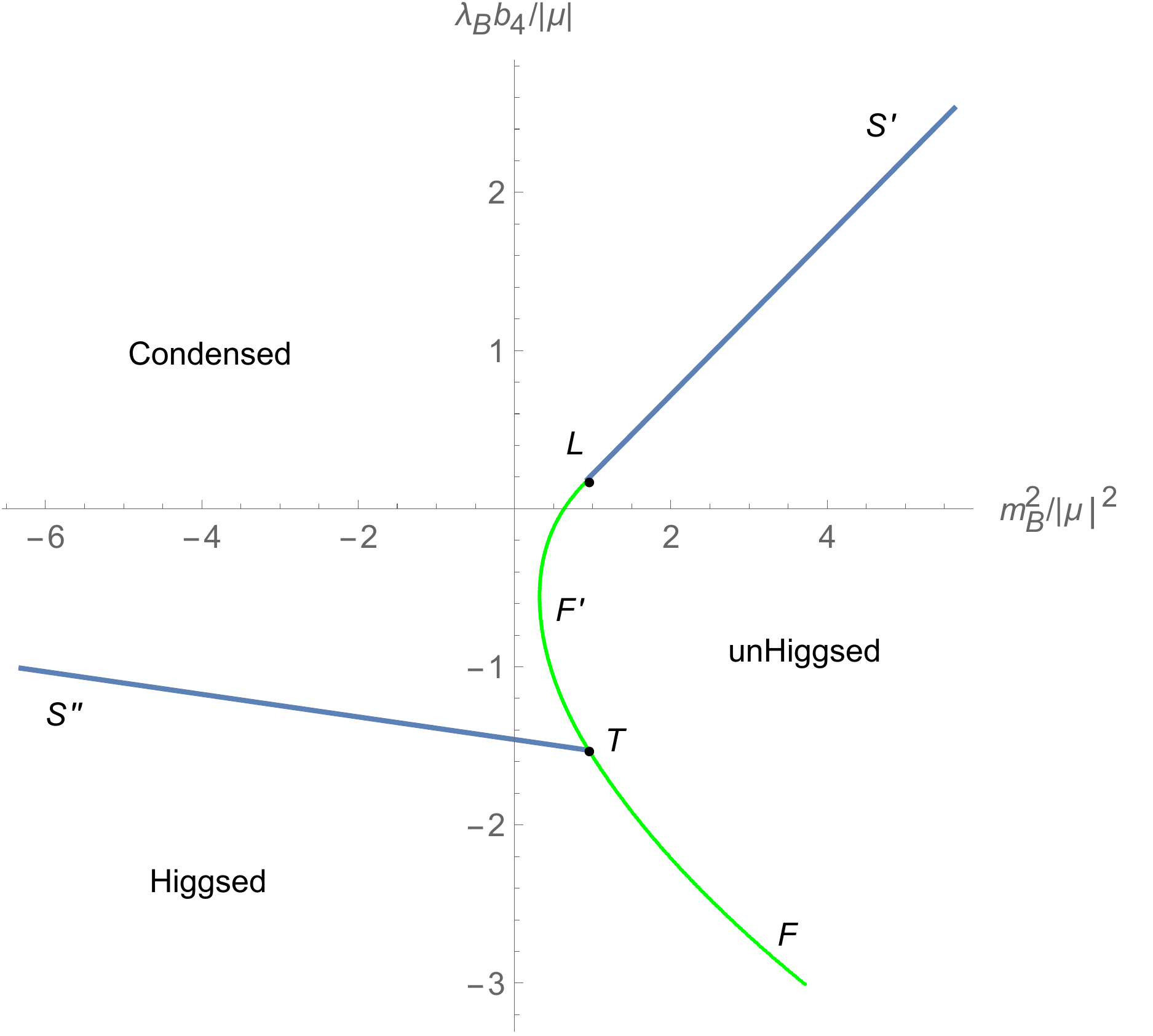}}
	\caption{The phase diagram of the regular boson theory for
          $|\lambda_B| = \tfrac{1}{4}$ and $x_6^B = 8$ such that it is
          in the stable range
          $0 < x_6^B < \tfrac{1}{2}(\phi_h + \phi_u)$. }
	\label{phasecaseC1}
\end{figure}

\begin{figure}[!h]
	\centering
	\scalebox{0.45}{\includegraphics{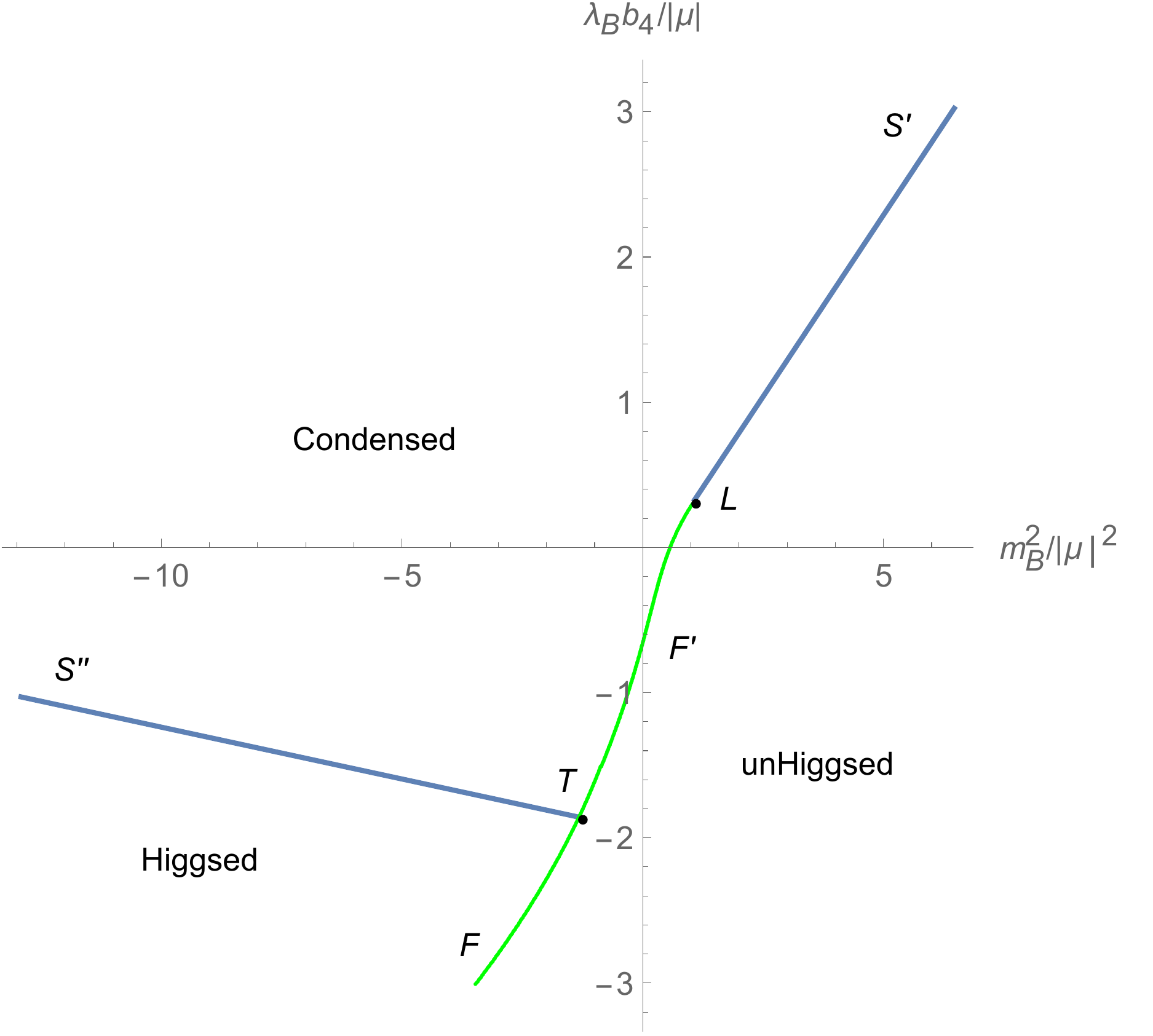}}
	\caption{The phase diagram of the regular boson theory for
          $|\lambda_B| = \tfrac{1}{4}$ and $x_6^B = 11$ such that it is
          in the stable range
          $\tfrac{1}{2}(\phi_h + \phi_u) < x_6^B < \phi_u$. }
	\label{phasecaseC2}
\end{figure}

To illustrate this fact, in this subsection we present three 
sample numerical plots, in Figures \ref{phasecaseB}, \ref{phasecaseC1}
and 
\ref{phasecaseC2} 
for the three subranges in the stable range of
$x_6^B$ for the choice of 't Hooft coupling $|\lambda_B| = 1 / 4$. The
schematic plots corresponding to these are in Figures
\ref{PhasediagB}, \ref{PhasediagC}(a) and \ref{PhasediagC}(b).

\newpage

\section{Discussion and Conclusions} \label{discu} 


Following the analysis of \cite{Geracie:2015drf}, in this paper 
we have presented completely explicit formulae 
\begin{equation} \begin{split} \label{occuptnconc}
\bar{n}_F({\e},q)&=\frac{1}{2}-\frac{1}{\pi|\lambda_F|}  \tan^{-1}\bigg(\frac{e^{\beta(\epsilon-q\mu)}-1}{e^{\beta(\epsilon-q \mu)}+1}\tan\frac{\pi|\lambda_F|}{2}\bigg)\ ,  \\
 \bar{n}_B({\e},q)&=\frac{1-|\lambda_B|}{2|\lambda_B|} - \frac{1}{\pi|\lambda_B|}  \tan^{-1}\bigg(\frac{e^{\beta(\epsilon-q \mu)}-1}{e^{\beta(\epsilon-q \mu)}+1}\cot\frac{\pi|\lambda_B|}{2}\bigg)\ .
 \end{split} 
\end{equation}
for the occupation numbers of fermionic and bosonic quasiparticle
states in large $N$ matter Chern-Simons theories at infinite volume as
a function of temperature and chemical potentials. These formulae
reduce to the famous formulae \eqref{occferm} and \eqref{stdbe} in the
weak coupling limit, and may be regarded as `anyonic' generalisations
of these (undergraduate) textbook formulae.  In our opinion the
formulae \eqref{occuptnconc} are very interesting, and deserve to be
understood and rederived from many points of view.

Specialising \eqref{occuptnconc} to zero temperature we have demonstrated that the runaway instability in
the Bose condensate phase of free (or large $N$ Wilson-Fisher) bosons
at values of the chemical potential larger than the thermal mass is
cured by the coupling to Chern-Simons gauge theory in large $N$ matter Chern-Simons theories. The final
stabilised Bose condensate is extremely simple: it can be viewed as a
theory of free bosonic quasiparticles, with each quasiparticle state
of energy less than $|\mu|$ occupied with an occupation number given
by \eqref{avon}. This regulated Bose condensate is the dual of an
effectively free Fermi sea of fermions coupled to an analogous
Chern-Simons gauge theory.

We find it fascinating that bosonic quasiparticle states appear to
obey a sort of modified exclusion principle, as a consequence of which
the Bose condensed phase is conceptually very similar to a Fermi sea
and should display many of its characteristic features. As in a Fermi
sea, in this phase every single particle state with energy less than
$|\mu|$ is fully occupied (i.e. has the maximum allowed occupation
number ${\bar n}$). As a consequence, excitations about this Bose
condensate are qualitatively similar to those about a Fermi sea. In
particular this Bose condensate hosts charged hole type
excitations. It also hosts uncharged particle hole pair excitations.
As in the case of the Fermi sea, when the particle carries momentum
${ \vec k_p}$ with $|\vec{k}_p|$ just larger than $k_F$ ($k_F$ is the
modulus of the momentum of the last occupied state) and the hole
carries momentum ${\vec k_h}$ with $|\vec{k}_h|$ just smaller than
$k_F$ then this particle-hole state carries nonzero momentum but
arbitrarily small energy. As in the case of a Fermi sea, the modulus
of the momentum of these particle hole excitations is always smaller
than $2k_F$. As a consequence, we expect to see $2k_F$ singularities -
similar to those in Fermi liquid theory - in correlation functions of
operators like the stress tensor that are built out of bilinears of
bosonic operators. The bosonic theory should also host collective
excitations of its condensate analogous to the zero sound of a Fermi
surface\footnote{We thank D. Tong for a discussion on this point.}.

It would be very interesting to understand whether and how the
modified `bosonic' exclusion principle manifests itself in dynamical
processes. Let us recall that for usual bosons
\begin{equation}\label{anne}
a^\dagger |n\rangle= 
\sqrt{n+1} |n+1\rangle
\end{equation}
(here $|n\rangle$ and $|n+1\rangle$ are the normalised states
corresponding to $n$ particles occupying a given single particle
energy level). The factor of $\sqrt{n+1}$ in this formula tells us
that bosons are gregarious; they preferentially transit to states that
are highly occupied. It would be fascinating to investigate what the
analogue of \eqref{anne} (if it exists) is for Chern-Simons coupled
bosons. At small $\lambda_B$ one would intuitively expect \eqref{anne}
to continue to apply for $n \ll {\bar n}$ (see \eqref{avon} for a
definition: recall ${\bar n}$ is large when $\lambda_B$ is
small). However \eqref{anne} will certainly be significantly modified
for values of $n$ of order ${\bar n}$.  The computation of scattering
processes at finite chemical potential and temperature (which may be
technically feasible at large $N$ along the lines of
\cite{Jain:2014nza}) could shed light on this question, and could also
make connection with Haldane's characterisation of statistics in terms
of occupation numbers \cite{Haldane:1991xg}\footnote{We thank
  D. Radicevic for bringing this paper, and its relevance to our
  current work, to our attention.}. The discussion of Section
\ref{bep} suggests that these S-matrices could also make contact with
$SU(N_B)_{k_B}$ representation theory.

The results of this paper have been obtained starting from expression
for the leading order in large $N$ (but exact to all orders in the 't
Hooft coupling) partition function of the system at finite temperature
and chemical potential. Explicit all-orders expressions for this
partition function had previously been derived at values of the
(modulus of the) chemical potential smaller than the quasiparticle
masses. In the current paper we have extended these known formulae to
values of the chemical potential larger than quasiparticle masses (see
Section \ref{mugcfe}).  We obtained the new free energy formulae of
this paper by analytically continuing previously obtained formulae in
$\mu$. It would be interesting to confirm these results from a direct
analysis of the path integral at values of the modulus of the chemical
potential larger than the thermal mass.

Our extended free energy formulae involved the addition of extra terms
to the free energies. These new terms qualitatively modify the
dynamics of the Bose condensate, stabilising it. Equivalent new terms
in the free energy expression for fermions are absent in the infinite
volume limit that we have focused on in this paper; as a consequence
the dynamics of the fermionic compressible phase in this limit is
essentially identical to that of a free Fermi sea.  Away from the
infinite volume limit, however, the modification to the free energy
does not vanish in other phases \cite{Jain:2013py} of the fermionic
theory (see subsection \ref{ofebugp}). This observation strongly
suggests that the thermodynamics of fermions in such phases differs
from that of a non-interacting Fermi sea; plausibly each single
particle fermion state has an occupation number less than unity in
such phases. This phenomenon deserves analysis, both from a direct
study of the relevant free energy formulae, as well as from a study of
the Schrodinger equation in the appropriate non-relativistic limits.

The fermionic and bosonic compressible phases described in this paper
can both be thought of as collections of quasiparticles which occupy
all states with energy smaller than the chemical potential with a
given occupation number (unity for the fermions and $\bar{n}$
\eqref{avon} for the bosons at zero temperature). The fact that each
of these quasiparticles each carry a definite spin\footnote{Note that
	this spin is a number and not a multiplet in $D=3$.} suggests that
each of these phases have extensive angular momenta (i.e.~have
definite angular momenta per unit area). In Appendix \ref{angmom}, we
have verified that this is indeed true of the free Fermi sea. It would
be interesting to compute the angular momentum per unit volume of all
the compressible phases studied in this paper as a function of the
chemical potential and the 't Hooft coupling. 
	
Since the results of this paper apply at every value of the chemical
potential $\mu$ and the quasiparticle mass $c_B$, they apply, in
particular, in the limit $\frac{|\mu| - c_B}{c_B} \ll 1$. In this
limit, the particles that make up the Bose condensate are all
non-relativistic and the Bose condensate described in this paper
should admit a description as a solution to the multi-particle
Schrodinger equation \cite{toappear1}. It would be very interesting to
reproduce the `bosonic exclusion principle' formula of Section 
\ref{bep} (also the more general formula \eqref{occuptnconc})  from this point of view. Such an analysis would likely furnish qualitatively new understanding.
	
The finite chemical potential phase diagram for the regular boson
theory, presented e.g.~in Figures \ref{PhasediagB} and
\ref{PhasediagC}, is rather intricate and has many interesting
features. At fixed values of the UV parameters, upon increasing the
chemical potential we always encounter phase transition indicating the
formation of a Bose condensate. For a range of values of the chemical
potential, this transition turns out to be of first order rather than
second order. It would be interesting to better understand the physics
of this switch of phase transitions between first and second order.

Finally, we have studied compressible phases by first taking
$N \to \infty$ and then taking $T \to 0$. In other words, the results
of this paper apply at temperatures that are much smaller than the
chemical potential but much larger than (for instance) the chemical
potential divided by any given positive power of $N$. It is possible
that, at still lower temperatures (temperatures of order of the
chemical potential divided by a positive power of $N$), new infrared
divergences invalidate the naive $1/N$ expansion (so that diagrams at
apparently different orders in $1/N$ all end up actually contributing
at the same order, see e.g.~\cite{Lee:2009epi}) leading to very
interesting low temperature dynamics. It would be very interesting to
investigate this further.

\section*{Acknowledgements}
We would like to especially thank A.~Dey, I.~Halder and S.~Jain for
several very useful discussions before and through the course of this
project. We would also like to thank K.~Damle, R.~Sensharma and
J.~McGreevy for useful discussions. We would also like to thank O. Aharony, S. Jain, A. Karch, G. Mandal,  J. McGreevy, D. Radicevic, T. Senthil, D. Tong,  A. Vishwanath and S. Wadia for comments on a preliminary draft of this manuscript. The work of S.~M., A.~M. and
N.~P. was supported by the Infosys Endowment for the study of the
Quantum Structure of Spacetime. Finally, we would like to acknowledge
our debt to the steady support of the people of India for research in
the basic sciences.

\appendix

\section{Bose-Fermi dualities at finite rank and level} \label{bfdfr}
In the present paper, we work in the 't Hooft large $N$ limit in which
the rank $N$ and level $\kappa$ are both taken to infinity with the
ratio $N / \kappa$ fixed. In this limit, the distinction between
$U(N)$ and $SU(N)$ gauge groups is lost. At finite $N$ and $\kappa$,
it is important to distinguish between $U(N)$ and $SU(N)$ groups and
Bose-Fermi duality works correctly only when the groups as well as the
corresponding levels are correctly chosen.

In this appendix we will carefully describe three distinct classes of
Chern-Simons theories: theories based on the $SU(N)$ gauge group and
two classes of $U(N)$ theories - the so-called Type II and Type I
theories - all of which become identical in the large $N$ limit. In
keeping with the focus of this paper, we will pay special attention to
the global symmetries of these theories. We will explain that though
the global symmetries of the $SU(N)$ and $U(N)$ theories are
structurally very different from each other, they become functionally
identical in the large $N$ limit.  We will also describe Bose-Fermi
duality conjectures at finite rank and level, keeping careful track of
all $U(1)$ factors \cite{Aharony:2015mjs}. The material presented in
this appendix is largely a review of well-known results (see
e.g.~\cite{Hsin:2016blu}), and is included here in order clearly
define all conventions and also in an attempt to make the current
paper self-contained.

\subsection{Pure Chern-Simons Theories}

\subsubsection{Conventions}

Throughout this paper we define the level $k$ of an $SU(N)$ pure
Chern-Simons theory to equal (upto sign) the level of its dual
`boundary' $SU(N)_k$ WZW theory.

We now turn to pure $U(N)$ Chern-Simons theories. Recall the
definition of the $U(N)$ group
\begin{equation}\label{modout}
U(N) = \frac{SU(N) \times U(1)}{\mbb{Z}_N}\ .
\end{equation}
In words, $U(N)$ has a simple $SU(N)$ factor and an abelian $U(1)$
factor, with the centre $\mbb{Z}_N$ of $SU(N)$ identified with a
corresponding $\mbb{Z}_N$ subgroup of the $U(1)$.\footnote{In the rest
  of this paper we often suppress writing the quotient by the diagonal
  $\mbb{Z}_N$ subgroup.} It follows that pure $U(N)$ Chern-Simons
theories are characterised by two levels, one each for the $SU(N)$ and
the $U(1)$ factors. We use the notation $U(N)_{k,k'}$ to denote these
levels where $k$ and $k'$ are defined as follows.
\begin{equation}
U(N)_{k,k'} \equiv  SU(N)_k \times U(1)_{Nk'}\ .
\end{equation}
$k$ is the level of the $SU(N)$ part of the gauge group as defined at
the beginning of this subsection. $Nk'$ is the level of the $U(1)$
part of the gauge group (normalised in the convention in which all
charged Wilson lines have integer charges). In fact, $Nk'$ is (upto
sign) the level of the `boundary' dual $U(1)$ WZW theory, normalised
so that all primary operators have integer charges. As we explain
later (see Section \ref{u1quant} below), the allowed values for $k'$
are
\begin{equation}\label{allkp}
  k'=k + m N\ ,
\end{equation} 
where $m$ is an arbitrary integer. 

In this appendix, we will be particularly interested in two $U(N)$
Chern-Simons theories, the Type II theories ($m=0$ in \eqref{allkp})
and Type I ($m={\rm sgn}(k)$ in \eqref{allkp})
\begin{equation}\label{ttcs}
  \text{Type II}:\ U(N)_{k, k}\ ,\quad \text{Type I}:\ U(N)_{k, \kappa}\ .
\end{equation}
where $\kappa$ is the so called `renormalized level', defined as
\begin{equation}\label{renbare}
\kappa = \sgn(k) (|k| + N)\ .
\end{equation}

The authors of \cite{Hsin:2016blu} have also studied another
interesting class of theories - we could call these Type I$'$ theories
- with $m=-{\rm sgn}(k)$. The properties of Type I$'$ theories are
analogous to those of the Type I theory (and Type I$'$ theories also
obey Bose-Fermi dualities closely analogous to those of Type I
theories). In this appendix we do not explicitly discuss Type I$'$
theories. However the interested reader will find it easy to modify
the discussion of Type I theories presented in this appendix to Type
I$'$ theories. Further, the authors of \cite{Radicevic:2016wqn}
generalise the dualities to the entire family of theories with $U(1)$
levels in \eqref{allkp}, of which the Type II, Type I and Type I$'$
theories are special cases.

We end this brief subsection with a notational remark. Following the
conventions of \cite{Hsin:2016blu}, dynamical gauge fields shall be
denoted by lowercase letters and background gauge fields by uppercase
letters. The $U(N)$ gauge field is generically denoted by $x_\mu$, and
it is often useful to split it into its traceless $SU(N)$ part and a
pure-trace $U(1)$ part:
\begin{equation}\label{splitnot}
  x_\mu = y_\mu + a_\mu \mathbb{1}_N\ ,\quad \tr\, y_\mu = 0\ ,
\end{equation}
where $\mathbb{1}_N$ is the $N \times N$ identity matrix.  We will
adopt the notation \eqref{splitnot} in the rest of this appendix. In
other words, we will always reserve the symbol $x_\mu$ for a general
$N \times N$ matrix valued gauge field, the symbol $y_\mu$ for the
traceless $N\times N$ matrix valued gauge field, and $a_\mu$ for the
$U(1)$ gauge field normalised so that $U(N)$ Wilson lines in the
fundamental / antifundamental of $U(N)$ carry charge $\pm 1$ under
$a_\mu$. The authors of \cite{Hsin:2016blu} work with the $U(1)$ gauge
field $\tr\, x_\mu = N a_\mu$ under which Wilson lines in the
fundamental / antifundamental representation of $U(N)$ carry a
fractional charge of $\pm 1 / N$.

\subsubsection{Lagrangians}\label{lagapp}

In this subsection we will explain how the levels for pure
Chern-Simons theories, defined in the previous subsection, enter the
Lagrangians for these theories. The discussion is very easy for the
$U(1)$ part of the Lagrangian. For a $U(N)_{k,k'}$ the Lagrangian for
the $U(1)$ factor is simply
\begin{equation}\label{lagpcu1}
  \mc{S}_{U(1)}[a] = \frac{\i k' N}{4\pi} \int d^3 x\, \epsilon^{\mu\nu\rho}\, a_\mu \partial_\nu a_\rho \ .
\end{equation}
This Lagrangian captures the dynamics of the $U(1)$ part of a
$U(N)_{k,k'}$ theory in every reasonable regulation scheme.

The discussion is a bit more complicated for the $SU(N)$ part of the
Lagrangian. In this case the value of the Chern-Simons path integral
depends on the regularization scheme used. The $SU(N)_k$ theory is
defined by the action
\begin{equation}\label{lagpcsn}
  \mc{S}_{SU(N)}[y] = \frac{\i c}{4\pi} \int d^3 x\, \epsilon^{\mu\nu\rho}\,\tr\left(y_\mu \partial_\nu y_\rho - \frac{2\i}{3} y_\mu y_\nu y_\rho\right)\ ,\quad (\text{given reg.~scheme})\ ,
\end{equation} 
where the value of $c$ depends on the regularization scheme
employed. It turns out $c=k$ when the path integral is regulated by
the addition of a Yang-Mills term in the Lagrangian with an
infinitesimal coefficient (a very large Yang-Mills coupling) - we
refer to this as the YM scheme. On the other hand $c=\kappa$ when the
path integral is regularized in the dimensional regularization scheme
(we call this the dim-reg scheme).

The Lagrangians presented in the main text of this paper are always
used in conjunction with the dimensional regularization scheme (see
\cite{Choudhury:2018iwf, Dey:2018ykx} for more details of the
regularization scheme)\footnote{The general rule is that the we get
  the same physical $SU(N)$ theory in the dim-reg and YM schemes if
  the coefficients $c$ in the actions used with these schemes differ
  by
$$c_{\text{dim-reg}}-c_{\text{YM-reg}}= {\rm sgn}(c_{\text{YM-reg}}) N .$$
\cite{Chen:1992ee} As far as $U(1)$ factors are concerned, we get the
same theories in the two regulation schemes if the two actions are
simply equal. The value of $c$ in the YM scheme equals the level of
the corresponding WZW theory.}.

We establish a bit of notation before we proceed. The actions may be
written very succinctly by employing differential forms. The $U(1)$ and
$SU(N)$ actions in \eqref{lagpcu1} and \eqref{lagpcsn} respectively
are written as
\begin{equation}
  \mc{S}_{U(1)}[a] = \frac{\i k' N}{4\pi} \int a \ud a\ ,\quad \mc{S}_{SU(N)}[y] = \frac{\i c}{4\pi} \int \tr\left(y \ud y - \frac{2\i}{3} y^3\right)\ ,
\end{equation}
where $a\ud a = d^3 x\, \epsilon^{\mu\nu\rho}\, a_\mu\partial_\nu a_\rho$
and $y^3 = d^3 x\, \epsilon^{\mu\nu\rho}\, y_\mu y_\nu y_\rho$.

We summarize the explicit Lagrangians in the different regularization
schemes below. The $SU(N)$ theory is defined as follows:
\begin{align}\label{sudr}
  \mc{S}_{SU(N)}[y] &=  \frac{\i \kappa}{4\pi} \int \tr\left(y \ud y - \frac{2\i}{3} y^3\right)\ ,\quad (\text{dim-reg})\ ,\nonumber\\
                    &= \frac{\i k}{4\pi} \int \tr\left(y \ud y - \frac{2\i}{3} y^3\right)\ ,\quad (\text{YM-reg})\ .
\end{align}
The Type I $U(N)_{k, \kappa}$ Chern-Simons theory is defined by the
Lagrangians
\begin{align}\label{tonedr}
  \mc{S}_{\rm Type\, I}[y,a] &=  \frac{\i \kappa}{4\pi} \int\tr\left(y\ud y - \frac{2\i}{3} y^3\right) + \frac{\i \kappa N}{4\pi} \int a \ud a\ ,\quad (\text{dim-reg})\ ,\nonumber\\
                             &= \frac{\i k}{4\pi} \int \tr\left(y \ud y - \frac{2\i}{3} y^3\right) +  \frac{\i\kappa N}{4\pi} \int a \ud a\ ,\quad (\text{YM-reg})\ .
\end{align}
Similarly, the Type II $U(N)_{k,k}$ theory is defined by the Lagrangians
\begin{align}\label{ttdr}
  \mc{S}_{\rm Type\, II}[y,a] &= \frac{\i\kappa}{4\pi} \int \tr\left(y \ud y - \frac{2\i}{3} y^3\right) + \frac{\i k N}{4\pi} \int  a \ud a\ ,\quad(\text{dim-reg})\ ,\nonumber\\
                              &= \frac{\i k}{4\pi} \int \tr\left(y \ud y - \frac{2\i}{3} y^3 \right) + \frac{\i k N}{4\pi} \int a \ud a\ ,\quad (\text{YM-reg})\ .
\end{align}

\subsubsection{Quantization of the $U(1)$ Chern-Simons levels}\label{u1quant}

The level of the gauge action for the $U(1)$ factor of a $U(N)$
Chern-Simons theory is generally of the form $Nk'$ with $k' = k + nN$
where $m$ is an arbitrary integer. In this subsubsection, we justify
this constraint on the $U(1)$ level.

The Lagrangian of a $U(N)_{k,k'}$ Chern-Simons theory is given by
\begin{equation}
  \mc{S}[y,a] = \frac{\i k}{4\pi}\int \tr\left(y\ud y - \frac{2\i}{3} y^3\right) + \frac{\i N k'}{4\pi} \int a \ud a\ .
\end{equation}
Rewriting the above Lagrangian in terms of the $U(N)$ gauge field
$x = y + a \mbb{1}_{N}$ in \eqref{splitnot}, we have
\begin{equation}\label{unxa}
  \mc{S}[x,a] = \frac{\i k}{4\pi}\int \tr\left(x\ud x - \frac{2\i}{3} x^3\right) + \frac{\i N (k'-k)}{4\pi} \int a \ud a\ .
\end{equation}
This action should be well-defined (and have all the physically
expected properties) for the $U(N)$ gauge field $x$.

Consider a $U(N)$ gauge configuration where only the gauge field
corresponding to one of the $U(1)$ subgroups of $U(N)$ is turned on,
say the first entry on the diagonal of the $U(N)$ gauge field. Suppose
this $U(1)$ gauge field (call it $b_\mu$) has one unit of spatial flux
and a constant time component $b_0$ with a holonomy of $2 \pi$ in the
time direction. This configuration is gauge equivalent (via a large
gauge transformation) to flux plus no holonomy; thus, it must yield
the same contribution to the path integral as this second
configuration. However, the action of the second configuration is
zero. Consequently the action of our original configuration must equal
an integral multiple of $2 \pi \i$. Using the fact that a $U(1)$ field
$b_\mu$ the above sort (namely with one unit of spatial flux and one
unit of timelike holonomy) has action\footnote{See footnote 15 of
  \cite{Witten:2015aoa} and the discussion above equation (5.15) of
  the lecture notes \cite{Tong:2016kpv} for an explanation of a tricky
  factor of $2$ here.} $$\frac{1}{4\pi} \int b\ud b = 2 \pi\ ,$$ it
follows that the first term in \eqref{unxa} evaluates to $2 \pi \i k$
which is $2 \pi \i$ times an integer as we required. We now turn to
the second term in \eqref{unxa}. Recall that the $U(1)$ gauge field is
related to the $U(N)$ gauge field as $$a = \frac{1}{N} \tr\, x\ .$$
Then, in the configuration above with only $b_\mu$ turned on, the
field $a_\mu$ is given by $\frac{1}{N} b_\mu$ and thus has $1/N$ units
of flux and a holonomy equal to $2 \pi /N$. It follows that the second
term in \eqref{unxa} evaluates to ${2 \pi \i}/{N^2}$. The total action
for the flux plus holonomy configuration is then
\begin{equation}\label{fluxphol}
  2\pi \i k + N(k'-k)\frac{2\pi\i}{N^2}\ .
\end{equation}
Thus, for this configuration to be $2\pi\i \times \text{integer}$ in
order to reproduce the same path integral as the configuration with
flux plus no holonomy, the quantity $\frac{k'-k}{N}$ must be an
integer. Thus, we have the condition
\begin{equation}\label{u1qt}
  k' = k + m N\ ,
\end{equation}
where $m$ is an arbitrary integer.

\subsubsection{Level-rank dualities}

Pure Chern-Simons theories are well known to enjoy level-rank
dualities (see e.g.~the relatively recent paper \cite{Hsin:2016blu}
and references therein). Level-rank duality asserts the equivalence of
two Type I $U(N)$ theories or of a Type II $U(N)$ theory and an
$SU(N)$ theory, when their ranks and levels are related as in Table
\ref{csdualitytab} below. Type I$'$ theories also enjoy analogous
level-rank dualities \cite{Hsin:2016blu} that we do not explicitly
enumerate here.
\begin{table}[!htbp]\centering
	\begin{tabular}{c|c}
          \toprule
          Type I-Type I & $U(N)_{k,\, \kappa } \longleftrightarrow U(|k|)_{\epsilon N,\, \epsilon \kappa }$ \\  \midrule
          Type II-${SU}(N)$ & $U(N)_{k,k} \longleftrightarrow SU(|k|)_{\epsilon N}$\\  \bottomrule
	\end{tabular}
	\caption{Level rank dualities between pure Chern-Simons
          theories. The quantity $\epsilon$ is given by
          $\epsilon=-{\sgn (k)}$. Theories on the same row are dual to
          each other. Here $\kappa= k + {\rm sgn}(k)N$. }\label{csdualitytab}
\end{table}

\subsection{Chern-Simons-matter theories}\label{couplemat}

\subsubsection{Conventions for levels and ranks}\label{clr}

In this paper we study Chern-Simons theories coupled to either scalars
or fermions in the fundamental representation of the gauge group. Like
their pure Chern-Simons counterparts, our Chern-Simons-matter theories
are labelled by the levels and ranks of the Chern-Simons gauge theory.

Throughout this appendix we will assume that the number of fundamental
scalars is smaller than $N$, and the number of fundamental fermions is
smaller than $|k|$. Under these conditions we adopt the following
convention for the definitions of the rank and level of a
Chern-Simons-matter theory. A Chern-Simons-matter theory with an
${SU}$ gauge group is said to have rank\footnote{The rank of the
  ${SU}(N)$ group is $N-1$ and \emph{not} $N$. However, we persist
  with this mild imprecision for the sake of brevity.} $N$ and level
$k$ if the following is satisfied: the low energy theory obtained by
massing up all scalars with a positive squared mass (or a positive
$m_B^{\rm cri}$ in the case of a critical boson theory) and by massing
up all fermion fields with a mass whose sign is the same as that of
the level $k$, is pure $SU(N)_k$ Chern-Simons theory. A similar
convention applies to the Type II $U(N)_{k,k}$ and the Type I
$U(N)_{k,\kappa}$ theories.

Now consider a Chern-Simons-matter theory coupled to $n_{\rm fer}$
fundamental fermions and $n_{\rm bos}$ fundamental boson fields. We
assume that all our matter fields are massive. However not all fields
need have masses of the same sign. Let
$n^{-}_{\rm fer} \leq n_{\rm fer}$ fundamental fermions have
(generic) masses whose sign is opposite to that of $k$. Also let
${n}_{\rm bos}^- \leq n_{\rm bos}$ fundamental bosons have (generic)
negative squared masses (or negative $m_B^{{\rm cri}}$ in the case of
the critical boson theory). Under these circumstances, the low energy
dynamics of the $SU(N)_k$, Type II $U(N)_{k,k}$ or Type I
$U(N)_{k,\kappa}$ theories is governed by pure
$SU(N_{\rm eff})_{k_{\rm eff}}$, Type II
$U(N_{\rm eff})_{k_{\rm eff},k_{\rm eff}}$ or Type I
$U(N_{\rm eff})_{k_{\rm eff},\kappa_{\rm eff}}$ Chern-Simons theories
with rank $N_{\rm eff}$ and level $k_{\rm eff}$ where
\begin{equation}\label{newenrank}
  N_{\rm eff}= N-{n}^-_{\rm bos}\ ,\quad  k_{\rm eff}=k-{n}_{\rm fer}^-\ ,\quad \kappa_{\rm eff} = k_{\rm eff} + \sgn(k_{\rm eff}) N_{\rm eff}\ .
\end{equation} 
The assumption spelt out at the beginning of the previous paragraph
ensures that $N_{\rm eff}$ is always positive and that $k_{\rm eff}$ has the same sign
as $k$.

The first of \eqref{newenrank} follows because the negative mass
fundamental scalars induce a Higgs mechanism which breaks the $SU(N)$
gauge group down to $SU(N_{\rm eff})$. The second of \eqref{newenrank}
follows because integrating out fermions shifts the low energy
Chern-Simons level in a manner that depends the sign of the masses of
the integrated-out fermions (more below).

\subsubsection{Lagrangians} 

Consider a Chern-Simons-matter theory as defined in the previous
subsubsection. Let us define\footnote{Note that ${\tilde k}$ is the
  average of the level of the low energy Chern-Simons theory
  \eqref{newenrank} obtained by massing up all fermions with a mass
  that has the same sign as $k$, and the level of the low energy
  Chern-Simons theory \eqref{newenrank} obtained by massing up all
  fermions with a mass that has the opposite sign as $k$.}
\begin{equation}\label{deftk}
{\tilde k}= k-{\rm sgn}(k)\frac{n_{\rm fer}}{2}\ ,
\quad {\tilde \kappa}= \kappa -{\rm sgn} (\kappa) \frac{n_{\rm fer}}{2}\ .
\end{equation} 
The gauge field Lagrangians for the $SU(N)$, Type II and Type I
Chern-Simons-matter theories are given by the Lagrangians in Section
\ref{lagapp} once we make the replacements
\begin{equation}\label{replacements}
  k \rightarrow {\tilde k}\ ,\quad \kappa \rightarrow {\tilde \kappa}\ .
\end{equation} 
Note that the `levels' that appear in the Lagrangians depend only on
the total number of fermionic fields and not on their masses. This is
what we would expect of a UV Lagrangian\footnote{One should be able to
  deform a UV CFT by mass terms of either sign, so the UV definition
  of CFT itself should be insensitive to the sign of the masses.}.

Within either the Yang-Mills or the dimensional regularization
schemes, integrating out a massive boson leaves all Chern-Simons
levels unchanged. On the other hand integrating out a massive fermion
of mass $m_F$ leaves the rank of Chern-Simons theory ($SU(N)$, Type II
or Type I) unchanged but shifts the coefficient $c$ (see
\eqref{lagpcsn}) of the Chern-Simons part of the
Lagrangian\footnote{The $U(1)$ levels are also shifted in the manner
  needed so as to keep the Type I or Type II theory respectively Type
  I or Type II.} according to the formula
\begin{equation}\label{shift}
  c \longrightarrow c + \frac{1}{2} \sgn(m_F)\ .
\end{equation}
After this renormalization, the effective IR level of the Chern-Simons
theory is given by
\begin{equation}\label{defkeff}
  k_{\rm eff}= k-n^{-}_{\rm fer}\ ,
\end{equation}
giving an explanation of \eqref{newenrank} (see around
\eqref{newenrank} for the definition of $n^{-}_{\rm fer}$).

\subsubsection{Level-rank duality} \label{lrd}

An $SU(N)_k$, Type II $U(N)_{k,k}$, Type I $U(N)_{k,\kappa}$
Chern-Simons-matter theory at rank $N$ and level $k$ with
$n_{\rm ferm}$ fundamental fermions and $n_{\rm bos}$ fundamental
bosons is conjectured to be dual to a Type II $U(N')_{k',k'}$,
$SU(N')$, Type I $U(N')_{k',\kappa'}$ Chern-Simons-matter theory
respectively with $n'_{\rm ferm}$ fundamental fermions and
$n'_{\rm bos}$ bosons where
\begin{equation}\label{dumap}
  k'=-{\rm sgn}(k) N\ ,\quad N'=|k|\ ,\quad n'_{\rm bos}=n_{\rm fer}\ ,\quad n'_{\rm fer}=n_{\rm bos}\ .
\end{equation} 
These conjectures are summarised in Table \ref{dualitytab} below
\begin{table}[!htbp]\centering
  \begin{tabular}{c|c|c}
    \toprule
    Type I~ - Type I & $U(N)_{k, \kappa}$ & $U(N')_{k',\kappa'}$ \\  \midrule
    Type II~ - $SU(N)$ & $U(N)_{k, k}$ &  $SU(N')_{k'}$ \\ \midrule
    $SU(N)$~- Type II  & $SU(N)_{k}$ & $U(N')_{k', k'}$ \\ \bottomrule
  \end{tabular}
  \caption{This table lists conjecturally dual pairs of
    theories. Theories in the same row of the table are conjecturally
    dual provided that the levels, ranks and the number of fermionic
    and bosonic fields are related as described in \eqref{dumap} this
    subsubsection. See earlier in this appendix for a detailed
    discussion of the conventions for the levels in
    Chern-Simons-matter theories.}
  \label{dualitytab}
\end{table}
Upon integrating out massive matter, the theories listed on the two
sides of the dualities listed in \ref{dualitytab} each reduce to pure
Chern-Simons theories of the same type (i.e. $SU(N)$, Type II or Type
I) as their parent matter theories, but with level and rank
$(N_{\rm eff}, k_{\rm eff})$ and $(N'_{\rm eff}, k'_{\rm eff})$
respectively. For every choice of matter masses it is easily verified
that
\begin{equation}\label{dumappcs}
  k'_{\rm eff}=-{\rm sgn}(k_{\rm eff}) N_{\rm eff}\ ,\quad N'_{\rm eff}=|k_{\rm eff}|\ .
\end{equation} 
As a consequence, these low energy topological theories can be
independently verified to be dual to each other using the known
level-rank dualities listed in table \ref{csdualitytab}. This
consistency check of the Chern-Simons-matter dualities was pointed out
in \cite{Aharony:2012nh}.

\subsection{Global Symmetries} \label{gs}

Each of the Chern-Simons-matter theories described above enjoys
invariance under a $U(1)$ global symmetry. In this subsection we will
describe the action of each of these symmetries by describing how each
of the Lagrangians above is modified when we turn on a background
field $C_\mu$ that couples to the global $U(1)$ symmetry\footnote{The
  chemical potential we have used in the main text of this paper is the imaginary
  holonomy of the gauge field $C$.}.

In all the theories studied in this paper, the Lagrangian for the
$SU(N)$ part $y$ of the dynamical gauge field is unaffected by the
coupling to $C$: there are no direct couplings between $C$ and any
term in the Lagrangian involving $y$. For this reason, in all the
Lagrangians presented in this section we omit the Lagrangian for the
$SU(N)$ gauge fields; the relevant terms continue to be given by
\eqref{tonedr}, \eqref{ttdr} with the replacements
\eqref{replacements}. However, as the $U(1)$ part of the dynamical
gauge field in the $U(N)$ theories does couple to $C$, we keep all
terms in the Lagrangian involving dynamical $U(1)$ gauge fields in the
formulae below. As these terms are the same in the dimensional
regularization and the Yang-Mills regularization schemes (see
\eqref{tonedr}, \eqref{ttdr}) the discussions of this section apply in
both schemes. In places where the $SU(N)$ Chern-Simons action is
relevant, we work with the Yang-Mills regularization scheme in this
subsection.

\subsubsection{$SU(N)$ theories}
The $SU(N)$ Chern-Simons-matter theories in this paper are coupled to
the background gauge field $C$ as follows. The matter fields that the
gauge fields in \eqref{SUNlag} interact with are each assigned a
charge $q$ under the gauge field $C$. The full Lagrangian of our
theory - including couplings of matter fields to $C_\mu$ - is obtained
as follows.  All derivatives acting on matter fields are modified to
$$D_\mu = \partial_\mu - \i q C_\mu\ .$$
In particular, for the Chern-Simons matter theories in this paper we
assign the charge $q=1$ to every fundamental field and $q=-1$ to every
antifundamental field. In other words, we replace the $SU(N)$ gauge
fields $y_\mu$ that appear in the Chern-Simons-matter Lagrangians
\eqref{cflag}, \eqref{cflag}, \eqref{cblag}, \eqref{rblag} by
\begin{equation}\label{Xrepl}
  z_\mu = y_\mu +  C_\mu\, \mbb{1}_N\ ,
\end{equation} 
(where $y_\mu$ is an $SU(N)$ matrix and $\mbb{1}_N$ is the $N\times N$
identity matrix). Note that the gauge field $z_\mu$ is a gauge field
for the $U(N)$ gauge group with its trace being the non-dynamical,
background gauge field $C$.

In addition to the terms coupling $C$ with matter above, our
Lagrangians are required to have a pure $C$ `contact' Lagrangian:
\begin{equation}\label{SUNlag}
  \mc{S}[\ldots;C] = \frac{\i {\tilde L}  }{4\pi} \int C \ud C\ ,
\end{equation}
where the ellipsis $\ldots$ depicts the dependence on the matter
fields and the $SU(N)$ gauge field which we have suppressed above. The
level ${\tilde L}$ must have certain integrality properties but also
has some arbitrariness which, when fixed, gives the precise definition
of the (contact terms in) the symmetry current correlators of the
theory. We pause now to explain in more detail the extent to which
${\tilde L}$ is fixed and the degree of its ambiguity.
 
From requirements of consistency, ${\tilde L}$ is constrained to take
the form
\begin{equation}\label{constoflid}
  {\tilde L}={\tilde k} N + L N^2\ .
\end{equation} 
To see why this is the case, let us integrate out all fermions which
we assume to be massive (for simplicity we assume we are in a phase
without any Higgsing due to the bosonic matter). Since the fermions
are minimally coupled to the dyanmical $SU(N)$ gauge field and the
background field $C$, this integrating-out procedure replaces
${\tilde k}$ with $k_{\rm eff}$ both in \eqref{constoflid} and in the
$SU(N)$ part of the Lagrangian. Restoring the $SU(N)$ part of the
lagrangian (and working in the Yang-Mills regularization scheme),
plugging in \eqref{constoflid} results in the effective low energy
Lagrangian
\begin{equation}\label{llelag}
  \mc{S}[\ldots,y;C] =\frac{\i k_{\rm eff}}{4\pi} \int \tr\left(y\ud y -
    \frac{2\i}{3} y^3\right) + \frac{\i N(k_{\rm eff} + L N) }{4\pi} \int C \ud C\ ,
\end{equation} 
or, in terms of the $U(N)$ gauge field $z = y + C \mbb{1}_N$ in
\eqref{Xrepl},
\begin{equation}\label{llelag1}
  \mc{S}[\ldots,z;C] =\frac{\i k_{\rm eff}}{4\pi} \int \tr\left(z\ud z -
    \frac{2\i}{3} z^3\right) + \frac{\i L N^2 }{4\pi} \int C \ud C\ ,\quad\text{with}\quad  C_\mu = \frac{1}{N}\tr\, z_\mu\ .
\end{equation} 
The above action is formally the same as the action for the $U(N)$
theory in \eqref{unxa} though the $U(1)$ field $C_\mu$ here is
non-dynamical as opposed to the dynamical $U(1)$ gauge field $a_\mu$
in \eqref{unxa}. The same arguments that led to the quantization rule
\eqref{u1qt} tell us that the level $L$ in \eqref{llelag} should be an
integer.

Recall that the gauge field configuration that was considered in
Section \eqref{u1quant} consisted of one unit of spatial magnetic flux
and one unit of holonomy around the time circle for the trace of the
$U(N)$ gauge field i.e. $\tr\, z_\mu = N C_\mu$. Thus, as in Section
\ref{u1quant}, the gauge field $C_\mu$ has a flux of $1 / N$ and a
holonomy of $2\pi / N$. The fact that the background field $C$ has
fractional flux on admissible gauge configurations tells us that it is
not normalised correctly to be viewed as a $U(1)$ gauge field. It
follows from the previous discussion that the correctly normalised
$U(1)$ gauge field is\footnote{Concretely, ${\tilde C}$ transforms
  under background global gauge transformations as
  ${\tilde C}_\mu \rightarrow {\tilde C}_\mu + \partial_\mu \alpha$
  where $\alpha$ is $U(1)$ valued, i.e.~obeys
  $\alpha \sim \alpha + 2 \pi$.}
\begin{equation}\label{rescalesu}
  {\tilde C}_\mu = N C_\mu\ .
\end{equation}
In terms of this field the low energy action \eqref{llelag} takes the
form
\begin{equation}\label{llelagt}
  \mc{S}[\ldots,z;\tilde{C}] = \frac{\i k_{\rm eff}}{4\pi} \int \tr\left(z \ud z - \frac{2\i}{3} z^3 \right) + \frac{\i L }{4\pi} \int {\tilde  C} \ud {\tilde  C}\ ,\quad\text{with}\quad {\tilde C}_\mu = \tr\, z_\mu\ .
\end{equation} 
Wording the action in terms of the $SU(N)$ gauge field $y_\mu$ and the
background field $C_\mu$ and suppressing the dependence on the $SU(N)$
gauge field as mentioned earlier, we find that the action is
\begin{equation}\label{SUNlagwr}
  \mc{S}[\ldots;\tilde{C}] = \frac{\i}{4\pi} \left(\frac{k_{\rm eff}}{N} + L\right) \int {\tilde C} \ud {\tilde C}\ .
\end{equation}
The above equation \eqref{SUNlagwr}, is, of course, just
\eqref{SUNlag} re-expressed in terms of the field ${\tilde C}$. Apart
from the fractional shift of $\frac{k_{\rm eff}}{N}$ that has its
origins in coupling with the $SU(N)$ part, $L$ is simply the level of
the $U(1)$ Chern-Simons action for the background field ${\tilde C}$.

There is another way to understand why ${\tilde C}$, rather than $C$,
is the correctly normalised $U(1)$ field. Under the field $C$ each
fundamental / antifundamental field carries charges
$\pm 1$.\footnote{Note that the above statement is accurate for
  fermions and only for scalar excitations in the unHiggsed phase of
  the bosonic theories. The $W$ boson excitations in the Higgsed phase
  carry charge $\frac{N_B}{N_B-1}$ under the gauge field $C$; see
  Section 2.9.7 of \cite{Choudhury:2018iwf}. It follows that
  $SU(N_B-1)$ baryons constructed out of $N_B-1$ $W$ bosons continue
  to carry charge $N_B$ under the gauge field $C$, or charge unity
  under the rescaled gauge field ${\tilde C}$.} However, fundamental /
antifundamental states are not $SU(N)$ gauge-invariant.  Every charged
operator that is $SU(N)$ gauge-invariant (and every charged state that
is $SU(N)$ gauge-invariant on $S^2$) is built out of $SU(N)$
Levi-Civita tensors and so carries charge that are quantized in units
of $N$ under the gauge field $C_\mu$ (see below for further discussion
of these \emph{baryonic} operators). On the other hand, these states
carry charges quantized in units of $1$ under the gauge field
${\tilde C}$. This explains why ${\tilde C}$, rather than $C$, is the
correctly normalised $U(1)$ gauge field.

\subsubsection{ Type II theories}
The $U(1)$ global symmetries of the Type II and Type I theories are of
the `topological' sort. As a consequence, the background field $C$
interacts only with the dynamical gauge fields of theory and not
directly with matter fields. Recall that the $U(1)$ part of the gauge
Lagrangian for Type II theory is
\begin{equation}\label{ttymmod}
  \mc{S}[a] = \frac{\i {\tilde k} N}{4\pi} \int a \ud a\ ,
\end{equation} 
where $a$ is the dynamical $U(1)$ gauge field and ${\tilde k}$ was
defined in \eqref{deftk}. After coupling with the background field
$C$, this `UV' action takes the form
\begin{equation}\label{ttymmodcoupold}
  \mc{S}[a;C] = \frac{\i {\tilde k} N}{4\pi} \int a \ud a  - \frac{\i k N}{2\pi} \int C \ud a  + \frac{\i L_{\rm II} k^2}{4\pi}  \int C \ud C\ .
\end{equation}
In the above, the gauge field $C$ couples to the topological current
\begin{equation}\label{jtypeII}
  J^\mu = \frac{kN}{2\pi} \epsilon^{\mu\nu\rho} \partial_\nu a_\rho = \frac{k}{2\pi} \epsilon^{\mu\nu\rho} \partial_\nu \tr\, x_\rho\ ,
\end{equation}
where $x_\mu$ is the dynamical $U(N)$ gauge field. The normalisation
of the current is such that for a monopole of flux $+1$ w.r.t.~one of
the $U(1)$ subgroups of $U(N_{})$, say the one with generator
$\text{diag}(1,0,\ldots,0)$, the electric charge corresponding to the
current \eqref{jtypeII} is $k$. More generally, the charges of all operators in this theory are quanized in units of $k$. As in the discussion around
\eqref{rescalesu}, it follows that $C_\mu$ in \eqref{ttymmodcoupold}
is not quite correctly normalised as a $U(1)$ gauge field, but that
the rescaled field
\begin{equation}\label{rescale}
{\tilde C}_\mu = k_{} C_\mu\ ,
\end{equation}
is indeed correctly normalised (note that the charge of monopole
operators is quantized in units of unity under the field
${\tilde C}_\mu$). The action \eqref{ttymmodcoupold} can thus be
rewritten as
\begin{equation}\label{redictwo}
\mc{S}_{\rm Type\, II}[a;C] = \frac{\i {\tilde k} N}{4\pi} \int a \ud
a\\  - \frac{\i N}{2\pi} \int {\tilde C} \ud a  + \frac{\i L_{\rm II} }{4\pi}
\int {\tilde C} \ud {\tilde C}\ ,
\end{equation}
As ${\tilde C}_\mu$ is a properly normalised $U(1)$ gauge field, it follows that  $L_{\rm II}$ must be an integer. 

Integrating out matter fields renormalizes ${\tilde k}$ to
$k_{\rm eff}$ and $N$ to $N_{\rm eff}$ according to \eqref{newenrank}
and we obtain a low-energy effective action of the form
\begin{equation}\label{ttymmodcoup}
  \mc{S}[a';C] = \frac{\i k_{\rm eff} N_{\rm eff}}{4\pi} \int a' \ud a'  - \frac{\i k N_{\rm eff}}{2\pi} \int C \ud a'  + \frac{\i L_{\rm II} k^2}{4\pi} \int C \ud C\ ,
\end{equation}
where $a'$ is the $U(1)$ gauge field given by $1 / N_{\rm eff}$ times
the trace of the $U(N_{\rm eff})$ gauge field. Note that this related
to the $U(1)$ gauge field part of the $U(N)$ gauge field $a$ by
$a' = \frac{N_{\rm eff}}{N} a$. Once again, we may re-express the
above action in terms of the correctly normalised ${\tilde C}$
\eqref{rescale}
\begin{equation}\label{ttymmodcoupm}
\mc{S}[a';C] = \frac{\i k_{\rm eff} N_{\rm eff}}{4\pi} \int a' \ud
a' - \frac{\i N_{\rm eff}}{2\pi} \int {\tilde C} \ud a'  + \frac{\i L_{\rm II} }{4\pi} \int {\tilde C} \ud {\tilde C}\ .
\end{equation}
The $a'$ equation of motion that follows from the above action
\eqref{ttymmodcoup} ensures that a single unit of $U(N_{\rm eff})$
flux traps $k_{\rm eff}$ fundamental fields while the $U(1)$ charge of
the single unit of flux is $k$. It follows that each fundamental field
effectively carries the charge $k / k_{\rm eff}$ under the field $C$
and the charge $1 / k_{\rm eff}$ under the correctly normalised
background field $\tl{C}$. Sometimes it will also be useful to define
another incorrectly normalised background field
$k_{\rm eff} C' = \tilde{C} = k C$ under which the fundamental field
charge $1$. In terms of this field $C'$, we have
\begin{equation}\label{ttymmodcoupnew}
\mc{S}[a';C'] = \frac{\i k_{\rm eff} N_{\rm eff}}{4\pi} \int a' \ud
a' - \frac{\i k_{\rm eff} N_{\rm eff}}{2\pi} \int C' \ud a'  + \frac{\i L_{\rm II} k_{\rm eff}^2}{4\pi} \int  C' \ud C'\ .
\end{equation}
Specialising to the case of the regular fermion and critical fermion
theories with one flavour of fermion with mass $m_F$, it follows that
the fundamental fermion effectively carries unit charge under $C$ when
$m_{F} k_F > 0$ but carries charge $\frac{|k_F|}{|k_F|-1}$ under this field when
$m_{F} k_F < 0$, matching the charge of scalar excitations and $W$
boson excitations under $SU(N)\, -\, {\rm Type~ II}$ duality.  In the
case of the critical boson or regular boson theories with one flavour
of boson, on the other hand, $k_{\rm eff} = k_B$ in both phases, so
the global charge of scalar excitations in the unHiggsed phase and the
$W$ bosons in the Higgsed phase are both unity. This charge assignment
is also consistent with $SU(N_F) - {\rm Type~ II}$ duality (recall
that $SU(N_F)$ fermions carry unit global charge for both signs of the
fermion mass).

\subsubsection{Type I theories}
The $U(1)$ part of the gauge Lagrangian for the Type I theory is
\begin{equation}\label{toneymmod}
  \mc{S}[a] =  \frac{\i {\tilde \kappa}  N}{4\pi} \int a \ud  a\ .
\end{equation}
(${\tilde \kappa}$ was defined in \eqref{deftk}). Upon coupling to the
background gauge field $C_\mu$, this action is modified in the UV to
\begin{equation}\label{toneymmodcoupold}
\mc{S}_{\rm Type\, I}[a;C] = \frac{\i { \tilde \kappa}_{}  N_{}}{4\pi} \int a
\ud a - \frac{\i\kappa N_{}}{2\pi} \int C \ud a + \frac{\i \kappa^2 L_{\rm I} }{4\pi}\int C \ud C\ ,
\end{equation}
The $U(1)$ global charge of a monopole of flux $+1$ w.r.t.~one of the
$U(1)$ subgroups of $U(N)$ carries charge ${\kappa}$, and so all
gauge-invariant operators in this theory carry global $U(1)$ charges
that are quantized in units of $\kappa$. It follows that $C$ in
\eqref{jtypeII} is not correctly normalised, and the correctly
normalised $U(1)$ field is the rescaled field
\begin{equation}\label{rescaleI}
  {\tilde C}_\mu = \kappa C_\mu\ .
\end{equation}
The coupling \eqref{toneymmodcoupold} can be rewritten in terms of the
field ${\tilde C}$ as
\begin{equation}\label{redicto}
\mc{S}[a;{\tilde C}] = \frac{\i { \tilde \kappa}  N_{}}{4\pi} \int a
\ud a - \frac{\i N}{2\pi} \int {\tilde C} \ud a + \frac{\i L_{\rm I} }{4\pi}
\int {\tilde C} \ud {\tilde C}\ .
\end{equation}
Since $\tilde{C}$ is a correctly normalised field, it follows that the
quantity $L_{\rm I}$ can be an arbitrary integer.

Accounting for the shifts in level and rank \eqref{newenrank} coming
from integrating out massive fermions and bosons, the action becomes
\begin{equation}\label{toneymmodcoup}
  \mc{S}[a';C] = \frac{\i { \tilde \kappa}_{\rm eff}  N_{\rm eff}}{4\pi} \int a'
  \ud a' - \frac{\i\kappa N_{\rm eff}}{2\pi} \int C \ud a' + \frac{\i \kappa^2 L_{\rm I} }{4\pi}
  \int C \ud C\ ,
\end{equation}
where, once again, $a'$ is the $U(1)$ gauge field associated to the
$U(N_{\rm eff})$ gauge field and is related to the $U(1)$ part $a$ of
the $U(N)$ gauge field as $a' = \frac{N}{N_{\rm eff}}a$. The above
action \eqref{toneymmodcoup} can also be rewritten in terms of
${\tilde C}$ yielding
\begin{equation}\label{toneymmodcouprw}
  \mc{S}[a';{\tilde C}] = \frac{\i { \tilde \kappa}_{\rm eff}  N_{\rm eff}}{4\pi} \int a'
  \ud a' - \frac{\i N_{\rm eff}}{2\pi} \int {\tilde C} \ud a' + \frac{\i L_{\rm I} }{4\pi}
  \int {\tilde C} \ud {\tilde C}\ ,
\end{equation}
Proceeding along the lines of the previous subsubsection we see that a
fundamental excitation carries an effective $U(1)$ charge of magnitude
$\frac{\kappa}{\kappa_{\rm eff}}$ under the gauge field $C$.  As
$\kappa$ and $\kappa_{\rm eff}$ both switch sign under duality,
charges of dual fundamental excitations match under
duality\footnote{More concretely, fundamental excitations carry unit
  charge in the Type I bosonic theories with unbroken gauge group and
  the Type I fermionic theories in which $m_F$ has the same sign as
  $k_F$. On the other hand fundamental excitations carry charge
  $\frac{|\kappa_F|}{|\kappa_F|-1}$ in Type I bosonic theories with
  broken gauge group or Type I fermionic theories in which the sign of
  $m_F$ is opposite to that of $k_F$.}. It is helpful to define
another incorrectly normalised background field $C'$ given by
$\kappa_{\rm eff} C' = \tilde{C} = \kappa C$ such that the fundamental
excitations have charge $1$ under $C'$. In terms of this field, the
action \eqref{toneymmodcouprw} becomes
\begin{equation}\label{toneymmodcoupmore}
  \mc{S}[a'; C'] = \frac{\i { \tilde \kappa}_{\rm eff}  N_{\rm eff}}{4\pi} \int a'
  \ud a' - \frac{\i\kappa_{\rm eff} N_{\rm eff}}{2\pi} \int  C' \ud a' + \frac{\i L_{\rm I} \kappa_{\rm eff}^2}{4\pi}
  \int  C' \ud  C'\ .
\end{equation}

\subsection{A precise statement of duality including background gauge
  fields}
We present the actions including Lagrangians for background gauge
fields for the dual pairs $SU(N)_k \longleftrightarrow U(N')_{k',k'}$
and $U(N)_{k,\kappa} \longleftrightarrow U(N')_{k',\kappa'}$. Recall
that the duality of ranks and levels are given by \eqref{dumap} which
we reproduce here for convenience:
\begin{equation}\label{dumapre}
  k' = -\sgn(k)N\ ,\quad N' = |k|\ .
\end{equation}

\subsubsection{$SU(N)$ and Type II}

The precise statement of the $SU(N)$\, - Type II duality is summarised
by the following two equivalent equations
\begin{equation}\label{cdo}
  \frac{\i {\tilde k} N}{4\pi} \int C \ud C
  \longleftrightarrow \frac{\i {\tilde k'} N'}{4\pi} \int a \ud
  a - \frac{\i N}{2\pi} \int { \tilde C} \ud a\ ,\quad\text{with}\quad {\tilde C} = k'C\ ,
\end{equation}
where we have again suppressed the gauge action for the $SU(N)$ and
$SU(N')$ gauge fields on the LHS and RHS respectively and the
dependence on the matter fields. Substituting $\tilde{C} = k' C$ on
the right hand side of \eqref{cdo}, we get
\begin{equation}\label{cd}
  \frac{\i {\tilde k} N}{4\pi} \int C \ud C \longleftrightarrow \frac{\i {\tilde k'} N'}{4\pi} \int a \ud a - \frac{\i N'k'}{2\pi} \int { C} \ud a\ .
\end{equation}
The above equations have the following meaning. They assert that the
path integral of a bosonic (resp.~fermionic) theory coupled to the
$SU(N)_k$ Chern-Simons theory on the LHS is the same as the path
integral of the fermionic (resp.~bosonic theory) coupled to the Type
II $U(N')_{k',k'}$ theory on the RHS.

Let us summarise \eqref{cd} and \eqref{cdo} in words. They assert that
the bosonic (resp. fermionic) $SU(N)_k$ theory with background level
$L=0$ (see \eqref{constoflid}) is dual to the fermionic
(resp.~bosonic) Type II $U(N')_{k',k'}$ theory with background level
$L_{\rm II}=0$ (see \eqref{redictwo}). More generally, the bosonic
(resp.~fermionic) $SU(N)$ theory with contact integer $L$ and the
fermionic (resp.~bosonic) Type II theory with contact integer
$L_{\rm II}=0$ are dual provided $L=L_{\rm II}$.

\subsubsection{Type I and Type I}

The precise statement of Type I\,-\,Type I duality is 
\begin{multline}\label{finconcsum}
  \frac{\i {\tilde \kappa} N}{4\pi} \int a \ud a -\frac{\i N
    \kappa}{2\pi} \int a \ud {H} \longleftrightarrow \\ \frac{\i
    {\tilde k'} N'}{4\pi} \int a' \ud a' -\frac{\i N' \kappa'}{2\pi}
  \int a' \ud {H} + \frac{ \i~ {\rm sgn}(\kappa')(\kappa')^2}{4
    \pi}\int { H} \ud {H}\ .
\end{multline}
Equivalently, in terms of the properly normalised $U(1)$
field\footnote{Note that fundamental fields carry unit charge under
  the field $H_\mu$ on both sides of the duality. Under the properly
  normalised gauge field ${\tilde H}$, on the other hand, fundamental
  fields carry charge $\frac{1}{\kappa}$ on the LHS of the duality,
  while fundamental fields on the RHS of the duality carry the
  opposite charge $\frac{1}{\kappa'}=-\frac{1}{\kappa}$. }
${\tilde H}= \kappa H=-\kappa' H$, we have
\begin{multline}\label{finconcsumn}
  \frac{\i {\tilde \kappa} N}{4\pi} \int a \ud a
  -\frac{\i N }{2\pi} \int a \ud {\tilde H} \longleftrightarrow
  \\ \frac{\i {\tilde k'} N'}{4\pi} \int a' \ud
  a' +\frac{\i N'}{2\pi} \int a' \ud \tilde{H} 
  + \frac{ \i~ {\rm sgn}(\kappa')}{4 \pi}\int {\tilde H} \ud {\tilde H}\ ,
\end{multline}
In words a bosonic (resp.~fermionic) Type I theory with rank and level
$(N,k)$ and with background level $L_{\rm I}=0$ is dual to the
fermionic (resp.~bosonic) Type I theory with rank and level $(N', k')$
and with background level $L'_{\rm I}={\sgn}(k')$. More generally, the
two Type I theories are dual to each other provided the background
levels on the LHS and RHS satisfy
$$L'_{\rm I}-L_{\rm I} ={\rm sgn}(k')=-{\rm sgn}(k)\ .$$

\subsection{Consistency check for conjectured dualities}

In this subsection we will present two consistency checks of the
dualities described in the previous subsection.  The consistency
checks presented below are essentially identical to those presented in
\cite{Hsin:2016blu}, and are included here for completeness, and also
to have a clear presentation of the dualities in the notation used in
this paper.

The consistency checks essentially involve the following operations on
both sides of a given duality. We promote the background field $C$ to
a dynamical gauge field $c$, add the coupling $\int H \ud c$ to a new
background gauge field and finally integrate over $c$ in the path
integral. If the old duality is consistent, we expect to recover
another duality with the correct actions including the correct
background levels on either side of the new duality.

\subsubsection{Self-consistency of $SU(N)$\,-\,Type II duality}
We start with an $SU(N)_k \longleftrightarrow U(N')_{k',k'}$
duality. It follows from \eqref{cdo} that
\begin{align}\label{cdn}
  \frac{\i {\tilde k} N}{4\pi} \int c \ud c -\frac{\i N}{2\pi} \int c
  \ud {\tilde H}
  &\longleftrightarrow \frac{\i {\tilde k'} N'}{4\pi}
    \int a \ud a - \frac{\i N'k'}{2\pi} \int {c} \ud a
    -\frac{\i N}{2\pi} \int c \ud {\tilde H}\ ,\nonumber\\
  &\longleftrightarrow \frac{\i {\tilde k'} N'}{4\pi} \int a \ud a - \frac{\i N'}{2\pi}\int {{\tilde c}} \ud a -\frac{\i\, {\rm sgn}(k')}{2\pi} \int {\tilde c} \ud {\tilde H}\ ,\nonumber\\
  &\longleftrightarrow \frac{\i {\tilde k'} N'}{4\pi} \int a \ud a\ ,\quad N'a = -{\rm sgn} (k') {\tilde H}\ ,\nonumber \\
  &\longleftrightarrow \frac{\i {\tilde k'} N'}{4\pi} \int H \ud
    H\ ,\quad N' H = \tilde{H} = -\sgn(k') N' a = k a \ .
\end{align}
We have started with \eqref{cdo}, renamed the gauge field $C_\mu$ as
$c_\mu$, added the same contact term\footnote{${\tilde H}$ is a
  properly normalised $U(1)$ gauge field; note that the integrality
  properties of the coefficients of all terms in the first line of
  \eqref{cdn} are consistent with this.}
$-\frac{\i N}{2\pi} \int c \ud {\tilde H}$ to both sides and then we
have path integrated over $c$ (i.e. made it a dynamical gauge
field). In going from the first to the second line of \eqref{cdn} we
have made the substitution $k' c = {\tilde c}$ and have used the fact
that $k'= {\rm sgn}(k') N$. In going from the second to the third line
we have integrated out the Lagrange multiplier field ${\tilde c}_\mu$
(note that the equation $N'a' = -{\rm sgn} (k') {\tilde H}$ is
sensible because both sides of the equation involved a properly
normalised $U(1)$ gauge field). In going from the third to the fourth
line we have substituted the incorrectly normalised background field
$H$ in place of the correctly normalised field $\tilde{H}$.

The final result of the manipulations in \eqref{cdn} is that
\begin{equation}\label{cdnfin}
  \frac{\i {\tilde k} N}{4\pi} \int c \ud c -\frac{\i N}{2\pi} \int c
  \ud {\tilde H} \longleftrightarrow \frac{\i {\tilde k'} N'}{4\pi} \int H \ud
  H\ .
\end{equation}
This is simply \eqref{cdo} with the replacement
$$(k,N) \leftrightarrow (k', N')\ .$$. Let us summarize. The starting point of the manipulations \eqref{cdn} was a duality in which the matter on the LHS is $SU(N)$ coupled with $L=0$ and the matter on the RHS is $U(N')_{k',k'}$ coupled with $L_{\rm II}=0$. The end point 
is the duality in which the matter on the RHS is $SU(N')$ gauged with
$L=0$ and the matter on the LHS is coupled to a $U(N)_{k,k}$ theory
with $L_{\rm II}=0$.

It follows that the discussion of this section can be regarded as a consistency
check of the proposed $SU(N)$\,-\,Type II duality.

\subsubsection{Type I\,-\,Type I from $SU(N)$\,-\,Type II}

As the starting point of the analysis of this section we use the
$SU(N)$\,-\,Type II duality \eqref{cd} with  contact term
$$\frac{ \i~ {\rm sgn}(k) N^2 }{4 \pi}\int C \ud C=-\frac{ \i~ {\rm sgn}(k') (k')^2 }{4 \pi}\int C \ud C\ ,$$ added to both sides\footnote{In other words the $SU(N)$ Type II duality with $L=L_{\rm II} = {\rm sgn}(k)=-{\rm sgn}(k')$.}, i.e.~with 
\begin{equation}\label{cdnew}
  \frac{\i {\tilde \kappa} N}{4\pi} \int C \ud C
  \longleftrightarrow \frac{\i {\tilde k'} N'}{4\pi} \int a \ud
  a  - \frac{\i N'k'}{2\pi} \int { C} \ud a 
  - \frac{ \i~ {\rm sgn}(k') (k')^2 }{4 \pi}\int C \ud C\ .
\end{equation}
Renaming $C$ as $c$, adding $-\frac{\i N}{2\pi} \int c \ud {\tilde H}$
to both sides of \eqref{cdnew} and then promoting $c$ to a dynamical
field (i.e. path integrating over it) we find
\begin{multline}\label{cdnnew}
  \frac{\i {\tilde \kappa} N}{4\pi} \int c \ud c -\frac{\i N}{2\pi}
  \int c \ud {\tilde H} \longleftrightarrow \\ \frac{\i {\tilde k'}
    N'}{4\pi} \int a \ud a - \frac{\i N'k'}{2\pi} \int {c} \ud a -
  \frac{ \i\, {\rm sgn}(k') (k')^2 }{4 \pi}\int c \ud c -\frac{\i\, {\rm
      sgn}(k') k'}{2\pi} \int c \ud {\tilde H}\ .
\end{multline}
The right hand side of \eqref{cdnnew} can be simplified as follows:
\begin{align}\label{cdnman}
  \text{RHS} &= \frac{\i {\tilde k'} N'}{4\pi} \int a \ud
  a - \frac{\i N'}{2\pi} \int {{\tilde c}} \ud a
  - \frac{ \i\, {\rm sgn}(k')}{4 \pi}\int {\tilde c} \ud {\tilde c}
  -\frac{\i\, {\rm sgn}(k')}{2\pi} \int {\tilde c} \ud {\tilde H}\ ,\nonumber\\
  &= \frac{\i {\tilde k'} N'}{4\pi} \int a \ud
    a + \frac{ \i\, {\rm sgn}(k')}{4 \pi}\int \left({\tilde H}+{\rm sgn}(k') N' a \right)  \ud \left({\tilde H} + {\rm sgn}(k') N' a \right)\ ,\nonumber\\
  &= \frac{\i {\tilde k'} N'}{4\pi} \int a \ud
    a + \frac{\i N'}{2\pi} \int a' \ud {\tilde H} 
    + \frac{ \i\, {\rm sgn}(k')}{4 \pi}\int {\tilde H} \ud {\tilde H}\ .
\end{align}
In going from \eqref{cdnnew} to the first line of \eqref{cdnman} we
have moved to the new integration variable ${\tilde c}=k'c$. In going
from the first to the second line of \eqref{cdnman} we have completed
the square in ${\tilde c}$ and have also use the fact that the path
integral
$$ -\frac{ \i~ {\rm sgn}(k')}{4 \pi}\int f \ud f\ ,\quad f={\tilde c} - {\tilde H} + {\rm sgn}(k') N' a'\ ,$$ 
is trivial and can be dropped\footnote{The fact that the path integral
  for a level $\pm 1$ $U(1)$ theory is trivial is essentially the
  statement that this theory is empty. This is familiar from many points of view. For instance, the  Hilbert space of this
  theory on the torus has only one state. The vacuum is the only
  primary operator of the WZW theory dual to this Chern-Simons
  theory.}  \cite{Hsin:2016blu}.  Moving to the background gauge field
$H_\mu$ defined by ${\tilde H} = \kappa H = -\kappa' H$, we get
\begin{equation}\label{finconc}
  \frac{\i {\tilde \kappa} N}{4\pi} \int c \ud c
  -\frac{\i N \kappa}{2\pi} \int c \ud {H}
  \sim \frac{\i {\tilde k'} N'}{4\pi} \int a \ud
  a -\frac{\i N' \kappa'}{2\pi} \int a \ud {H} 
  + \frac{ \i~ {\rm sgn}(\kappa')(\kappa')^2}{4 \pi}\int { H} \ud {H}\ ,
\end{equation} 
where now, the dynamical gauge field $c$ on the left hand side is the
$U(1)$ part of the $U(N)$ gauge field of a Type I theory. Similarly,
the gauge field $a$ on the right hand side, which was originally the
$U(1)$ gauge field in a Type II theory, now becomes a $U(1)$ gauge
field of a Type I theory due to the change in its level. Thus,
\eqref{finconc} is in perfect agreement with the  statement of
the Type I\ - Type I duality \eqref{finconcsum}, including the correct
relative values of the background levels $L_I$ and $L'_I$. Starting
from the $SU(N)$\ - Type II duality we have thus been able to derive the
Type I\ - Type I duality.

\subsection{Charged gauge-invariant operators}\label{ops}

As we have mentioned in the main text in Section \ref{symop}, the
spectrum of gauge invariant operators in the theories under study
include `single-sum' operators (operators that are constructed from
the gauge contractions of derivatives acting on a fundamental and
other derivatives acting on an antifundamental) whose dimension
remains finite in the large $N$ limit and more complicated operators
whose dimension goes to infinity in the large $N$ limit. In each case
the single sum operators are uncharged under the global $U(1)$
symmetry. The only charged operators in the theory are the more
complicated operators referred to above, whose nature we now describe.

\subsubsection{$SU(N)$ theories}\label{SUNop}

The complicated operators for $SU(N)$ theories are baryonic in nature.
Baryon operators are given by the contraction of $N$ fundamental
fields with the colour Levi-Civita tensor (and in general, with
different numbers of derivatives acting on each of the $N$
fields). There are many distinct baryon operators, labelled by the $N$
distinct `letters' (the structure of the derivative operators acting
on the fundamental fields) used to construct the operator.

In an $SU(N)$ theory coupled to a single fundamental boson, the baryon
operators are all bosonic in nature i.e.~they have integral spins (all
arising from the derivatives in the operator in the case of scalars
and also additionally from the spin of the $W$ bosons in the Higgsed
phase). On the other hand, in an $SU(N)$ theory coupled to a single
fundamental fermion, the baryon operators are bosonic when $N$ is even
and fermionic when $N$ is odd.

Baryon operators built out of fundamental fields carry charge 
$N$ under the gauge field $C_\mu$ in \eqref{SUNlag}, but 
carry unit charge under the gauge field ${\tilde C}_\mu$ 
defined in \eqref{rescalesu}. 

\subsubsection{Type II $U(N)$ theories}\label{TypeIIop}

In the case of Type II $U(N)_{k,k}$ theories, the more complicated
operators are monopole operators. A basic monopole operator consists
of unit magnetic flux for trace of the $U(N)$ gauge field. The
Chern-Simons coupling endows the operator with a electric gauge charge
which is in the `totally symmetric' $SU(N)$ colour representation with
$|k|$ boxes and a charge of $k$ for the $U(1)$ factor of the gauge
group (i.e. this representation carries charge $k$ under the gauge
field $a_\mu$ in \eqref{ttymmod}). In order to make this gauge
invariant, we have to attach $|k|$ copies of the fundamental (or
antifundamental) field in the colour representation
\begin{equation}\label{fsrep}
  \underbrace{\yng(2) \cdots \yng(1)}_{|k|~\text{boxes}}\ ,
\end{equation}
(and in general, different numbers of derivatives acting on each of
the $|k|$ fields). As in the case of the baryon operators, there are
many distinct monopole operators labelled by the $|k|$ distinct
`letters' (derivatives acting on fundamental fields).

The spin of a monopole operator is given by the sum of the intrinsic
spin $s_{\rm int}$ and the statistical spin $s_{\rm stat}$
\begin{equation}
  s = s_{\rm int} + s_{\rm stat}\ .
\end{equation}
The intrinsic spin is the sum of the spin of all the letters that go
into the construction of the operator. The `statistical spin' is given
by the formula
\begin{equation}\label{statspin}
  s_{\rm stat}= -\frac{C_2(R)}{\kappa} - \frac{Q^2}{2Nk}\ ,
\end{equation} 
where $C_2(R)$ is the quadratic Casimir of the $SU(N)$ representation
$R$ with Young diagram \eqref{fsrep} and $Q$ is the $U(1)$ charge of
this representation. It is not difficult to verify that
\cite{toappear1}
\begin{equation}\label{casimir}
  C_2(R) = \frac{1}{2}\left(k^2 - |k|  + N|k| - \frac{k^2}{N}\right)\ ,\quad Q = |k|\ ,\quad\text{and hence}\quad s_{\rm stat} = -\frac{k}{2}\ .
\end{equation} 
In the case of the theory with only bosonic matter the intrinsic spin
$s_{\rm int}$ of this operator is always integral (and comes only from
the derivatives for scalar excitations and also additionally from the
$W$ bosons in the Higgsed phase). On the other hand, the statistical
spin $s_{\rm stat}$ given in \eqref{casimir} is integral or
half-integral depending on whether $k$ is even or odd. It follows that
the monopole operator has integer spin (and hence bosonic) or
half-integer spin (and hence fermionic) depending on whether $k$ is
even or odd respectively.

In the case of the theory with only fermionic matter, the intrinsic
spin of the monopole operator is integral or half-integral depending
on whether $k$ is even or odd respectively. The statistical spin of
the operator continues to be integral or half-integral depending on
whether $k$ is even or odd respectively. Thus, monopole operators in
Type II theories coupled to fermions are always bosonic.

Note that this maps perfectly to the statistics of the baryon
operators of the dual $SU(N')$ theories with $N' = |k|$. When the Type
II $U(N)$ theory has only bosonic (resp.~fermionic) matter, the
$SU(N')$ theory has only fermionic (resp.~bosonic) matter. In the case
that the Type II theory was coupled to fermionic matter, the monopole
operator is bosonic when $k$ is even and fermionic when $k$ odd. The
same is true of the baryon operators for the $SU(N')$ theory with
bosonic matter since $N' = |k|$. Similarly, the Type II theory with
fermionic matter has only bosonic monopole operators, which is in
follows that, in this case, the in perfect agreement that the baryon
operators of the $SU(N')$ theory with bosonic matter are always
bosonic.

The monopole operator discussed in this subsubsection carries 
charge $k$ under the gauge field $C_\mu$, and charge unity 
under the rescaled gauge field ${\tilde C}_\mu$. 

We re-emphasize that that the monopole operator defined above has
half-integer spin in the theory with bosonic matter when $k$ is an odd
integer. This tells us that the spectrum on $S^2$ of Chern-Simons
theories coupled to bosonic fundamental matter includes fermions when
$k$ is odd. It follows that Chern-Simons theories coupled to bosons in
the fundamental representation are secretly fermionic theories, and
are well-defined only on manifolds with a specified spin structure.

The observations of the previous paragraph are a reflection of the
following facts about pure Chern-Simons theory \cite{Hsin:2016blu}.
Despite the fact that the monopole operator described in this
subsection appears to transform in the representation \eqref{fsrep},
it is actually gauge invariant (the gauge charges of the matter fields
\eqref{fsrep} are screened by the monopole flux). This screening is a
reflection of the fact that the only representation that appears in
the completely symmetric fusion of $k$ fundamental primaries of WZW is
the identity representation. The authors of \cite{Hsin:2016blu} noted
that Wilson lines in the representation \eqref{fsrep} are `invisible'
in the sense of monodromy, but are not quite nothing because they
carry half integral spin. These otherwise invisible fermionic lines
have to be added to the spectrum, and make pure Chern Simons theory at
odd $k$ a theory that is only well defined on a spin manifold with a
specified spin structure.

\subsubsection{Type I $U(N)$ theories}\label{TypeIop}

The complicated operators in Type I $U(N)_{k,\kappa}$ are monopole
operators. The electric gauge charge of these operators under the
$U(1)$ factor of the $U(N)$ gauge group is now $\kappa$ rather than
$k$, while the $SU(N)$ representation is still the totally symmetric
representation with $|k|$ boxes. The $SU(N)$ representation can be
neutralised by attaching $|k|$ fundamental (or antifundamental) fields
in the representation \eqref{fsrep}, but this is not sufficient to
neutralise the $U(1)$ charge. The solution is to take $N$ additional
fundamental fields but now transforming in a gauge singlet (in other
words, a baryon-type operator). In other words, we have the following
representations of the matter field attachments:
\begin{equation}
  \underbrace{\yng(2) \cdots \yng(1)}_{|k|~\text{boxes}}\ \oplus\ \text{\footnotesize{$N$ boxes}} \begin{cases} \ytableausetup
{mathmode, boxsize=1.2em}
\begin{ytableau}
\\
\\
\none[\vdots]  \\
\\
\end{ytableau}  \end{cases}
\end{equation}
The intrinsic spins of these operators get a contribution from the
totally symmetric $|k|$ number of fundamental fields and an additional
contribution from the totally antisymmetric baryon-type attachment
with $N$ fundamental fields. The statistical spin is now given by the
formula
\begin{equation}\label{statspinn}
s_{\rm stat} = -\frac{C_2(R)}{\kappa} - \frac{Q^2}{2(N\kappa)}\ ,
\end{equation} 
with $C_2(R)$ being the quadratic Casimir for the representation
\eqref{fsrep} whose value is displayed in \eqref{casimir}, and $Q$ is
the $U(1)$ electric charge of the monopole operator which is given by
$\kappa$. Plugging these into \eqref{statspinn}, we get
\cite{toappear1}
\begin{equation}\label{statspinnfin}
  s_{\rm stat} = -\sgn(k) \frac{|k| + 1}{2}\ .
\end{equation}
In the theory with bosonic matter, the baryon-type attachment as well
as the fundamental fields in the totally symmetric representation
\eqref{fsrep} are bosonic and hence have integral intrinsic spin. The
statistical spin is half-integral (resp.~integral) when $k$ is even
(resp.~odd). Thus, the monopole operator is a fermion when $k$ is even
and a boson when $k$ is odd.

In the theory with fermionic matter, the baryon-type attachment has
half-integral (resp.~integral) intrinsic spin when $N$ is odd
(resp.~even) and the $|k|$ fundamental fields in the totally symmetric
representation have half-integral (resp.~integral) spin when $k$ is
odd (resp.~even). The statistical spin is integral
(resp.~half-integral) when $k$ is odd (resp.~even). Thus the
contribution to the total spin from the $|k|$ fundamental fields in
the totally symmetric representation and the statistical spin is
always half-integral for any value of $k$. Thus, the monopole operator
in fermionic theories are bosons when $N$ is odd and fermions when $N$
is even.

Since $|k|$ is mapped to $N'$ and bosonic matter is mapped to
fermionic matter under the Bose-Fermi duality, it is straightforward
to check that the above statements are consistent with duality. The
monopole operator discussed in this subsubsection carries charge
$\kappa$ under the gauge field $C_\mu$, and charge unity under the
rescaled gauge field ${\tilde C}_\mu$.

As in the previous subsection, the formula \eqref{statspinnfin} tells us that even  Type I theories coupled to bosonic matter have fermions in their spectrum, 
and so are well defined only on spin manifolds (and need a spin structure for their definition) when $k$ is even.
As in the previous subsection this fact is closely related to the fact that pure Type I Chern Simons theories have 
otherwise invisible fermionic line operators in their spectrum when $k$ is even \cite{Hsin:2016blu}. 
 
\subsection{$SU(N)$ and $U(N)$ theories at finite chemical potential}\label{usu}

In this brief subsection we will explain that even though the global
symmetries for Type II and Type I theories are structurally very
different from that of the $SU(N)$ theory, it is nonetheless the case
that a finite chemical potential ensemble for Type II and Type I
theories are identical to their $SU(N)$ counterparts in the large $N$
limit.

Consider the Type II theory \eqref{ttymmodcoup} in the presence of a
nontrivial background gauge field $C_\mu$. In this section, we work
with the incorrectly normalised field $C' = \frac{k}{k_{\rm eff}} C$
as defined above equation \eqref{ttymmodcoupnew} such that the
fundamental excitations of a particular phase of the theory have unit
charge. We reproduce the action here for convenience:
\begin{equation}\label{chempottypeII}
\mc{S}[a';C'] = \frac{\i k_{\rm eff} N_{\rm eff}}{4\pi} \int a' \ud
a' - \frac{\i k_{\rm eff} N_{\rm eff}}{2\pi} \int C' \ud a'  + \frac{\i L_{\rm II} k_{\rm eff}^2}{4\pi} \int  C' \ud C'\ .
\end{equation}
The equation of motion for the dynamical $U(1)$ gauge field $a'_\mu$
is
\begin{equation}\label{eomam}
  \frac{\i N_{\rm eff} k_{\rm eff}}{2\pi} \epsilon^{\mu\nu\rho} (\partial_\nu a'_\rho - \partial_\nu C'_\rho) =  J^\mu\ ,
\end{equation}
where $J^\mu$ is the matter $U(1)$ current that couples to the
dynamical gauge field $a_\mu$. The solution to \eqref{eomam} clearly
takes the form
\begin{equation}\label{aset}
a'_\mu= C'_\mu + {\tilde a}_\mu\ ,
\end{equation} 
where ${\tilde a}_\mu$ satisfies\footnote{In order to get
  \eqref{asetn}, we have used the fact that, in Euclidean space,
  $\epsilon_{\alpha \beta \rho} \epsilon^{\mu\nu \rho} =
  \delta^{\mu}_{\alpha} \delta^{\nu}_{\beta} - \delta^{\mu}_{\beta}
  \delta^{\nu}_{\alpha}$ with the choice
  $\epsilon_{123}=\epsilon^{123}=+1$.}
\begin{equation}\label{asetn} 
  \frac{\i N_{\rm eff} k_{\rm eff}}{4\pi} \epsilon^{\mu\nu\rho} {\tilde f}_{\nu\rho} \equiv \frac{\i N_{\rm eff} k_{\rm eff}}{2\pi} \epsilon^{\mu\nu\rho}\partial_\nu {\tilde a}_\rho = J^\mu\ .
\end{equation} 
Turning on a chemical potential $\mu$ amounts to setting
$C'_0=\i \mu$. It follows that
\begin{equation}\label{effectiveass}
a_\nu= \i \mu ~\delta_{\nu 0} + {\tilde a}_\nu\ .
\end{equation} 

Had we been dealing with an $SU(N)$ rather than a Type II theory,
$a_\mu$ would have been the background field and
$a_\nu= \i \mu ~\delta_{\nu 0}$ would have corresponded to turning on
a chemical potential $\mu$. The term ${\tilde a}_\nu$ on the RHS of
\eqref{effectiveass} is therefore captures the difference between
turning on a chemical potential of magnitude $\mu$ in the Type II
theory and the $SU(N)$ theory.

In order to understand the nature and magnitude of this difference, we
note that turning on a chemical potential of order unity gives rise to
a charge density of order $N_{\rm eff}$. Two things immediately
follow by consulting \eqref{asetn}. First, ${\tilde a}_\nu$
corresponds to the gauge field for a uniform magnetic field, at least
approximately since the charge density on the RHS of \eqref{asetn} is
approximately uniform. In other words, the effect of turning on a
chemical potential for the Type II theory is similar to turning on
both a chemical potential and a background magnetic field for the
$SU(N)$ theory. Second, the magnitude of this magnetic field scales
like ${1}/{k_{\rm eff}}$. When $k_{\rm eff}$ is of order unity, this
magnetic field plays a key role in the dynamics of the Type II theory
at finite chemical potential. On the other hand, in the 't Hooft large
$N$ limit, this magnetic field is parametrically small and can be
ignored.

Of course the discussion of a chemical potential in the Type I theory
is very similar; in this case ${\tilde a}_\nu$ is the field strength
corresponding to a magnetic field whose magnitude is of order
$\frac{1}{\kappa_{\rm eff}}$. It follows that there is effectively no
difference between $SU(N)$, Type II and Type I theories at finite
chemical potential at leading order in the 't Hooft large $N$ limit.

\section{Duality of off-shell free energies and occupation numbers}\label{dualityapp}

We show that the off-shell free energies of the appropriate fermionic
and bosonic theories map to each other under the duality map. This
requires the mapping of levels and ranks \eqref{level-rankmap}
\begin{equation}\label{applevelrankmap}
N_{B}=|\kappa_F|-N_{F}\ ,\quad \kappa_B = -\kappa_F\ ,
\end{equation}
the identification of the UV parameters \eqref{rfcbmap} for the
quasi-fermionic theories
\begin{equation}\label{apprfcbmap}
  \lambda_B m_B^{\rm cri} = - m_F\ ,
\end{equation}
and \eqref{rbcfmap} for the quasi-bosonic theories
\begin{equation}\label{apprbcfmap}
  x_6^B = x_6^F\ ,\quad b_4 = y_4\ ,\quad m_B^2 = y_2^2\ ,
\end{equation}
the map between the holonomy distributions \eqref{holonomymap}
\begin{equation}\label{appholonomymap}
  |\lambda_B|\rho_B(\pi-\alpha) + |\lambda_F| \rho_F(\alpha) = \frac{1}{2\pi}\ ,
\end{equation}
and the following identification of the off-shell variables
\eqref{offshvarmap}:
\begin{equation}\label{appoffshvarmap}
  c_B = c_F\ ,\quad \lambda_B \tl\cS = \lambda_F \tl\cC - \tfrac{1}{2} \sgn(\lambda_F) c_F\ ,\quad 2\lambda_B \sigma_B = -\frac{4\pi\zeta_F}{\kappa_F}\ .
\end{equation}
We divide the checking of duality into two parts. Note that the
holonomy distributions appear only in one term in each of the free
energies of the two bosonic and two fermionic theories. We first check
that the terms independent of the holonomy distribution in the free
energies map to each other under duality. We then separately check
that the terms involving the holonomy distributions map to each other
as well.

It is easy to check that the terms independent of the holonomy in the
free energies of the critical boson and regular fermion map to each
other provided we make the duality identifications
\eqref{applevelrankmap}, \eqref{apprfcbmap} and \eqref{appoffshvarmap}
(see e.g. around equations A.6 in
\cite{Choudhury:2018iwf}). Similarly, it is easy to check (see
e.g. \cite{Dey:2018ykx}) that that the regular boson and critical
fermion free energies \eqref{RBoffshellfe} and \eqref{CFoffshellfe})
map to each other once we make the duality identifications
\eqref{applevelrankmap}, \eqref{apprbcfmap} and
\eqref{appoffshvarmap}. Next, we check the duality of the
holonomy-dependent term in the free energies.

\subsection{Duality of the off-shell free energy when
  \texorpdfstring{$c_B > |\mu|$}{cB>mu}}

It is useful to introduce the following notation for the holonomy
integral in the last line of the each of the bosonic (see equations
\eqref{CB2voffshellfe} and \eqref{RBoffshellfe}) and fermionic (see
equations \eqref{RF2voffshellfe} and \eqref{CFoffshellfe}) free
energies.
\begin{align} \label{defscs}
\mathcal{C}_1({\epsilon}, |{\mu}|) & \equiv \int_{-\pi}^{\pi} d\alpha \ \rho_F(\alpha) \   \left(\log\big(1+e^{-\hat{\epsilon}-|\hat{\mu}|-\i\alpha }\big)+\log\big(1+e^{-\hat{\epsilon}+ |\hat{\mu}|+\i\alpha }\big)  \right)\ ,\nonumber \\
\cS_1({\epsilon},|{\mu}|) &\equiv  \int_{-\pi}^{\pi} d\alpha \ \rho_B(\alpha) \ \left(\log\big(1-e^{-\hat{\epsilon}-|\hat{\mu}|-\i\alpha }\big)+\log\big(1-e^{-\hat{\epsilon}+ |\hat{\mu}|+\i \alpha }\big)  \right)\ ,
\end{align}
where, $\cC_1$ is the holonomy integral in the last line of the free
energies for both the regular (in \eqref{CFoffshellfe}) and critical fermion (in \eqref{RF2voffshellfe}) while $\cS_1$ is
the holonomy integral for both the regular (in \eqref{RBoffshellfe}) and critical bosonic (in \eqref{CB2voffshellfe})
theories. It is then easy to verify using the map \eqref{appholonomymap}
between the holonomy distributions that
\begin{equation}\label{relmcms} 
|\lambda_B| \mathcal{S}_1({\epsilon}, |{\mu}|) + |\lambda_F|  \mathcal{C}_1({\epsilon}, |{\mu}|) = 0\quad{\rm when}\quad { \epsilon} >|{\mu}|\ .
\end{equation} 
To verify \eqref{relmcms}, we start with the expression for
$\mathcal{S}_1({\epsilon}, |{\mu}|)$ presented in the second line of
\eqref{defscs}, substitute $\rho_B$ in terms of $\rho_F$ using
\eqref{appholonomymap} and make the change of variables
\begin{equation}\label{changeofvar}
  \alpha= \pi -{\tilde \alpha}\ , \quad{ \rm when}\ {0<\alpha<\pi}\ ,\qquad {\rm and}\qquad \alpha = -\pi +{\tilde \alpha}\ ,\quad{\rm when}\ {-\pi <\alpha<0}\ .
\end{equation}
We find
\begin{equation}\label{mcS1andmcC1reln}
  \mathcal{S}_1({\epsilon}, |{\mu}|)= \frac{\lambda_F}{\lambda_B}  \mathcal{C}_1({\epsilon}, |{\mu}|)+ \frac{1}{|\lambda_B|} 
  \int_{-\pi}^{\pi}  \frac{d\alpha}{2\pi} \  \left(\log\big(1-e^{-\hat{\epsilon}-|\hat{\mu}|-\i\alpha }\big) + \log\big(1-e^{-\hat{\epsilon}+ |\hat{\mu}|+\i \alpha }\big)  \right)\ .
\end{equation}
When ${\epsilon}>|{\mu}|$, the integral in the second term on the RHS
vanishes. One can see this by power-series-expanding each of the
logarithms in $e^{-\hat{\epsilon} \pm (|\hat{\mu}| +\i \alpha)}$
(which is less than $1$ since $\hat{\epsilon} >|\hat{\mu}|$) and
integrating term by term. Alternatively, we may convert the integral
over $\alpha$ into a contour integral which nowhere intersects the
branch cut of the $\log$ because ${\epsilon} > {\mu}$, and use
Cauchy's theorem. We have thus established \eqref{relmcms}.

There is a further integral of $\cC_1$ or $\cS_1$ over the energy
$\epsilon$ in each of the bosonic and fermionic free energies
\eqref{CB2voffshellfe}, \eqref{RBoffshellfe}, \eqref{CFoffshellfe} and
\eqref{RF2voffshellfe}. The range of the $\epsilon$ integral is from
$\hat{c}_B$ to infinity for the bosons and from $\hat{c}_F$ to
infinity for the fermions. Since we work with values of $\mu$ such
that $c_B > |\mu|$ and $c_F > |\mu|$, it follows that all the values
of ${\epsilon}$ that appear in these integrals obey the condition
${\epsilon}> |{\mu}|$ and hence \eqref{relmcms} applies. Thus, the
terms with the holonomy integrals also map to each other under
Bose-Fermi duality.

In summary, it follows that the off-shell free energies of the regular
fermion and critical boson theories map to each other under duality
and similarly for the critical fermion and regular boson theories. It
follows that the expressions for the full thermal free energies map to
each other under duality, at least when $|\mu| \leq c_B$ and
$|\mu|\leq c_F$. In the next subsection, we check the duality of the
modified off-shell free energies in the situation when $|\mu| > c_B$
or $|\mu| > c_F$.

\subsection{Duality of the modified off-shell
  free energies for \texorpdfstring{$c_B < |\mu|$}{cB<mu}}

The holonomy dependent parts of the modified bosonic free energies
\eqref{RBoffshellfecorn} and \eqref{CB2voffshellfecorn}
\begin{multline}\label{recastb}
\cS'_1({ \epsilon}, |\mu|) =\int_{-\pi}^{\pi} d\alpha \ \rho_B(\alpha) \ \left(\log(1-e^{-\hat{\e} + |\hat{\mu}| + \i \a})+\log(1-e^{-\hat{\e} - |\hat{\mu}| - \i \alpha})\right) \\ - 2 \pi \Theta(|{ \mu}|-{ \epsilon} ) \int_{e^{{\hat \epsilon} -|{\hat \mu}|} }^1 dx \ \frac{{\tilde \rho}_B(x)}{x}\ ,
\end{multline}
and those of the modified fermionic free energies
\eqref{RF2voffshellfecorn} and \eqref{CFoffshellfecorn} are
\begin{multline}\label{recastf}
\cC'_1({ \epsilon}, |\mu|)  =\int_{-\pi}^{\pi} d\alpha \ \rho_F(\alpha) \ \left(\log(1 + e^{-\hat{\e} + |\hat{\mu}| + \i \a})+\log(1 + e^{-\hat{\e}-|\hat{\mu}|-\i \alpha})\right) \\ - 2 \pi \Theta(|{ \mu}|-{ \epsilon} ) \int_{e^{{\hat \epsilon} - |{\hat \mu}|} }^1 dx \ \frac{{\tilde \rho}_F(-x)}{x}\ .
\end{multline}
Similar to the discussion in the previous subsection, it is clearly
sufficient to show that
\begin{equation}\label{mapmcS1mcC1mod}
  |\lambda_B|\ \mc{S}_{1}'({\e},|{\mu}|) + |\lambda_F| \  \mc{C}_{1}'({\e},|{\mu}|) = 0\ ,
\end{equation}
in order to establish that \eqref{recastb} and \eqref{recastf} map to
each other under duality. We will now proceed to verify
\eqref{mapmcS1mcC1mod}. We need the duality relations between the
holonomies $\rho_B$, $\rho_F$ \eqref{appholonomymap} and the analytically
continued holonomies $\tl\rho_B$, $\tl\rho_F$ \eqref{tilderel} which
we reproduce below:
\begin{equation} \label{rhorho} 
|\lambda_B| \rho_B(\alpha) +|\lambda_F| \rho_F(\pi - \alpha) =\frac{1}{2\pi}\ ,\quad
|\lambda_B| \tl{\rho}_B(z) +|\lambda_F| \tl{\rho}_F(-z) =\frac{1}{2\pi}\ .
\end{equation}
It is then not difficult to verify that the LHS of
\eqref{mapmcS1mcC1mod} simplifies to
\begin{equation}\label{extratermsduality} 
\int_{-\pi}^{\pi} \frac{d\alpha}{2\pi} \ \left(\log\big(1-e^{-\hat{\e}+|\hat{\mu}|+\i \a}\big)+\log\big(1-e^{-\hat{\e}-|\hat{\mu}|-\i \alpha}\big)\right) -\Theta(|{ \mu}|-{ \e} ) \int_{e^{{\hat \epsilon} -{|\hat \mu}|} }^1  \ \frac{dx}{x}\ .
\end{equation}
The above equation can be rewritten using the variable change
$z = e^{\i\alpha}$ as
\begin{equation}\label{reer} 
\oint_C \frac{dz }{2\pi \i z } \ \left(\log\Big(1-z\, e^{-\hat{\e}+|\hat{\mu}|}\big)+\log\big(1-z\, e^{-\hat{\e}-|\hat{\mu}|}\big)\right) -\Theta(|{ \mu}|-{ \e} ) \int_{e^{{\hat \epsilon} -{|\hat \mu}|} }^1  \ \frac{dx}{x}\ ,
\end{equation}
where the contour $C$ runs over the unit circle. The second logarithm
in the first term of \eqref{reer} is analytic inside the unit circle
(its branch cut starts outside the unit circle) and so, by Cauchy's
theorem, the contour integral involving this logarithm vanishes. It
follows that \eqref{reer} simplifies to
\begin{equation}\label{reersimp} 
  \oint_C \frac{d z}{2\pi\i z} \ \log\big(1- z\,e^{-\hat{\e}+|\hat{\mu}|}\big) -\Theta(|{ \mu}|-{ \e} ) \int_{e^{{\hat \epsilon} -{|\hat \mu}|} }^1  \ \frac{dx}{x}\ .
\end{equation}
Now the contour integral involving the logarithm in \eqref{reersimp}
does not vanish, but receives contributions only from the
discontinuity across the cut of this logarithm.  Evaluating this
discontinuity we find that \eqref{reersimp} becomes
\begin{equation}\label{simpextrazfinal} 
\Theta(|{ \mu}|-{ \e} ) \int_{e^{{\hat \epsilon} -{|\hat \mu}|} }^{1}  \ \frac{dx}{x} -\Theta(|{\hat \mu}|-{\hat \e} ) \int_{e^{{\hat \epsilon} -{|\hat \mu}|} }^{1}  \ \frac{dx}{x} = 0\ ,
\end{equation}
establishing \eqref{mapmcS1mcC1mod}.

\subsection{Duality of the occupation numbers}

We will now examine how the occupation numbers
$\bar{n}_B({\e},q)$ (as computed in \eqref{occuptnBfin})
and $\bar{n}_F({\e},q)$ (as computed in
\eqref{occuptnFfin}) map to each other under duality. We have
\begin{align}\label{dualmanip}
&|\lambda_B| \bar{n}_B({\e},q) -
|\lambda_F| \bar{n}_F({\e},q)\nonumber \\
&=  \int_{-\pi}^{\pi} d\a \ |\lambda_B| \rho_B(\a) \  \frac{1}{e^{\hat{\e} - q \hat{\mu} -\i \a }-1} -
\int_{-\pi}^{\pi} d\a \ |\lambda_F| \rho_F(\a) \  \frac{1}{e^{\hat{\e} - q \hat{\mu} -\i \a }+1}\nonumber  \\
&\qquad + 2 \pi \Theta( q { \mu}-{ \e}) \Big( |\lambda_B|  {\tilde \rho}_B(e^{-|{\hat \mu}|+{\hat \epsilon}}) + |\lambda_F|  {\tilde \rho}_F(-e^{-|{\hat \mu}|+{\hat \epsilon}}) \Big)\ ,\nonumber \\
&= \frac{1}{2\pi} \int_{-\pi}^{\pi} d\a \  \frac{1}{e^{\hat{\e} - q \hat{\mu} -\i \a }-1} + \Theta( q { \mu}-{ \epsilon})\ ,\nonumber\\
&=0\ .
\end{align}
We have used 
\begin{equation}
|\lambda_B| \tl{\rho}_{B}(z) +|\lambda_F| \tl{\rho}_{F}(-z) =\frac{1}{2\pi}  \ ,
\end{equation}
to argue that the third line of \eqref{dualmanip} evaluates to the
second term on the fourth line of the same equation.  We have also
broken up the integral in the second term in the second line (the term
involving ${\rho}_F$) into an integral from $-\pi$ to $0$ and $0$ to
$\pi$, and then made the change of variables \eqref{changeofvar} and
used $e^{\pm \i \pi q}= -1 $ and \eqref{rhorho} to demonstrate that
the second line of \eqref{dualmanip} reduces to the first term on the
fourth line of the same equation. In going from fourth line to the
last line of \eqref{dualmanip}, we have used
\begin{equation}
  \int_{-\pi}^{\pi} \frac{d\a}{2\pi} \  \frac{1}{e^{\hat{\e} - q\hat{\mu} -\i \a q}-1}  = - \Theta(q{\mu}-{\e})\ .
\end{equation}
It thus follows from \eqref{dualmanip} that
\begin{equation}\label{ovn}
N_B  \bar{n}_B({\e},q)=  N_F \bar{n}_F({\e},q)\ .
\end{equation}

\section{Curve sketching}\label{curveplot}

In this appendix we study a toy model which is governed the following
quantum effective potential
\begin{equation} \label{tmpot} 
  U_{\rm eff}(\sigma) = \left\{\renewcommand{\arraystretch}{2}\begin{array}{cc} A \sigma^3 + B \sigma^2 + C \sigma &\qquad \sigma < 0\ , \\ (A + a) \sigma^3 + (B + b) \sigma^2 + C \sigma &\qquad \sigma > 0\ ,\end{array}\right.
\end{equation}
Note that $U_{\rm eff}(\sigma)$ is analytic (and polynomial) everywhere
away from $\sigma=0$. At $\sigma=0$, $U_{\rm eff}(\sigma)$ is
continuous and once-differentiable, but its second and third
derivatives are discontinuous.

We designate the $\sigma < 0$ branch in \eqref{tmpot} as the `$\m$'
branch and the $\sigma > 0$ branch as the `$\p$' branch. Whenever a
local minimum exists in the $\m$ or $\p$ branch and we choose our
theory to be in that vacuum, the theory is said to be in the $\m$ or
$\p$ phase.

The potential $U_{\rm eff}(\sigma)$ has 5 parameters, $A$, $B$, $C$,
$a$ and $b$. We will also impose the condition $a>0$ \footnote{Simply
  because this condition always turns out to be true in the case of
  physical interest to us, see Section \ref{localpot}.} but allow all
the other four parameters to have either sign\footnote{Under the
  change of variables $\sigma \rightarrow \alpha \sigma$ (with
  $\alpha>0$), these 5 variables scale as
$$ A \rightarrow \alpha^3 A\ ,\quad a \rightarrow \alpha^3 a\ ,\quad B \rightarrow \alpha^2 B\ ,\quad b \rightarrow \alpha^2 b\ ,\quad C \rightarrow \alpha C $$
so in actuality our toy model \eqref{tmpot} has 4 continuous
parameters.}. In this section we will study the phase diagram of the
model \eqref{tmpot} at fixed values of $A$, $a$, $b$, but as functions
of the two parameters $B$ and $C$. Thus, the phase diagram will be two
dimensional, with $C$ parametrizing the $x$-axis and $B$ parametrizing
the $y$-axis.

It turns out that the phase
diagrams in question are qualitatively different depending on the
signs of $A$, $A+a$ and $b$. As a consequence we will present
$2^3 - 2$ different phase diagrams, one for each of the \emph{allowed}
signs for these three `parameters' (since we have $a > 0$, the case
$A > 0$, $A + a < 0$ with either sign of $b$ is excluded from the
analysis).

The sign of $A$ (resp.~$A+a$) decides the boundedness of the potential
for large negative (resp.~positive) $\sigma$. We thus consider the
three cases separately\footnote{When one of the $\sigma < 0$ or
  $\sigma > 0$ branches in \eqref{tmpot} corresponds to the condensed
  branch of the regular boson effective potential \eqref{finalu} which
  exists only for a finite range of $\sigma$, the question of
  boundedness is not very relevant since $\sigma$ is not allowed to
  grow too large.}:
\begin{equation}
  \text{Case I:}\ A > 0\ ,\ A+a > 0\ ,\quad   \text{Case II:}\ A < 0\ ,\ A+a > 0\ ,\quad  \text{Case III:}\ A < 0\ ,\ A+a < 0\ .
\end{equation}
Cases I and III are unbounded below (for negative $\sigma$ and
positive $\sigma$ respectively) while Case II is stable. Within each
case, we consider two subcases corresponding to $b > 0$ and $b < 0$.

\subsection{Some analytic features of the phase diagram} \label{phasean}

The extremization equation of the effective potential,
$U_{\rm eff}'(\sigma)=0$, is separately quadratic both when $\sigma<0$
and $\sigma>0$. The discriminants of these two quadratic equations are
given by
\begin{equation}\label{disc}
  D_- = B^2 - 3 A C\ ,\quad D_+ = (B+b)^2 - 3 (A+a) C\ .
\end{equation}
and will play a key role in our analysis below. When these
discriminants are negative, there are no extrema in the corresponding
branch of the potential. The $B$-axis i.e.~the $C = 0$ line is tangent
to the discriminant curves $D_-$ and $D_+$ in \eqref{disc}.

\begin{figure}[htbp]
      \begin{center}
        \scalebox{1}{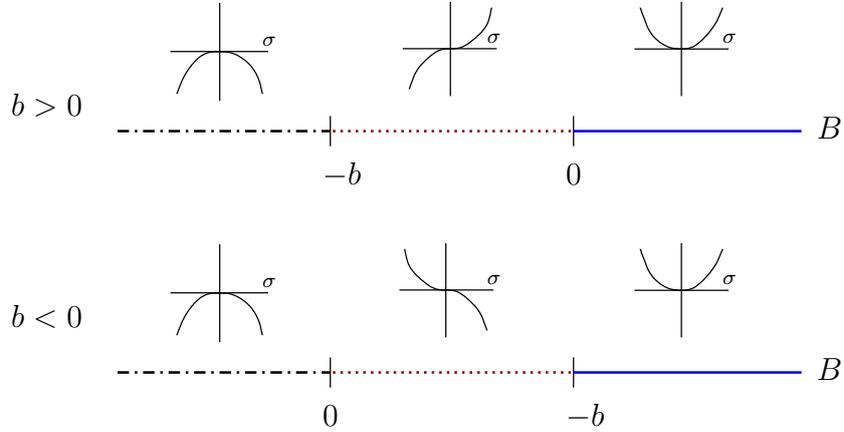}
        \caption{Plots of the potential $U_{\rm eff}(\sigma)$ at
          $C = 0$ and at different values of the parameter $B$ (the
          $x$-axis of the figure). The blue solid line indicates that
          the potential has a minimum at $\sigma = 0$, the red dotted
          line indicates that it has a point of inflection and the
          black dashed-dotted line indicates that it has a maximum.}
        \label{Czero}
      \end{center}
    \end{figure}

The $C = 0$ line in the phase diagram is very interesting
analytically. On the $C = 0$ line, there are extrema in both the $\m$
and $\p$ branches since the discriminants $D_-$ and $D_+$ in
\eqref{disc} are both positive. The extrema for the $\m$ and $\p$
branches are at
\begin{align}
  \m\ \text{branch}&:\quad \sigma = 0\ ,\quad \sigma = - \frac{2 B}{3 A}\ ,\nonumber\\
  \p\ \text{branch}&:\quad \sigma = 0\ ,\quad \sigma = - \frac{2 (B+b)}{3 (A+a)}\ .
\end{align}
The extremum at $\sigma = 0$ is always a legitimate one for both
branches. However, the other extremum must be discarded if it occurs
outside the range of validity $\sigma \lessgtr 0$ for the $\m$ or $\p$
branches. The second derivative of the potential in either branch at
the extremum $\sigma = 0$ is given by
\begin{equation}
  \m\ \text{branch}:\quad U''_{\rm eff}(0) = 2 B\ ,\quad \p\ \text{branch}:\quad U''_{\rm eff}(0) = 2 (B+b)\ .
\end{equation}
Based on the above facts, we deduce three different kinds of
transitions in the behaviour of the potential \eqref{tmpot} depending
on the range of $B$. We give the details separately for $b > 0$ and
$b < 0$. The potential is sketched in Figure \ref{Czero} for each of
the different cases.

\begin{enumerate}
\item $b > 0$
  \begin{enumerate}
  \item $B > 0$: The full potential has a minimum at $\sigma = 0$
    i.e.~both branches of the potential have a minimum at
    $\sigma = 0$.

  \item $-b < B < 0$: A maximum occurs in the $\m$ branch at
    $\sigma = 0$ while a minimum occurs in the $\p$ branch. Thus, the
    potential has a point of inflection at $\sigma = 0$.

  \item $B < -b$: The full potential has a maximum at $\sigma = 0$
    since both branches of the potential have a maximum at
    $\sigma = 0$.
  \end{enumerate}

\item $b < 0$
  \begin{enumerate}
  \item $B > -b$: Both branches have a minimum at $\sigma = 0$ and
    hence the full potential has a minimum at this point.

  \item $0 < B < -b$: A minimum occurs in the $\m$ branch at
    $\sigma = 0$ while a maximum occurs in the $\p$ branch. Thus, the
    potential has a point of inflection at $\sigma = 0$.

  \item $B < 0$: Both branches of the potential have a maximum at
    $\sigma = 0$, thus giving rise to a maximum for the full potential
    at $\sigma = 0$.
  \end{enumerate}
\end{enumerate}

So far, we have looked at the behaviour on the potential by setting
$C = 0$ and varying $B$. Suppose we now broaden our focus to a small
neighbourhood of the $C = 0$ line. To begin with, note that the slope
of the potential $U_{\rm eff}(\sigma)$ at $\sigma = 0$ is given by $C$
in both branches of the potential. Let us study the potential as we
take $C$ through zero from positive to negative values by choosing $B$
to be in one of the three ranges discussed previously. When $B$ is
such that we are on the blue line in Figure \ref{Czero}, a local
minimum crosses from the $\m$ branch to the $\p$ branch. When we
choose $B$ to be on the black line in Figure \ref{Czero}, a local
maximum crosses from the $\p$ branch to the $\m$ branch.

Finally, when we choose $B$ such that we are on the red line segment,
two kinds of behaviour are possible depending on the sign of
$b$. First, consider the case $b > 0$. The potential is monotonic near
the origin when $C > 0$, has a point of inflection at $C = 0$ and
develops a local maximum in the $\m$ branch and a local minimum in the
$\p$ branch for $C < 0$. Second, when $b < 0$, the potential starts
with a local minimum in the $\m$ branch and a local maximum in the
$\p$ branch for $C > 0$, has a point of inflection at $C = 0$ and is
monotonic near $\sigma = 0$ for $C < 0$.

We stress that the above description applies to only the local
behaviour of the potential near $\sigma = 0$, the junction of the two
branches. In particular, there may be other local extrema of the
potential as one goes deeper into either of its branches. A full
analysis of the potential is required in order to understand its
global structure, which is what we turn to next.

\subsection{Case I: $A > 0$, $A+a > 0$}
The potential is bounded below for $\sigma > 0$ and unbounded below
for $\sigma < 0$. We have the following cases:
  \begin{enumerate}
  \item $C > 0$: Displayed in Figures \ref{CaseI}(a)-(d). The
    potential in the $\m$ branch is monotonic when $B < 0$ or ($B > 0$
    with $D_- < 0$) and has extrema when ($B > 0$ with $D_- >
    0$). Similarly, the potential is monotonic in the $\p$ branch when
    $B + b > 0$ or ($B + b < 0$ with $D_+ < 0$) and has extrema when
    ($B + b < 0$ with $D_+ > 0$). Then, we have the following
    possibilities.
    \begin{enumerate}
    \item Either $B < 0$ or ($B > 0$ with $D_- < 0$); either
      $B + b > 0$ or ($B + b < 0$ with $D_+ < 0$): Monotonic in both
      $\p$ and $\m$ branches.

    \item ($B > 0$ with $D_- > 0$); either $B + b > 0$ or ($B+b < 0$
      with $D_+ < 0$): Two extrema present in the $\m$ branch, no
      extremum in the $\p$ branch.

    \item Either $B < 0$ or ($B > 0$ with $D_- < 0$); ($B + b < 0$
      with $D_+ > 0$): No extremum in the $\m$ branch, two extrema
      present in the $\p$ branch.

    \item ($B > 0$ with $D_- > 0$); ($B + b < 0$ with $D_+ > 0$):
      Maximum-minimum pairs in both Higgsed and unHiggsed
      phases. Depending on the relative magnitudes of the parameters
      $B$ and $C$, the minimum in one of the $\p$ or $\m$ branches
      will be deeper and hence more dominant. A first order transition
      is possible when the dominant minimum switches from one branch
      to the other. We depict the two situations in Figure
      \ref{CaseI}(d).
    \end{enumerate}

  \item $C < 0$: In Figure \ref{CaseI}(e). There is always a minimum
    in the $\m$ branch and a maximum in the $\p$ branch, irrespective
    of the values of the parameter $B$.
    
    \begin{figure}[htbp]
      \begin{center}
        \scalebox{0.8}{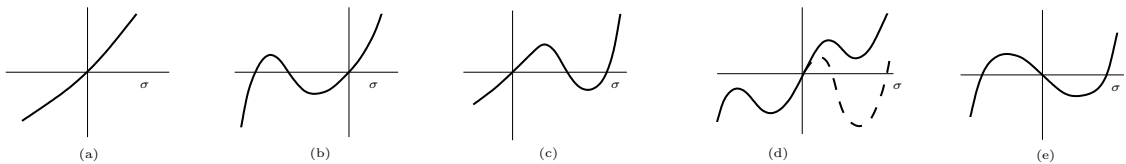}
        \caption{Effective potential for Case I: $A > 0$, $A + a > 0$.}
        \label{CaseI}
      \end{center}
    \end{figure}
\end{enumerate}

\begin{figure}[htbp]
  \begin{center}
    \scalebox{0.8}{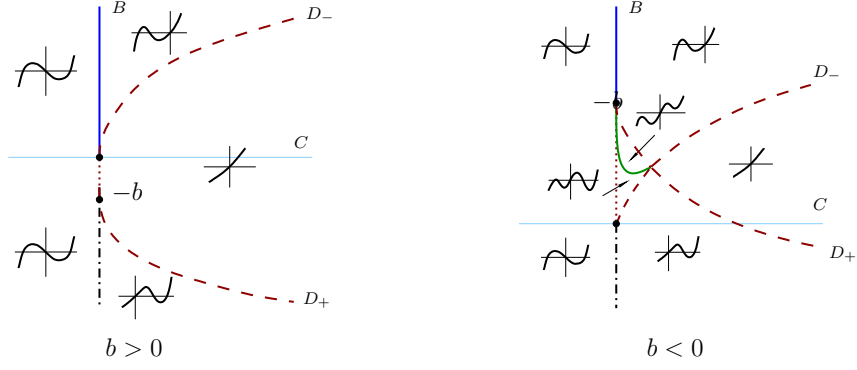}
    \caption{Detailed structure of the potential $U_{\rm eff}(\sigma)$
      \eqref{tmpot} for Case I: $A > 0$, $A + a > 0$. The solid blue
      (resp.~dashed-dotted black) lines signify a local minimum
      (resp.~local maximum) crossing from one branch to the other
      through $\sigma = 0$. The red dashed lines $D_+$ and $D_-$
      signify the (dis)appearance of new extrema in the potential
      (cf.~\eqref{disc}). The dotted red line segment between $B = 0$
      and $B= -b$ signifies a local maximum in one branch and a local
      minimum in the other meeting at $\sigma = 0$. The solid green
      line is a first order phase transition line which signifies the
      presence of two equal local minima of the potential, one in each
      branch.}
    \label{PhaseI}
  \end{center}
\end{figure}

\begin{figure}[htbp]
  \begin{center}
    \scalebox{0.8}{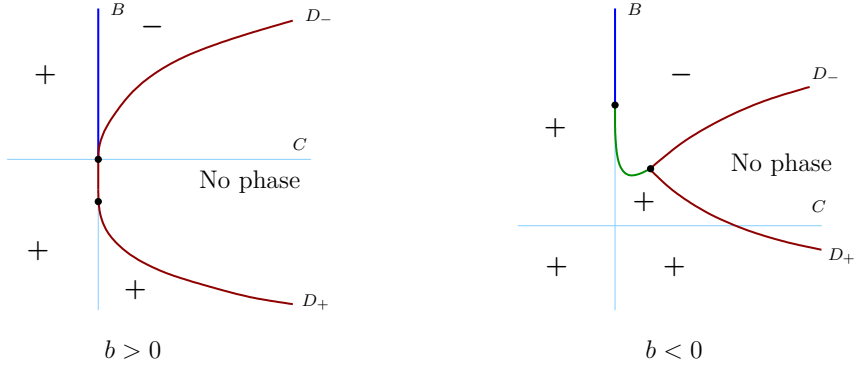}
    \caption{Phase diagram for Case I: $A > 0$, $A + a > 0$. The solid
      blue line is a second order phase transition between the $\m$
      and $\p$ phases. The solid red lines demarcate regions of
      monotonic potential. The solid green line is a first order phase
      transition line between the $\m$ and $\p$ phases.}
    \label{diagI}
  \end{center}
\end{figure}

\subsection{Case II: $A < 0$, $A + a > 0$}
The potential is bounded below for both $\sigma < 0$ and $\sigma > 0$.
\begin{enumerate}
\item $C > 0$, $B$ either sign
  \begin{enumerate}
  \item either $B + b > 0$ or ($B + b < 0$ with $D_+ < 0$): Minimum in
    $\m$ branch, monotonic in $\p$ branch. Shown in Figure
    \ref{CaseII}(a).

  \item ($B + b < 0$ with $D_+ > 0$): Minimum in $\m$ branch, a
    maximum-minimum pair in the $\p$ branch. Shown in Figure
    \ref{CaseII}(b). The minimum in the $\p$ branch could also be lower
    than the minimum in the $\m$ branch though it is not depicted in
    Figure \ref{CaseII}(b).
  \end{enumerate}

\item $C < 0$, $B + b$ either sign
  \begin{enumerate}
  \item $B > 0$ or ($B < 0$ with $D_- < 0$): Monotonic in the $\m$
    branch, minimum in the $\p$ branch. Shown in Figure
    \ref{CaseII}(c).

  \item ($B < 0$ with $D_- > 0$): Minimum in the $\p$ branch, a
    maximum-minimum pair in the $\m$ branch.  Shown in Figure
    \ref{CaseII}(d). The minimum in the $\m$ branch could also be
    lower than the minimum in the $\p$ branch though it is not
    depicted in Figure \ref{CaseII}(d).

  \end{enumerate}

\begin{figure}[htbp]
  \begin{center}
    \scalebox{1}{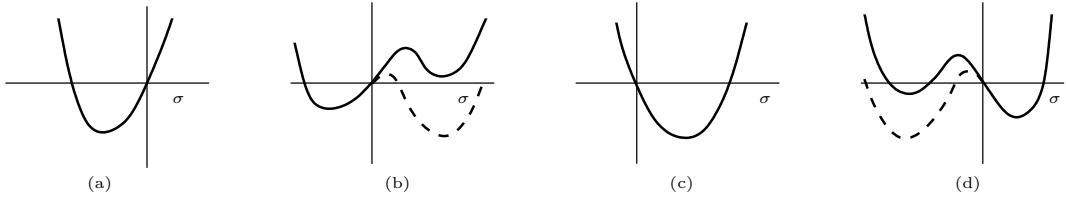}
    \caption{Effective potential for Case II: $A < 0$, $A + a > 0$.}
    \label{CaseII}
  \end{center}
\end{figure}
\end{enumerate}

\begin{figure}[htbp]
  \begin{center}
    \scalebox{0.8}{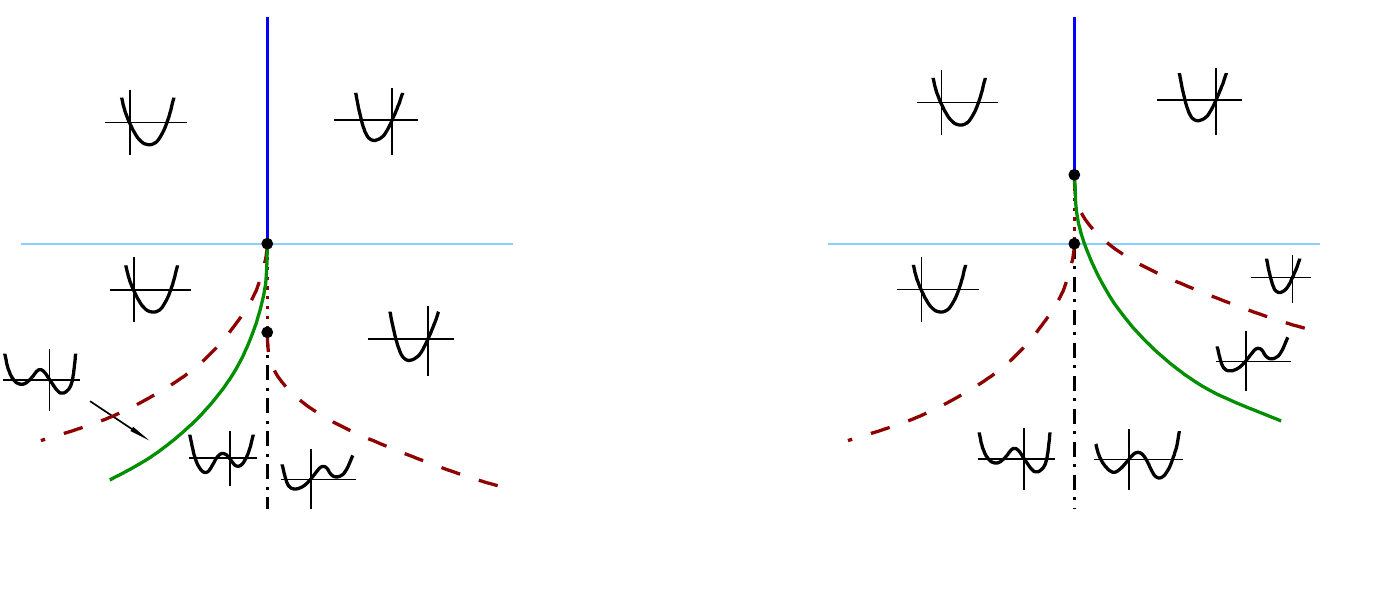}
    \caption{Detailed structure of the potential for Case II: $A < 0$,
      $A + a > 0$. Refer to the caption of Figure \ref{CaseI} for
      details on the various marked lines.}
    \label{PhaseII}
  \end{center}
\end{figure}

\begin{figure}[htbp]
  \begin{center}
    \scalebox{0.8}{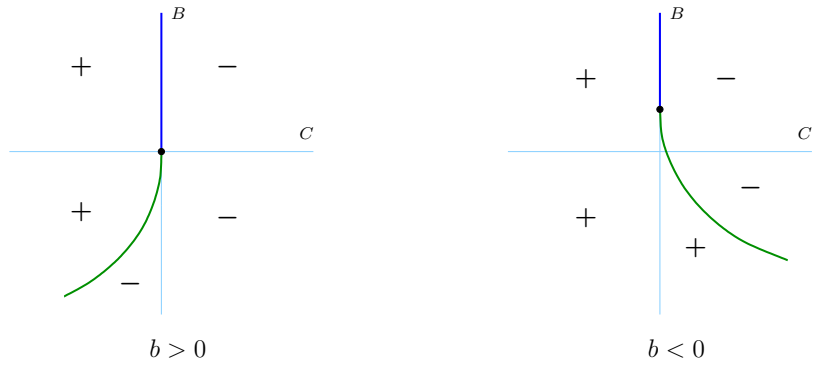}
    \caption{Phase diagram for Case II: $A < 0$, $A + a > 0$. The
      solid blue line is a second order phase transition between the
      $\m$ and $\p$ phases. The solid green line is a first order
      phase transition line between the $\m$ and $\p$ phases.}
    \label{diagII}
  \end{center}
\end{figure}

\subsection{Case III: $A < 0$, $A + a < 0$}
In this case, the effective potential is bounded below for
$\sigma < 0$ and unbounded below for $\sigma > 0$.

\begin{enumerate}
\item $C < 0$. The potential in the $\m$ branch is monotonic when
  $B > 0$ or ($B < 0$ with $D_- < 0$) and has two extrema when
  ($B < 0$ with $D_- > 0$). Similarly, the potential in the $\p$
  branch is monotonic when $B + b < 0$ or ($B + b > 0$ with $D_+ < 0$)
  and has two extrema when ($B + b > 0$ with $D_+ > 0$). We now
  consider the following possibilities.
  \begin{enumerate}
  \item Either $B + b < 0$ or ($B + b > 0$ with $D_+ < 0$); either
    $B > 0$ or ($B < 0$ with $D_- < 0$): Potential is monotonic.

  \item Either $B + b < 0$ or ($B + b > 0$ with $D_+ < 0$); ($B < 0$
    with $D_- > 0$): Two extrema in the $\m$ branch, monotonic in the
    $\p$ branch.

  \item ($B + b > 0$ with $D_+ > 0$); either $B > 0$ or ($B < 0$ with
    $D_- < 0$): Two extrema in the $\p$ branch, monotonic in the $\m$
    branch.

  \item ($B + b > 0$ with $D_+ > 0$); ($B < 0$ with $D_- > 0$):
    Maximum-minimum pairs in both branches. The minimum in one of the
    branches is typically lower than the other. There will be a first
    order phase transition when the dominant minimum switches from one
    branch to the other. We have shown the two possibilities in Figure
    \ref{CaseIII}(d).
  \end{enumerate}

\item $C > 0$: There is always a minimum in $\m$ branch and a maximum
  in the $\p$ branch, irrespective of the value of the parameter $B$.

\begin{figure}[htbp]
  \begin{center}
    \scalebox{0.8}{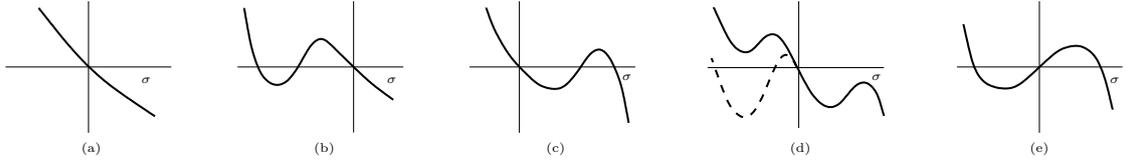}
    \caption{Effective potential for Case III: $A < 0$, $A + a < 0$.}
    \label{CaseIII}
  \end{center}
\end{figure}
\end{enumerate}

\begin{figure}[htbp]
  \begin{center}
    \scalebox{0.8}{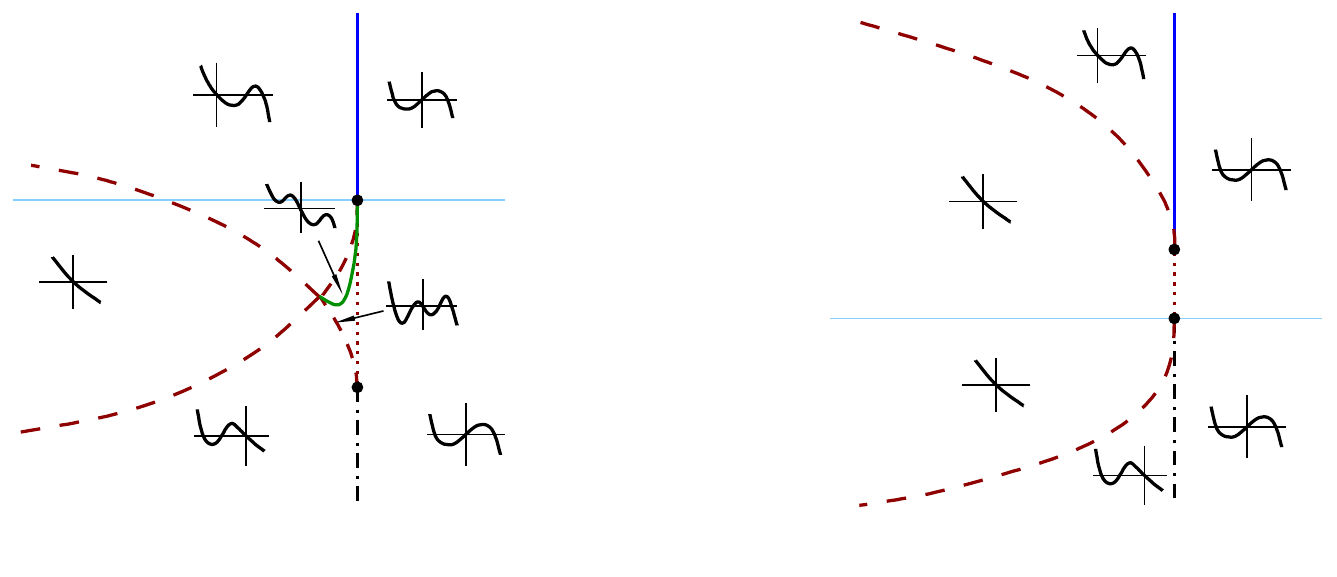}
    \caption{Detailed structure of the potential for Case III: $A < 0$,
      $A + a < 0$. Refer to the caption of Figure \ref{CaseI} for
      details on the various marked lines.}
    \label{PhaseIII}
  \end{center}
\end{figure}

\begin{figure}[!htbp]
  \begin{center}
    \scalebox{0.8}{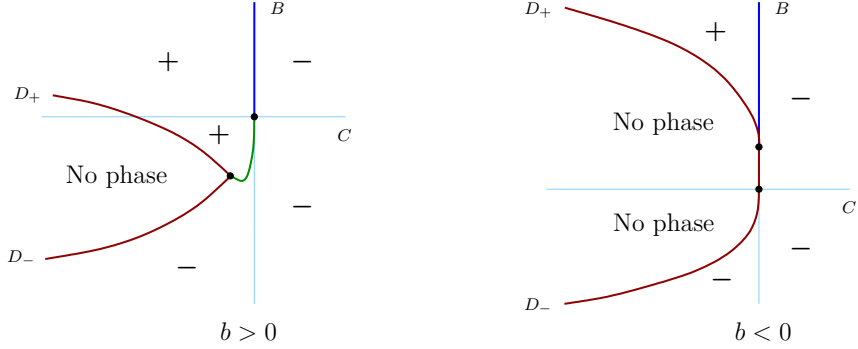}
    \caption{Phase diagram for Case III: $A < 0$, $A + a < 0$. Refer
      to the caption of Figure \ref{diagI} for details on the
      various marked lines.}
    \label{diagIII}
  \end{center}
\end{figure}

\section{The phase diagram of the regular boson theory with an
  unstable potential}\label{unstable}

In this section, we present the analysis of the effective potential
\eqref{finalu} for the unstable ranges of $x_6^B$ viz.~Case A and Case
D in \eqref{x6case}:
\begin{equation}
  \text{Case A}:\  x_6^B < \phi_h\ ,\quad \text{Case D}:\ \phi_u < x_6^B\ .
\end{equation}
As we have remarked earlier, since the potential is unbounded below
either for large and positive $\sigma_B$ (Case A) or for large and
negative $\sigma_B$ (Case D), the phases that may exist are metastable
at best. However, we present the calculations simply because it is
informative and for the sake of completeness. The analysis of the
current section may also be useful in the study of the $\mc{N} = 2$
supersymmetric Chern-Simons theory with one fundamental chiral
multiplet in the presence of chemical potential\footnote{The phase
  diagram of this theory with zero chemical potential was computed in
  \cite{Dey:2019ihe}. The main strategy in that paper was to use the
  results for the effective potential for the regular boson to
  understand the local behaviour of the effective potential for the
  $\mc{N} = 2$ theory near its non-analytic points (analogous to what
  we do in this paper). The local behaviour of the potential was
  sometimes decided by the unstable cases of the regular boson, though
  the final global potential was stable. The global stability of the
  potential essentially meant that the unboundedness of the local
  potential was replaced by a very deep minimum in the global
  potential which was situated outside the range of applicability of
  the local potential.}.

Recall that we work only at leading order in the large $N$ limit in
which case the parameter $x_6^B$ is exactly marginal. However, this is
no longer the case when when the subleading corrections in $1 / N$ for
the $\beta$-function of $x_6^B$ are taken into account
\cite{Aharony:2018pjn}. The current analysis may be useful in this
situation since the fixed points of $x_6^B$ may well end up being in
one of the above unstable ranges (see \cite{Aharony:2018pjn} for more
details).

\subsection{Case A:
  \texorpdfstring{$x_6^B < \phi_h$}{x6<phi-h}}\label{CaseA}
For this range of $x_6^B$, both local expressions of the potential
\eqref{potl} and \eqref{potr} are unbounded below for large and
positive $\sigma_B'$ and $\sigma_B''$ respectively since the quantity
$A$ satisfies $A < 0$ and $A +a < 0$ for both \eqref{potl} and
\eqref{potr}. Thus, Case III of Appendix \ref{curveplot} applies to
the potential near both non-analytic points. We must remark that the
unboundedness of \eqref{potl} is not very serious since it occurs in
the condensed branch which is eventually cut off at
$\sigma_B' = 1 / |\lambda_B|$.

\begin{figure}[!ht]
  \begin{center}
    \scalebox{0.8}{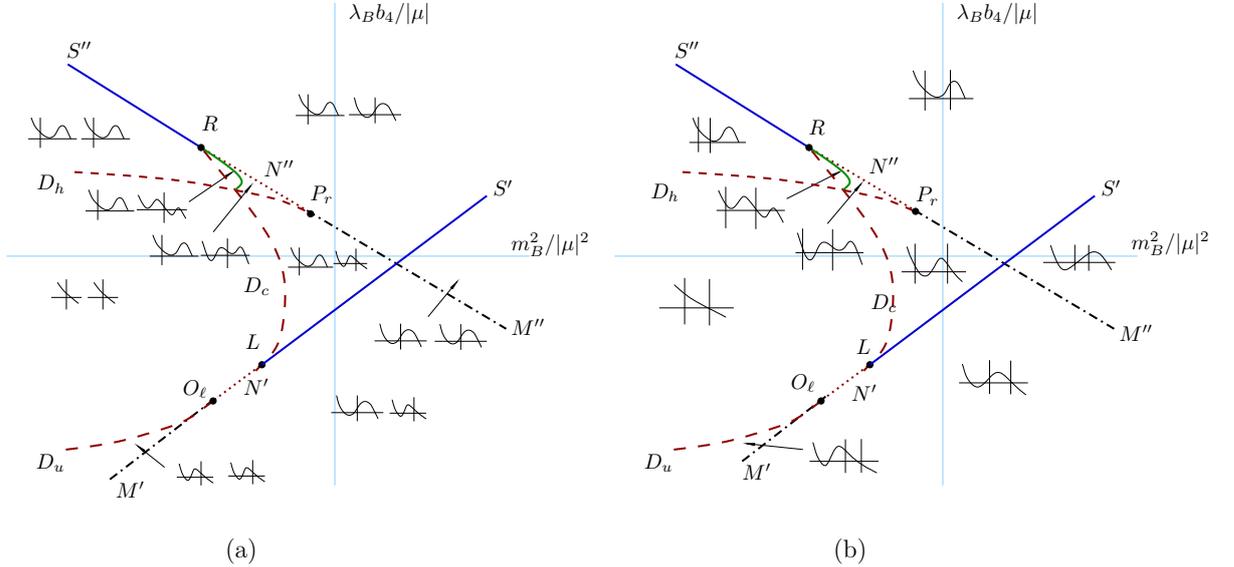}
    \caption{Structure of the potential for $x_6^B < \phi_h$. Refer to
      Sections \ref{nearell} and \ref{nearr} for the labels of the
      various points and lines. We have restored the $|\mu|$
      dependence of the phase diagram by rescaling $m_B^2$ and
      $\lambda_B b_4$ by appropriate powers of $|\mu|$.}
    \label{PhaseplotA}
  \end{center}
\end{figure}

We plot the special lines corresponding to the potential near both
non-analytic points in Figure \ref{PhaseplotA}. The segments of $D_u$,
$D_c$ and $D_h$ which are present in this phase diagram are specified
by the following conditions:
\begin{align}
  D_u: &\quad \lambda_B b_4\ \leq\ \tfrac{3}{4} \lambda_B^2 (x_6^B - \phi_u)\ ,\qquad   D_h: \quad -\tfrac{3}{4}(2-|\lambda_B|)(x_6^B - \phi_h)\ \leq\ \lambda_B b_4\ ,
\end{align}
\begin{multline}
D_c: \quad \tfrac{3}{4}\lambda_B^2 (x_6^B - \phi_u) - (1-|\lambda_B|)\ \leq\ \lambda_B b_4\ \leq \\ -\tfrac{3}{4}|\lambda_B|(2-|\lambda_B|)(x_6^B - \phi_h) + \frac{|\lambda_B|(1-|\lambda_B|)}{2-|\lambda_B|}\ .
\end{multline}
For each region demarcated by the special lines in Figure
\ref{PhaseplotA}(a), we display the shape of the potential in the
neighbourhood of the points $\sigma_B = \ell$ and $\sigma_B = r$ as a
small inset figure (with the plot on the left (resp.~right)
corresponding to $\ell$ (resp.~$r$)). In Figure \ref{PhaseplotA}(b),
we combine the information from the plots near $\ell$ and $r$ to give
a global shape of the potential in each region demarcated by the
special lines. Note that one is able to obtain this information just
by comparing the common region of the two plots and without any
detailed quantitative analysis of the potential.

\begin{figure}[!ht]
  \begin{center}
    \scalebox{0.8}{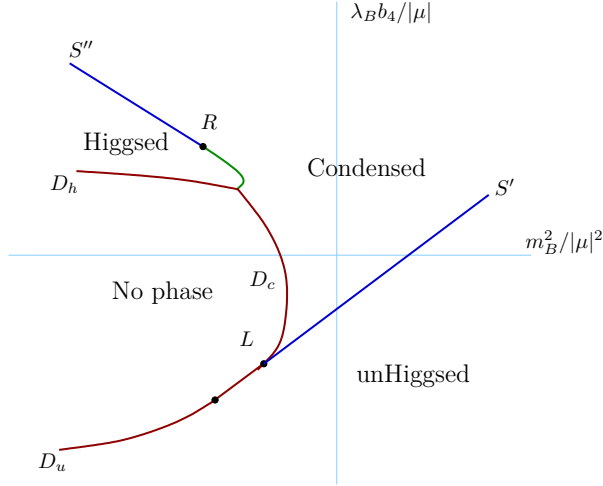}
    \caption{Phase diagram for $x_6^B < \phi_h$. The blue lines are
      second order transitions, the green line is a first order
      transition while the red lines demarcate regions of runaway
      potential from the regions where metastable phases exist. Recall
      that the potential is unstable and all the phases displayed
      above are metastable at best. We have restored the $|\mu|$
      dependence of the phase diagram by rescaling $m_B^2$ and
      $\lambda_B b_4$ by appropriate powers of $|\mu|$.}
    \label{PhasediagA}
  \end{center}
\end{figure}

Since the potential is unbounded from below for $\sigma_B \gg 0$, all
the phases for this range of $x_6^B$ are metastable at best. We show
this ``phase'' diagram in Figure \ref{PhasediagA}.

\subsection{Case D: \texorpdfstring{$\phi_u < x_6^B$}{phi-u<x6}}
\begin{figure}[!ht]
  \begin{center}
    \scalebox{0.8}{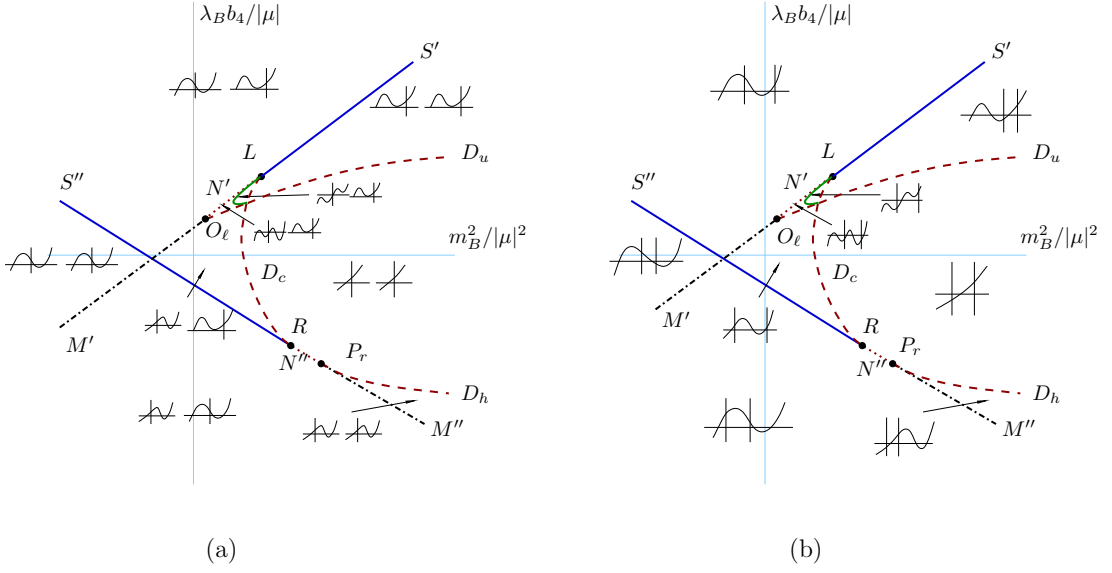}
    \caption{Structure of the potential for $\phi_u < x_6^B$. Recall
      that the effective potential is unbounded from below in this
      case so that the `phases' displayed here are dominant metastable
      phases when they exist. To the right of the red curve the
      effective potential is monotonic so no metastable phases
      exist. The green curve is a first order transition (between
      dominant metastable phases) while the blue curves are second
      order transitions between such phases. We have restored the
      $|\mu|$ dependence of the phase diagram by rescaling $m_B^2$ and
      $\lambda_B b_4$ by appropriate powers of $|\mu|$.}
    \label{PhaseplotD}
  \end{center}
\end{figure}

\begin{figure}[!ht]
  \begin{center}
    \scalebox{0.8}{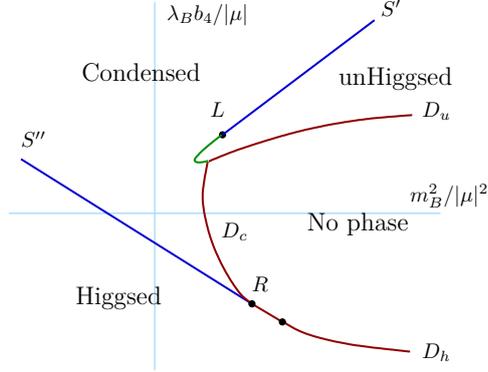}
    \caption{Phase diagram for $\phi_u < x_6^B$.  Recall that the
      effective potential is unbounded from below in this case so that
      the `phases' displayed here are dominant metastable phases when
      they exist. To the right of the red curve the effective
      potential is monotonic so no meta-table phases exist. The blue
      lines are second order transitions and the green curve is a
      first order transition between metastable phases. To the `right'
      of the red curve no metastable phases exist. We have restored
      the $|\mu|$ dependence of the phase diagram by rescaling $m_B^2$
      and $\lambda_B b_4$ by appropriate powers of $|\mu|$.}
    \label{PhasediagD}
  \end{center}
\end{figure}

Case I of Appendix \ref{curveplot} applies to the behaviour of the
potential near both non-analytic points. The conditions on
$\lambda_B b_4$ which specify the segments of the discriminant
parabolas that appear in the phase diagram are
\begin{align}
  D_u: &\quad \lambda_B b_4\ \geq\ \tfrac{3}{4} \lambda_B^2 (x_6^B - \phi_u)\ ,\qquad   D_h: \quad \lambda_B b_4 \ \leq\ -\tfrac{3}{4}(2-|\lambda_B|)(x_6^B - \phi_h)\ ,
\end{align}
\begin{multline}
D_c: \quad -\tfrac{3}{4}|\lambda_B|(2-|\lambda_B|)(x_6^B - \phi_h) + \frac{|\lambda_B|(1-|\lambda_B|)}{2-|\lambda_B|} \ \leq\ \lambda_B b_4\ \leq \\ \tfrac{3}{4}\lambda_B^2 (x_6^B - \phi_u) - (1-|\lambda_B|)\ .
\end{multline}
The detailed structure of the potential is given as small inset plots
in Figure \ref{PhaseplotD}(a) in each region demarcated by the special
lines of the two local potentials \eqref{potl} and \eqref{potr}. The
global structure of the potential is displayed in each of these
regions in Figure \ref{PhaseplotD}(b). Since the potential is
unbounded below in this range of $x_6^B$, the phases in this case are
metastable and are displayed in Figure \ref{PhasediagD}.

\newpage 

\section{Preliminary discussions on angular momentum}\label{angmom}

The $2+1$ dimensional bosons and fermions that we study in the main
text of this paper do not appear in spin multiplets but instead each
carry the same spin in the rest frame. For this reason we expect a
macroscopic collection of these particles to carry angular momentum
that is extensive (i.e. scales like the volume).

In this appendix, we will demonstrate that - at least with one
reasonable definition of the angular momentum - this expectation is
indeed correct for the Fermi sea of a collection of free fermions of unit charge in
$2+1$ dimensions. We leave the definition and computation of the
angular momentum of thermodynamical ensembles of Chern-Simons-matter
theories to future work\footnote{One additional complication that the
  Chern-Simons gauge field would introduce is the presence of WZW
  boundary degrees of freedom, which, by themselves, could carry
  extensive angular momentum.}.

\subsection{The free Dirac equation and its symmetries}
The free Dirac equation
\begin{equation} \label{eoms}
(\g^\mu \D_\mu + m_F )\Psi(x^\a)=0\ ,
\end{equation}
follows from the action 
\begin{equation}\label{lags}
  \mc{S}= \i \int d^3x \ \bar{\Psi} \left( \g^\mu \partial_{\mu} +m_F \right)\Psi\ .
\end{equation}
We make the following choice of gamma matrices 
\begin{equation}\label{gammat}
  \gamma^0 \equiv \g^t = \i \sigma^3\ ,\quad \gamma^1 \equiv \g^x = \sigma^1\ ,\quad \gamma^2 \equiv \g^y
  =\sigma^2\ .
\end{equation}
Note that
\begin{equation}\label{goi}
\gamma^0 \gamma^i=  - \epsilon^{ij} \gamma_j\ ,
\end{equation} 
where $\epsilon^{12}= - \epsilon^{21} = 1$. 

\subsubsection{Symmetries of the Lagrangian} 

\textbf{Parity}: It is easy to check that the field redefinition
\begin{equation}\label{varcch}
  \chi(x,y,t) = \gamma^y \Psi(x,-y,t) \equiv (\mc{P}^y \cdot \Psi)(x,y,t)\ ,
\end{equation} 
turns the Dirac action \eqref{lags} into
\begin{equation}\label{lagsmo}
  \mc{S}= \i \int d^3x \ \overline{\chi} \left( \g^\mu \partial_{\mu} -m_F \right)\chi\ .
\end{equation}
In other words the parity operation \eqref{varcch} (corresponding to
$y \rightarrow -y$) flips the sign of the mass term. Equivalently, the
parity operation combined with a flip in the sign of the fermion mass
is a symmetry of the system.

\noindent \textbf{Charge conjugation}: In a similar vein, it is easy
to check that the field redefinition
\begin{equation}\label{varcchn}
  \chi(x,y,t)= \gamma^x \Psi^*(x,y,t) \equiv (\mc{C} \cdot \Psi)(x,y,t)\ ,
\end{equation} 
turns the Dirac action \eqref{lags} into
\begin{equation}\label{lagsmc}
  \mc{S}= \i \int d^3x \ \overline{\chi} \left( \g^\mu \partial_{\mu} +m_F \right)\chi\ .
\end{equation}
In other words, complex conjugation \eqref{varcchn} preserves the
Dirac action and is a symmetry of the system.

\noindent \textbf{Time reversal}: Finally, note that the field
redefinition
\begin{equation}\label{varcchnn}
  \chi(x,y,t)= \gamma^0 \Psi(x,y,-t) \equiv (\mc{T}' \cdot \Psi)(x,y,t)\ ,
\end{equation}
turns the Dirac Lagrangian \eqref{lags} into
\begin{equation} \label{lagsmt}
\mc{S}= -\i \int d^3x \ \overline{\chi} \left( \g^\mu \partial_{\mu} -m_F \right)\chi 
\end{equation} 
The `time reversal transformation' $\mc{T}'$
\eqref{varcchnn}\footnote{More precisely \eqref{varcchnn} is a
  ${\cal C} \cal{T}$ transformation.} and a flip in the sign of the
fermion mass together flips the overall sign of the Lagrangian and so
preserves the Dirac equation.

\subsubsection{The Hamiltonian, angular momentum and
  symmetries} \label{has}

The (single-particle) Dirac Hamiltonian
\begin{equation}\label{Hamilt}
  H (m_F) = \i \gamma^0 \gamma^i \partial_i + \i m_F \gamma^0 \ .
\end{equation} 
The single-particle angular momentum operator is given by
\begin{equation}\label{smao} 
  J=\i \epsilon^{ij} x_i \partial_j +\i\frac{\gamma^0}{2} = \i (x \partial_y - y \partial_x) +\i \frac{\gamma^0}{2}=  \i \partial_\phi - \frac{\sigma^3}{2} \ .
\end{equation} 
Note that
\begin{equation}\label{comrel}
  [H,J]= -\gamma^0 \gamma_i \epsilon^{ik}
  \partial_k - \gamma^0 \gamma^i \gamma^0 \partial_i = \epsilon_{ij}
  \epsilon^{ik} \gamma^j \partial_k - \gamma^i \partial_i = 0\ ,
\end{equation}
and
\begin{equation}\label{jcomrel}
[J,x^k]=\i  \epsilon_{ik} x^i\ ,\quad [J, \gamma^k]= \i \epsilon_{mk} \gamma^m\ .
\end{equation}
As expected, $x^i$ and $\gamma^i$ rotate the same way under rotations
(both are vectors in the spatial plane). It follows that the operators
$x^i\gamma_i$ and $\gamma^0$ both commute with $J$.

We now study the interplay between solutions of the Dirac equation and
the discrete symmetries discussed in the previous subsubsection.

First, let us look at parity. The analysis around \eqref{lagsmo} tells
us that if $\Psi(x,y, t)$ is a solution of the Dirac equation then
$\gamma^y \Psi(x,-y, t)$ is also a solution of the same equation with
flipped mass. Applying this result to solutions of the form
$\psi(x,y)e^{-\i Et}$, we conclude that if $\psi(x,y)$ is an
eigenstate of the Dirac Hamiltonian $H(m_F)$ with energy $E$, then
$\gamma^y \psi(x,-y)$ is an eigenstate of the Dirac Hamiltonian
$H(-m_F)$ with the same energy. This is easy to check directly. Let
$\psi(x,y)$ be an eigenstate of $H(m_F)$ with energy $E$ i.e.

\begin{equation}\label{eig}
  \left( \i \gamma^0 \gamma^i \partial_i +\i m_F \gamma^0 \right) 
  \psi(x,y)= E \psi(x,y) \ .
\end{equation}
Then it is easily verified that
\begin{equation}\label{eigm}
  \left( \i \gamma^0 \gamma^i \partial_i - \i m_F \gamma^0 \right) 
  \gamma^y \psi(x,-y)= E \gamma^y \psi(x,-y)\ .
\end{equation}
Then, the above discussion can be phrased in terms of the parity
operator $\mc{P}^y$ as
\begin{equation}
  \mc{P}^y H(m_F) = H(-m_F) \mc{P}^y\ .
\end{equation}
Again, it is easily verified that\footnote{If $\Psi(x,y,t)$ is an
  eigenstate of the angular momentum operator \eqref{smao} with
  eigenvalue $j$ i.e.
\begin{equation}
J \Psi(x,y,t) = j \Psi(x,y,t)\ ,
\end{equation}
then, 
\begin{equation}
J \big( \gamma^y \Psi(x,-y,t) \big) = - j \big( \gamma^y \Psi(x,-y,t) \big) 
\end{equation}
}
\begin{equation}\label{jonn} 
  J \left( \gamma^y \psi(x,-y) \right) 
  = - \gamma^y \left( J \psi(\alpha, \beta) \right)|_{\alpha=x,~\beta=-y}\ .
\end{equation}
The above equation can be recast in terms of the parity operator $P^y$
as
\begin{equation}
  (J \mc{P}^y \psi)(x,y) = - (\mc{P}^y J \psi)(x,y)\ ,\quad\text{i.e.}\quad J \mc{P}^y = - \mc{P}^y J\ .
\end{equation}
It follows that parity operator flips angular momentum; it maps an
angular momentum eigenstate with eigenvalue $j$ to an eigenstate of
eigenvalue $-j$. Consider a simultaneous eigenstate of the Dirac
Hamiltonian $H(m_F)$ and the angular momentum operator $J$ with
eigenvalues $E$ and $j$ respectively. Then the parity operator
$\mc{P}^y$ maps this eigenstate to an eigenstate of $H(-m_F)$ and $J$
with eigenvalues $E$ and $-j$ respectively.

We can repeat this discussion for the charge conjugation operation.
It follows from the discussion around \eqref{lagsmc} that if
$\psi(x,y)e^{-\i Et}$ is a solution to the Dirac equation with mass
$m_F$, then $\gamma^x \psi^*(x,y)e^{\i Et}$ is also a solution to the
Dirac equation with mass $m_F$. In other words, if $\psi(x,y)$ is an
eigenstate of $H(m_F)$ with energy $E$ then $\gamma^x \psi^*(x,y)$ is
another eigenstate of $H(m_F)$ but with energy $-E$. Again this is
easy to check explicitly. Given that \eqref{eig} holds, it follows
that
\begin{equation}\label{eigmm}
  \left( \i \gamma^0 \gamma^i \partial_i + \i m_F \gamma^0 \right) 
  \gamma^x \psi^*(x,y)=-E \gamma^x \psi^*(x,y)\ ,\quad\text{i.e.}\quad \mc{C} H(m_F) = -H(m_F) \mc{C}\ .
\end{equation}
Similarly, it is easy to verify that\footnote{More explicitly, if
  $\Psi(x,y,t)$ is an eigenstate of the angular momentum operator
  \eqref{smao} with eigenvalue $j$, then, it is easy to convince
  oneself that
\begin{equation}
J \big( \gamma^x \Psi^*(x,y,t) \big) = - j \big( \gamma^x \Psi^*(x,y,t) \big) 
\end{equation}
}
\begin{equation}\label{jonnn} 
  J \left( \gamma^x \psi^*(x,y) \right)  = - \gamma^x \left( J \psi(x, y) \right)^*\ ,\quad\text{i.e.}\quad J \mc{C} = -\mc{C} J\ .
\end{equation}
In summary, if the state $\psi$ has energy $E$ and angular momentum
$J$, then the state $\gamma^x \psi^*$ has energy $-E$ and angular
momentum $-J$.
 
Finally, the discussion around \eqref{lagsmt} tells us that if
$\psi(x,y)e^{-\i Et}$ is a solution to the Dirac equation with mass
$m_F$, then $\gamma^0 \psi(x,y)e^{\i Et}$ is also a solution to the
Dirac equation with mass $-m_F$. It follows that if $\psi(x,y)$ is an
eigenstate of the Dirac Hamiltonian $H(m_F)$ with energy $E$, then
$\gamma^0 \psi(x,y)$ is another eigenstate of $H(-m_F)$ with energy
$-E$. Again this is easy to check explicitly:

\begin{equation}
  \left( \i \gamma^0 \gamma^i \partial_i +\i m_F \gamma^0 \right) 
  \psi(x,y,t)= \i \partial_t \psi(x,y,t) \ .
\end{equation}

\begin{equation}\label{eigmmm}
  \i \gamma^0\left(  \gamma^i \partial_i - m_F  \right) 
  \gamma^0 \psi(x,y)=- E \gamma^0 \psi(x,y) \ ,\quad\text{i.e.}\quad \mc{T}' H(m_F) = - H(-m_F) \mc{T}'\ .
\end{equation}
Also it is easy to verify that\footnote{In other words, if
  $\Psi(x,y,t)$ is an eigenstate of the angular momentum operator
  \eqref{smao} with eigenvalue $j$, then, it follows that
	\begin{equation}
	J \big( \gamma^0 \Psi(x,y,-t) \big) =  j \big( \gamma^0 \Psi(x,y,-t) \big) 
	\end{equation}
}
\begin{equation}\label{jonnnn} 
  J \left( \gamma^0 \psi(x,y) \right) = \gamma^0 \left( J \psi(x, y) \right)\ ,\quad\text{i.e.}\quad J \mc{T}' = \mc{T}' J\ .
\end{equation}
In summary, if a simultaneous eigenstate $\psi$ of $H(m_F)$ and $J$
has energy $E$ and angular momentum $j$ respectively, then the state
$\gamma^0 \psi$ is a simultaneous eigenstate of $H(-m_F)$ and $J$ with
energy $-E$ and angular momentum $j$ respectively.

\subsection{Boundary conditions at infinity}

Since the angular momentum of a Fermi sea on a truly infinite space
will turn out to be infinite, we need an IR regulator to get a
sensible answer. We adopt the following strategy.  Let us define the
radial spatial variable $r$ and the unit position vector by
\begin{equation}\label{rdef}
r^2=x^2+y^2\ ,\quad {\hat x}^i= \frac{x^i}{r}\ .
\end{equation} 
We study the Dirac on the cut off plane $r \leq R$, subject to one of
the two the boundary conditions
\begin{equation}\label{bcs}
  \gamma^i {\hat x}^i \psi = \pm \psi\quad\text{at}\quad r = R\ .
\end{equation} 
The boundary conditions with the $\pm$ signs in \eqref{bcs} define
two distinct Hilbert Spaces $\mc{H}^+$ and $\mc{H}^-$ respectively.

We will now discuss the interplay between the boundary conditions
\eqref{bcs} and the symmetry operations \eqref{varcch},
\eqref{varcchn} and \eqref{varcchnn}. Consider solutions
$\psi_\pm \in \mc{H}^\pm$ that obeys the boundary conditions
\eqref{bcs}. It is easy to check that
\begin{equation} \label{paritybcs} \begin{split}
&\gamma^i {\hat x}^i \big( \gamma^y \psi_\pm(x,-y) \big) = \mp \gamma^y \psi_\pm(x,-y) \\
&\gamma^i {\hat x}^i \big( \gamma^x \psi_\pm^*(x,y) \big) = \pm \gamma^x \psi_\pm^*(x,y) \\
&\gamma^i {\hat x}^i \big( \gamma^0 \psi_\pm(x,y) \big) = \mp \gamma^0 \psi_\pm(x,y) \\
\end{split} 
\end{equation}
Thus, parity and time reversal map $\mc{H}^+$ to $\mc{H}^-$ and vice
versa (recall that these two operations also flip the sign of the mass
parameter $m_F$). On the other hand the charge conjugation operation
acts separately within $\mc{H}^+$ and $\mc{H}^-$.

\subsection{A convenient definition of the angular momentum of the
  Fermi sea}

Let $J^\pm(\mu, m_F)$ be the angular momentum of the Fermi sea with
chemical potential $\mu$ of fermions with mass $m_F$ in the Hilbert
Space $\mc{H}^\pm$. The fact that the charge conjugation operation
maps every state of energy $E$ and angular momentum $j$ in $\mc{H}^+$
(resp.~$\mc{H}^-$) to another state of energy $-E$ and angular
momentum $-j$ in $\mc{H}^+$ (resp.~$\mc{H}^-$) (and the fact that a
antiparticle is the removal of a particle in a state of negative
energy\footnote{In the first quantized formalism in which we work in
  this Appendix, the fermion has both positive and negative energy
  states.  The vacuum of the theory at $\mu=0$ is the Dirac Sea in
  which all negative energy states are filled and all positive energy
  states are unfilled. The positive $\mu$ Fermi sea is the state in
  which all negative energy states continue to be filled, but in
  addition, positive energy states with energies less than $\mu$ are
  also filled. At negative $\mu$, on the other hand, no positive
  energy states are filled and, in addition, some negative energy
  states - those with $\epsilon > \mu$ - are also now empty (the other
  negative energy states continue to be filled). In contrast to this
  Appendix, in the main text we have adopted the second quantized
  viewpoint in which all fermions states have positive energy, but
  there are two kinds of fermion states; those corresponding to
  positive charge `particles' (these map to the positive energy states
  of the Dirac Sea picture) and those corresponding to negative charge
  `antiparticles' (these map to the removal of negative energy states
  in the Dirac Sea picture). } and therefore has minus the charges of
the negative energy state in question) tells us that the spectrum of
particle and antiparticle states are exactly isomorphic. For every
particle state of charge 1, energy $E$ and angular momentum $j$, there
is a corresponding antiparticle state of charge $-1$, energy $E$ and
angular momentum $j$. In other words
\begin{equation}\label{jpmf}
J^\pm(\mu, m_F)= J^\pm(-\mu, m_F) \ .
\end{equation}
The parity operator tells us that for every eigenstate of $H(m_F)$ of
energy $E$ and angular momentum $j$ in $\mc{H}^+$ there is a
corresponding eigenstate of $H(-m_F)$ with energy $E$ and angular
momentum $-j$ in $\mc{H}^-$. It follows that
\begin{equation}\label{jpms}
J^\pm(\mu, m_F)= - J^\mp(\mu, -m_F) \ .
\end{equation}
Finally, the time reversal operation tells us that for every
eigenstate of $H(m_F)$ with energy $E$ and angular momentum $j$ in
$\mc{H}^+$ there is a corresponding eigenstate of $H(-m_F)$ with
energy $-E$ and angular momentum $j$ in $\mc{H}^-$. It follows that
\begin{equation}\label{jpmt}
J^\pm(\mu, m_F)= - J^\mp(-\mu, -m_F) \ .
\end{equation}
Clearly \eqref{jpmt} carries no new information, but follows from
\eqref{jpmf} and \eqref{jpms}.

A symmetric definition of the angular momentum - one that we will
adopt in the rest of this appendix - is
\begin{equation}\label{angmomdef} 
  J(\mu, m_F) = \frac{J^+(\mu, m_F) + J^-(\mu, m_F)}{2} \ .
\end{equation} 
It follows immediately from \eqref{jpmf}, \eqref{jpms} and
\eqref{jpmt} that
\begin{equation}\label{symo} 
J(\mu, m_F) = J(-\mu, m_F)=-J(\mu, -m_F)=-J(-\mu, -m_F)\ .
\end{equation}

\subsection{Solutions of the Dirac equation}

With the choice of $\gamma$ matrices \eqref{gammat}, it is not
difficult to verify that a basis of regular solutions to the Dirac
equation in polar coordinates is given - upto normalisation - by
\begin{equation}\label{gensoln}
\Psi(r,\phi,t) = \begin{pmatrix} 
e^{\i n \phi} \ J_{|n|}(pr)  \\ 
- {\rm sgn}(n) \eta \sqrt{\frac{|E|+ \eta m_F}{|E|-\eta m_F}} \ e^{\i (n+1)\phi}\ J_{|n+1|}(pr) 
\end{pmatrix} e^{-\i \eta |E|t}
\end{equation} 
where $n$ is an integer, positive or negative and the energy is given
by $E=\eta|E|=\eta \sqrt{p^2+m_F^2}$. $\eta=\pm 1$ for
positive/negative energy solutions, respectively \footnote{More
  explicitly the solutions are given in terms of positive integers $n$
  by
	\begin{equation}\label{gensolnp}
	\Psi(r,\phi,t) = \begin{pmatrix} 
	e^{\i |n| \phi} \ J_{|n|}(pr)  \\ 
	-\eta \sqrt{\frac{|E|+\eta m_F}{|E|-\eta m_F}} \ e^{\i (|n|+1)\phi}\ J_{|n|+1}(pr) 
	\end{pmatrix} e^{-\i \eta |E|t}
	\end{equation} 
	and 
	\begin{equation}\label{gensolnm}
	\Psi(r,\phi,t) = \begin{pmatrix} 
	e^{-\i |n| \phi} \ J_{|n|}(pr)  \\ 
	\eta \sqrt{\frac{|E|+ \eta m_F}{|E|-\eta m_F}} \ e^{\i (-|n|+1)\phi}\ J_{|n|-1}(pr) 
	\end{pmatrix} e^{-\i \eta |E|t}
	\end{equation}  }.
Explicitly, the positive energy solutions are given by 
\begin{equation}\label{pesoln}
  \Psi(r,\phi,t) = \begin{pmatrix} 
    e^{\i n \phi} \ J_{|n|}(pr)  \\ 
    - {\rm sgn}(n) \sqrt{\frac{|E|+  m_F}{|E|-m_F}} \ e^{\i (n+1)\phi}\ J_{|n+1|}(pr) 
\end{pmatrix} e^{-\i  |E|t}
\end{equation} 
And the negative energy solutions are given by 
\begin{equation}\label{nesoln}
\Psi(r,\phi,t) = \begin{pmatrix} 
e^{\i n \phi} \ J_{|n|}(pr)  \\ 
 {\rm sgn}(n) \sqrt{\frac{|E|- m_F}{|E|+ m_F}} \ e^{\i (n+1)\phi}\ J_{|n+1|}(pr) 
\end{pmatrix} e^{\i |E|t}
\end{equation} 
Together with the integer $n$, the solution \eqref{gensolnm} is
parametrized by the continuous radial momentum $p$. The momentum $p$
determines the energy of the solution via
\begin{equation}\label{ensol}
  E^2=p^2+m_F^2\ .
\end{equation}
The integer $n$ determines the angular momentum of the solution via
the equation
\begin{equation}\label{ansol}
  j = - \Big( n+\frac{1}{2} \Big) \ .
\end{equation}

Let $\psi_{E, j, m_F}$ denote the eigenfunction of the Dirac
Hamiltonian $H(m_F)$, energy $E$ and angular momentum $j$ ($E$ and $j$
both range over positive and negative values in our notation). The
explicit form of $\psi_{E, j, m_F}$ is easily read off from
\eqref{gensoln}, \eqref{ensol} and \eqref{ansol}. As a check on our
algebra we have verified that, these explicit solutions obey
\begin{align}
  &\gamma^y \psi_{E,j, m_F}(r,-\phi) \propto \psi_{E,-j, -m_F}(r,\phi)\ ,\nonumber \\
  & \gamma^x \psi^*_{E,j, m_F}(r,\phi) \propto \psi_{-E,-j, m_F}(r,\phi)\ ,\nonumber \\
  & \gamma^0 \psi_{E,j, m_F}(r,\phi) \propto \psi_{-E,j, -m_F}(r,\phi) \ .
\end{align}
as predicted on general grounds in Section \ref{has}.

\subsection{Boundary conditions on the solutions}

In writing \eqref{gensoln} we have not yet imposed any boundary
conditions at $r=R$. Turning to the boundary conditions \eqref{bcs}
now, it is easy to check that the solution \eqref{gensoln} belongs to
$\mc{H}^\pm$ provided
\begin{equation}\label{beghp}
\chi^\pm_{n,\eta }(p) \equiv  J_{|n|+{\rm sgn}(n)}(pr) \pm {\rm sgn}(n) \eta \sqrt{\frac{|E|-\eta m_F}{|E|+\eta m_F}} \ J_{|n|}(pr) =0\ .
\end{equation} 
Explicitly, the boundary condition equations for positive energy
solutions are given by
\begin{equation}\label{beghppe}
\chi^\pm_{n,+}(p) \equiv  J_{|n|+{\rm sgn}(n)}(pr) \pm {\rm sgn}(n) \sqrt{\frac{|E|- m_F}{|E|+ m_F}} \ J_{|n|}(pr) =0\ ,
\end{equation} 
and for negative energy solutions 
\begin{equation}\label{beghpne}
\chi^\pm_{n,-}(p) \equiv  J_{|n|+{\rm sgn}(n)}(pr) \mp {\rm sgn}(n) \sqrt{\frac{|E|+ m_F}{|E|- m_F}} \ J_{|n|}(pr) =0\ .
\end{equation} 

\subsection{Computation of the angular momentum} 

We wish to enumerate the states in both $\mc{H}^+$ and $\mc{H}^-$
graded by their angular momenta that contribute to the Fermi sea of
chemical potential $\mu$. In the computations presented in the rest of
this appendix we will assume that $\mu>0$ so that we need to deal with
only particles with positive energy states rather than `antiparticles' with
positive energy (i.e.~states in which particles of negative energy are
removed or rather, absent). As we know that $J^\pm(\mu, m_F)$ is an
even function of $\mu$ (see \eqref{jpmf}), this restriction results in
no loss of generality.

We proceed with our computation as follows. Let $N_n^\pm(\mu, m_F)$
denote the number of eigenstates of the Dirac Hamiltonian $H(m_F)$ in
the Hilbert Space $\mc{H}^\pm$, with discrete label $n$ and with
energy less than $\mu$. These are states such that
$$p^2 +m_F^2 < \mu^2\ ,$$
i.e with $p<p_F$ where $p_F$ is defined by the equation 
\begin{equation}\label{fermom} 
p_F^2 +m_F^2 = \mu^2\ .
\end{equation} 
It follows that 
\begin{equation}\label{jpform} 
J^\pm (\mu, m_F)= - \sum_{n=-\infty}^\infty \Big(n+\frac{1}{2} \Big)  \ N^\pm_n(\mu, m_F)\ , 
\end{equation} 
and so (using \eqref{angmomdef})
\begin{equation}\label{jpformtot} 
J(\mu, m_F)=  - \frac{1}{2} \left( \sum_{n=-\infty}^\infty \Big(n+\frac{1}{2} \Big) \  N^+_n(\mu, m_F)  + \sum_{n=-\infty}^\infty \Big( n+\frac{1}{2} \Big) \ N^-_n(\mu, m_F) \right) \ .
\end{equation}

\subsubsection{Rough estimate of angular momentum}

The angular momentum $J(\mu,m_F)$ is a function of $R$. We are
interested in computing it at large $R$ while keeping only those terms
that are at least extensive i.e.~that scale like $R^2$ or faster. The
following rough estimate gives a sense of the scales involved. It is
not too difficult to verify that, at large $n$, the Bessel function
$J_n(x)$ is everywhere well approximated by a WKB approximation (e.g., see Appendix E.2 in \cite{Bhattacharyya:2016nhn}) as follows. Let
\begin{equation}
  n_{*}=\sqrt{4n^2-1}\ ,\quad f(n,x) = \frac{n_{*}}{2}\sin^{-1}\frac{n_{*}}{2x} + \sqrt{x^2-\frac{n_{*}^2}{4} } \ .
\end{equation}
Then, we have
\begin{equation}\label{WKBbesself}
  J_{n}(x)  =\left\{\renewcommand{\arraystretch}{2.8}\begin{array}{cc} \frac{1}{\sqrt{2\pi x} }   \frac{1}{\big(\frac{n_{*}^2}{4x^2}-1\big)^{1/4}}  \left( \frac{n_{*}}{2x} - \sqrt{\frac{n_{*}^2}{4x^2}-1} \right)^{n_{*}/2}  e^{\sqrt{\frac{n_{*}^2}{4}-x^2} } \ , & \quad  0< x\ll \frac{n_{*}}{2}-n_{*}^{1/3}\ , \\
                                                       \sqrt{\frac{2}{x}} \ \Big(\frac{n_{*}}{4}\Big)^{1/6}  Ai\Big( -(4/n_{*})^{1/3}  \big(x-\frac{n_{*}}{2}\big) \Big) \ , & \quad  \frac{n_{*}}{2}-\frac{n_{*}}{6} \ll x \ll  \frac{n_{*}}{2}+\frac{n_{*}}{6}\ ,  \\
                                                       \sqrt{\frac{2}{ \pi x}} \frac{1}{(1-\frac{n_{*}^2}{4x^2})^{1/4}} \cos  \bigg(   \frac{\pi}{4} +\frac{n_{*}\pi}{4}- f(n,x) \bigg)\ , & \quad x\gg \frac{n_{*}}{2}+n_{*}^{1/3}\ .\end{array}\right.
\end{equation}
The first line in \eqref{WKBbesself} is the WKB approximation to this
function in the classically disallowed region. The second line in
\eqref{WKBbesself} is the Airy form\footnote{Here, $Ai(x)$ is the
  standard Airy function of the first kind.} that occurs in the WKB
approximation in the transition from the allowed to the disallowed
region. The last line is the WKB approximation in the classically
allowed region. Equation \eqref{WKBbesself} leads to the following
rough approximation (valid at leading order in $R$ assuming $n$ has a
generic value, i.e. $n \sim R$)
\begin{equation}\label{thetfun}
  N^\pm_{|n|}(\mu, m_F) = \frac{1}{\pi} \ \Theta(p_FR-|n|) \ \left( |n| \sin^{-1}\left(\frac{|n|}{p_FR}\right) +\sqrt{(p_FR)^2-|n|^2} -\frac{|n|\pi }{2} \right)\ .
\end{equation} 
We see that $N^\pm_{|n|} \sim R$. Also, the number of $n$ values that
contribute (before the $\Theta$ function in \eqref{thetfun} kills the
contribution) is also of order $R$. Given that generic $n \sim R$, it
follows that the contribution of positive $n$ to the sums
\eqref{jpform} and \eqref{jpformmod} is of order $R^3$, and so is
super-extensive.

This super-extensive term is killed by a cancellation between terms of
positive and negative angular momenta, as we now explain. To see this,
note that the sum in \eqref{jpform} can be reorganised in three useful
ways, each of which we now list:
\begin{align}\label{jpformmod} 
  &J^\pm (\mu, m_F)\nonumber\\
  &= - \sum_{n=0}^\infty \left(n+\frac{1}{2}\right) \left( N^\pm_n(\mu, m_F) -  N^\pm_{-n-1}(\mu, m_F) \right)\ ,\nonumber \\
&= - \sum_{n=0}^\infty n \left( N^\pm_n(\mu, m_F) -  N^\pm_{-n-2}(\mu, m_F) \right) - \frac{1}{2}\sum_{n=0}^\infty \left( N^\pm_n(\mu, m_F) -  3N^\pm_{-n-2}(\mu, m_F) \right)\ ,\nonumber \\
&= \frac{1}{2}N_{0}^{\pm}(\mu,m_F) -  \sum_{n=0}^\infty n
\left( N^\pm_n(\mu, m_F) -  N^\pm_{-n}(\mu, m_F) \right) - \frac{1}{2} \sum_{n=0}^\infty \left( N^\pm_n(\mu, m_F) +  N^\pm_{-n}(\mu, m_F) \right).
\end{align}
In particular note that the difference between $N$ values in the first
line in \eqref{jpformmod} is of order unity rather than order $R$ (it
is easy to check that this quantity vanishes when $m_F=0$ using the
boundary condition equations \eqref{beghp}). This establishes that the
angular momentum is actually of order $R^2$ rather than $R^3$.

\subsection{Angular momentum in the non-relativistic limit}

The utility of the second line of \eqref{jpformmod} is that it allows
us to see that the angular momentum of the Fermi sea is simply
$\frac{1}{2}$ times the number of occupied single particle states in
the non-relativistic limit at positive mass $m_F > 0$. To see this, we
note that, for large enough $n$, it follows from \eqref{thetfun} that
$N_{n}^{\pm} =N_{-n-2}^{\pm} $. And so, we get from the second line of
\eqref{jpformmod}
\begin{equation}\label{jpformnr} 
J^\pm (\mu, m_F) =\sum_{n=0}^\infty  N^\pm_n(\mu, m_F) \ .
\end{equation} 
It follows that the total angular momentum \eqref{jpformtot} is given by 
\begin{equation}\label{jtotnr} 
J(\mu, m_F) = \frac{1}{2} \sum_{n=0}^\infty  \Big( N^{+}_n(\mu, m_F) +N^{-}_n(\mu, m_F) \Big) 
\end{equation} 
and that, in this limit, the angular momentum is exactly half of the
charge of the Fermi sea.

To see this explicitly in formulas, we note that at large $n$, we can
replace this summation by an integral and use \eqref{thetfun} to get
\begin{equation}\label{jtotnrintpm} 
J(\mu, m_F) =\frac{p_FR}{\pi} \int_{0}^{1}   \Big( \alpha \sin^{-1}\alpha  +\sqrt{1-\alpha^2} -\frac{\pi \alpha}{2} \Big)~d\alpha 
\end{equation} 
where, we have defined $\alpha = n/(p_FR)$ for brevity. Performing the
integral, the final result is
\begin{equation}\label{jtotnrpmF} 
J(\mu, m_F) =\frac{1}{8} (p_FR)^2 
\end{equation}
This result is exactly the $1/2$ times number of occupied
single-particle states\footnote{This can be easily obtained from the
  phase-space counting,
  $ \int_{p\leq p_F} \frac{\mc{V}_2
    d^2p}{(2\pi)^2}=\frac{(p_FR)^2}{4}$, where, $\mc{V}_2=\pi R^2$ is
  the volume of the two dimensional space with a radial cutoff at
  $R$.} in the non-relativistic limit at positive mass $m_F > 0$.

In the same way, the third line allows us to see that the angular
momentum of the Fermi sea is simply $-\frac{1}{2}$ times the number of
occupied single-particle states in the non-relativistic limit at
negative mass $m_F < 0$. The argument is similar to the positive mass
case. The final answer in this case is given by
\begin{equation}\label{jtotnrnmF}
  J(\mu, m_F) =-\frac{1}{8} (p_FR)^2 \ .
\end{equation}
Combining \eqref{jtotnrpmF} and \eqref{jtotnrnmF}, we see that in the
non-relativistic limit
\begin{equation}\label{jtotnrf} 
  J(\mu, m_F) =\frac{\sgn(m_F)}{2} \frac{(p_FR)^2}{4}\ .
\end{equation}
Note, in particular, that this angular momentum is extensive. 

\subsection{General form of angular momentum}

Away from the non-relativistic limit the angular momentum of the
Fermi sea, defined in this appendix, takes the form
\begin{equation}\label{angu}
  J(\mu,m_F) = \frac{1}{2} \frac{(p_FR)^2}{4}  h\left(\frac{m_F}{|\mu|}\right)\ ,
\end{equation} 
where $h(x)$ is a yet to be carefully computed function which has the
properties
\begin{equation}\label{hval}
h(-1)=-1\ ,\quad h(0)=0\ ,\quad h(1)=1\ .
\end{equation}
A very crude estimation - one that makes approximations that we have
not attempted to systematically justify - suggests
\begin{equation}\label{hest}
h(x) =\frac{2x}{1+|x|}.
\end{equation} 
We leave the verification or improvement of \eqref{hest} (a relatively
easy exercise) and its generalisation to nonzero 't Hooft coupling (a
more interesting exercise) to future work.

\bibliography{cp}\bibliographystyle{JHEP}

\end{document}